\documentclass[12pt,preprint]{aastex}
\usepackage{amssymb}
\usepackage{graphicx}
\usepackage{grffile}
\usepackage{subfigure}
\usepackage{subfloat}
\usepackage{url}
\usepackage{float}
\usepackage{verbatim}
\usepackage{epsfig}
\usepackage{color}
\usepackage{captcont}
\usepackage{natbib}
\usepackage{comment}

\definecolor{orange}{rgb}{1,0.5,0}

\newcommand{\etal}{et~al.~}

\newcommand{\kms}{\ifmmode\,{\rm km}\,{\rm s}^{-1}\else km$\,$s$^{-1}$\fi}
\newcommand{\be}{\begin{equation}}
\newcommand{\ee}{\end{equation}}
\newcommand{\bea}{\begin{eqnarray}}
\newcommand{\eea}{\end{eqnarray}}

\newcommand{\magarc}{\ifmmode {{{{\rm mag}~{\rm arcsec}}^{-2}}}
             \else {{{mag}$~${arcsec}$^{-2}$}}
             \fi}
\def \spose#1{\hbox to 0pt{#1\hss}}
\def \ltsim{\mathrel{\spose{\lower 3pt\hbox{$\sim$}}
     \raise 2.0pt\hbox{$<$}}}
\def \gtsim{\mathrel{\spose{\lower 3pt\hbox{$\sim$}}
     \raise 2.0pt\hbox{$>$}}}
\def\eq#1{Equation~(\ref{eq:#1})}
\def\se#1{Section~\ref{sec:#1}}
\def\ap#1{Appendix~\ref{sec:#1}}
\def\Fig#1{Figure~\ref{fig:#1}}
\def\Table#1{Table~\ref{fig:#1}}

\def\ifm#1{\relax\ifmmode#1\else$\mathsurround=0pt #1$\fi}
\def\kms{\ifmmode\,{\rm km}\,{\rm s}^{-1}\else km$\,$s$^{-1}$\fi}

\def\ltsima{$\; \buildrel < \over \sim \;$}
\def\lsim{\lower.5ex\hbox{\ltsima}}
\def\gtsima{$\; \buildrel > \over \sim \;$}
\def\gsim{\lower.5ex\hbox{\gtsima}}

\def\C28{\rm C_{28}}

\def\pmb#1{\setbox0=\hbox{#1}%
\kern-.025em\copy0\kern-\wd0 \kern.05em\copy0\kern-\wd0
\kern-.025em\raise.0433em\box0}

\def \littlemm{\ifmmode{\scriptscriptstyle m }
     \else{\hbox{$\scriptscriptstyle m $ }}\fi}
\def \topemm{\raise .9ex \hbox{\littlemm}}
\def \magpoint{\hbox to 2pt{}\rlap{\hskip -.5ex
     \topemm}.\hbox to 2pt{}}

\def \magarc {mag arcsec$^{-2}$}
\def \Teff {$T_{\rm eff}$}
\def \logg {$\log g$}
\def \lbfgsb {{\fontfamily{pcr}\selectfont L-BFGS-B}}
\def \fmincon {{\fontfamily{pcr}\selectfont fmincon}}

\usepackage{lscape}
\slugcomment{Accepted for publication at MNRAS}

\shorttitle{Stellar Content of Galactic Globular Glusters}
\shortauthors{Barber \etal 2013}

\begin{document}


\title{Validation of optimised population synthesis through mock spectra
and Galactic globular clusters}

\author{Christopher Barber\footnote{Current address: University of Victoria,
 Dept. of Physics and Astronomy, Victoria, BC Canada}, St\'{e}phane Courteau,
 Joel C. Roediger}
\affil{Queen's University, Dept. of Physics, Engineering Physics and
Astronomy, Kingston, ON Canada}
\and
\author{Ricardo Schiavon\footnote{Current address: Astrophysics Research Institute, Liverpool John Moores University, 146 Brownlow Hill, Liverpool, L3 5RF, United Kingdom}}
\affil{Gemini Observatory, 670 N. A'Ohoku Place, Hilo, HI 96720, USA}

\email{cbar@uvic.ca, courteau@astro.queensu.ca, jroediger@astro.queensu.ca,
R.P.Schiavon@ljmu.ac.uk}


\begin{abstract}

Optimised population synthesis provides an empirical method to extract the relative mix of stellar evolutionary stages and the distribution of atmospheric parameters within unresolved stellar systems, yet a robust validation of this method is still lacking.  We here provide a calibration of population synthesis via non-linear bound-constrained optimisation of stellar populations based upon optical spectra of mock stellar systems and observed Galactic Globular Clusters (GGCs).  The MILES stellar library is used as a basis for mock spectra as well as templates for the synthesis of deep GGC spectra from \citet{S05}. Optimised population synthesis applied to mock spectra recovers mean light-weighted stellar atmospheric parameters to within a mean uncertainty of 240 K, 0.04 dex, and 0.03 dex for \Teff, \logg, and [Fe/H], respectively. We use additional information from HST/ACS deep colour-magnitude diagrams (CMDs) from \citet{Ata} and literature metallicities to validate our optimisation results on GGCs. Decompositions of both mock and GGC spectra confirm the method's ability to recover the expected mean light-weighted metallicity in dust-free conditions ($E(B-V) \lsim 0.15$) with uncertainties comparable to evolutionary population synthesis methods. Dustier conditions require either appropriate dust-modelling when fitting to the full spectrum, or fitting only to select spectral features. We derive light-weighted fractions of stellar evolutionary stages from our population synthesis fits to GGCs, yielding on average a combined $25\pm6$ per cent from main sequence and turnoff dwarfs, $64\pm7 $ per cent from subgiant, red giant and asymptotic giant branch stars, and $15\pm7 $ per cent from horizontal branch stars and blue stragglers. Excellent agreement is found between these fractions and those estimated from deep HST/ACS CMDs. Overall, optimised population synthesis remains a powerful tool for understanding the stellar populations within the integrated light of galaxies and globular clusters.

\end{abstract}



\section{INTRODUCTION}\label{sec:Intro}
Fundamental information on the physical processes which dominate the
formation and evolution of galaxies can be gleaned from the study of
their stellar populations.  Early studies of galaxies' stellar content relied on the
technique of \textit{population synthesis}, wherein the integrated spectrum
of a galaxy is decomposed into a sum of suitably-weighted spectra of
individal stars with known basic properties such as temperature, surface
gravity, and metallicity \citep{spinrad, faber, oconnell, pickles}.
In order to obtain information such as age or star formation history,
stellar evolutionary models should then be applied subsequent to the decomposition.
Advances in our knowledge of stellar evolution have however enabled
a new technique for stellar population analyses, so-called
\textit{evolutionary population synthesis} \citep{Renzini81,Buzzoni89, Brucha93,Maraston98}.
By combining individual stellar spectra with isochrones and an initial
mass function, integrated spectra of entire galaxies (or stellar clusters)
can then be constructed over a wide range of ages and metallicities. 
Thus, evolutionary population synthesis folds in evolutionary models
as part of the spectral decomposition. The power of the latter technique
to reduce the degrees of freedom in stellar population analyses resulted
in the demise of the former.

However, various uncertainties affecting current stellar evolution models
\citep[e.g. thermally-pulsating asymptotic giant branch, horizontal branch,
and blue straggler stars;][]{Maraston07,ConroyGunn09}, especially pertaining
to the brightest phases of stellar evolution, are cause for concern in any
application of evolutionary population synthesis.  For instance, the presence
of blue horizontal branch stars in elliptical galaxies, coupled to the lack of
a predictive theory for their origins, can be misinterpreted as their having
recently experienced a burst of star formation \citep{Maraston&Thomas00}.
Population synthesis, on the other hand, should in principle be free from
the outshining effect that uncertain phases of stellar evolution have on
stellar population analyses.  We therefore expect this technique to be able
to unravel at once the contributions of all stars to the integrated light
of a stellar system.

In light of these issues and the availability of exceptional computing power,
we wish to re-examine and validate the population synthesis optimisation
method as a means to provide estimates of the stellar content of unresolved
systems independent of any assumed stellar evolution model.
And while the latter is true, it should still be noted that the coverage in
age, metallicity, and surface gravity of the adopted stellar basis may have
a significant impact on the final spectral decompositions \citep{Koleva2008}.

Before applying the population synthesis technique to galaxy spectra,
it must be rigorously tested on data samples for which the underlying
stellar contents are known.  In this paper, we use both mock and observational data 
to test the technique under realistic (imperfect) observing conditions.
Owing to the existence of deep integrated spectra
\citep[hereafter S05]{S05} and colour-magnitude diagrams \citep[hereafter CMDs;][]{Ata}
as well as high-resolution star-by-star spectroscopic
abundances (\citealt{Harris}; \citealt{Roediger2014}, and references
therein) for them, Galactic globular clusters (GGCs) provide
the best astrophysical test bed for population synthesis.

Specifically, we test the ability of the technique to
recover the luminosity-weighted distributions of stellar atmospheric parameters
(effective temperature, surface gravity, metalliticy, and colour), as well as the
contributions of various evolutionary phases to the integrated light of GGCs
which can be directly validated with corresponding estimates based on CMDs.
In doing so, we identify regions of parameter space (both observational and physical)
where the technique may fail to reproduce known stellar system data.
It should be noted that, without the inclusion of a stellar evolutionary model, population
synthesis cannot derive evolutionary properties of stellar systems such as age
and star formation histories; properties that we do not attempt to measure in this paper.

To our knowledge, no study of this kind has addressed the reliability of population
synthesis methods using both mock and real data as constraints.
\citet{Koleva2008} performed a similar study to our own using {\it evolutionary}
population synthesis on the same S05 GGC spectra used here. We will show below that our
population synthesis method is just as reliable in the determination of GGC metallicity
despite the lack of stellar evolution modelling.  

The computational engine central to our numerical decompositions
is a non-linear bound-constrained optimisation.  Similar spectral
decompositions of stellar systems via constrained optimisation
have been applied before \citep[e.g.][]{Macarthur09, CidFernandes05, Walcher06}.
Other inversion methods used to fit stellar spectra or SSP models
to integrated spectra have also been reported recently by, e.g. \citet{Vergely2002},
\citet{Moultaka05}, \citet{Ocvirk06}, \citet{tojeiro07}, \citet{Koleva2008}, and \citet{koleva2009}.

The organization of the paper is as follows.  In \se{libs}, we present our
libraries of GGC and stellar spectra. In \se{Optimisation}, we describe
the optimisation algorithm used for decomposing mock and observed
integrated spectra into sums of individual stellar spectra,
as well as establish the typical level of random and systematic errors
inherent to any given decomposition.  We then apply our algorithm in
\se{popsynth} to the GGC integrated spectra of S05 to determine the
fractional contributions (by light) of stellar parameters (effective
temperature, surface gravity and metallicity) for each cluster.
These fractions in turn yield estimates of each cluster's light-weighted
metallicity and stellar evolutionary budget. \se{met} and \se{CMD}
present a comparison against independent constraints.  Our discussion
and conclusions are presented in \se{conc}, followed by two appendices.
\ap{AppendixA} includes our reconstruction of each GGC spectrum analyzed
in this paper. Finally, deep CMDs for the 24 S05 GGCs in common with the
HST/ACS database of \citet{Ata}, along with their breakdown into the various
stellar evolutionary zones, are presented and compared with CMDs derived
from population synthesis in \ap{AppendixC}.


\section{SPECTRAL LIBRARIES}\label{sec:libs}

The GGC integrated spectra that we wish to model through population
synthesis methods come from the S05 library.  S05 obtained deep,
drift-scan optical spectra of the cores of 41 GGCs using the R-C
spectrograph on the Blanco 4-m telescope at the Cerro-Tololo-Interamerican
Observatory. These data combine the merits of broad spectral coverage
(${\sim}3350-6430$~\AA), intermediate spectral resolution (FWHM ${\sim}3.1$~\AA)
and high signal-to-noise (50 $<$ S/N/\AA\ $<$ 240 at 4000~\AA).  Further details
on the observational setup or reduction methods for these spectra can
be found in S05.

\begin{figure}
  \begin{center}
    \includegraphics[width=\textwidth]{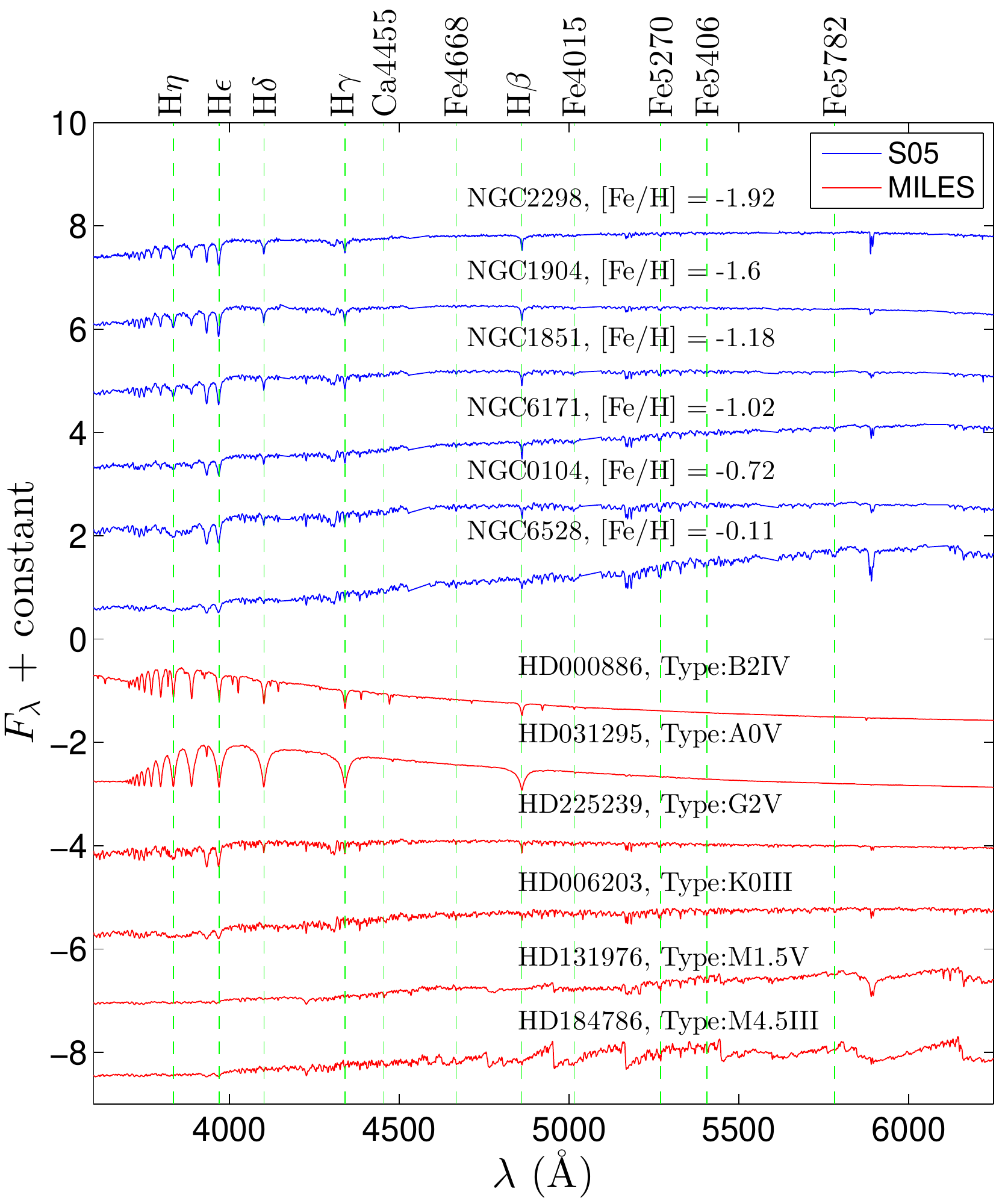}
  \end{center}
  \caption{Representative spectra from the S05 and MILES spectral
   libraries, shown on an arbitrary flux scale.  Vertical green dashed lines
   denote the locations of major (atomic) absorption features listed in
   \citet{Worthey}, with the label of each feature provided at the figure
    top. The S05 and MILES spectra are stacked vertically in order of
    increasing metallicity ([Fe/H]) and effective temperature, respectively.}
  \label{fig:global}
\end{figure}

Population synthesis typically takes advantage of an extensive library of stellar
spectra in order to reproduce the integrated spectrum of a (normally) unresolved
stellar system.   To reconstruct the S05 spectra, we use the Medium-resolution
Isaac Newton Telescope stellar library (\citealt{MILES}; hereafter MILES), which
consists of empirical spectra for 985 stars obtained at the 2.5-m Isaac Newton
Telescope.  The comparable wavelength range (${\sim}3525-7500$~\AA) and
resolution (FWHM=2.54 $\pm$ 0.08~\AA; \citealt{milesII,falcon}) of these
data and S05 makes the MILES library ideal for our purposes\footnote{Note
that adopting a different stellar library might yield different results.
However, the investigation of differences between popular stellar libraries
is beyond the scope of this paper.}.
\Fig{global} shows sample spectra from both the S05 and MILES libraries,
where the former (shown in blue) span the full metallicity range of the
S05 library, while the latter (shown in red) span the full spectral range
of the MILES library.  A simple by-eye comparison of the continuum shapes
from these two examples already suggests that GGC spectra are predominantly
generated by cool (i.e., old) G-, K- and M-type stars rather than hot
(i.e., young) O-, B- and A-type stars as expected given the mean age
($\gsim 10$ Gyr) of globular clusters. We will indeed confirm this
impression in \se{GGCsynthesis}.

The successful decomposition of the GGC spectra hinges on the accurate
characterization of both the wavelength calibration and spectral resolution
of the S05 and MILES libraries. While the MILES spectra have a constant
full width at half maximum resolution throughout their wavelength
range \citep[FWHM$_{\rm M}$ = 2.54~\AA; ][]{falcon}, that of the S05
spectra (FWHM$_{\rm S}$) varies with wavelength according to,
\be
\mathrm{FWHM_{\rm S}}(\lambda) = 15.290 - 6.079 \times 10^{-3} \lambda +
9.472 \times 10^{-7} \lambda^2 - 4.395 \times 10^{-11} \lambda^3
\label{eq:FWHMExpectedGGC}
\ee
where $\lambda$ is the wavelength in \AA.  This function gives
FWHM$_{\rm S} \simeq$ 3.1 \AA\ at the central wavelengths
(${\sim}4500-5500$ \AA) of the S05 spectra and roughly 3.6 \AA\ 
at their edges ($\sim$3350 \AA\ and 6430 \AA). The MILES spectra were 
therefore degraded by convolving with a Gaussian kernel to match the
resolution of S05.  A mean difference of 0.05 $\pm$ 0.13~\AA\
between the centroids from S05 and MILES was measured, implying that
the wavelength calibrations of the two libraries are consistent to within
the random error, which is more than sufficient for our needs.

\subsection{Stellar basis}\label{sec:basis}

We must now select a stellar spectral basis for our optimisation code.
This basis would ideally comprise as many stellar types as possible whilst
keeping the number of spectra manageable, lest our code be made
computationally prohibitive.  We first trimmed the original MILES library
to exclude ``exotic'' stars (e.g. binaries, emission line objects), highly
reddened stars (with $E(B-V) > 0.1$), stars exhibiting spectral peculiarities,
and stars with unknown distance measurements. Stars belonging
to globular clusters were also removed to avoid possible
contamination in their spectra and because their metallicities are
assigned the mean of the entire globular cluster. All remaining stars
were manually inspected for emission lines and other possible spectral
blemishes.  In the end, we were left with a ``restricted'' library
of 774 stars from which to construct our stellar basis.

\begin{figure}
  \begin{center}
    \includegraphics[width=\textwidth]{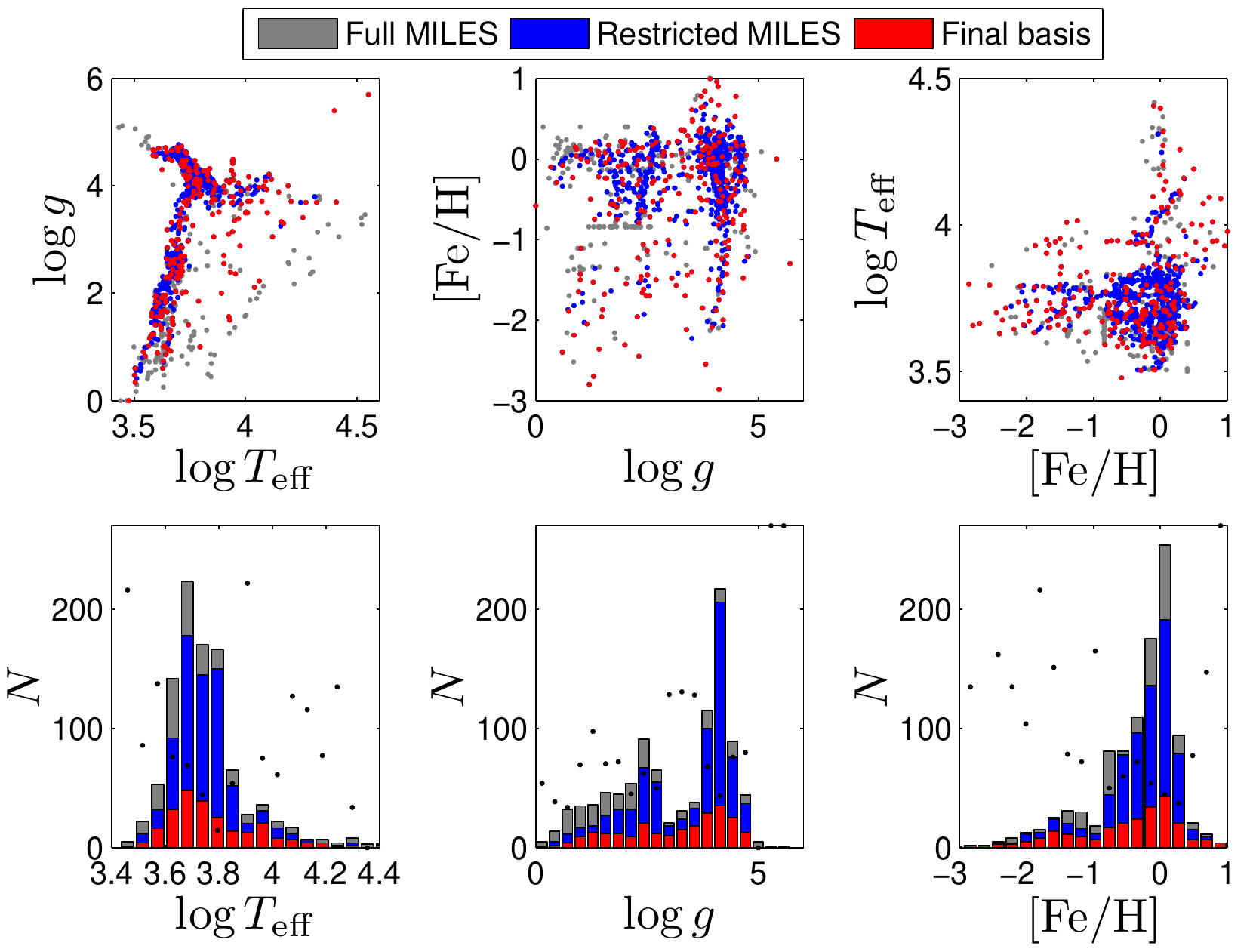}
  \end{center}
  \caption{(Top panels) Bivariate distributions of the atmospheric
parameters
   \Teff, \logg~and [Fe/H] for the full MILES library, as well as two of its
   subsets (see \se{basis} for details). (Bottom panels) Distribution of
   each atmospheric parameter in the full MILES library and its two
   subsets. Black points above each bin indicate the coverage of the
   final basis relative to the full MILES library, scaled from 0 to 1 in
  each panel from bottom to top respectively.}
  \label{fig:basisParams}
\end{figure}

The range of physical parameters spanned by the stars in the restricted
MILES library is highlighted in \Fig{basisParams}.  For each MILES star,
the atmospheric parameters \Teff, \logg\ and [Fe/H] were extracted
from~\citet{Prugniel}, where \Teff\ is the effective temperature (in
K), log$(g)$ is the surface gravity (in cgs units), and [Fe/H]
is the metallicity.

Each star in the full MILES library is shown as a grey dot while the restricted
set is shown as blue dots in the top row of \Fig{basisParams}.  The group of
stars in the middle panel at [Fe/H]$=-0.84$ are all members of the globular
cluster NGC 6838 and are thus ignored.  Many of the high {\Teff}- low {\logg}
stars were also removed in light of strong emission lines in their spectra.

To minimize degeneracies and safeguard against computationally expensive
optimisations, a subset of our restricted stars was chosen such that each member
had to be above a minimum distance from all others in the 3-dimensional
parameter space of \Teff, \logg\ and [Fe/H].  To ensure that each parameter
was sufficiently sampled, a minimum (arbitrary) distance of 1/17th of the
size of the parameter space was chosen. That distance corresponds to 0.07, 0.35,
and 0.24 in $\log$ \Teff, \logg\ and [Fe/H] respectively.  The fact that this
spacing is larger than the typical errors in these parameters prevents degeneracies
in our basis. We have verified that choosing a smaller distance (and thus a larger
basis) does not affect our results.  Our final stellar basis of 242 stars, hereafter ``final basis'',
is shown with red points in \Fig{basisParams}. Fine sampling of the final basis
is shown with histograms in the bottom panels of \Fig{basisParams}, following
the same colour scheme as its top panels. A black dot also indicates the fraction
of MILES stars used for the final basis in each bin. The final basis includes stars
which nicely sample the parameter space defined by the entire MILES library.
Because the distributions are not uniform, any random error in our derived
atmospheric parameters will likely bias the results toward the medians
of these distributions.


\section{Optimisation Method}\label{sec:Optimisation}
\subsection{The Algorithm}\label{sec:Code}

To disentangle the light-weighted stellar populations that make up the
integrated spectra of GGCs, we set up an optimisation method in MATLAB
which follows closely that presented in~\citet{Macarthur09}.
The overall optimisation problem can be summarized as trying to find the
global minimum of a merit function $f(\vec{x})$ subject to linear
constraints $a < x_i < b$.  The merit function to be minimized is defined as,
\be
\mathrm{Merit}= \left({\frac{1}{N-M}\sum\limits^N_{i=1} R_i^2}
\right)^{\frac{1}{2}},
\label{eq:merit}
\ee
where $N$ is the number of data points included in the fit, $M$ is the
number of stellar templates in the basis, and
\be
R_i = w_i(G_i - S_i),
\label{eq:Ri}
\ee
where $G_i$ is the flux of the S05 GGC spectrum at wavelength
$\lambda_i$, $w_i$ is the weight of the $i^{th}$ pixel, and $S_i$
is the modelled GGC flux at $\lambda_i$ given by,
\be
S_i= \sum\limits^M_{j=1} x_jF_{ji}.
\label{eq:Si}
\ee
Here $F_{ji}$ denotes the flux at the $i^{th}$ wavelength of the $j^{th}$
template, and $x_j$ is the relative contribution of the $j^{th}$ template to
the synthesized spectrum.  We define the weights per pixel
$w_i = m_i \sigma_i^{-2}$, where $\sigma_i$ is the error on $G_i$ and $m_i$
can take a value of 0 or 1 to mask out blemishes in the GGC spectra (see S05).
We take $\sigma_i^2$ as the variance given by the ratio of the SED to its S/N spectrum.

No templates can acquire a negative contribution to the synthesized spectrum
as this would be unphysical.  That is,
\be
x_j \geq 0, ~~~ \forall~ j = 1 .. M
\label{eq:lowerbound}
\ee

As in~\citet{Macarthur09}, we have used the \lbfgsb~optimisation algorithm to
minimize the merit function.  This algorithm is a limited-memory quasi-Newton
code designed for the problem of bound-constrained optimisation.  The advantage
of \lbfgsb\ is that it does not require the full Hessian matrix at each
iteration, but rather uses an approximation based on earlier iterations.
This approximation is very useful in problems like ours wherein hundreds
of independent variables may exist and
computing the full Hessian can be computationally expensive and unnecessary.
A MATLAB implementation of \lbfgsb\ was obtained from
{\tt http://www.cs.ubc.ca/~pcarbo/lbfgsb-for-matlab.html}.

We also compare in the next section a similar optimisation routine, called
\fmincon, which is inherent to the MATLAB package.  The latter uses an
``interior-point'' optimisation method.  It will be shown below to perform
poorly relative to \lbfgsb\ in the case of realistic (noisy) data.

\subsection{Tests with Mock Spectra}\label{sec:mock}

We must first test the internal accuracy of \lbfgsb\ and \fmincon\
in the context of our optimisation problem.  To do so, mock composite
spectra are created by summing individual stellar spectra from our
final basis which can then be applied to either code.
More specifically, a mock spectrum is created by first selecting
ten stars from the final basis at random and giving
them randomly-assigned positive weights $p_i$, subject to the constraint
$\sum\limits^{10}_{i=1}p_i = 1$ with each $p_i > 0.05$; all other stars
in the basis were given $p_i = 0$.  The basis spectra are then multiplied
by their $p_i$ and summed to produce the mock spectrum.  The figure
of merit of our test is its ability to recover the correct $p_i$ for all of
the stars in our final basis at once (including those with $p_i = 0$).

In order to simulate realistic conditions, Gaussian noise is added to
the spectrum up to a desired $S/N$ ratio per pixel. The two optimisation codes
can then be tested with the same mock spectra in order to compare the
recovered $p_i$.  A given optimisation is run ten times for each $S/N$
ratio, varying the set of ten stars in the mock spectrum and their $p_i$ each time.  The $S/N$
is varied over the range [10,10000] and the case of no additional noise
is also considered. Such a large range is considered to show the very
slow convergence of FMINCON with increasing $S/N$.

Initial conditions are required for these optimisation routines.
By fitting to the same mock spectrum repeatedly with different
initial conditions each time, we find that the optimisation results
are largely insensitive to the initial conditions. Therefore for this
and all subsequent decompositions, we use equal weights for
all the basis stars as initial conditions for the optimisation code.

\begin{figure}
  \begin{center}
    \includegraphics[width=\textwidth]{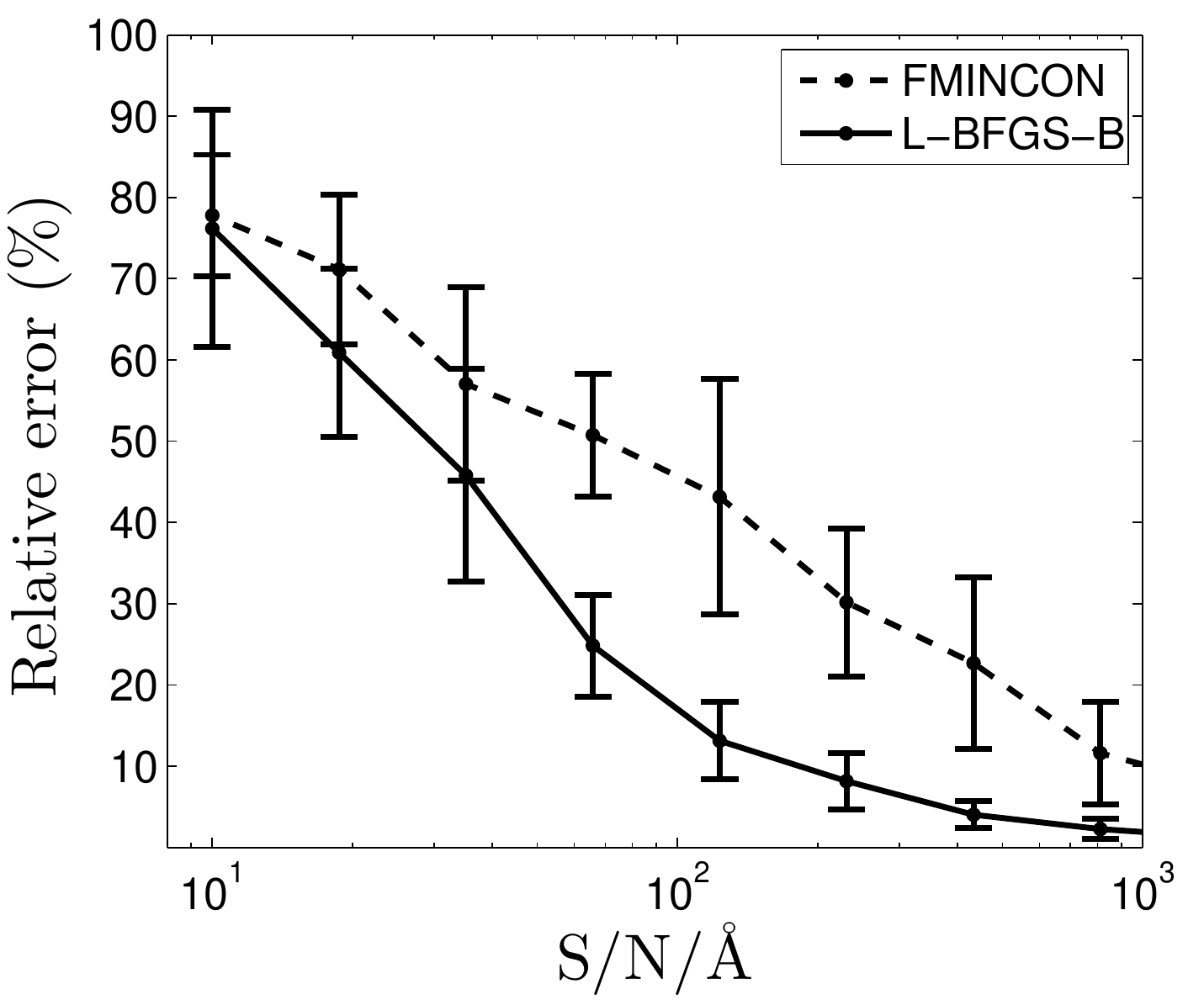}
  \end{center}
  \caption{Relative error between the recovered and assigned weights
   in the decompositions of mock spectra as a function of $S/N$.  Error bars
   represent the rms spread from ten realisations of the spectra at each
   $S/N$ value.  The dashed and solid lines correspond to the two optimisation
   algorithms described in the text (see legend).  The relative error at each
   $S/N$ value corresponds to the absolute difference between the fit and
   expected weights, divided by the expected weights and averaged over
   ten trials.}
  \label{fig:fakeInBasisErrors}
\end{figure}

For each $S/N$ value, the relative residuals between the ten non-zero
($p_i\neq 0$) mock spectrum weights and their corresponding fit weights
are computed and averaged over the ten weights and ten tests.
These averaged relative residuals are shown as a function of $S/N$
in \Fig{fakeInBasisErrors}. When no noise is added to the artificial spectrum,
\fmincon\ and \lbfgsb\ both converge to within $0.1$ per cent of the correct result.
However, in the presence of noise, \lbfgsb\ proved far more reliable and
converged faster.  In light of these results, we only use \lbfgsb\ for
subsequent optimisations.  Since the S05 spectra have $S/N$ levels
in the range [50,240], \Fig{fakeInBasisErrors} shows that relative errors
of $\sim$20 per cent should be assigned to our derived light-weighted
fractions of individual stars. However, this error may likely result from
degeneracies in our final basis. We show later in this section that the
error in the recovered atmospheric parameters is much smaller than
20 per cent.

\begin{figure}
  \begin{center}
    \includegraphics[width=\textwidth]{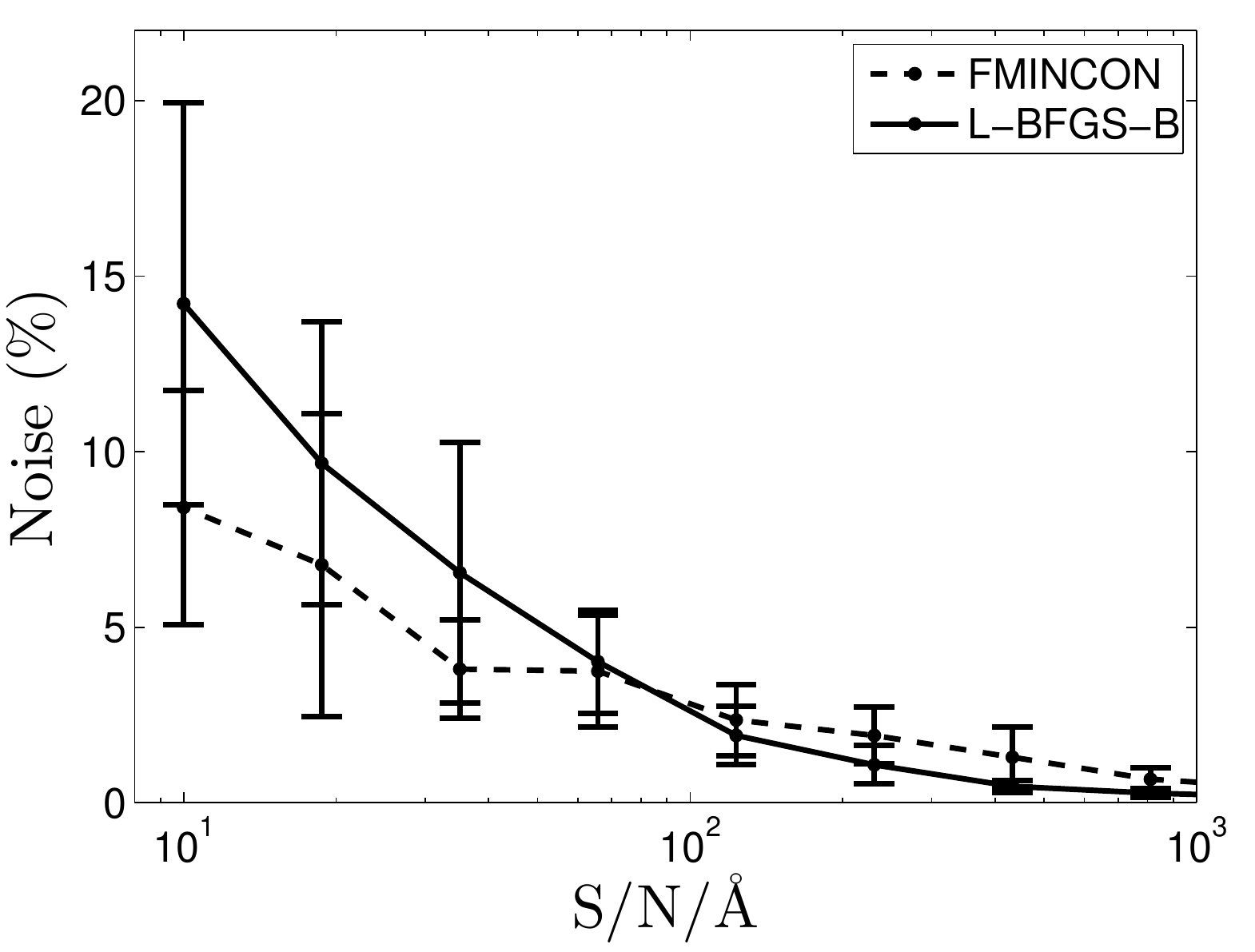}
  \end{center}
  \caption{As in \Fig{fakeInBasisErrors} but here showing the maximum noise
   level in the mock spectra decompositions.
   The noise at each $S/N$ per \AA\ was computed by taking the
   maximum fit weight expected to be zero in each trial run
   and averaging over all 10 trials for that $S/N$ ratio.}
  \label{fig:fakeInBasisSTD}
\end{figure}

In assessing the internal errors of our optimisation procedure, false
positives may also contribute to the random noise involved in the
decomposition results.  We test for this by computing for each run
the maximum fractional contribution (by light) of those stars whose
weights are assigned to be zero in the mock spectra. This maximum
deviation from zero is averaged over all tests and is shown as a
function of $S/N$ in \Fig{fakeInBasisSTD}.  Considering again the $S/N$ range
of the S05 spectra, we infer that the random noise level for all subsequent
fits should be less than 5 per cent.  Therefore, any fractional
contributions above 10 per cent will be significantly different from random noise
at the 2$\sigma$ level.

A similar test to the one above uses mock spectra constructed instead with
stars {\sl external} to the final basis but included in our restricted
MILES subsample (see \se{basis}). The final basis is thus fully independent
from these mock spectra.  We run our code on five different mock spectra
constructed this way.  No noise is added here since the noise intrinsic
to the spectra is now independent between these two datasets.

\begin{figure}
  \begin{center}
    \includegraphics[width=0.9\textwidth]{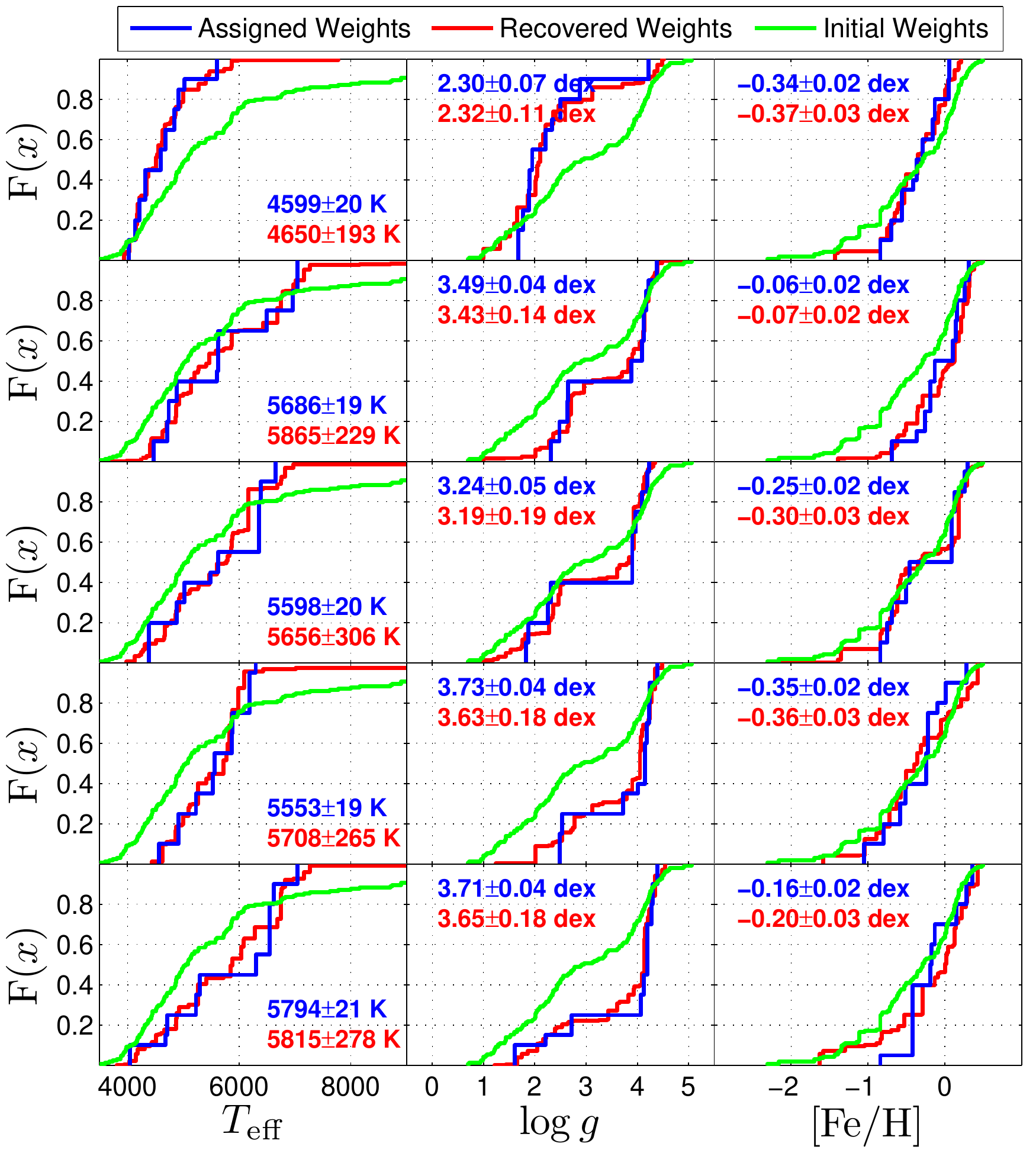}
  \end{center}
  \caption{Comparison of the cumulative distribution functions (CDFs)
   in the space of the stellar parameters \Teff, \logg, and [Fe/H]
   obtained from decompositions of mock spectra constructed from
   the MILES database but external to our final basis.  The blue CDFs
   correspond to the mock spectra parameters derived from randomly
   {\it assigned} weights; the red CDFs are based on parameters
   {\it recovered} by our \lbfgsb\ optimisation code.
   The green histograms represent the initial conditions for the
   optimisation program (i.e., all stars given equal weighting).  Each row
   corresponds to a different mock spectrum, while each column represents a
   different atmospheric parameter. The light-weighted mean atmospheric
   parameters for mock and fit spectra are also shown in blue and red text respectively.}
  \label{fig:fake_kstests}
\end{figure}

Since the assigned and recovered weights for stars cannot be directly
compared in these tests (the parent samples are different), we evaluate
the performance of the \lbfgsb\ routine through cumulative distribution
functions (CDFs) of the resulting luminosity fractions which comprise
the mock and fitted spectra, as a function of their atmospheric parameters.
The CDFs are shown in \Fig{fake_kstests} where the assigned and fitted
weights are
represented by the blue and red curves, respectively.  For reference,
the CDFs obtained if all of the final basis stars are given equal weights
(initial conditions of the optimisation routine) are shown as green curves.
The overall qualitative impression from \Fig{fake_kstests}, with the red
and blue CDFs being very similar, is that \lbfgsb\ is quite robust at
recovering the right mix of stellar parameters for the mock spectra.
Mean light-weighted atmospheric parameters were computed for each test
and compared between assigned and recovered spectra. These means
can be seen in blue and red text in each panel respectively. The mock
and recovered values agree within the uncertainties, which are computed
from the error in the atmospheric parameters from \citet{Prugniel} in
addition to an assumed error of 20 per cent in each derived light-weighted
fraction. The mean residuals between assigned and recovered parameters
are 240 K, 0.04 dex, and 0.03 dex for \Teff, \logg, and [Fe/H], respectively.
Thus, under these ideal (mock) conditions, population synthesis is excellent
at recovering atmospheric parameters from integrated spectra.

Since our analysis depends on matching the full SED of a stellar system,
and not strictly on line indices, a blue-ward depression of the continuum
due to dust extinction would clearly affect our ability to recover the right
stellar mix.  Line indices are indeed largely impervious to dust extinction
effects \citep{MacArthur2005} but the latter is not true for full SED
fitting.
In order to quantify the influence of dust reddening on our spectral
optimisations, mock spectra created from stars external to the main spectral
basis were reddened according to the Milky Way extinction law
\citep{Cardelli1989}.

\begin{figure}
  \begin{center}
    \includegraphics[width=\textwidth]{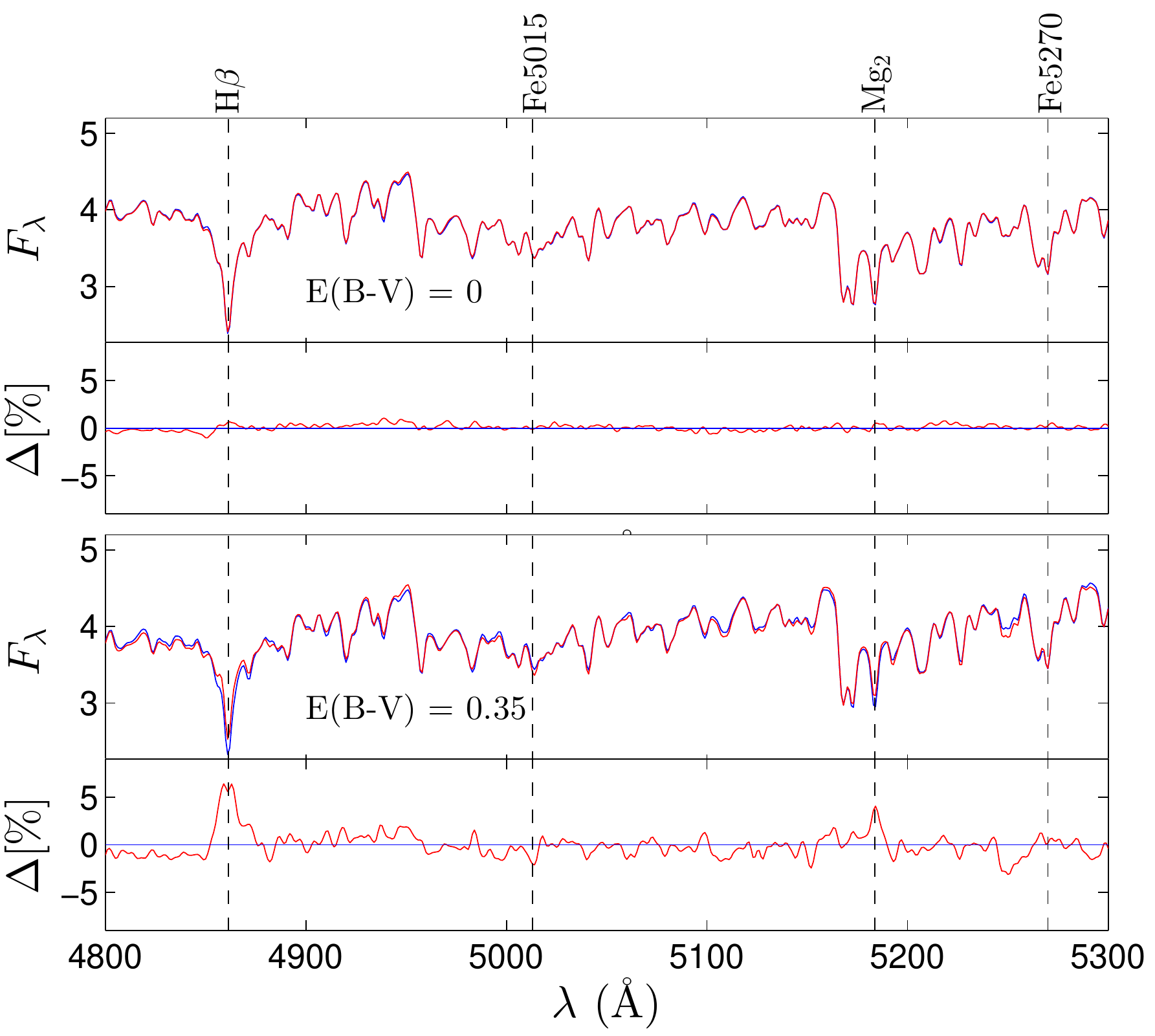}
  \end{center}
  \caption{Fits to mock spectra before and after reddening (top and bottom
   sets of panels, respectively). The mock and fit spectra are shown in blue
   and red, respectively. Residuals are shown below the spectra.
   The mock spectrum in the bottom panel was reddened with $E(B-V) = 0.35$.
   Important spectral features are indicated above the plot.}
  \label{fig:fake_reddened_spectra}
\end{figure}

The optimisation is performed on both reddened and non-reddened mock
spectra through ten random realisations. \Fig{fake_reddened_spectra}
shows the fit to a sample mock spectrum before and after reddening.
In all cases, results were the same: fits to the non-reddened spectra
are excellent, but application of a reddening law with $E(B-V) = 0.35$
to the mock spectrum prior to optimisation causes a poorer fit.  The
highest discrepancies arise at prominent absorption features such as
Mg$_2$, Fe5250, and H$\beta$ whose strengths depend strongly
on metallicity and age.

\begin{figure}
  \begin{center}
    \includegraphics[width=\textwidth]{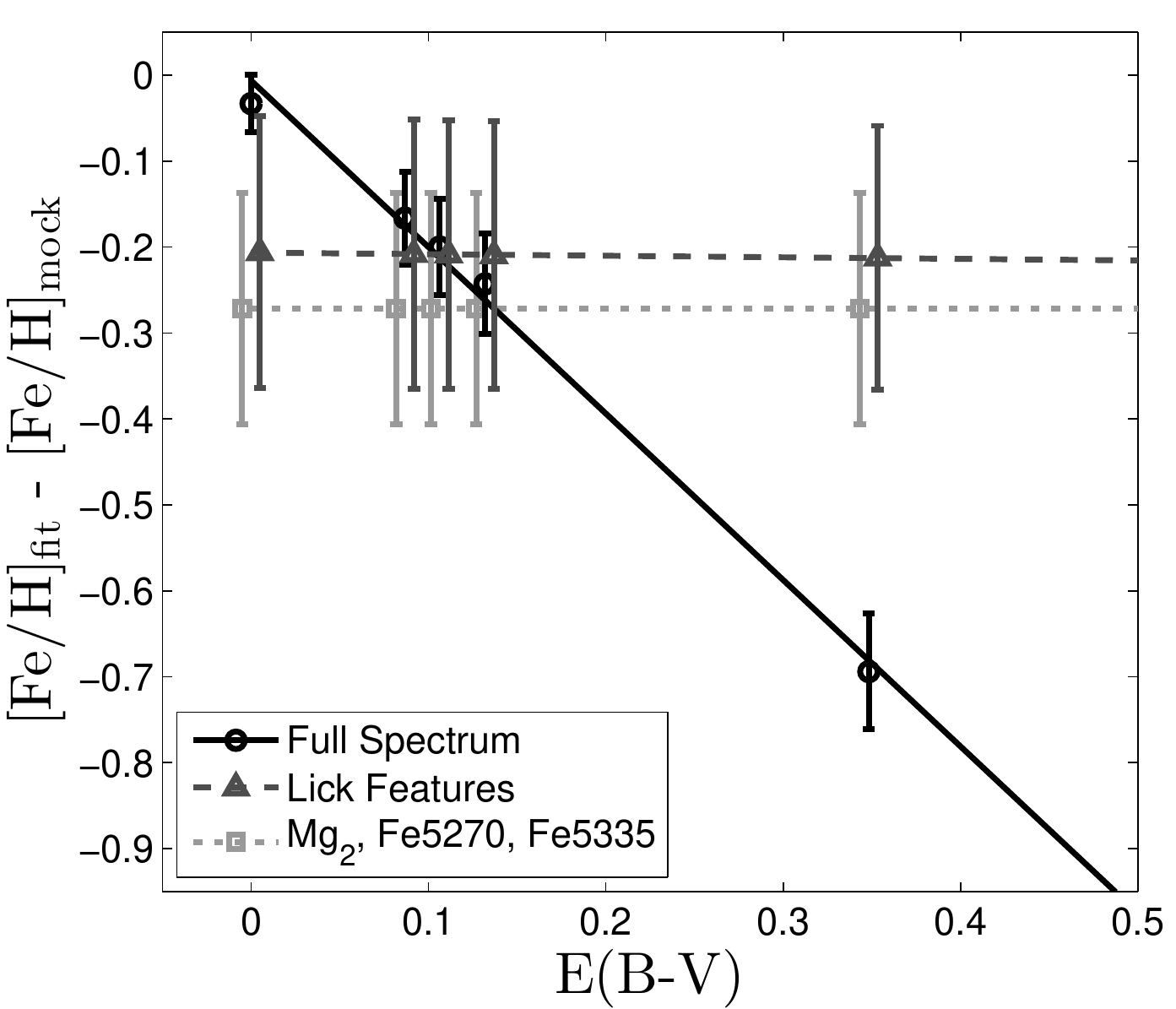}
  \end{center}
  \caption{Residuals between the light-weighted metallicities of mock
   spectra ([Fe/H]$_{\rm mock}$) and those of our population synthesis fits
   ([Fe/H]$_{\rm fit}$) as a function of reddening applied to the mock spectra
   prior to fitting. Fits to the full spectrum, Lick features only, and only
   three metallicity-sensitive features are shown as circles, triangles, and
   squares respectively. Error bars indicate the rms dispersion over 10 tests.
   Linear fits to the data are also shown with different line types (see
   legend). For clarity, the data points for the fits to Lick or metallic
   features are offset in $E(B-V)$ by $\pm 0.005$ respectively.}
  \label{fig:fake_reddened_metallicities}
\end{figure}

To investigate the effect of reddening on the recovered metallicity,
we compute a light-weighted metallicity of the mock and fit spectra,
and average the difference over the ten realisations. This process
is repeated with a range of reddening values.
In \Fig{fake_reddened_metallicities}, we show the metallicity residuals
between the fit and mock spectra as a function of applied reddening with
circular points.  Error bars indicate the rms dispersion over all ten realisations.
When no reddening is applied, we recover the original metallicity accurately,
with a mean $\Delta$[Fe/H]$ = -0.018 \pm 0.023$.  When the mock spectra
are reddened, we recover a lower than expected metallicity, an effect
which grows linearly with $E(B-V)$.

In order to control this trend, we attempt to remove the effect
of reddening from the reddened mock spectra by subtracting their
continua. This procedure is performed by fitting and dividing by
linear pseudo-continua around 13 prominent absorption features
(hereafter ``Lick features'') as in \citet{Worthey}. The continuum
level defined by the spectral region outside each feature is set to zero, 
leaving us with only 453 non-zero pixels per spectrum compared with
2770 in the full-spectrum case.
We remove the continua this way for both the reddened and non-reddened
MILES spectra (including the basis) and rerun the decompositions.  Results
of this test are shown as triangular points in \Fig{fake_reddened_metallicities}.
With reddening effects removed from the spectra, we measure [Fe/H]
consistently to within an error of 0.4 dex. Above $E(B-V) = 0.15$,
the derived metallicities are consistently better than when fitting to
the (reddened) full spectrum indicating that this dereddening task was
successful. However, for weakly reddened spectra, the fit to the full
spectrum results in a more accurate value of [Fe/H]. In the weakly-reddened
regime, one thus ought to fit to as many pixels as possible in order
to maximize the accuracy of the derived parameters in a stellar population.

Note that the derived [Fe/H] values in this test tend to be underestimated.
Since the metallicities of these mock spectra are generally higher (-0.5 $\leq$ [Fe/H] $\leq$ 0)
than the mean metallicity of the basis ([Fe/H] $\sim -0.5$), random
errors are expected to on average bias the average to underestimate the metallicities
of these mock spectra, as observed.  Indeed, we have checked that forcing the mock spectra
to have [Fe/H] $ < -1$ yields an {\it overestimate} in metallicity
of 0.1 to 0.2 dex for highly reddened mock spectra.

In an attempt to increase our metallicity sensitivity, an additional
test was performed where we instead fit only to three metallicity-sensitive
spectral features: Mg$_2$, Fe5270, and Fe5335.  Results of this test
are shown as square points in \Fig{fake_reddened_metallicities}. Again,
no trend with reddening is observed.  However, the error in [Fe/H]
estimates is now larger due to the even lower number (122) of pixels
remaining in the spectra.

Our tests confirm both the code's ability to recover the correct
metallicity in dust-free conditions and to fail in the presence of dust.
Full-spectrum population synthesis of integrated stellar spectra thus
ill-advised for highly-reddened (e.g. low latitude) stellar systems without
appropriate dust modelling.


\section{Population Synthesis of Galactic Globular Clusters}
\label{sec:popsynth}

\subsection{Population Synthesis of the S05 GGC Spectra}
\label{sec:GGCsynthesis}

The \lbfgsb\ optimisation code was applied to the S05 GGC spectra using our final basis
of 242 stars. Because the stars in the final basis are sampled mostly from the solar neighbourhood,
and thus have similar alpha abundance patterns to the GGCs, we do not expect alpha-enhancement
to be a concern \citep[see][]{Schiavon2007}.  As in \se{mock}, we have also verified that our results
are insensitive to initial conditions when fitting to GGC data. Therefore we use equal weighting
for each basis spectrum as initial conditions.

In light of our tests in \se{mock} and the fact that many of our GGC spectra suffer from Galactic extinction, we have implemented a reddening correction in \eq{Ri} by multiplying $S_i$ by a Cardelli reddening law of the form $e^{-\tau (\lambda)}$, where:

\be
\tau(\lambda) = R_V E(B-V) \left( a(x)+\frac{b(x)}{R_V}\right) \frac{1}{1.086}.
\label{eq:CardelliRedLaw}
\ee

and $a(x)$ and $b(x)$ are defined in \citet{Cardelli1989}. We adopt $R_V=3.1$ for the Milky Way. For each GGC, we varied $E(B-V)$ in \eq{CardelliRedLaw} in steps of 0.02 from 0 to 1. The fit with the lowest value of the Merit function qualified as our best.

Comparisons of the synthesized and original spectra, along with the
residuals of the fit, are shown in the electronic version of \ap{AppendixA} for all 41 GGCs.  The
residuals were computed by taking the difference between the fitted
and original spectra and normalizing by the mean value of the latter.
$\chi^2$/dof values (hereafter $\chi^2$) are also shown for each
decomposition. As in \citet{Macarthur09}, these values are computed
as the ratio between the actual and expected variance in the fit.
Most of our fits are excellent, such as those
of NGC 1851 and NGC 1904 with $\chi^2 \sim 1$, and demonstrate
the applicability of our method.

\begin{figure}
  \begin{center}
    \subfigure{\includegraphics{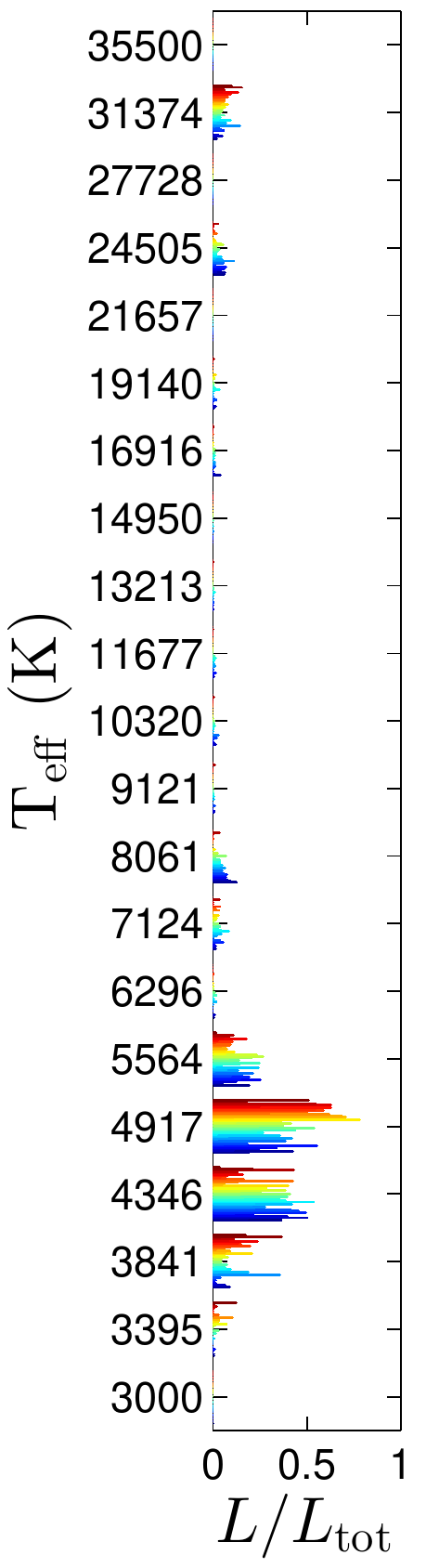}}
    \subfigure{\includegraphics{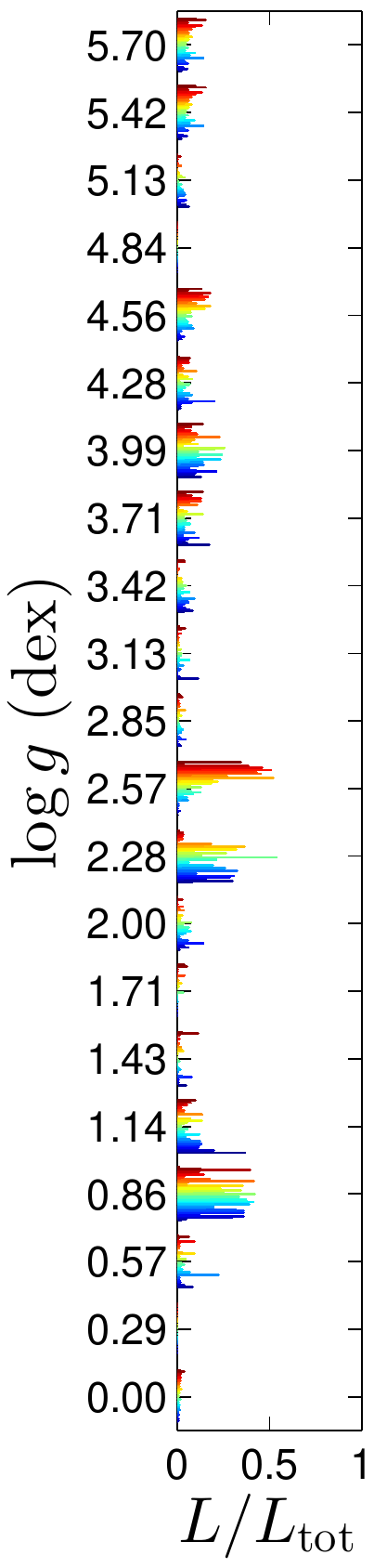}}
    \subfigure{\includegraphics{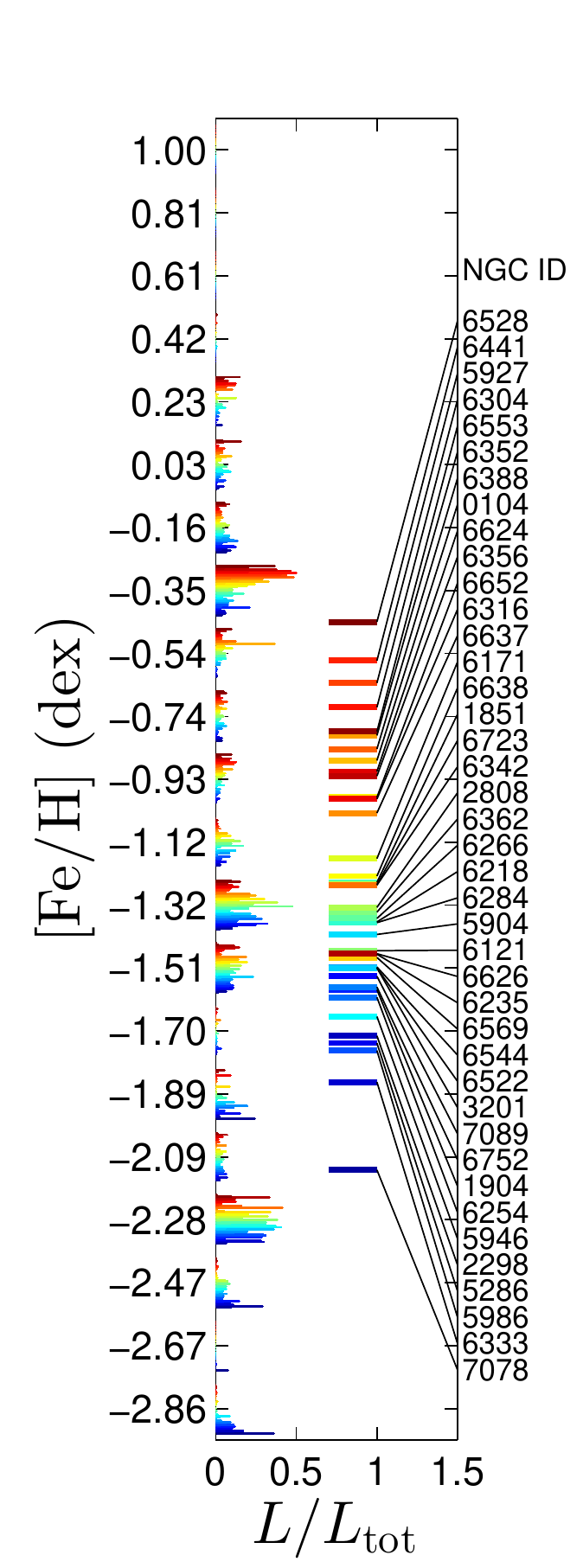}}
  \end{center}
  \caption{Optimised fractional contributions of final basis stars
   to each of the 41 S05 GGC spectra versus total
   cluster $V-$band luminosity, binned by either \Teff, \logg\, or [Fe/H].
   Each colour strip represents the best final fit of a separate GGC,
   organized from low (bluer) to high (redder) metallicity.  The values
   on the Y-axis correspond to the lower bounds of each atmospheric
   parameter bin. The derived light-weighted metallicity of each cluster
   is also indicated on the right side of the right panel, with a black line
connecting each NGC ID with its metallicity.}
  \label{fig:param_Conts}
\end{figure}

The luminosity fractions of different stellar types which make up the total
light of our GGCs, as inferred from our optimisation code, are binned
by \Teff, \logg, and [Fe/H] and plotted in \Fig{param_Conts}.
Each colour bar in this figure represents the fit to a different cluster,
ordered by metallicity.
Recall that the relative uncertainty assigned to the weight of a given star
is $\sim$20 per cent and that the absolute noise level is $\sim$5 per cent (see
\se{mock}). However in this figure, the mean error in the magnitude of each
colour bar will be less than that for individual stars.
As a conservative upper limit, we still assign these same 20 per cent errors
to the colour bars in this figure.  Therefore all light fractions in this
figure contributing less than 5 per cent may be considered noise.

In the \Teff~regime, most of the contribution ($\sim$90 per cent) to the
total light of the GGCs comes from stars with \Teff $\sim$4000-6000~K.
Interestingly, two very hot stars ($2.4$ and $3.1\times10^4$~K) together
make a small $\sim$5 per cent contribution to all the fits.

The fact that our modelling of these old systems requires very hot blue
stars in all cases is certainly odd, and for some it might be a numerical artifact.
This is especially true for GGCs such as 47 Tuc and NGC 6652 which are not
expected to harbour such hot stars.  Not surprisingly, the contribution from
these hot stars in these clusters is less than our expected 5 per cent random error.
Thus we may be overfitting these spectra in an effort to numerically minimise
the residuals.  However, in the cases where the hot star contribution does exceed
5 per cent, the decomposition would then be physically motivated; that is, these
clusters are known to have a significant population of hot (e.g. blue horizontal
branch and blue straggler) stars.  Such is the case for e.g. NGC 6752 and 7089
(see \Table{lumFracTableCMD} and \ap{AppendixC}).

We find that these hot stars are necessary in order to obtain an accurate fit to the
integrated spectra. Indeed, setting their light-weighting to zero after the
optimisation yields a poor fit. 
The need for hot stars in similar fits was also reported by \citet{Koleva2008}
who found that adding a contribution of very hot stars ($T_{\rm eff}$ = 6000 to 20 000 K)
to their fits to the S05 spectra improved the match considerably, especially
for clusters with strong blue horizontal branches.

The largest mean contributions to the total light in the \logg\ regime
come from high surface gravity main sequence stars ($\sim 35 $ per cent between
\logg\ = 3.42 and 4.84 dex) and low surface gravity red giant stars
($\sim 33 $ per cent between \logg\ = 0.5 and 1.43 dex). A slightly smaller
contribution of $\sim 28 $ per cent on average comes from intermediate
surface gravity stars with \logg\ between 2.28 and 2.85 dex.

In the [Fe/H] regime, contributions come from a wide range of values.
For clarity, the mean light-weighted metallicity of each GGC is indicated on the right side of 
the right panel. GGCs with low (high) metallicity have larger contributions from
low (high) metallicity stars, as expected.  \se{met} provides a more detailed
comparison between these derived metallicities and current literature estimates.

\begin{figure}
  \begin{center}
    \includegraphics[width=\textwidth]{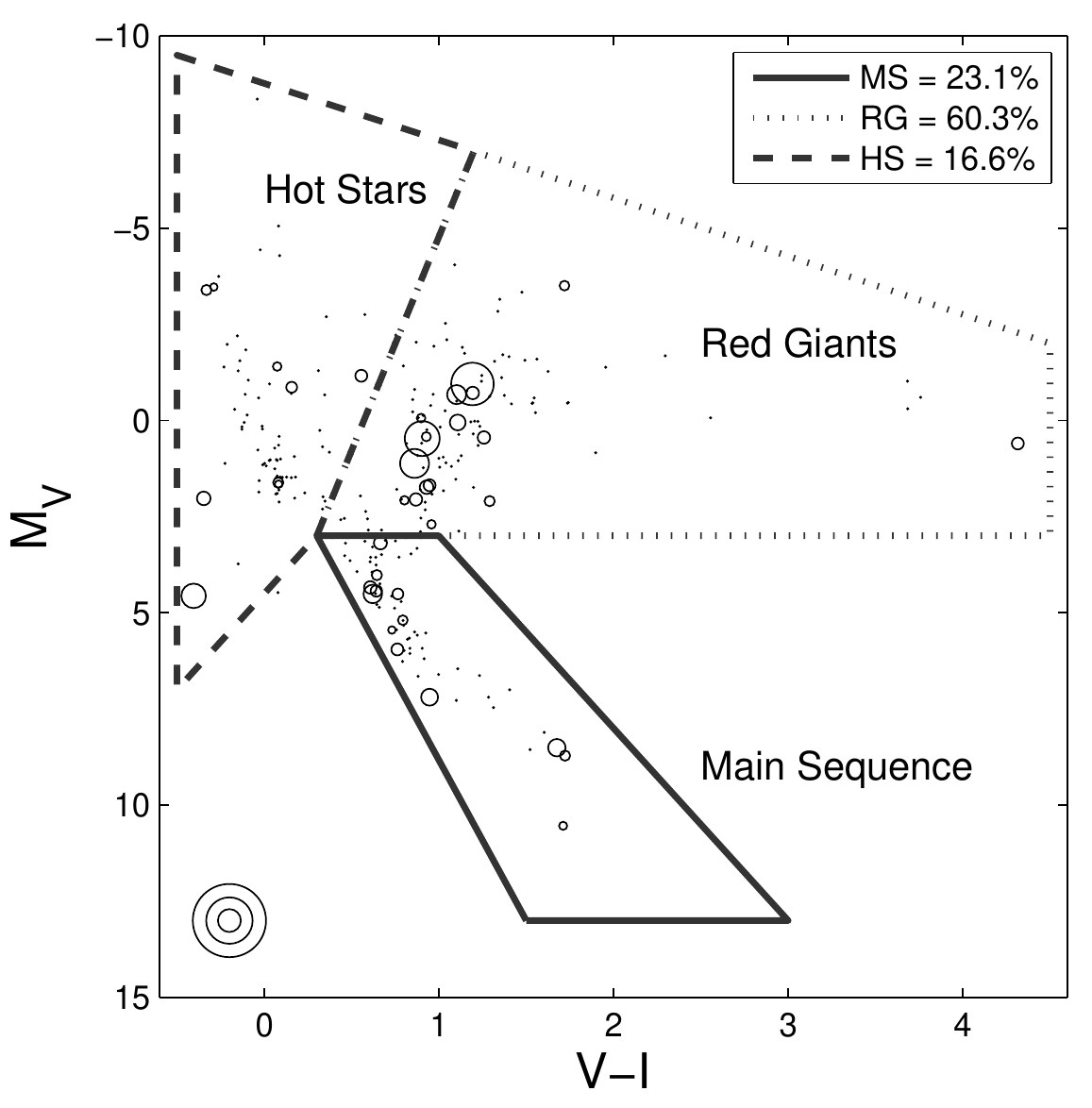}
  \end{center}
  \caption{Mean fractional contributions (by light) of MILES stars
   to the spectra of all 41 GGCs plotted on a colour magnitude
   diagram, as indicated by the area covered by each point. Concentric
   circles at the bottom left corner indicate contributions of 5 per cent, 20 per cent,
   and 50 per cent. The legend indicates the total contribution from 
   hot stars (HS), main sequence (MS), and red giant branch (RG) stars.}
  \label{fig:fit_CMD}
\end{figure}

For a better visualisation of these results, \Fig{fit_CMD} shows the
fractional contributions by light of MILES stars to the GGC spectra,
averaged over all 41 fits to our GGCs.  Relative
contributions are directly proportional to the area of each circle.
For reference, concentric circles are drawn in the bottom left corner to 
indicate contributions of 5 per cent, 20 per cent, and 50 per cent respectively.
Rough stellar evolutionary zones have been sketched to investigate the
contribution of various phases of stellar evolution. Because of the coarse
resolution of the MILES library in (V-I) - $M_V$ space, only three zones
have been drawn: Lower main sequence (MS), red giant branch (RG), and hot stars (HS).
We find that roughly 60, 23, and 17 per cent of the integrated light of GGCs
comes from each zone, respectively. As we shall see
in \se{CMD}, these evolutionary phase fractions agree well with the
stellar fractions inferred from GGC CMDs, thus lending our
optimisation method further support.

\subsection{Metallicity Comparisons}\label{sec:met}

\begin{figure}
  \begin{center}
    \includegraphics[width=\textwidth]{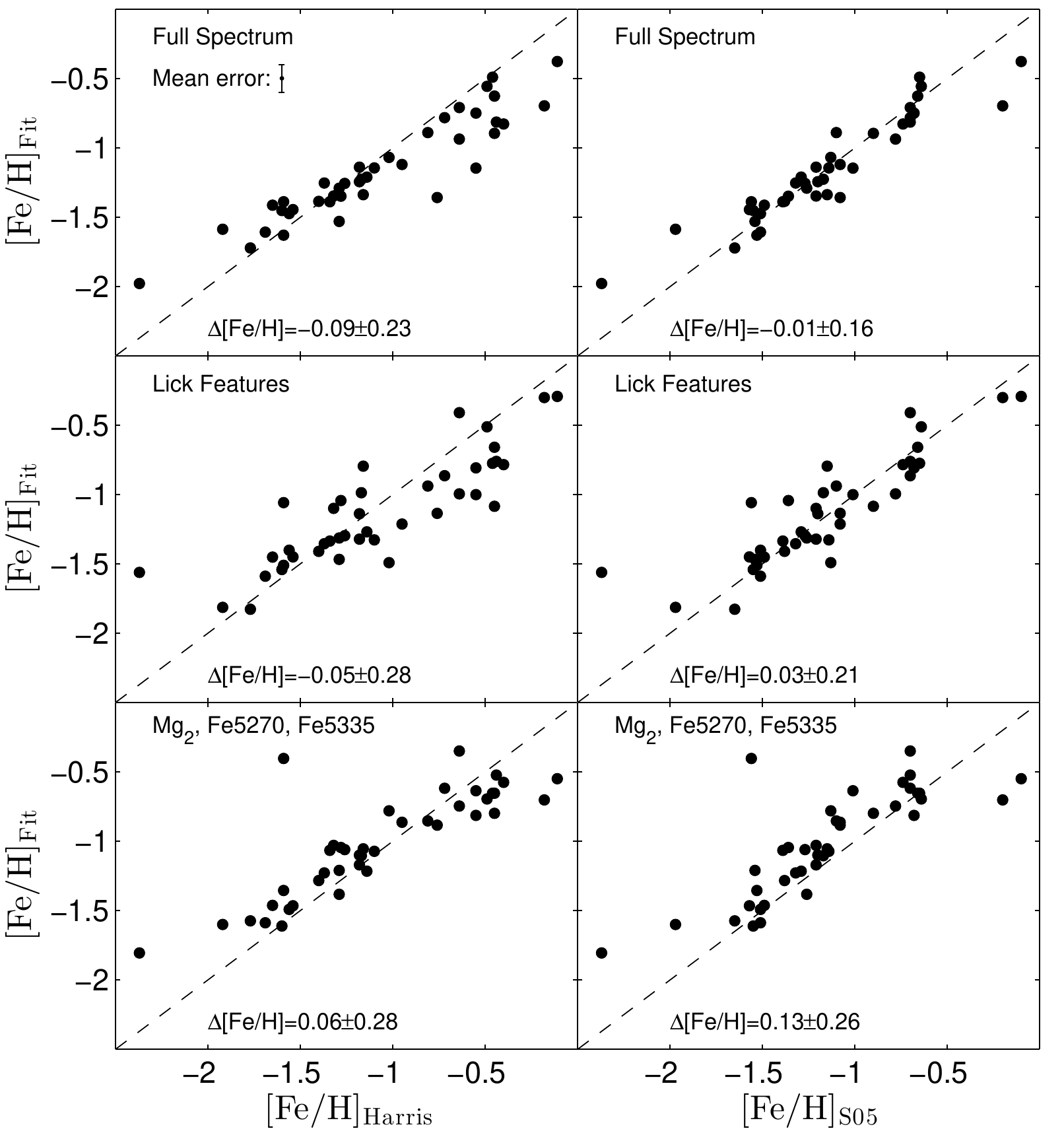}
  \end{center}
  \caption{Comparison between metallicities derived from our optimisation
   synthesis of S05 GGC spectra and those from the~\citet{Harris} catalogue
   (left column) or S05 (right column).
   1:1 relations are shown as dashed lines.
   The top row shows results from
   population synthesis using the full spectrum. The middle
   and bottom panels show the results of fitting only
   continuum-subtracted Lick features (thirteen generally
   prominent and three metallicity-sensitive features, respectively).
   Mean and rms metallicity offsets are indicated in each panel.}
   \label{fig:fehVsFeh}
\end{figure}

We can also test our code's ability to reproduce
known GGC metallicities. \Fig{fehVsFeh} compares literature values of
[Fe/H] with those obtained through our decompositions. The error bar
in the top-left panel represents the nominal uncertainty in the fitted
values, based on our mock spectra decompositions (see \se{mock}).

The top-left panel of \Fig{fehVsFeh} shows a comparison between the
metallicities from \citet[][circa 2010]{Harris} and fits to our GGCs.
The mean and rms offset between our fits and literature values
($\Delta$[Fe/H]) are indicated in each panel.  Between [Fe/H] of -1.5 and -1,
the two datasets are in remarkable agreement. However, we tend to
underestimate and overestimate [Fe/H] at higher and lower metallicity, respectively. 

\begin{figure}
  \begin{center}
    \includegraphics[width=\textwidth]{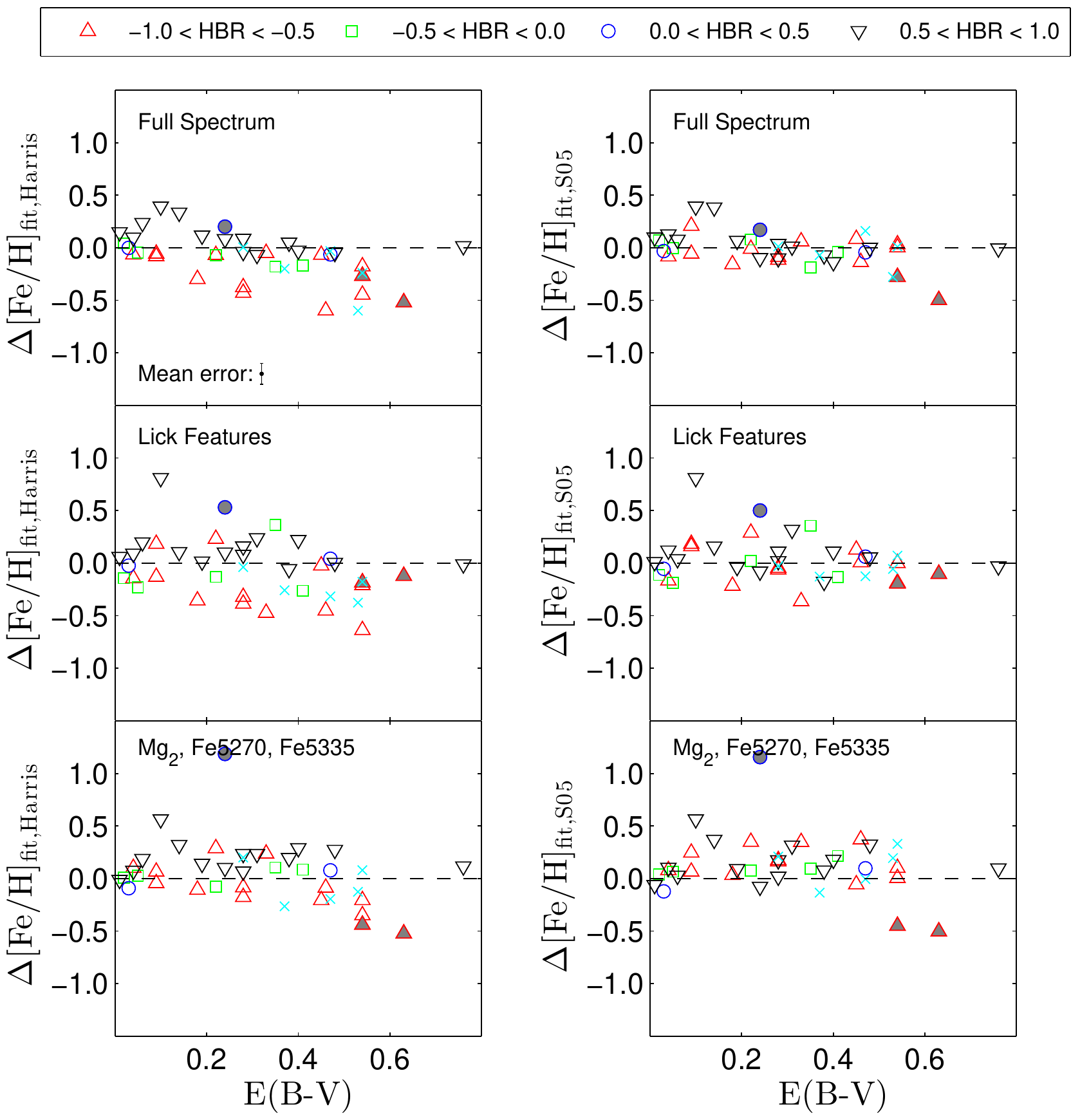}
  \end{center}
  \caption{
  Residuals between metallicities derived from our optimisation
   synthesis of S05 GGC spectra and metallicities taken from
   the~\citet{Harris} catalogue, as a function of $E(B-V)$ (also
   taken from~\citealt{Harris}). Poor fits ($\chi^2 > 5$) are indicated
   with gray-filled markers. Points are styled according to horizontal branch ratio (HBR; see legend), 
while clusters without an HBR estimate are shown as cyan x's.}
   \label{fig:HarrisresidVsreddening}
\end{figure}

This discrepancy at high metallicity may stem from the fact that metal-rich GGCs are mostly
located near the Milky Way mid-plane and suffer more extinction than
those lying above it. The top-left panel of \Fig{HarrisresidVsreddening}
shows the difference between our inferred (light-weighted) mean metallicity
values and those taken from the 2010 edition of the~\citet{Harris}
catalogue, as a function of colour excess $E(B-V)$.  Here, poor fits
($\chi^2>5$) are indicated with filled gray markers. Data points are
also coloured according to horizontal branch ratio (HBR), defined as
HBR = (B-R)/(B+V+R), where B, V, and R are the number of stars bluer
than, within, and redder than, the RR lyrae region of an HR diagram,
respectively \citep{Harris}. Thus a low HBR indicates a higher fraction
of red-to-blue stars.  For GGCs with low (negative) HBR and high
reddening ($E(B-V) > 0.15$), we underestimate [Fe/H] relative to
the \citet{Harris} values. This trend corroborates our results
of mock tests with reddening in \se{mock}.  In this first pass,
it would appear that our Cardelli reddening correction was 
unsuccessful; however, we show below that this is not the case.

\citet{Cezario13} also estimate [Fe/H] for the S05 GGCs via spectrum
synthesis with the modelling program {\it ULySS} using for a basis SSP
models based on MILES rather than the (empirical) MILES itself. When
comparing with the [Fe/H] values of \citet{Carretta2009}, they achieve
$\Delta$[Fe/H] =$ -0.15 \pm 0.17$.  Note that comparing with \citet{Carretta2009}
instead of \citet{Harris} does not change our results, which is expected since the
Harris catalogue is largely based on that of \citet{Carretta2009}. Thus, relative
to \citet{Cezario13}, our method yields a smaller net offset but similar scatter
with $\Delta$[Fe/H] =$ -0.09 \pm 0.23$.

In order to investigate and ultimately remove any reddening dependence,
we subtract the continua of both the GGC and template MILES spectra
using the same method as in \se{mock}, and rerun the decompositions. 
Note that this time we do not include the Cardelli reddening law coefficient
as in fits with the full spectrum. The middle rows of Figures \ref{fig:fehVsFeh}
and \ref{fig:HarrisresidVsreddening} show the result of this exercise.
The left middle panel of \Fig{fehVsFeh} indicates that the discrepancy
between [Fe/H] derived from our decompositions and \citet{Harris} for
highly reddened GGCs is largely unchanged, and now exhibits more scatter.

Despite our continuum subtraction, the trend with reddening is still observed.
Furthermore, we still find a lower than expected [Fe/H] for GGCs with negative HBR,
but satisfactory agreement for those with positive HBR.
This discrepancy may be partially explained by
the fact that the 2010 version of the \citet{Harris} catalogue
uses a different metallicity scale than does \citet{Prugniel}
for the MILES stars. The \citet{Harris} values have been
converted to a new scale defined by \citet{Carretta2009},
while Prugniel's scale agrees with that used by \citet{MILES},
whom in turn used \citet{SKC} as a reference for their [Fe/H] values.
Thus, rather than using the current \citet{Harris} catalogue,
we compare our derived [Fe/H] values with those compiled (and
derived) by S05.  Their [Fe/H] values are taken from
\citet{KraftIvans2003} or \citet{CarrettaGratton1997},
or extracted using their own spectroscopic methods.

The right-hand column of \Fig{fehVsFeh} shows a comparison between
the [Fe/H] values derived from our decompositions and those reported
by S05. The top-right panel of this figure shows that our values for
clusters with high [Fe/H] are in much better agreement with S05
than with \citet[][2010 version]{Harris}, even when fitting to the
full spectrum. \citet{Cezario13} also find better agreement when
comparing with S05 values ($\Delta[{\rm Fe/H}] = -0.05 \pm 0.16$), 
whereas we find $\Delta[{\rm Fe/H}]= -0.01 \pm 0.16$. Thus our errors
in [Fe/H] are likely not driven by our inability to measure [Fe/H],
but rather due to systematic uncertainties in the reference library.

Our scatter is consistent with a similar study by \citet{Koleva2008}, who
performed population synthesis on the S05 GGC spectra using single stellar
population models based on the MILES library. They report a standard
deviation of 0.17 dex between their determined [Fe/H] and those from S05.
It is remarkable that we can achieve a similar result using only stellar
spectra from MILES as a basis, without any evolutionary modelling.

Despite our corrections, we still seem to overestimate [Fe/H] at the low
metallicity end ([Fe/H] $< -2$) when comparing to S05. This discrepancy
might be due to incompleteness in our basis for stars with [Fe/H] $< 2$, as
is shown in the right-hand column of \Fig{basisParams}.  In an attempt
to correct for this, we allow all MILES stars with [Fe/H] $< 2 $ into
our basis regardless of peculiarities, and rerun the decompositions
but no improvement is found.  This is not surprising since the entire
MILES library itself contains few stars below this metallicity.

For completeness we show the result of fitting to only the Lick features
in the middle-right panel (again without a Cardelli reddening correction).
While no trend with reddening is observed, the scatter is much higher
than when fitting to the full spectrum. The same panel in \Fig{HarrisresidVsreddening}
shows that the residual trend with $E(B-V)$ has disappeared and shows lower scatter
than when comparing with \citet{Harris} [Fe/H] values, indicating that this task was
indeed successful.

Rather than using all of the prominent features in our continuum-subtracted
spectra, we now attempt to further improve the recovered metallicities by
considering only three of the most metallicity-sensitive lines:
Mg$_2$, Fe5250, and Fe5335. These results are displayed in the bottom
panels of Figures \ref{fig:fehVsFeh} and \ref{fig:HarrisresidVsreddening}.
In this case comparisons between our fit [Fe/H] and those of \citet{Harris}
and S05 show good agreement.  This method however {\it overestimates}
metallicities for most of the GGCs regardless of our choice of [Fe/H] library
comparison. This is likely because we err toward the mean of the distribution
of atmospheric parameters in the basis ([Fe/H] $= -0.5$; see \se{mock}) 
which is higher than the [Fe/H] of most of our GGCs.
Indeed, for [Fe/H] $>-0.5$, we tend to underestimate [Fe/H] when
comparing with \citet{Harris} values, a phenomenon consistent with
our tests with mock spectra.

We thus conclude that, in terms of measuring light-weighted metallicities
of highly reddened stellar systems accurately, fitting to the full spectrum
with the inclusion of an appropriate reddening law yields the best results.
However, in the absence of a reliable reddening correction, fitting to many
spectral features with the continuum subtracted is a viable alternative.
In the absence of significant reddening, fitting to the full spectrum yields
the most accurate and precise metallicities.

\subsection{GGC Population Synthesis through Colour-Magnitude Diagrams}
\label{sec:CMD}

A complementary approach to testing population synthesis results is through
CMDs with which luminosity, colour, and stellar evolutionary phase distributions
of member stars for each GGC can readily be extracted.  From the luminosity
and colour data, we can create a light-weighted colour distribution for each
CMD.  Thus, if colour information of the MILES stars can be obtained,
we can directly compare the light-weighted stellar fractions resulting
from our GGC spectral decompositions with those extracted from CMDs.

Accurate CMD data for our GGCs are available from the GGC HST/ACS survey ~\citep{Ata}.
This database contains $V-$ and $I-$band magnitudes of individual stars in 66 GGCs.
Twenty-four of these ACS GGCs overlap with the S05 database, thus providing
ample data for comparison.

\subsubsection{Stellar Evolutionary Stages}\label{sec:CMDevolution}

In order to obtain luminosity fractions of the many stellar
evolutionary stages in globular clusters, we have defined specific
CMD zones which correspond to these phases.
The carefully delineated stellar evolutionary phase zones
are shown for the 24 GGCs that overlap with~\citet{Ata}
in \ap{AppendixD}.

For each GGC, we then compute the total stellar light
contained in each region and normalize those values
by the total luminosity of the cluster, in the $V$-band.  Stars lying outside
of the evolutionary zones are ignored. The evolutionary
phase luminosity fractions from CMDs are listed in \Table{lumFracTableCMD}. 
For a discussion on the luminosity fractions derived from these CMDs, see \ap{AppendixB}.


\begin{table}
\caption{Luminosity fractions of various stellar evolutionary
 phases for the 24 GGCs in common between \citet{S05} and \citet{Ata}.
 Fractions are derived via colour-magnitude diagrams.}
\centering
\begin{tabular}{lccccccc}
\\ [-1ex]
\hline \hline \\[-2ex]
& \multicolumn{6}{c}{Luminosity fraction} \\
NGC ID & MS & TO & SGB & RGB+AGB & RHB & BHB & BS\\ 
\hline
\\ [-1ex]
0104 & 0.13 & 0.10 & 0.05 & 0.56 & 0.15 & 0.00 & 0.01\\
1851 & 0.14 & 0.09 & 0.03 & 0.58 & 0.11 & 0.04 & 0.01\\
2298 & 0.13 & 0.09 & 0.05 & 0.61 & 0.02 & 0.09 & 0.01\\
2808 & 0.14 & 0.09 & 0.06 & 0.60 & 0.07 & 0.02 & 0.01\\
3201 & 0.15 & 0.09 & 0.04 & 0.59 & 0.06 & 0.06 & 0.01\\
5286 & 0.16 & 0.09 & 0.05 & 0.60 & 0.02 & 0.07 & 0.00\\
5904 & 0.16 & 0.09 & 0.04 & 0.59 & 0.03 & 0.07 & 0.01\\
5927 & 0.14 & 0.12 & 0.06 & 0.52 & 0.14 & 0.00 & 0.01\\
5986 & 0.15 & 0.09 & 0.06 & 0.59 & 0.03 & 0.07 & 0.01\\
6121 & 0.15 & 0.12 & 0.05 & 0.51 & 0.10 & 0.07 & 0.01\\
6171 & 0.17 & 0.12 & 0.04 & 0.52 & 0.11 & 0.03 & 0.01\\
6218 & 0.16 & 0.12 & 0.06 & 0.57 & 0.00 & 0.07 & 0.01\\
6254 & 0.18 & 0.10 & 0.04 & 0.61 & 0.00 & 0.06 & 0.01\\
6304 & 0.19 & 0.11 & 0.04 & 0.47 & 0.14 & 0.01 & 0.02\\
6352 & 0.16 & 0.13 & 0.06 & 0.43 & 0.18 & 0.01 & 0.03\\
6362 & 0.13 & 0.10 & 0.05 & 0.56 & 0.09 & 0.05 & 0.01\\
6388 & 0.11 & 0.08 & 0.05 & 0.59 & 0.15 & 0.01 & 0.01\\
6441 & 0.11 & 0.06 & 0.04 & 0.55 & 0.20 & 0.02 & 0.01\\
6624 & 0.13 & 0.12 & 0.06 & 0.49 & 0.17 & 0.01 & 0.02\\
6637 & 0.13 & 0.11 & 0.04 & 0.56 & 0.14 & 0.01 & 0.01\\
6652 & 0.12 & 0.11 & 0.05 & 0.56 & 0.13 & 0.02 & 0.02\\
6723 & 0.16 & 0.11 & 0.05 & 0.53 & 0.09 & 0.05 & 0.01\\
6752 & 0.15 & 0.10 & 0.07 & 0.61 & 0.00 & 0.06 & 0.00\\
7089 & 0.17 & 0.10 & 0.05 & 0.59 & 0.02 & 0.07 & 0.01\\ \hline
Mean & 0.15 & 0.10 & 0.05 & 0.56 & 0.09 & 0.04 & 0.01\\
$\sigma$ & 0.02 & 0.01 & 0.01 & 0.05 & 0.06 & 0.03 & 0.01\\ [1ex]
\hline
\end{tabular}
\label{fig:lumFracTableCMD}
\end{table}

In order to compare the evolutionary zone fractions between the ACS CMDs and those derived from population synthesis, we redrew coarse zones on the ACS CMDs that match those boundaries shown
in \Fig{fit_CMD}. The resulting light fractions are listed in \Table{lumFracTableComp} along with
those from populations synthesis. 
The light fractions from CMDs come solely from $V$-band luminosities, while those from
our optimisation are measured across our entire wavelength range from $\sim 3500$
to $6000$ \AA\ which overlaps with the $B$ and $V$ bands. We do not expect this slight
discrepancy to affect our results significantly. Indeed, the photometrically-derived results 
are in good statistical agreement with those obtained from our spectroscopic decompositions. 
For instance, the MS stars in CMDs make up $17 \pm 3$ per cent
of the total light, while the mean dwarf light fraction in our
spectral decompositions is $22 \pm 6$ per cent.  For red giant stars, we
find a contribution of $64 \pm 7 $ per cent from our spectral decompositions
compared to the sum ($69 \pm 6 $ per cent) obtained from CMDs. The light
fraction of hot stars also agrees well, with a fit
fraction of $15 \pm 7 $ per cent compared with $14 \pm 6 $ per cent from CMDs.
The uncertainty in these light fractions was
computed by summing in quadrature the rms noise of the luminosity
fractions for the aforementioned stellar evolutionary zones. 
The magnitude cut in \se{CMDcolor} was not employed here; we have checked 
that doing so did not significantly affect these results.


\begin{table}
\caption{Luminosity fractions of various stellar evolutionary
 phases for the 24 GGCs in common between \citet{S05} and \citet{Ata}.
 Fractions are derived via colour-magnitude diagrams.}
\centering
\begin{tabular}{lccccccc}
\\ [-1ex]
\hline \hline \\[-2ex]
 & \multicolumn{3}{c}{CMDs} & \multicolumn{3}{c}{Spectrum synthesis} \\
NGC ID&MS&RG&HS&MS&RG&HS \\
\hline
\\ [-1ex]
0104&0.22&0.77&0.01&0.26&0.69&0.05\\
1851&0.17&0.71&0.12&0.29&0.62&0.10\\
2298&0.20&0.62&0.18&0.04&0.72&0.24\\
2808&0.16&0.77&0.06&0.25&0.64&0.10\\
3201&0.21&0.69&0.10&0.20&0.58&0.22\\
5286&0.21&0.65&0.14&0.09&0.65&0.27\\
5904&0.15&0.72&0.13&0.30&0.54&0.17\\
5927&0.14&0.68&0.18&0.28&0.62&0.10\\
5986&0.19&0.67&0.13&0.08&0.64&0.27\\
6121&0.19&0.63&0.18&0.15&0.63&0.22\\
6171&0.19&0.63&0.18&0.17&0.69&0.14\\
6218&0.20&0.69&0.12&0.22&0.57&0.21\\
6254&0.16&0.75&0.09&0.16&0.62&0.21\\
6304&0.15&0.79&0.05&0.24&0.66&0.10\\
6352&0.17&0.74&0.09&0.21&0.72&0.06\\
6362&0.22&0.62&0.17&0.24&0.68&0.08\\
6388&0.11&0.73&0.16&0.16&0.73&0.11\\
6441&0.16&0.67&0.17&0.24&0.63&0.13\\
6624&0.15&0.75&0.10&0.24&0.69&0.07\\
6637&0.13&0.66&0.22&0.24&0.70&0.07\\
6652&0.13&0.58&0.29&0.31&0.61&0.08\\
6723&0.17&0.61&0.21&0.34&0.55&0.11\\
6752&0.17&0.71&0.12&0.20&0.58&0.21\\
7089&0.17&0.70&0.13&0.27&0.55&0.18\\  \hline
Mean&0.17&0.69&0.14&0.22&0.64&0.15\\
$\sigma$&0.03&0.06&0.06&0.06&0.07&0.07\\ [1ex]
\hline
\end{tabular}
\label{fig:lumFracTableComp}
\end{table}

\begin{figure}
  \begin{center}
    \includegraphics[width=\textwidth]{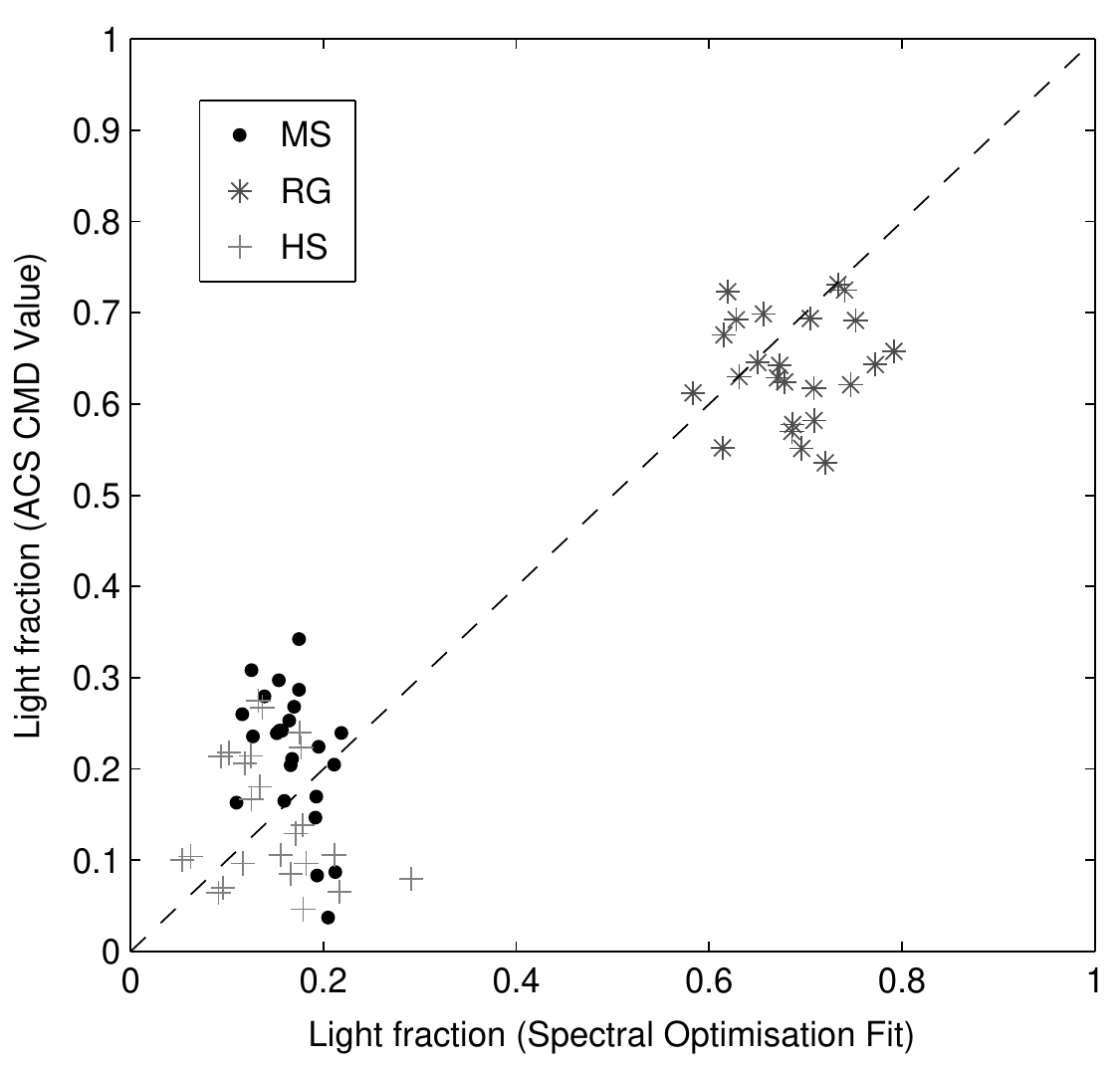}
  \end{center}
  \caption{Light fractions of various evolutionary stages from the ACS CMDs as a function of those obtained by our population synthesis model. Fractions from main sequence (MS), red giant (RG), and ``hot star'' (HS) regions are shown as circles, stars, and crosses, respectively. The 1:1 relation is shown as a dashed line.}
  \label{fig:lightfracs_ACS_vs_fit}
\end{figure}

Regarding the fractions for individual clusters, we note that the
HS fractions derived spectroscopically do not always agree with those from CMDs. 
This can be seen in \Fig{lightfracs_ACS_vs_fit} where we show the light fractions of
MS, RG, and HS from the ACS CMDs against our fitted results. While the agreement
is generally very good, the scatter is large. The latter may result from the crudeness
of our boundaries. For instance, our RG zone (see \Fig{lightConts}) may contain
some RHB stars that truly belong in the HS zone. This is especially true when
the RHB is concentrated close to the RGB in the CMD (e.g. NGC 104, NGC 6441).
Conversely, some MILES stars that fall in our HS zone but lie near
the boundary may in fact belong to the RG zone (e.g. NGC 6752, NGC 7089).
Thus in order to obtain accurate evolutionary phase fractions, the
CMD zone boundaries must be carefully defined. See \ap{AppendixC}
for a visual representation on a per-cluster basis.

The test above is yet another important validation of our numerical
optimisation spectral decomposition method.

\subsubsection{Comparison of (V-I) Distributions}\label{sec:CMDcolor}

In order to compare the colour distributions based on our
spectral decompositions and extracted from CMDs, we extract
$V-I$ colours for MILES stars from the SIMBAD astronomical
database\footnote{\tt http://simbad.u-strasbg.fr/simbad/}.
Because the latter lacks $I-$band magnitudes for many MILES stars,
and short of any other catalogue with colour information, we use the
colour interpolator of ~\citet{WortheyLee} to convert the atmospheric
parameters \Teff, \logg, and [Fe/H] for our final basis stars into
$V-I$ colours.

A check on the reliability of the calculated values is performed
by comparing the Worthey \& Lee estimates with measured values from
SIMBAD (if available).  Modulo a small group of 5 discrepant
stars (none of which are included in our final basis), the overall agreement
is excellent.  The mean difference in $V-I$ between the two datasets
for stars in our basis is $-0.03$ with an rms dispersion of $0.11$.
We thus verify the reliability of the Worthey $\&$ Lee calculator, as those
authors also demonstrated in their paper. 

\begin{figure}
  \begin{center}
    \includegraphics[width=\textwidth]{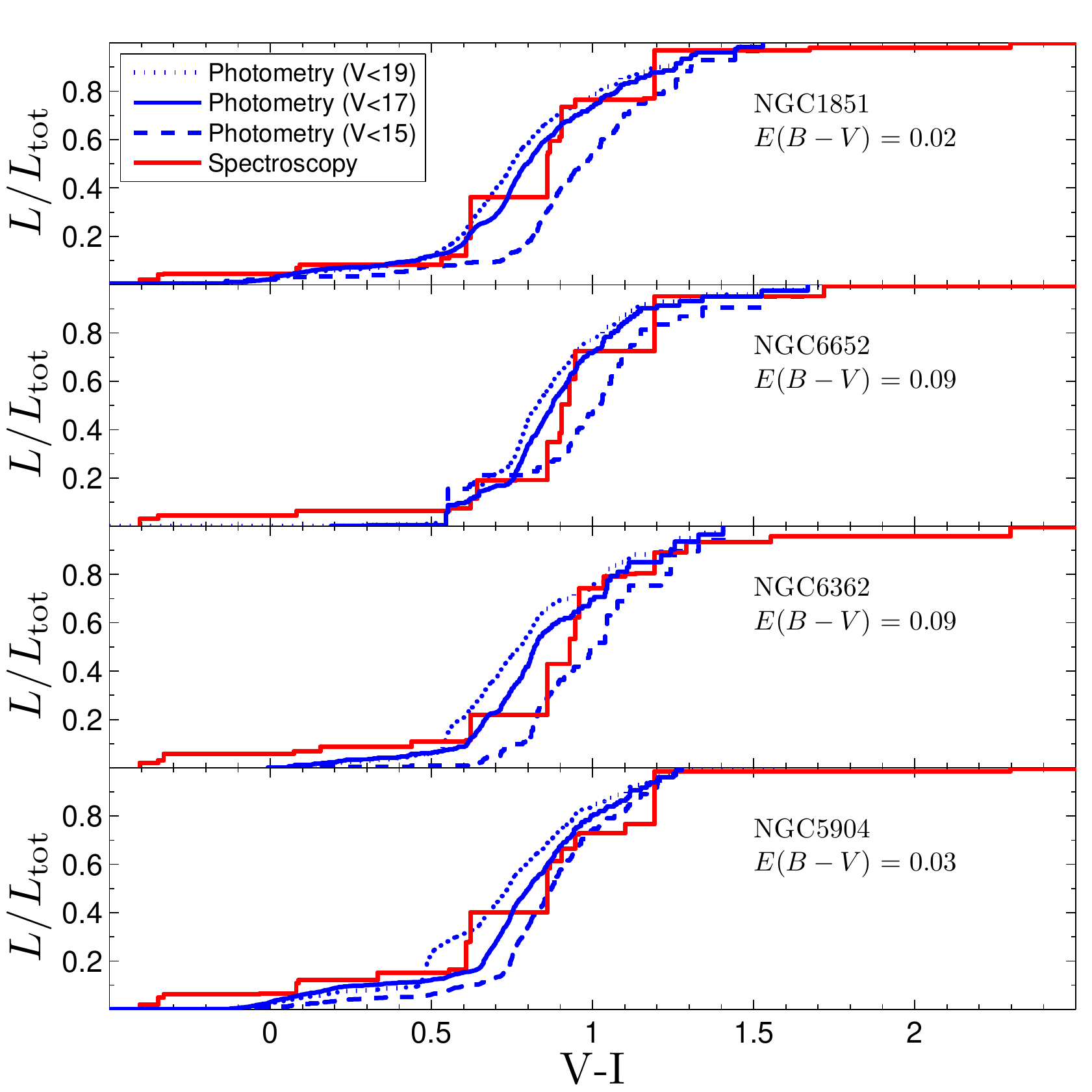}
  \end{center}
  \caption{CDFs of the $V-I$ distributions obtained from our decompositions
   (red) compared to those from CMDs (blue) for a select sample of GGCs with
   low reddening.  The CMD CDFs are shown for a range of the $V-$band
   magnitude cut-offs (see upper-left corner legend), as described in the
   text.  The best match is obtained for a magnitude cut-off of $V<17$.}
  \label{fig:CMD_Vthresh}
\end{figure}

\begin{figure}
  \begin{center}
    \includegraphics[width=\textwidth]{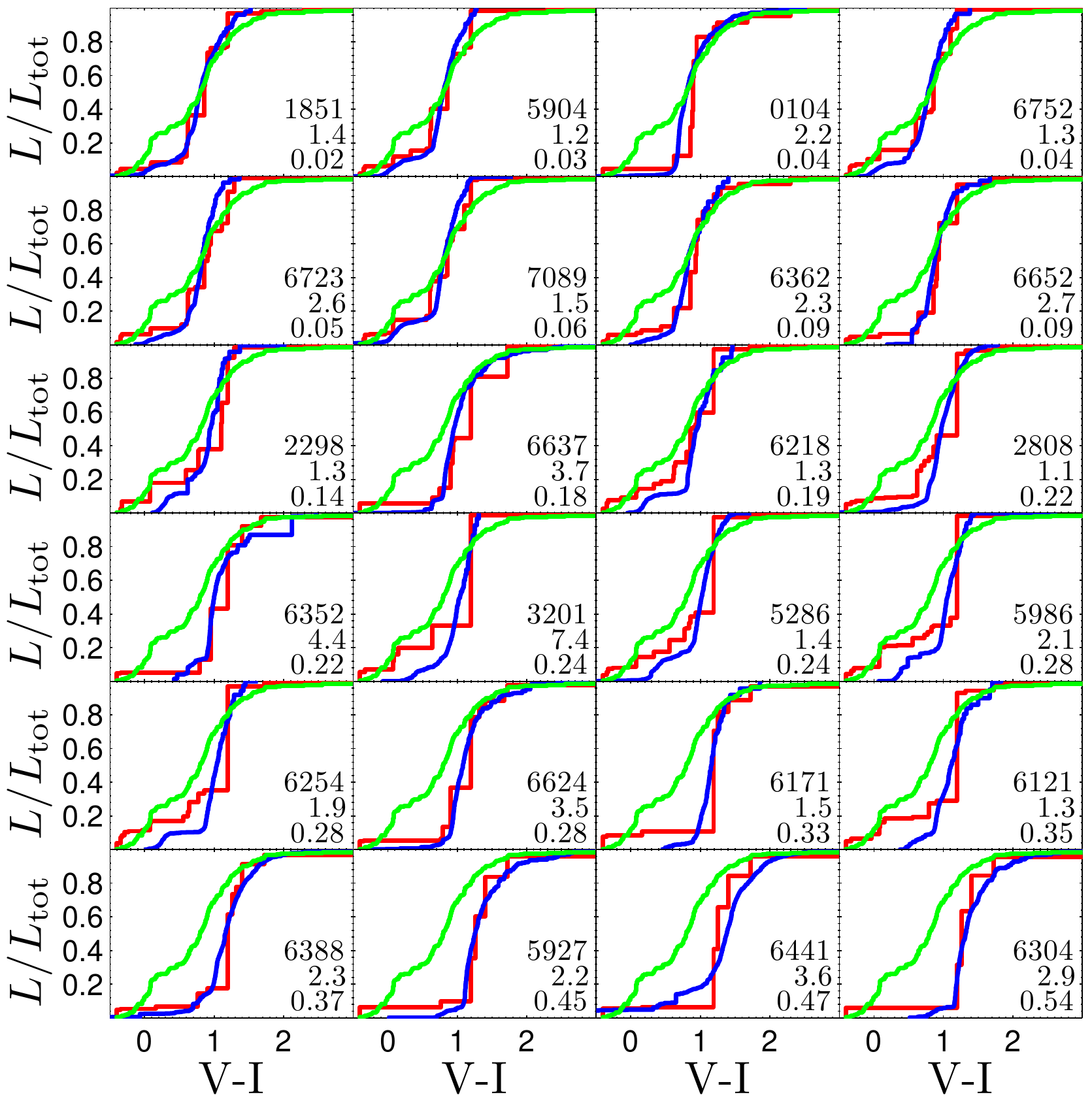}
  \end{center}
  \caption{CDFs of the $V-I$ distributions obtained from our spectral
   optimisations (red), compared with those derived from CMDs (blue),
   for all the GGCs overlapping between the HST ACS~\citep{Ata} and
   the S05 libraries.  The colour distribution corresponding to equal
   weight for each star (i.e., initial conditions) is shown in green.
   NGC ID, $\chi^2$ value, and $E(B-V)$ from \citet{Harris} are
   displayed at the bottom right of each panel (top to bottom,
   respectively).}
  \label{fig:CMD_all}
\end{figure}

The $V-I$ distributions from our spectral fits can now be compared
with those extracted from the CMDs.  This comparison uses the same
method described in \se{mock} for the mock spectra created from stars
lying outside of our final basis.  

A potential concern in this comparison is that our optimisation method may be insensitive
to faint, cool stars that contribute little to the total integrated light in the GGC spectra.
Such stars may strongly affect gravity-sensitive or molecular absorption features which
only make up a few out of the thousands of pixels in the spectra, so fitting to the full
spectrum may miss them. In an attempt to correct for this, we compute light fractions
using only stars in the CMD brighter than some magnitude limit\footnote{This exercise
fully ignores the relative number of stars of a given luminosity. Clearly, the integrated
spectrum is representative of the {\it total} brightness of stars of a given luminosity.}.
We tested $V$-band magnitude cut-off values ranging from $V=$ 14 to 22 for each
CMD-spectral optimisation comparison and found that the best match is found for
an empirically-derived limit of $V<17$.  \Fig{CMD_Vthresh} shows this comparison
for three values of $V-$band magnitude cutoff, where the CMD and optimisation
CDFs are shown in blue and red, respectively.  This threshold should not be adopted
universally; it is merely representative of the sensitivity of our particular spectroscopic
set-up. 

The CDFs for the ACS clusters versus their $V-I$ distribution are shown
in \Fig{CMD_all}, with the clusters arranged from the top-left to
bottom-right corners in order of increasing $E(B-V)$.  As gauged visually,
the match between the CMD and optimisation CDFs is superb for most clusters.

This test further supports the applicability of our population synthesis method.
However, some clusters, such as NGC 3201, 5986, and 6254, show large discrepancies
in the bluer regions. Since these clusters have reasonably good fits (see \ap{AppendixA}),
this discrepancy may be the result of unknown method degeneracies.


\section{Conclusions}\label{sec:conc}

We have tested the method of non-linear optimisation (population synthesis)
to decompose the integrated spectra of stellar systems into distributions
of fundamental stellar parameters.  To this end, we have used the spectral
MILES library \citep{MILES} to construct a suitable basis for the optimisation.
Our decomposition method was tested on mock spectra constructed from the
spectral library basis, yielding relative uncertainties of 20 per cent in the
light fractions and absolute total light noise levels of 5 per cent or less,
for a given star.

The stellar atmospheric parameters \Teff, \logg, and [Fe/H] of mock spectra
constructed from stars outside our stellar basis were also extracted
with great accuracy by our optimisation technique. The mean errors between
mock and fitted atmospheric parameters are 240 K, 0.04 dex, and 0.03 dex for
\Teff, \logg, and [Fe/H], respectively.

Having established the reliability of our spectral decomposition method,
we applied our code to the individual integrated spectra of 41 Galactic
globular clusters from the collection of \citet{S05}.  These spectra were
decomposed into relative fractions (by light) of a suitably chosen basis
of stellar spectra from the MILES library.  

The light-weighted GGC metallicities obtained from population synthesis agree well with
those of \citet{Harris} and our own literature compilation \citep{S05} when
an appropriate reddening model is included in the decomposition. In the
absence of such a model,  good agreement for highly reddened clusters
is only found when the continuum is removed (i.e. only prominent absorption
features enter the fit). 

The decompositions based on our optimised population synthesis were
compared with CMD data of the 24 S05 GGCs which overlap with those from
\citet{Ata}. Our CMD analysis yielded light-weighted luminosity fractions
for various stellar evolutionary stages as well as their $V-I$ colour
distributions. We found superb agreement between these quantities and
the luminosity fractions derived from our population synthesis optimisations.
The extracted spectroscopic luminosity fractions are reported in the abstract 
and in \Table{lumFracTableComp} are compared against similar values from CMDs.

Overall, we find the technique
of numerical optimisation to be a reliable tool for extracting the mean metallicity
and light fractions of stellar populations in unresolved stellar systems.  Some
caveats pertaining to the depth of the spectroscopic data and the line of sight
reddening must be taken into consideration.

\begin{acknowledgments}

We thank Ata Sarajedini for kindly providing CMD data files
from the ACS Globular Cluster Survey.  The referee is also
thanked for excellent comments that substantially improved
the manuscript.  SC and JCR acknowledge financial support
from the National Science and Engineering Council of Canada
through a Discovery Grant and a postgraduate scholarship, respectively.
R.P.S. acknowledges support from Gemini Observatory, which is operated by 
the Association of Universities for Research in Astronomy, Inc., on behalf of 
the international Gemini partnership of Argentina, Australia, Brazil, Canada, Chile, 
and the United States of America. The referee is also thanked for 
excellent comments that greatly improved the manuscript.

\end{acknowledgments}

\bibliographystyle{mn2e} 
\bibliography{paper} 

\begin{thebibliography}{}

\bibitem[\protect\citeauthoryear{{Bruzual} \& {Charlot}}{{Bruzual} \&
  {Charlot}}{1993}]{Brucha93}
{Bruzual} G.,  {Charlot} S.,  1993, \apj, 405, 538

\bibitem[\protect\citeauthoryear{{Buzzoni}}{{Buzzoni}}{1989}]{Buzzoni89}
{Buzzoni} A.,  1989, \apjs, 71, 817

\bibitem[\protect\citeauthoryear{{Cardelli}, {Clayton} \& {Mathis}}{{Cardelli}
  et~al.}{1989}]{Cardelli1989}
{Cardelli} J.~A.,  {Clayton} G.~C.,    {Mathis} J.~S.,  1989, \apj, 345, 245

\bibitem[\protect\citeauthoryear{{Carretta}, {Bragaglia}, {Gratton}, {D'Orazi}
  \& {Lucatello}}{{Carretta} et~al.}{2009}]{Carretta2009}
{Carretta} E.,  {Bragaglia} A.,  {Gratton} R.,  {D'Orazi} V.,    {Lucatello}
  S.,  2009, \aap, 508, 695

\bibitem[\protect\citeauthoryear{{Carretta} \& {Gratton}}{{Carretta} \&
  {Gratton}}{1997}]{CarrettaGratton1997}
{Carretta} E.,  {Gratton} R.~G.,  1997, \aaps, 121, 95

\bibitem[\protect\citeauthoryear{{Cenarro}, {Peletier},
  {S{\'a}nchez-Bl{\'a}zquez}, {Selam}, {Toloba}, {Cardiel},
  {Falc{\'o}n-Barroso}, {Gorgas}, {Jim{\'e}nez-Vicente} \&
  {Vazdekis}}{{Cenarro} et~al.}{2007}]{milesII}
{Cenarro} A.~J.,  {Peletier} R.~F.,  {S{\'a}nchez-Bl{\'a}zquez} P.,  {Selam}
  S.~O.,  {Toloba} E.,  {Cardiel} N.,  {Falc{\'o}n-Barroso} J.,  {Gorgas} J.,
  {Jim{\'e}nez-Vicente} J.,    {Vazdekis} A.,  2007, \mnras, 374, 664

\bibitem[\protect\citeauthoryear{{Cezario}, {Coelho}, {Alves-Brito}, {Forbes}
  \& {Brodie}}{{Cezario} et~al.}{2013}]{Cezario13}
{Cezario} E.,  {Coelho} P.~R.~T.,  {Alves-Brito} A.,  {Forbes} D.~A.,
  {Brodie} J.~P.,  2013, \aap, 549, A60

\bibitem[\protect\citeauthoryear{{Cid Fernandes}, {Mateus}, {Sodr{\'e}},
  {Stasi{\'n}ska} \& {Gomes}}{{Cid Fernandes} et~al.}{2005}]{CidFernandes05}
{Cid Fernandes} R.,  {Mateus} A.,  {Sodr{\'e}} L.,  {Stasi{\'n}ska} G.,
  {Gomes} J.~M.,  2005, \mnras, 358, 363

\bibitem[\protect\citeauthoryear{{Conroy}, {Gunn} \& {White}}{{Conroy}
  et~al.}{2009}]{ConroyGunn09}
{Conroy} C.,  {Gunn} J.~E.,    {White} M.,  2009, \apj, 699, 486

\bibitem[\protect\citeauthoryear{{Faber}}{{Faber}}{1972}]{faber}
{Faber} S.~M.,  1972, \aap, 20, 361

\bibitem[\protect\citeauthoryear{{Falc{\'o}n-Barroso},
  {S{\'a}nchez-Bl{\'a}zquez}, {Vazdekis}, {Ricciardelli}, {Cardiel}, {Cenarro},
  {Gorgas} \& {Peletier}}{{Falc{\'o}n-Barroso} et~al.}{2011}]{falcon}
{Falc{\'o}n-Barroso} J.,  {S{\'a}nchez-Bl{\'a}zquez} P.,  {Vazdekis} A.,
  {Ricciardelli} E.,  {Cardiel} N.,  {Cenarro} A.~J.,  {Gorgas} J.,
  {Peletier} R.~F.,  2011, \aap, 532, A95+

\bibitem[\protect\citeauthoryear{{Harris}}{{Harris}}{1996}]{Harris}
{Harris} W.~E.,  1996, \aj, 112, 1487

\bibitem[\protect\citeauthoryear{{Koleva}, {Prugniel}, {Bouchard} \&
  {Wu}}{{Koleva} et~al.}{2009}]{koleva2009}
{Koleva} M.,  {Prugniel} P.,  {Bouchard} A.,    {Wu} Y.,  2009, \aap, 501, 1269

\bibitem[\protect\citeauthoryear{{Koleva}, {Prugniel}, {Ocvirk}, {Le Borgne} \&
  {Soubiran}}{{Koleva} et~al.}{2008}]{Koleva2008}
{Koleva} M.,  {Prugniel} P.,  {Ocvirk} P.,  {Le Borgne} D.,    {Soubiran} C.,
  2008, \mnras, 385, 1998

\bibitem[\protect\citeauthoryear{{Kraft} \& {Ivans}}{{Kraft} \&
  {Ivans}}{2003}]{KraftIvans2003}
{Kraft} R.~P.,  {Ivans} I.~I.,  2003, \pasp, 115, 143

\bibitem[\protect\citeauthoryear{{MacArthur}}{{MacArthur}}{2005}]{MacArthur2005}
{MacArthur} L.~A.,  2005, \apj, 623, 795

\bibitem[\protect\citeauthoryear{{MacArthur}, {Gonz{\'a}lez} \&
  {Courteau}}{{MacArthur} et~al.}{2009}]{Macarthur09}
{MacArthur} L.~A.,  {Gonz{\'a}lez} J.~J.,    {Courteau} S.,  2009, \mnras, 395,
  28

\bibitem[\protect\citeauthoryear{{Maraston}}{{Maraston}}{1998}]{Maraston98}
{Maraston} C.,  1998, \mnras, 300, 872

\bibitem[\protect\citeauthoryear{{Maraston}}{{Maraston}}{2007}]{Maraston07}
{Maraston} C.,  2007, in {Combes} F.,  {Palou{\v s}} J.,  eds, IAU Symposium
  Vol.~235 of IAU Symposium, {Stellar Population Models}.
pp 52--56

\bibitem[\protect\citeauthoryear{{Maraston} \& {Thomas}}{{Maraston} \&
  {Thomas}}{2000}]{Maraston&Thomas00}
{Maraston} C.,  {Thomas} D.,  2000, \apj, 541, 126

\bibitem[\protect\citeauthoryear{{Moultaka}}{{Moultaka}}{2005}]{Moultaka05}
{Moultaka} J.,  2005, \aap, 430, 95

\bibitem[\protect\citeauthoryear{{O'Connell}}{{O'Connell}}{1976}]{oconnell}
{O'Connell} R.~W.,  1976, \apj, 206, 370

\bibitem[\protect\citeauthoryear{{Ocvirk}, {Pichon}, {Lan{\c c}on} \&
  {Thi{\'e}baut}}{{Ocvirk} et~al.}{2006}]{Ocvirk06}
{Ocvirk} P.,  {Pichon} C.,  {Lan{\c c}on} A.,    {Thi{\'e}baut} E.,  2006,
  \mnras, 365, 46

\bibitem[\protect\citeauthoryear{{Pickles}}{{Pickles}}{1985}]{pickles}
{Pickles} A.~J.,  1985, \apj, 296, 340

\bibitem[\protect\citeauthoryear{{Prugniel}, {Vauglin}, {I.} \&
  {Koleva}}{{Prugniel} et~al.}{2011}]{Prugniel}
{Prugniel} P.,  {Vauglin} {I.}   {Koleva} M.,  2011, VizieR Online Data
  Catalog, 353, 19165

\bibitem[\protect\citeauthoryear{{Renzini}}{{Renzini}}{1981}]{Renzini81}
{Renzini} A.,  1981, Annales de Physique, 6, 87

\bibitem[\protect\citeauthoryear{{Roediger}, {Courteau}, {Graves} \&
  {Schiavon}}{{Roediger} et~al.}{2014}]{Roediger2014}
{Roediger} J.~C.,  {Courteau} S.,  {Graves} G.,    {Schiavon} R.~P.,  2014,
  \apjs, 210, 10

\bibitem[\protect\citeauthoryear{{S{\'a}nchez-Bl{\'a}zquez}, {Peletier},
  {Jim{\'e}nez-Vicente}, {Cardiel}, {Cenarro}, {Falc{\'o}n-Barroso}, {Gorgas},
  {Selam} \& {Vazdekis}}{{S{\'a}nchez-Bl{\'a}zquez} et~al.}{2006}]{MILES}
{S{\'a}nchez-Bl{\'a}zquez} P.,  {Peletier} R.~F.,  {Jim{\'e}nez-Vicente} J.,
  {Cardiel} N.,  {Cenarro} A.~J.,  {Falc{\'o}n-Barroso} J.,  {Gorgas} J.,
  {Selam} S.,    {Vazdekis} A.,  2006, \mnras, 371, 703

\bibitem[\protect\citeauthoryear{{Sarajedini}}{{Sarajedini}}{1992}]{sara92}
{Sarajedini} A.,  1992, PhD thesis, Yale University., New Haven, CT.

\bibitem[\protect\citeauthoryear{{Sarajedini}, {Bedin}, {Chaboyer}, {Dotter},
  {Siegel}, {Anderson}, {Aparicio}, {King}, {Majewski}, {Mar{\'{\i}}n-Franch},
  {Piotto}, {Reid} \& {Rosenberg}}{{Sarajedini} et~al.}{2007}]{Ata}
{Sarajedini} A.,  {Bedin} L.~R.,  {Chaboyer} B.,  {Dotter} A.,  {Siegel} M.,
  {Anderson} J.,  {Aparicio} A.,  {King} I.,  {Majewski} S.,
  {Mar{\'{\i}}n-Franch} A.,  {Piotto} G.,  {Reid} I.~N.,    {Rosenberg} A.,
  2007, \aj, 133, 1658

\bibitem[\protect\citeauthoryear{{Schiavon}}{{Schiavon}}{2007}]{Schiavon2007}
{Schiavon} R.~P.,  2007, \apjs, 171, 146

\bibitem[\protect\citeauthoryear{{Schiavon}, {Rose}, {Courteau} \&
  {MacArthur}}{{Schiavon} et~al.}{2005}]{S05}
{Schiavon} R.~P.,  {Rose} J.~A.,  {Courteau} S.,    {MacArthur} L.~A.,  2005,
  \apjs, 160, 163

\bibitem[\protect\citeauthoryear{{Soubiran}, {Katz} \& {Cayrel}}{{Soubiran}
  et~al.}{1998}]{SKC}
{Soubiran} C.,  {Katz} D.,    {Cayrel} R.,  1998, \aaps, 133, 221

\bibitem[\protect\citeauthoryear{{Spinrad} \& {Taylor}}{{Spinrad} \&
  {Taylor}}{1971}]{spinrad}
{Spinrad} H.,  {Taylor} B.~J.,  1971, \apjs, 22, 445

\bibitem[\protect\citeauthoryear{{Tojeiro}, {Heavens}, {Jimenez} \&
  {Panter}}{{Tojeiro} et~al.}{2007}]{tojeiro07}
{Tojeiro} R.,  {Heavens} A.~F.,  {Jimenez} R.,    {Panter} B.,  2007, \mnras,
  381, 1252

\bibitem[\protect\citeauthoryear{{Vergely}, {Lan{\c c}on} \&
  {Mouhcine}}{{Vergely} et~al.}{2002}]{Vergely2002}
{Vergely} J.-L.,  {Lan{\c c}on} A.,    {Mouhcine} 2002, \aap, 394, 807

\bibitem[\protect\citeauthoryear{{Walcher}, {B{\"o}ker}, {Charlot}, {Ho},
  {Rix}, {Rossa}, {Shields} \& {van der Marel}}{{Walcher}
  et~al.}{2006}]{Walcher06}
{Walcher} C.~J.,  {B{\"o}ker} T.,  {Charlot} S.,  {Ho} L.~C.,  {Rix} H.-W.,
  {Rossa} J.,  {Shields} J.~C.,    {van der Marel} R.~P.,  2006, \apj, 649, 692

\bibitem[\protect\citeauthoryear{{Worthey}, {Faber}, {Gonzalez} \&
  {Burstein}}{{Worthey} et~al.}{1994}]{Worthey}
{Worthey} G.,  {Faber} S.~M.,  {Gonzalez} J.~J.,    {Burstein} D.,  1994,
  \apjs, 94, 687

\bibitem[\protect\citeauthoryear{{Worthey} \& {Lee}}{{Worthey} \&
  {Lee}}{2011}]{WortheyLee}
{Worthey} G.,  {Lee} H.-c.,  2011, \apjs, 193, 1

\end{thebibliography}


\clearpage

\appendix

\section{APPENDIX A}\label{sec:AppendixA}

In this Appendix, we show the fits to our 41 GGC spectra from our
population synthesis code. The S05 spectra and their optimisation fits
are shown in blue and red, respectively.  Goodness of fit is indicated
by $\chi^2$/dof values and data-model residuals are shown in the inset
below each spectrum. To estimate $\chi^2$/dof, the errors in each GGC
spectrum were computed pixel-to-pixel by dividing the spectrum by
the S/N ratio spectra provided by S05.  Metallicity and reddening
values from the 2010 version of \citet{Harris} and from our fits are also indicated
on each panel.


\begin{figure}[H]
  \begin{center}
    \includegraphics[width=\textwidth]{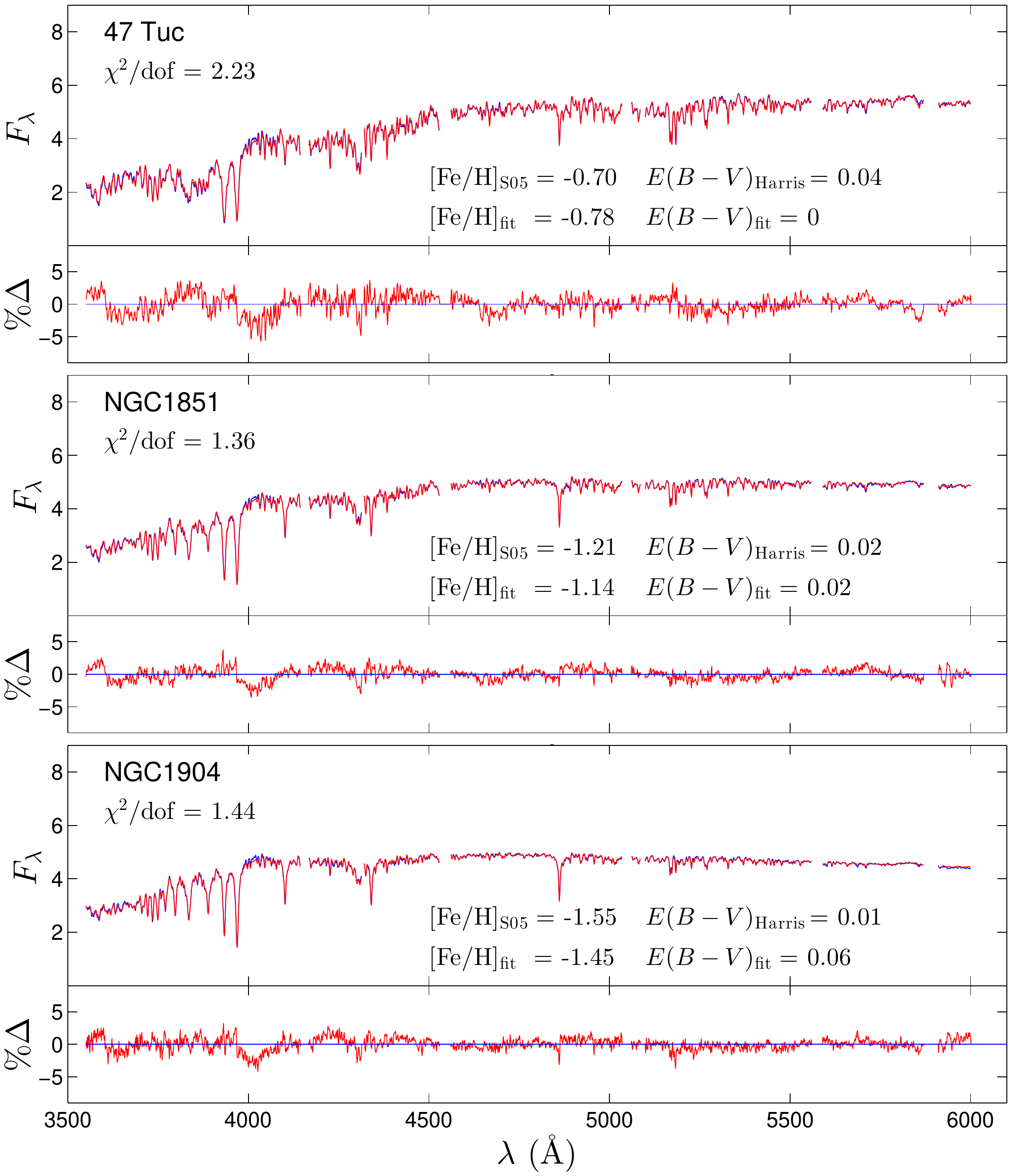}
  \end{center}
  \captcont{Spectral optimisation fits to GGC spectra.}
  \label{fig:allFits}
\end{figure}

\clearpage
\begin{figure}
  \begin{center}
    \includegraphics[width=\textwidth]{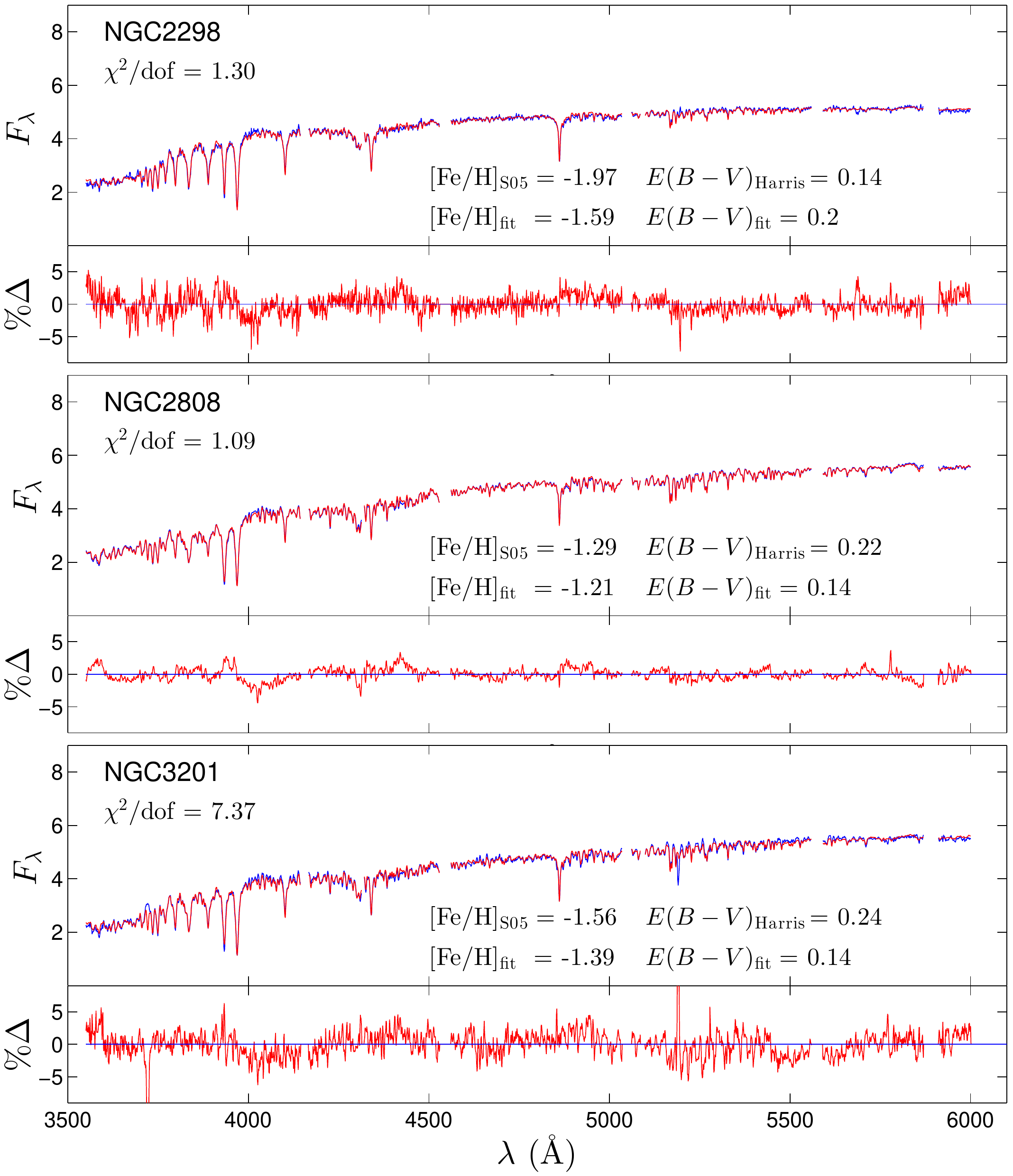}
  \end{center}
  \captcont{ .../continued }
\end{figure}

\clearpage
\begin{figure}
  \begin{center}
    \includegraphics[width=\textwidth]{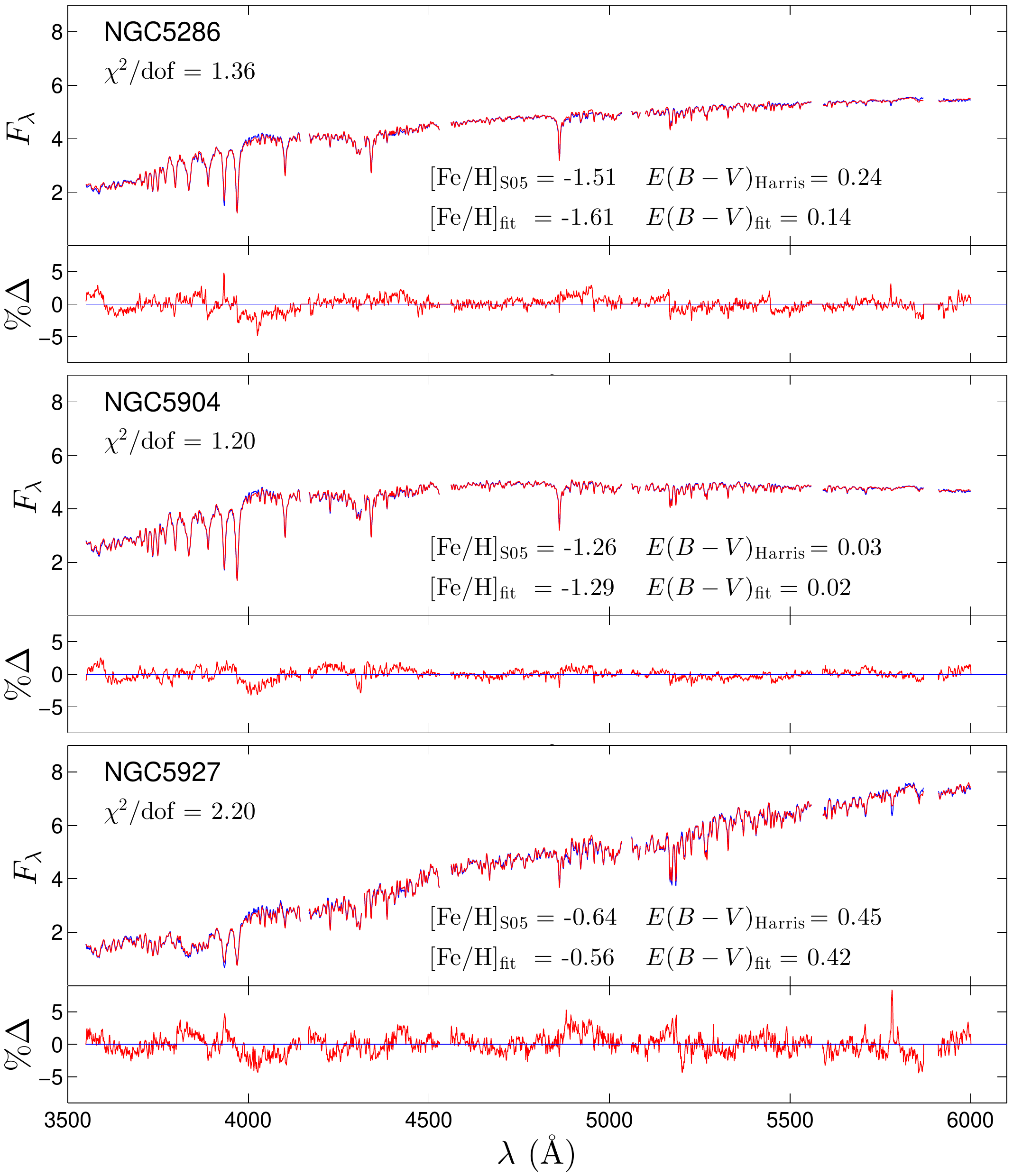}
  \end{center}
  \captcont{ .../continued }
\end{figure}

\clearpage
\begin{figure}
  \begin{center}
    \includegraphics[width=\textwidth]{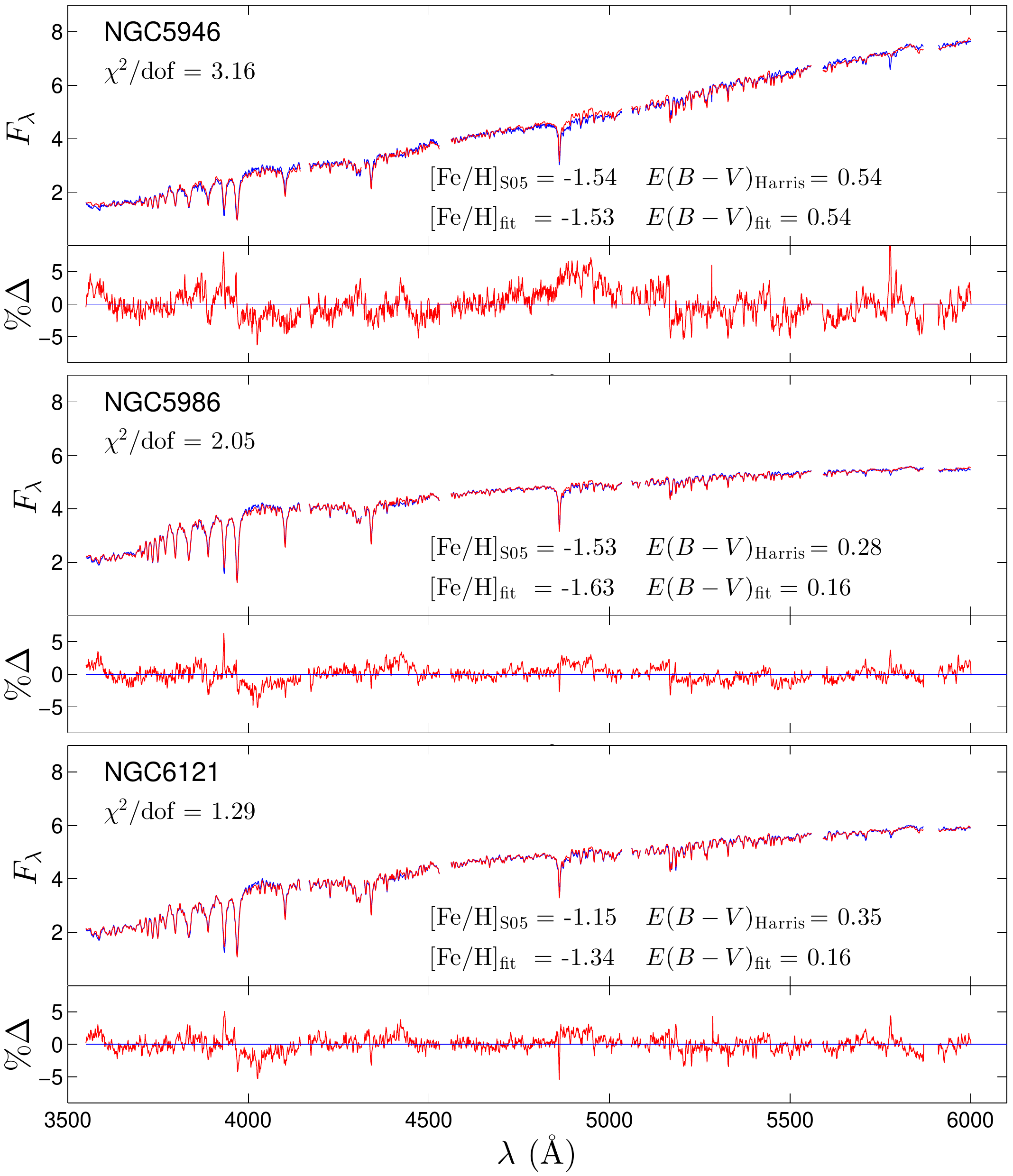}
  \end{center}
  \captcont{ .../continued }
\end{figure}

\clearpage
\begin{figure}
  \begin{center}
    \includegraphics[width=\textwidth]{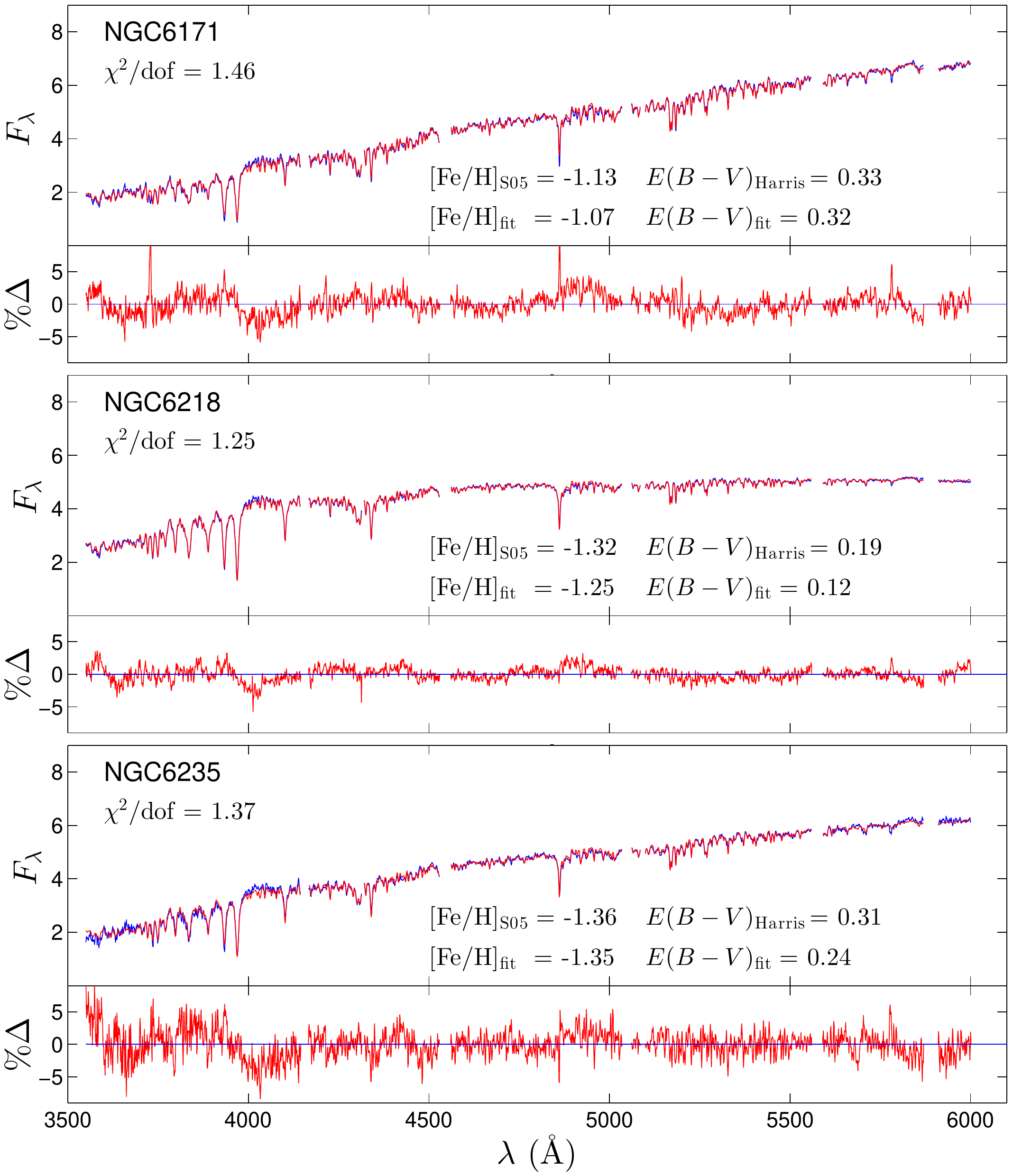}
  \end{center}
  \captcont{ .../continued }
\end{figure}

\clearpage
\begin{figure}
  \begin{center}
    \includegraphics[width=\textwidth]{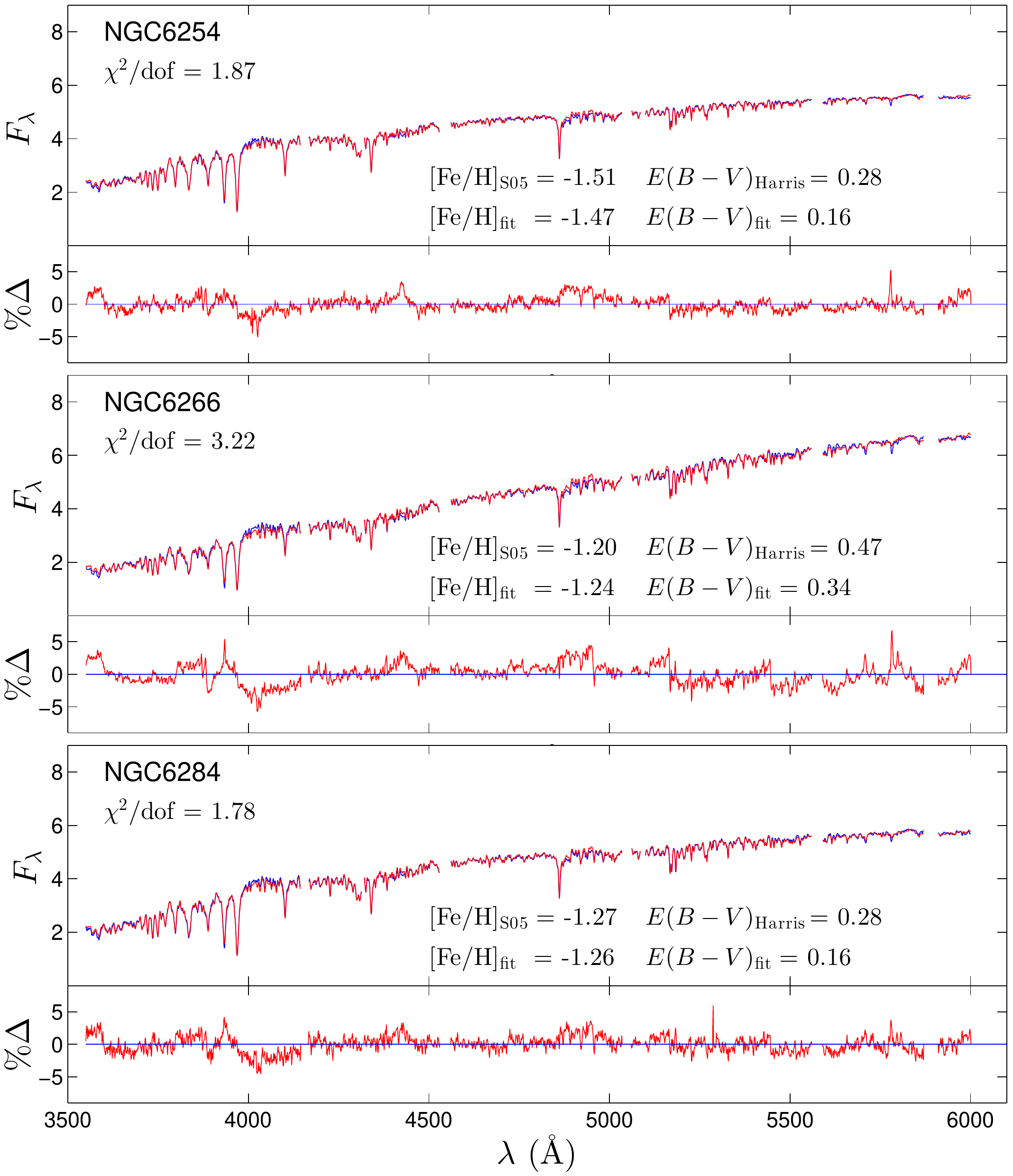}
  \end{center}
  \captcont{ .../continued }
\end{figure}

\clearpage
\begin{figure}
  \begin{center}
    \includegraphics[width=\textwidth]{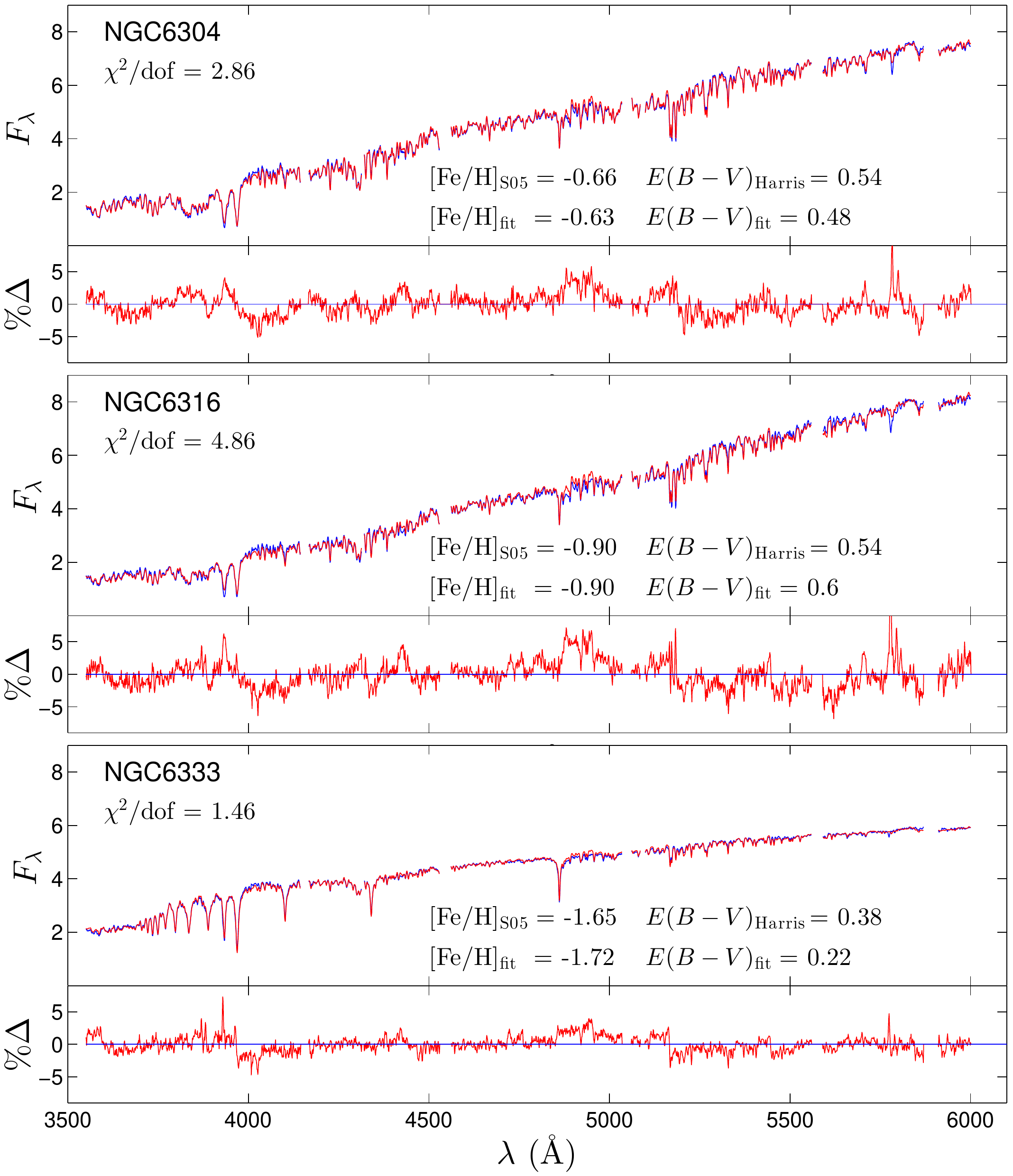}
  \end{center}
  \captcont{ .../continued }
\end{figure}

\clearpage
\begin{figure}
  \begin{center}
    \includegraphics[width=\textwidth]{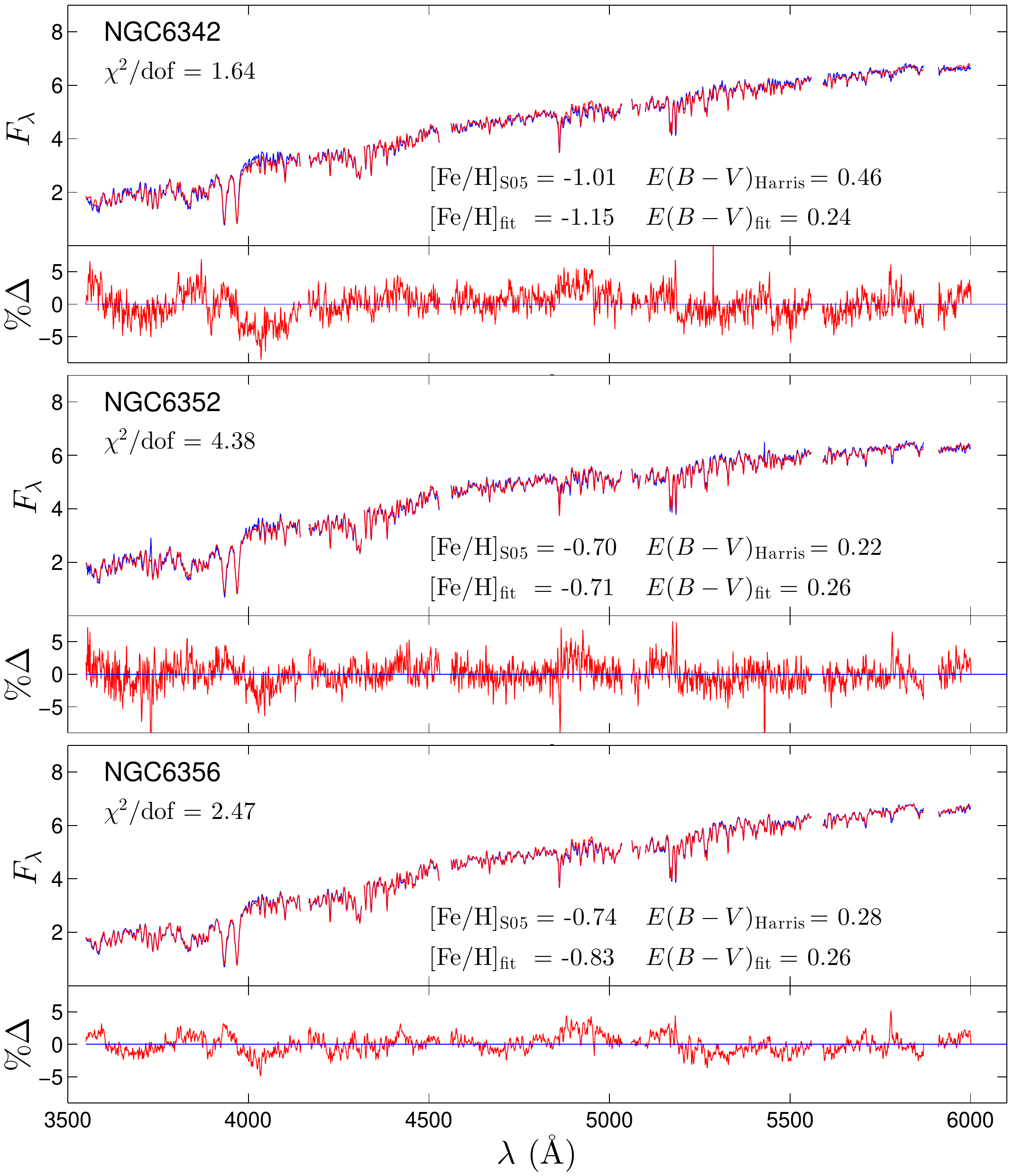}
  \end{center}
  \captcont{ .../continued }
\end{figure}

\clearpage
\begin{figure}
  \begin{center}
    \includegraphics[width=\textwidth]{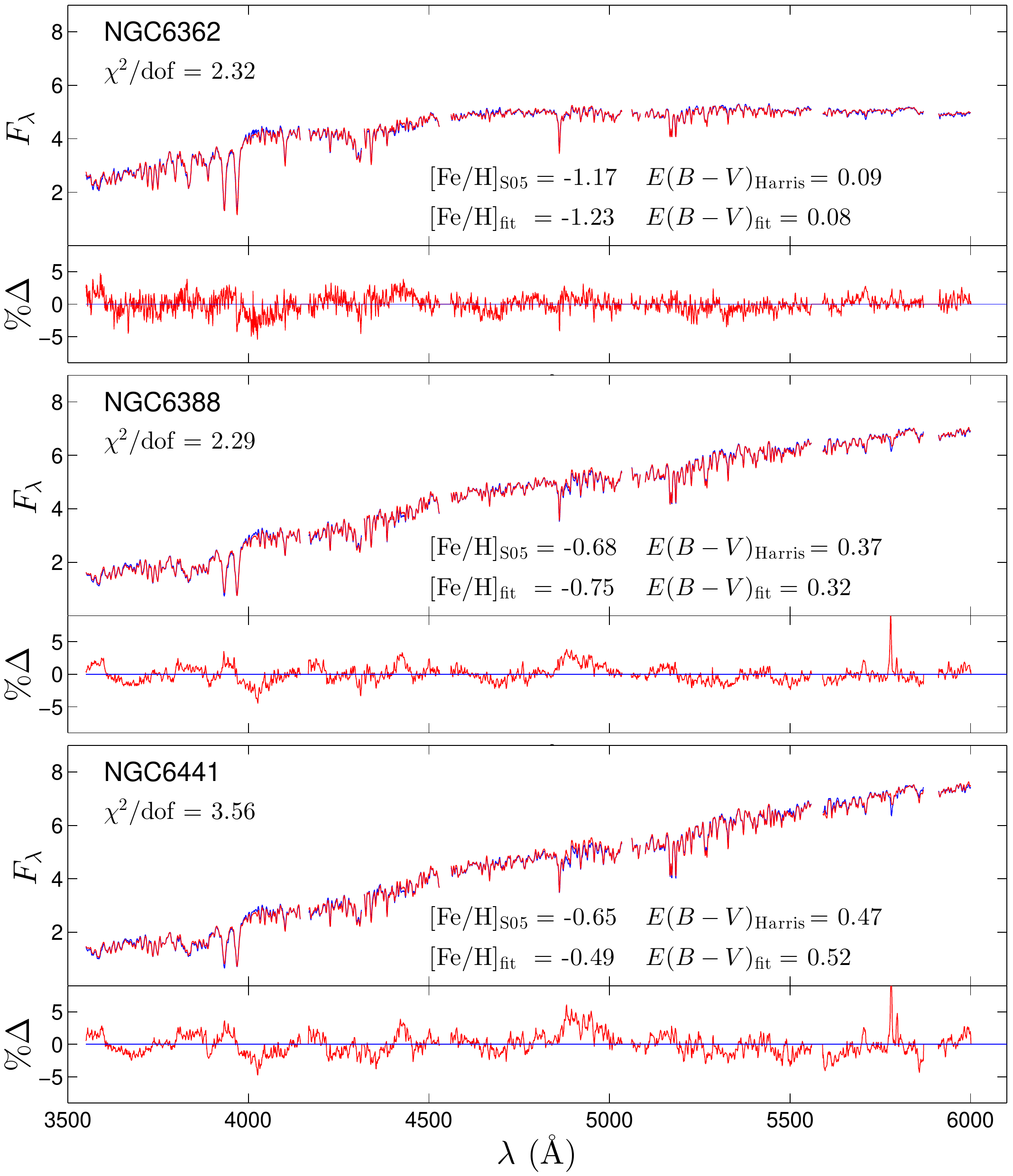}
  \end{center}
  \captcont{ .../continued }
\end{figure}

\clearpage
\begin{figure}
  \begin{center}
    \includegraphics[width=\textwidth]{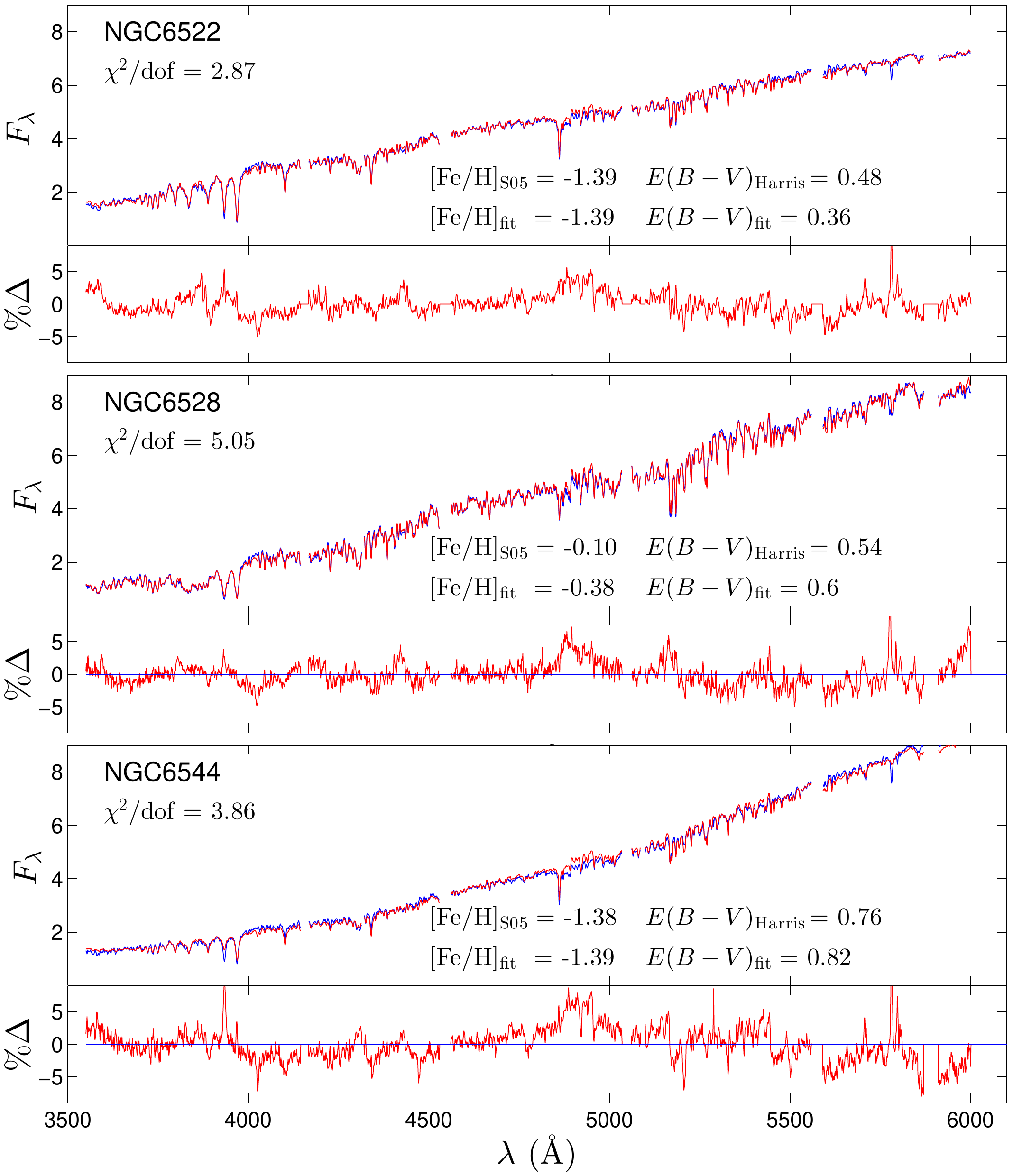}
  \end{center}
  \captcont{ .../continued }
\end{figure}

\clearpage
\begin{figure}
  \begin{center}
    \includegraphics[width=\textwidth]{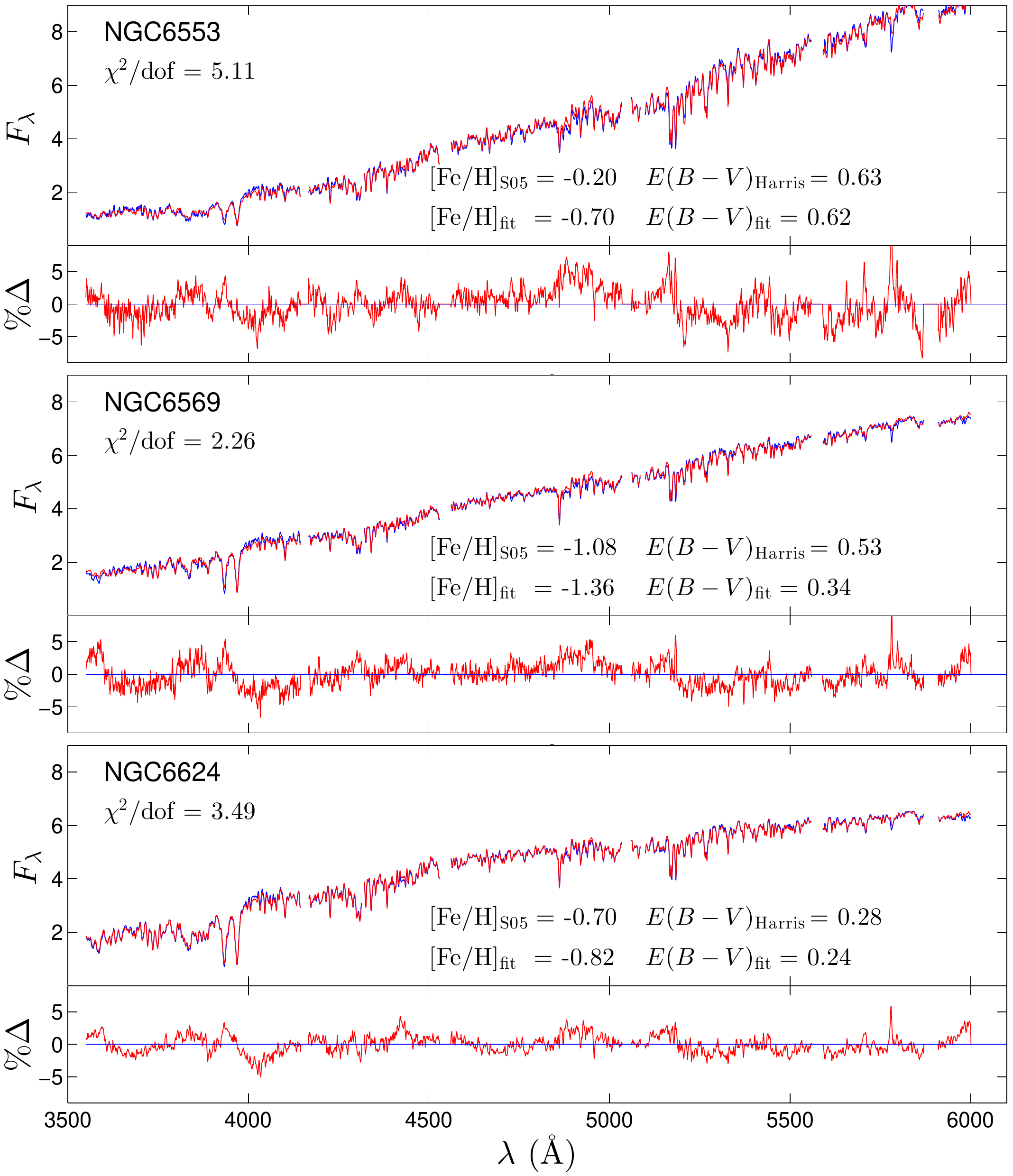}
  \end{center}
  \captcont{ .../continued }
\end{figure}

\clearpage
\begin{figure}
  \begin{center}
    \includegraphics[width=\textwidth]{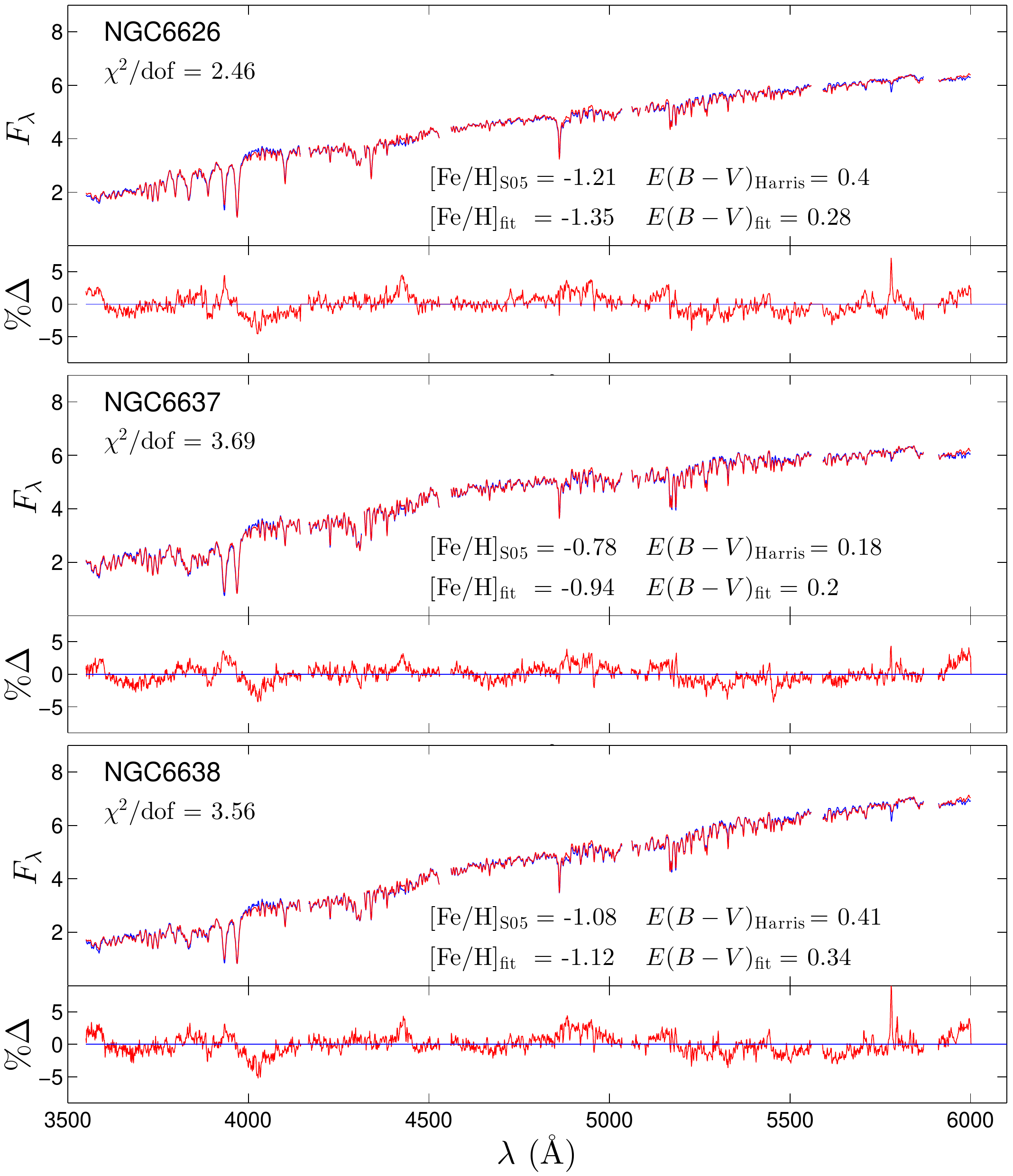}
  \end{center}
  \captcont{ .../continued }
\end{figure}

\clearpage
\begin{figure}
  \begin{center}
    \includegraphics[width=\textwidth]{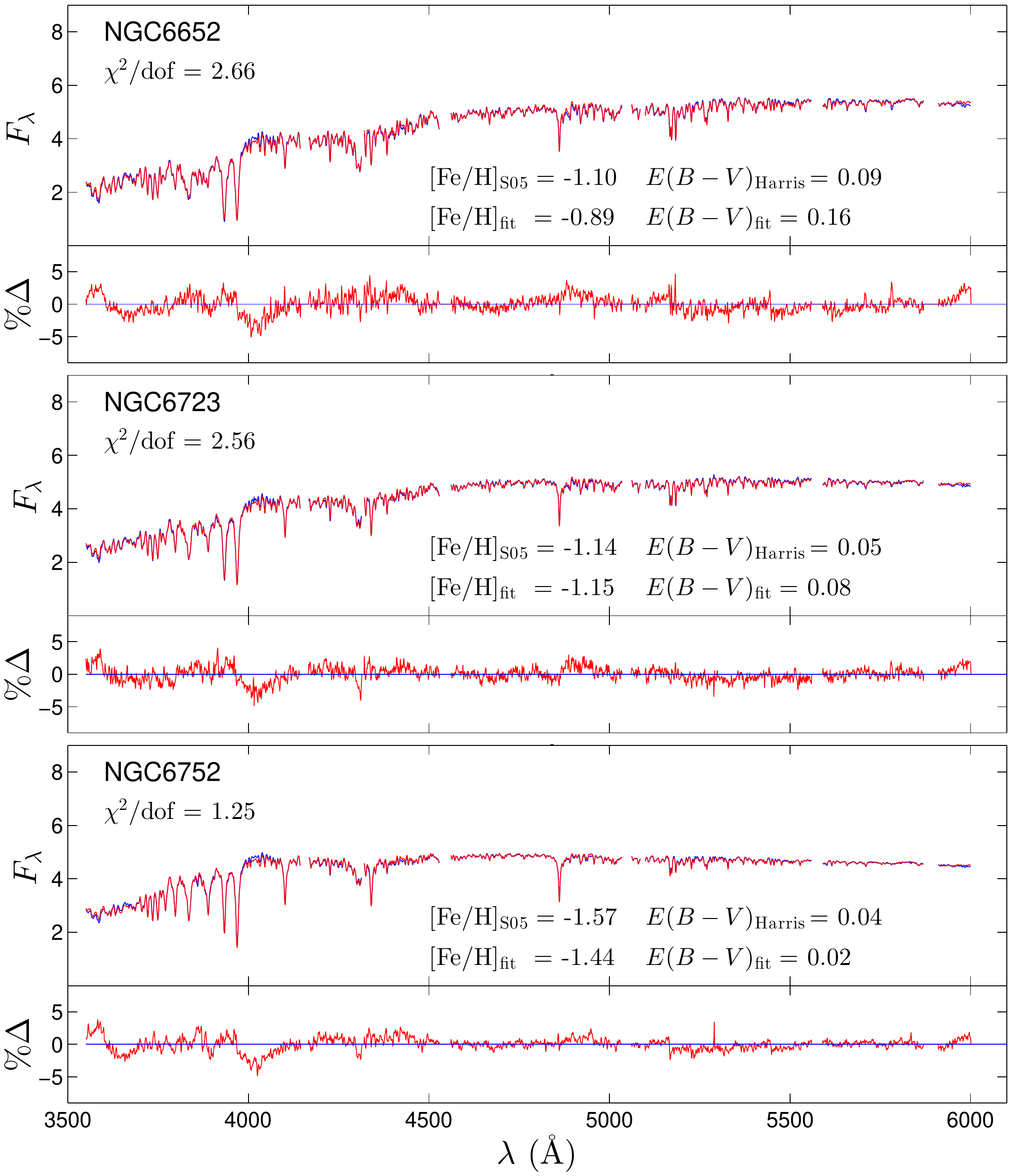}
  \end{center}
  \captcont{ .../continued }
\end{figure}

\clearpage
\begin{figure}
  \begin{center}
    \includegraphics[width=\textwidth]{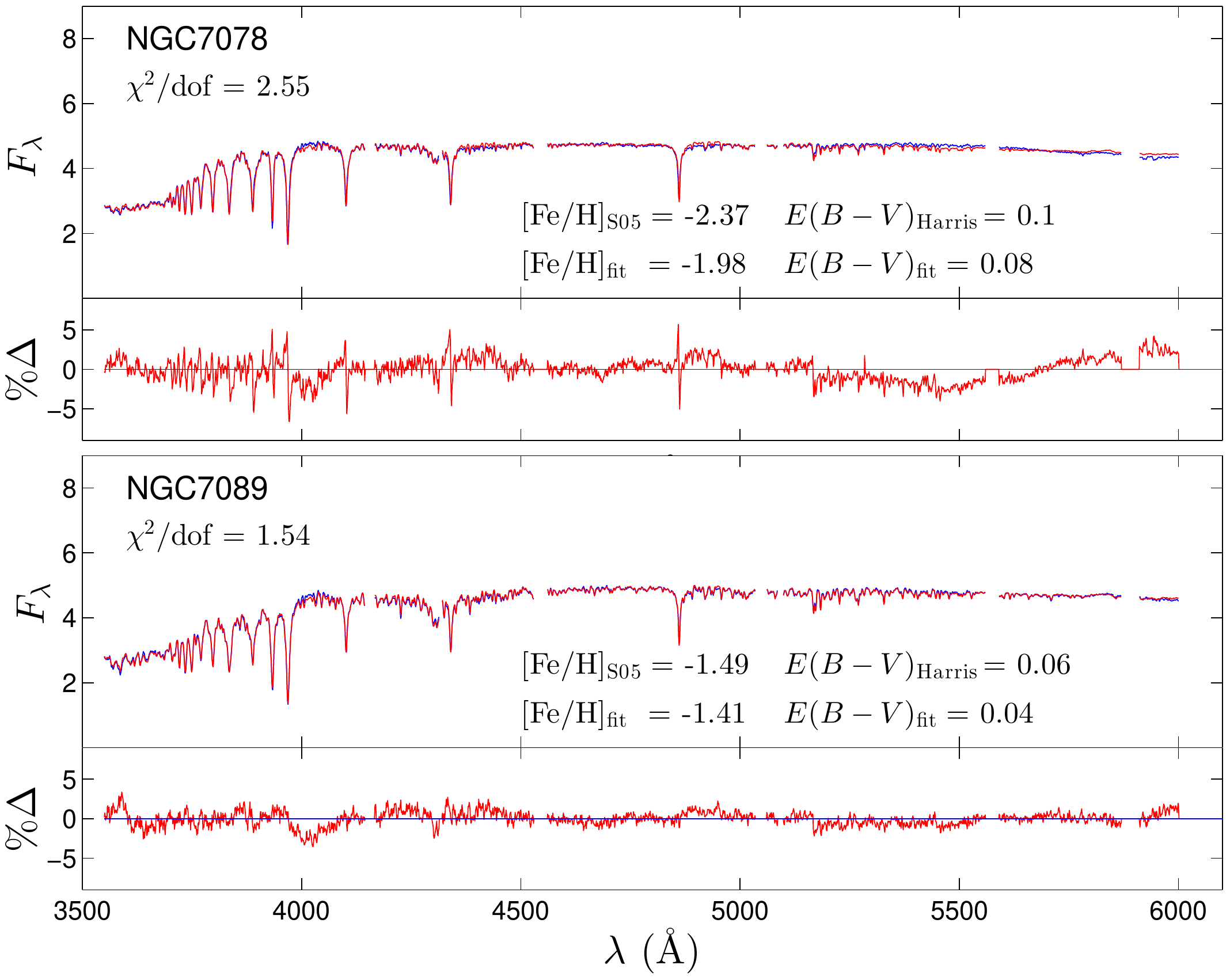}
  \end{center}
  \caption{ .../continued }
\end{figure}

\clearpage

\section{APPENDIX B}\label{sec:AppendixB}

\begin{figure}[H]
  \begin{center}
    \includegraphics[width=0.8\textwidth]{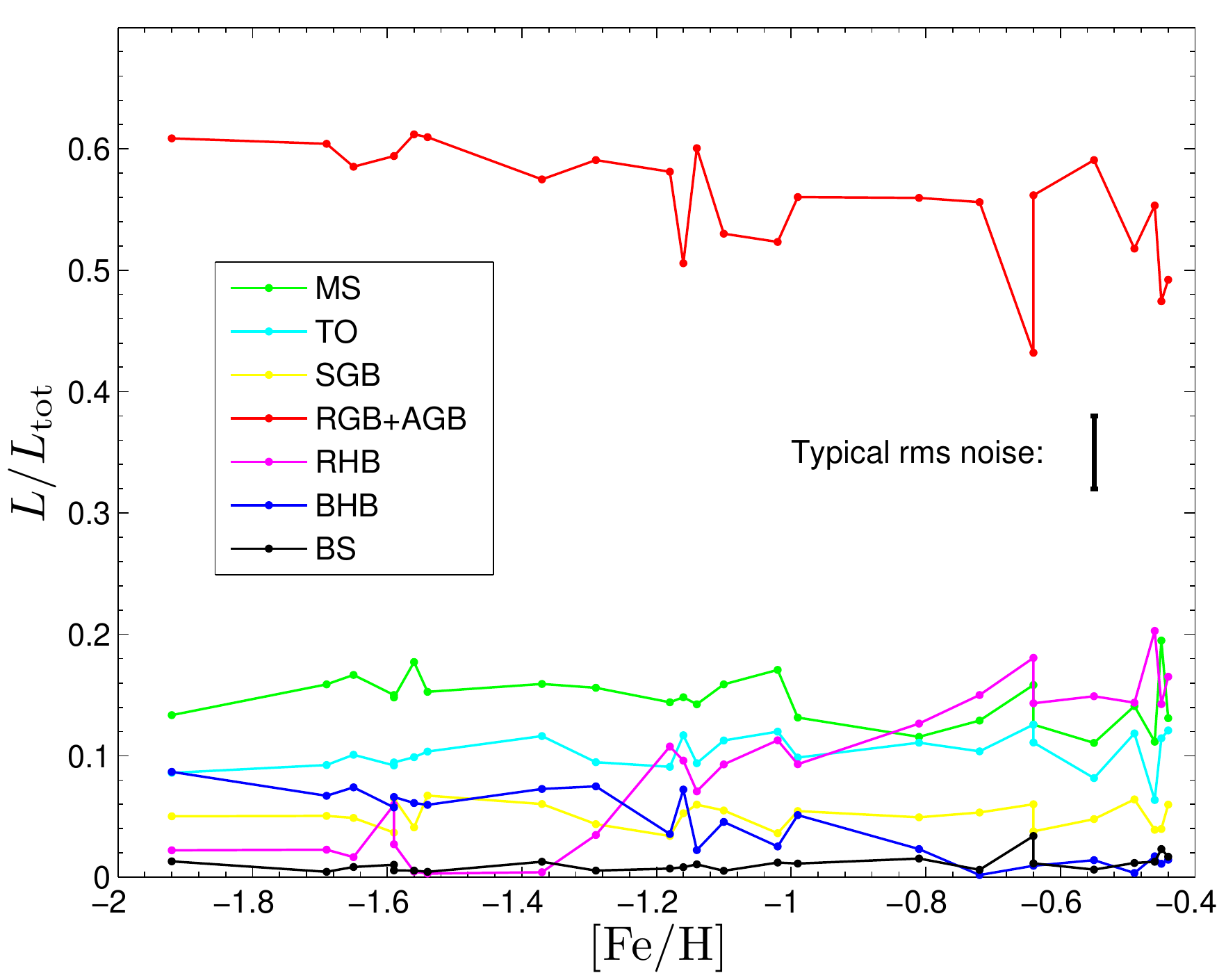}
  \end{center}
  \caption{Fractional contributions of different stellar evolutionary phases
   to the total light of the 24 GGCs from \citet{Ata} overlapping with S05,
   as a function of cluster metallicity. Evolutionary phases include
   turn-off (TO), subgiant branch (SGB), red giant and asymptotic giant branch
   (RGB+AGB), red horizontal branch (RHB), blue horizontal branch (BHB), and
   blue stragglers (BS). The rms noise in the luminosity fractions of turn-off
   stars is shown on the figure as a representative uncertainty.}
  \label{fig:lightConts}
\end{figure}

The $V$-band luminosity fractions of the many stellar evolutionary phases
for the 24 ACS GGC clusters are displayed graphically in \Fig{lightConts}.
The light fractions of main sequence (MS), turn-off (TO), subgiant branch (SGB),
 and blue straggler (BS) stars remain fairly constant with [Fe/H].  Conversely,
the red and asymptotic giant branch (RGB and AGB respectively) light
fractions seem to decrease with increasing GGC metallicity.
The fairly smooth and real increase and decrease of the red and blue
horizontal branch (RHB and BHB, respectively)
light fractions with metallicity, especially from [Fe/H] = -1.4 to -1,
is a well-known dependence of the horizontal branch morphology.  Note
that our data would only be weakly sensitive to any second-order
dependence of HB morphology \citep{sara92}.

\clearpage

\section{APPENDIX C}\label{sec:AppendixC}

In this Appendix, we present colour-magnitude diagrams for the 24 GGCs
of \citet{Ata} which coincide with those studied by S05. Stages
of stellar evolution are indicated by distinct boxes while the per cent
contribution of each stage to the total luminosity of the cluster is
provided in the legend. The uncertainty in the displayed luminosity
fractions is estimated at 6 per cent, based on the rms noise in the luminosity
fractions of the red giant stars.

Also included are CMDs of the same clusters, constructed through our
optimised decompositions of their integrated spectra. Point sizes
indicate the relative light-weighted contribution to the integrated
spectra.

\begin{figure}[H]
  \centering
    \subfigure{
\includegraphics[width=0.7\textwidth]{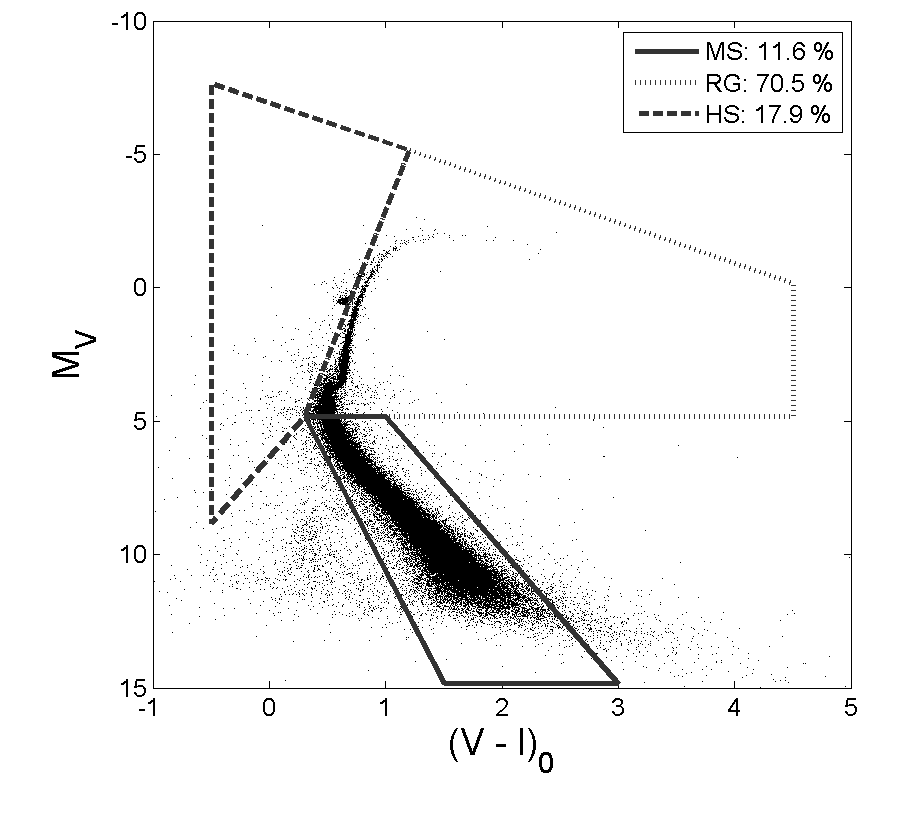}}
    \subfigure{
\includegraphics[width=0.655\textwidth]{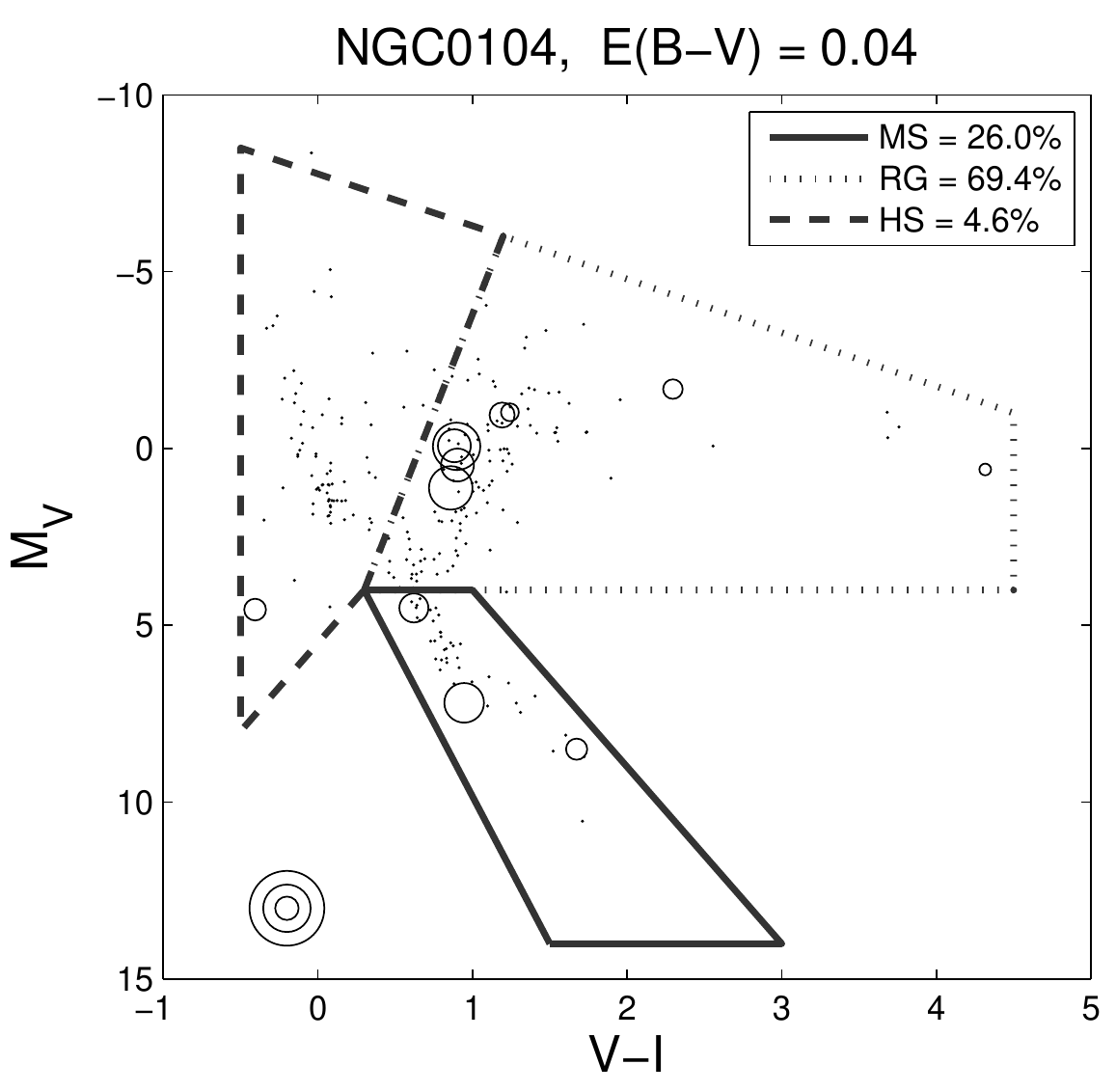}}
  \captcont{Colour-magnitude diagrams of GGCs; results are displayed as in \Fig{fit_CMD}. The top and bottom panels correspond to CMDs from \citet{Ata} and MILES, respectively.  }
  \label{fig:CMDsReal}
\end{figure}

\clearpage
\begin{figure}
  \centering
    \subfigure{
\includegraphics[width=0.75\textwidth]{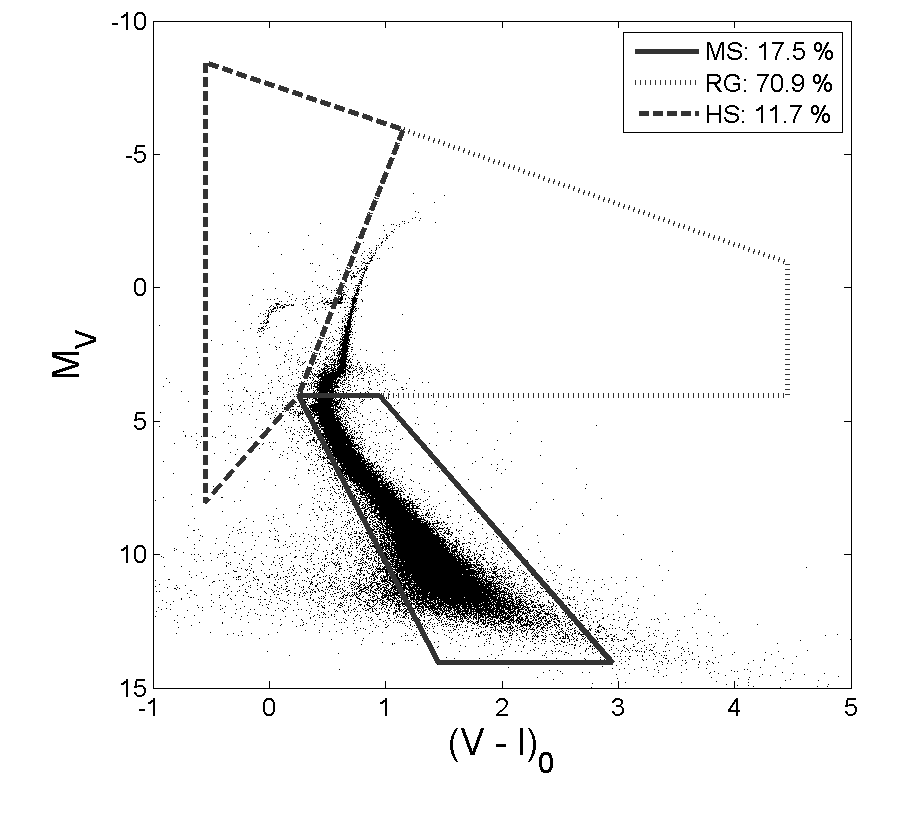}}
    \subfigure{
\includegraphics[width=0.7\textwidth]{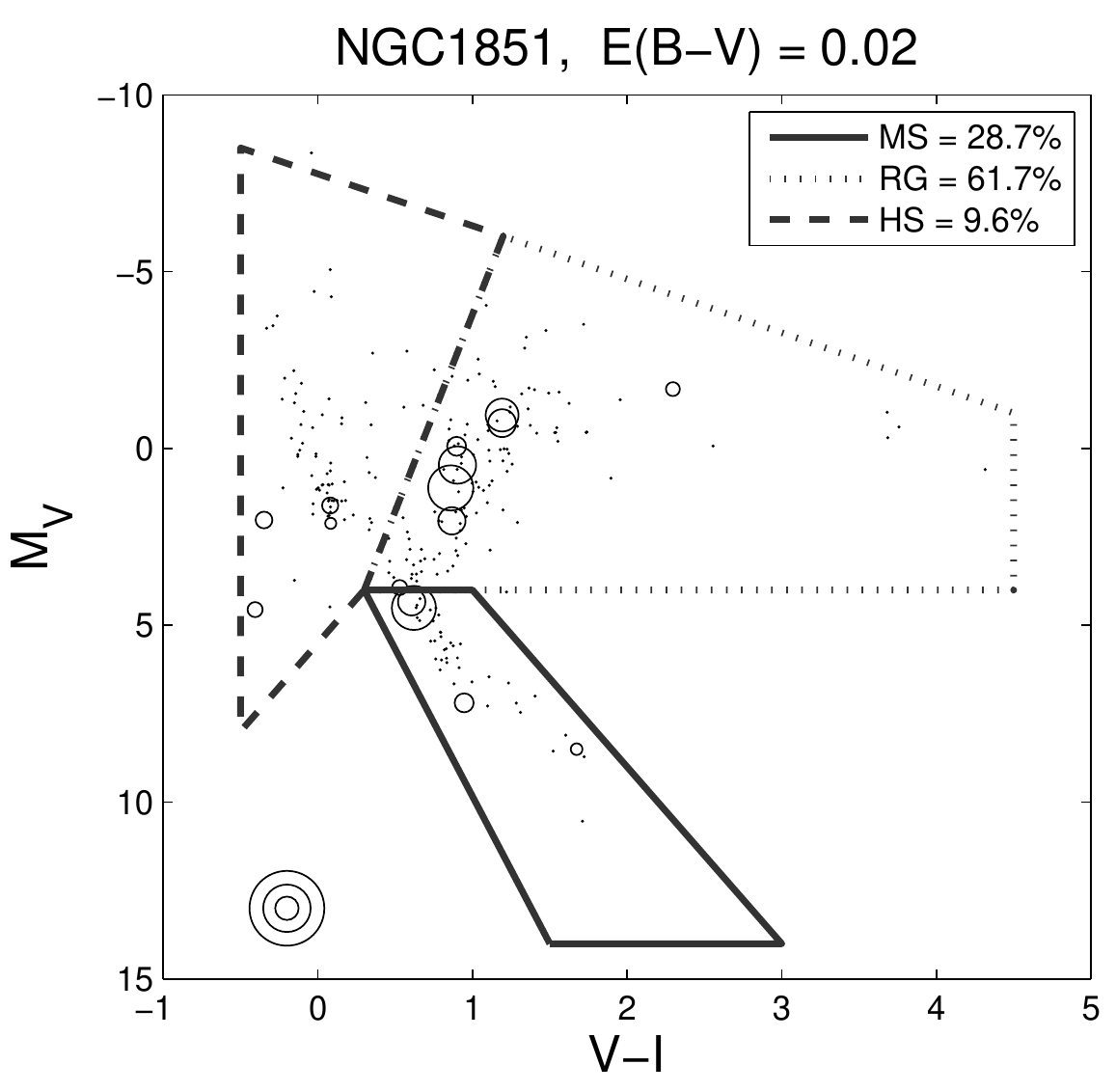}}
  \captcont{.../continued}
\end{figure}

\clearpage
\begin{figure}
  \centering
    \subfigure{
\includegraphics[width=0.75\textwidth]{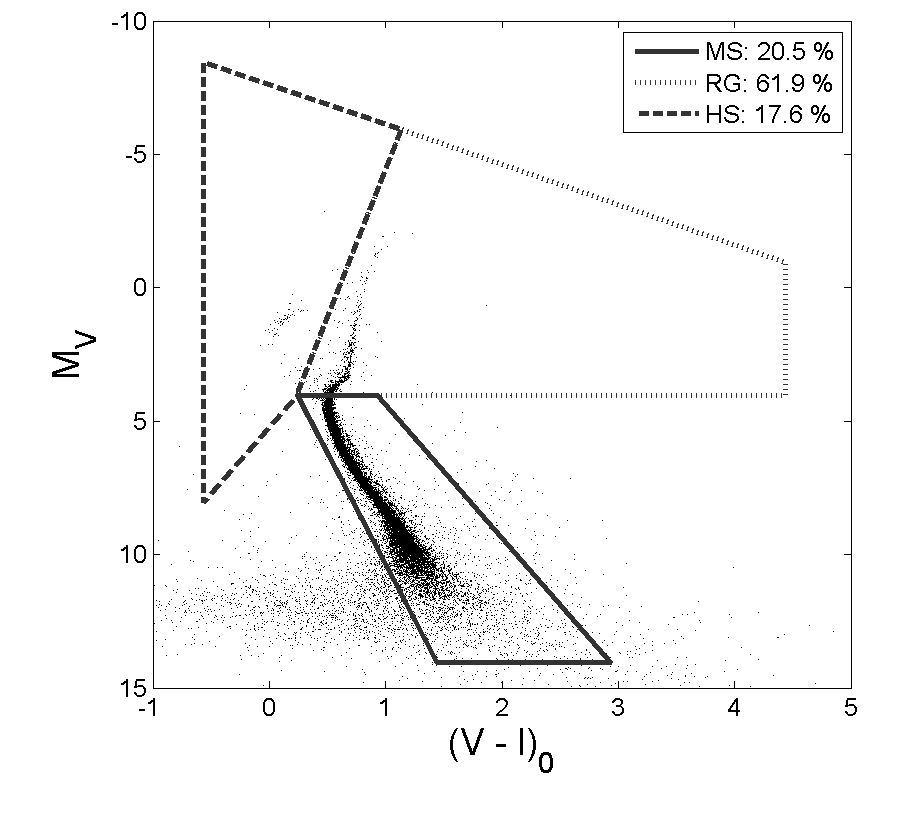}}
    \subfigure{
\includegraphics[width=0.7\textwidth]{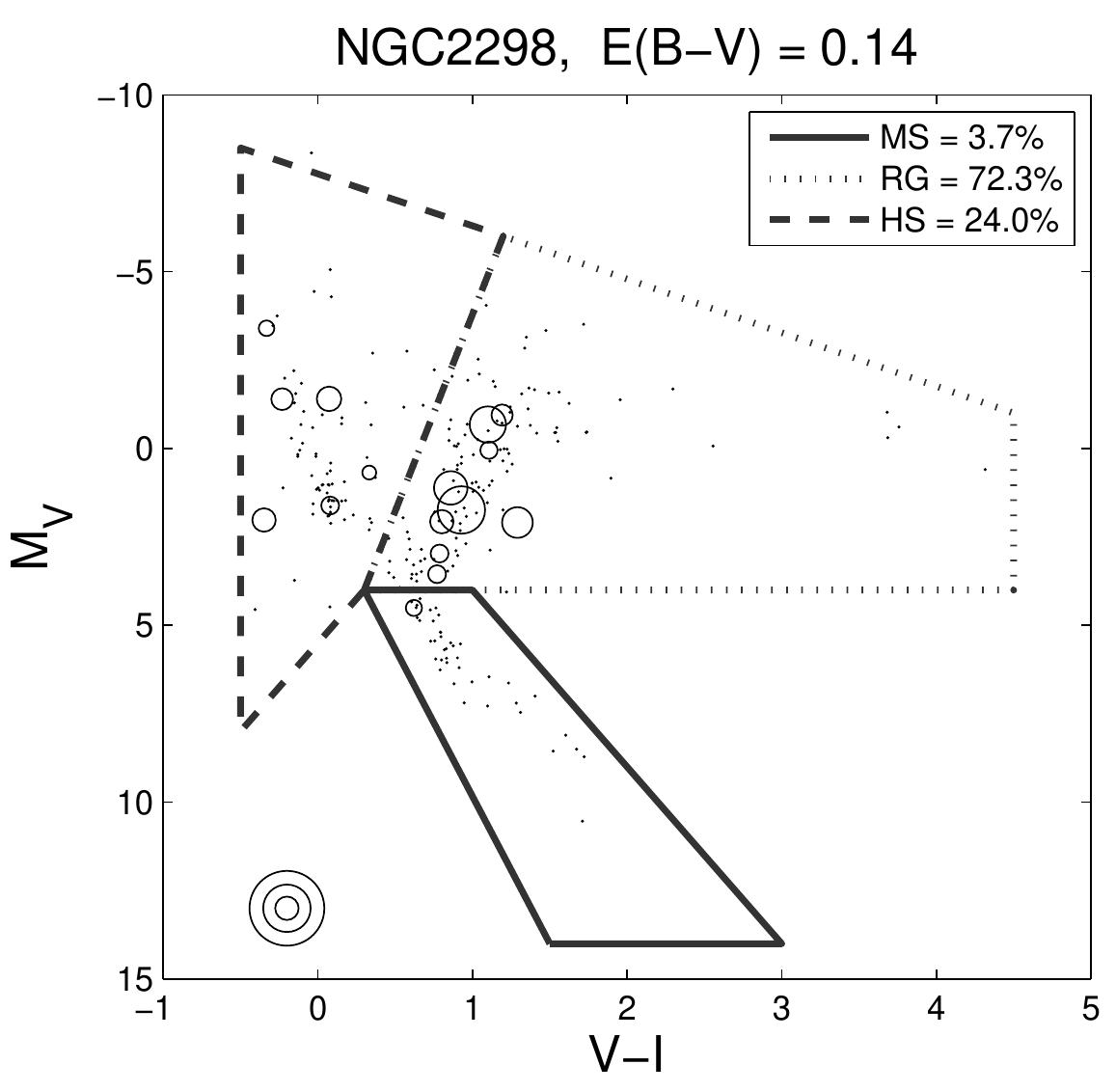}}
  \captcont{.../continued}
\end{figure}

\clearpage
\begin{figure}
  \centering
    \subfigure{
\includegraphics[width=0.75\textwidth]{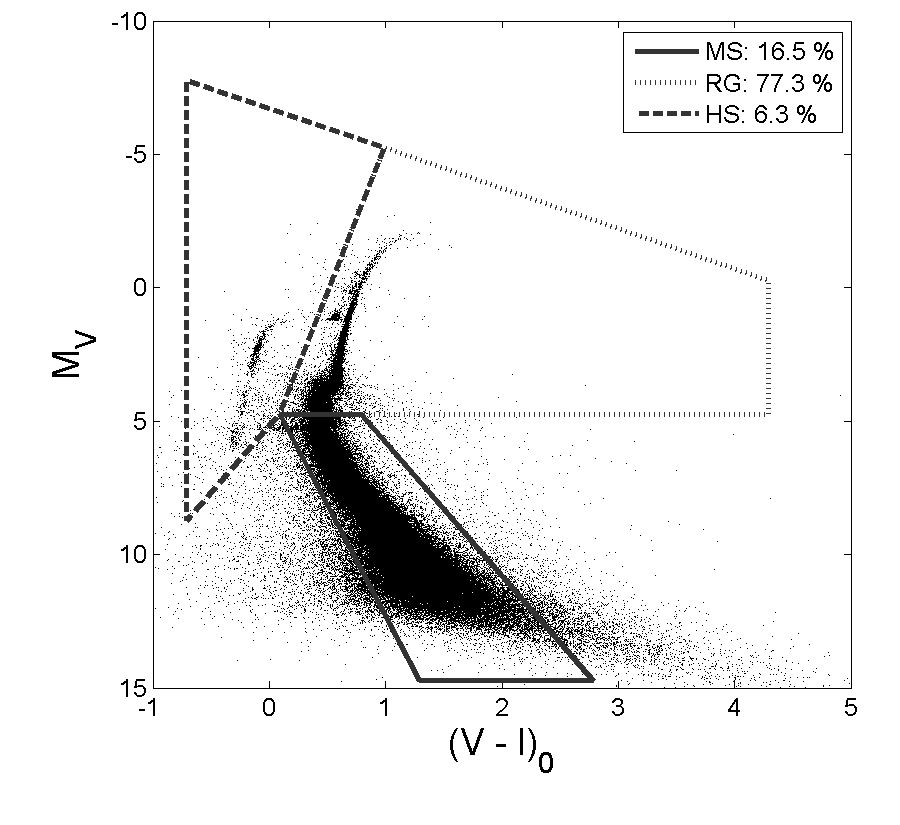}}
    \subfigure{
\includegraphics[width=0.7\textwidth]{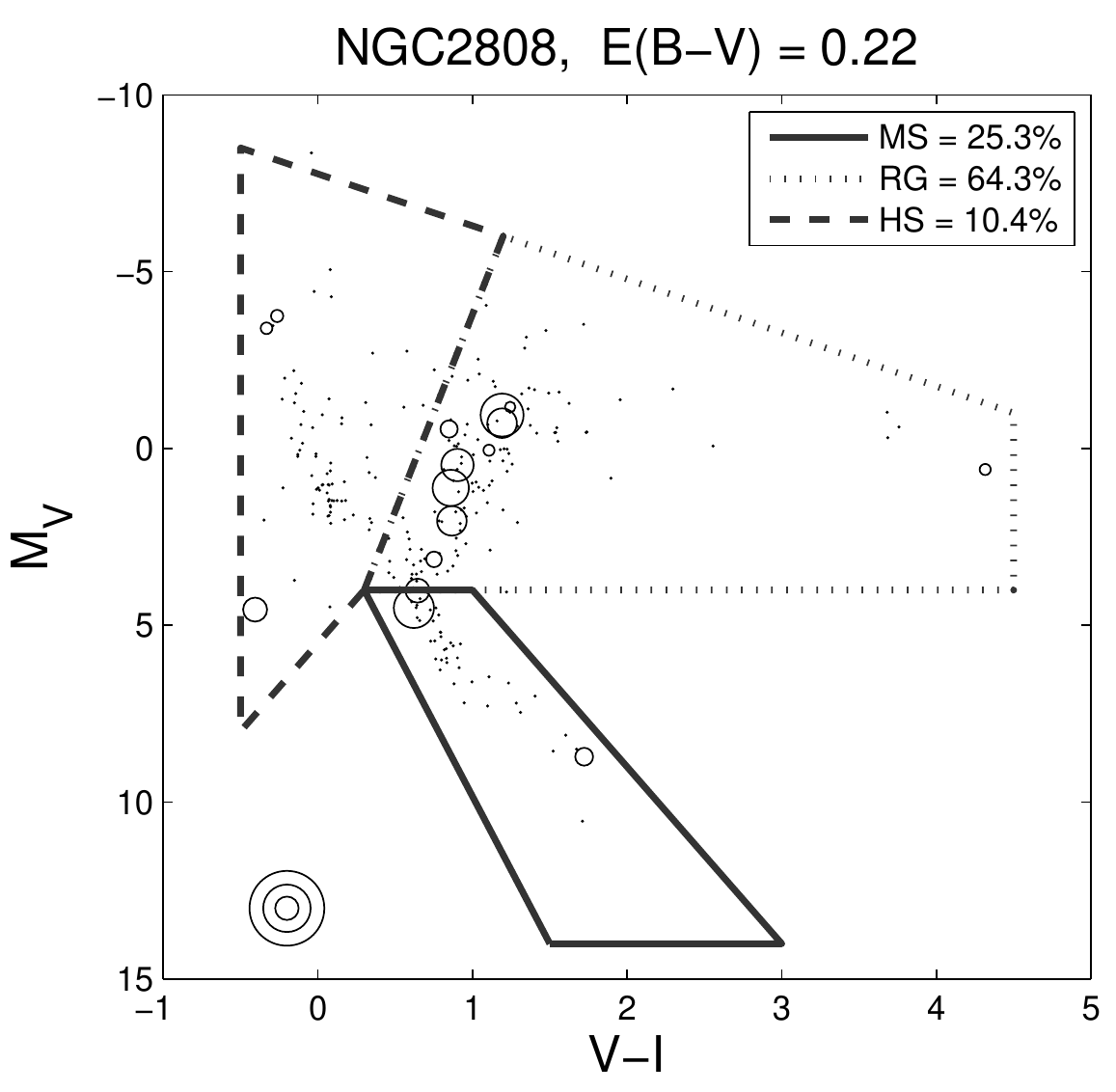}}
  \captcont{.../continued}
\end{figure}

\clearpage
\begin{figure}
  \centering
    \subfigure{
\includegraphics[width=0.75\textwidth]{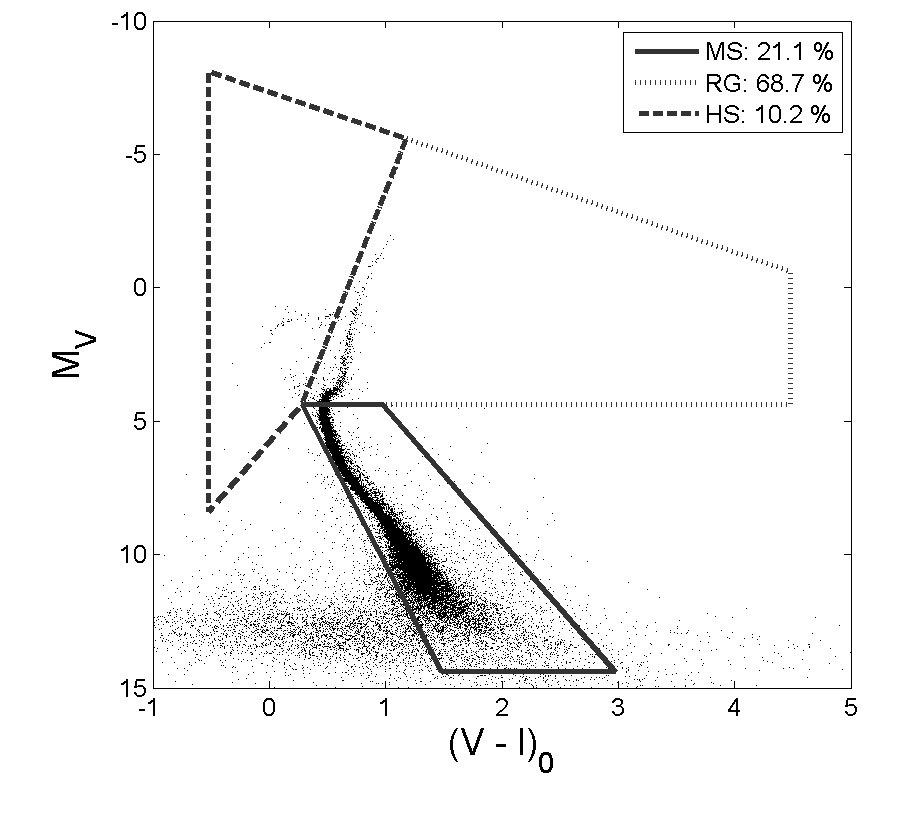}}
    \subfigure{
\includegraphics[width=0.7\textwidth]{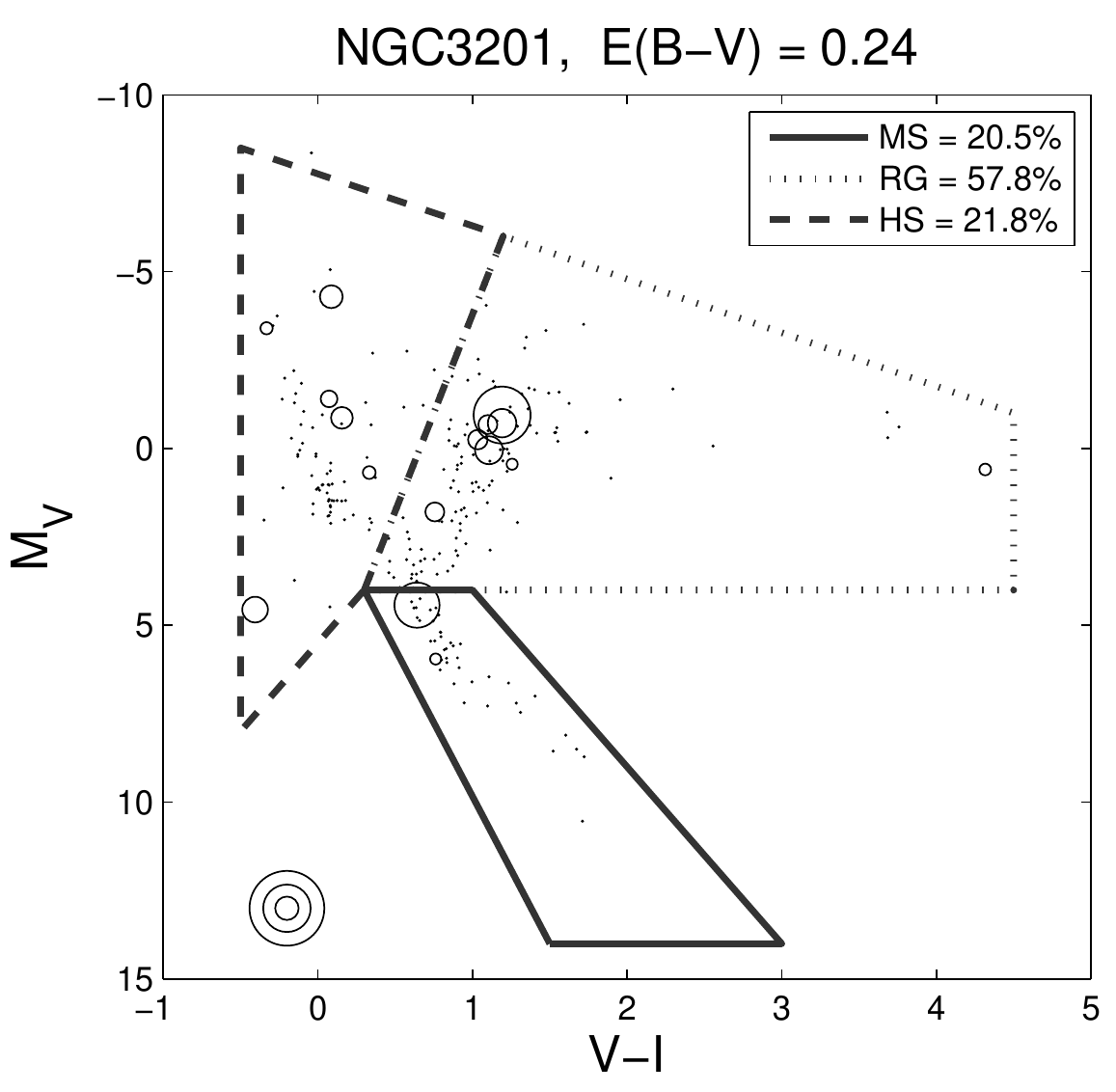}}
  \captcont{.../continued}
\end{figure}

\clearpage
\begin{figure}
  \centering
    \subfigure{
\includegraphics[width=0.75\textwidth]{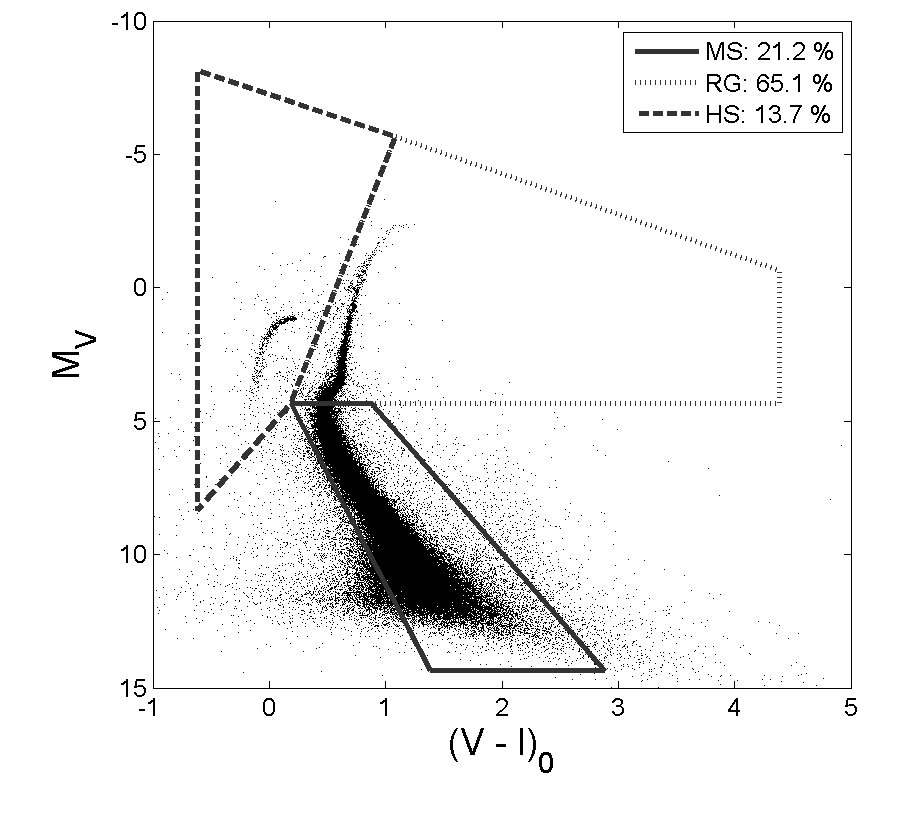}}
    \subfigure{
\includegraphics[width=0.7\textwidth]{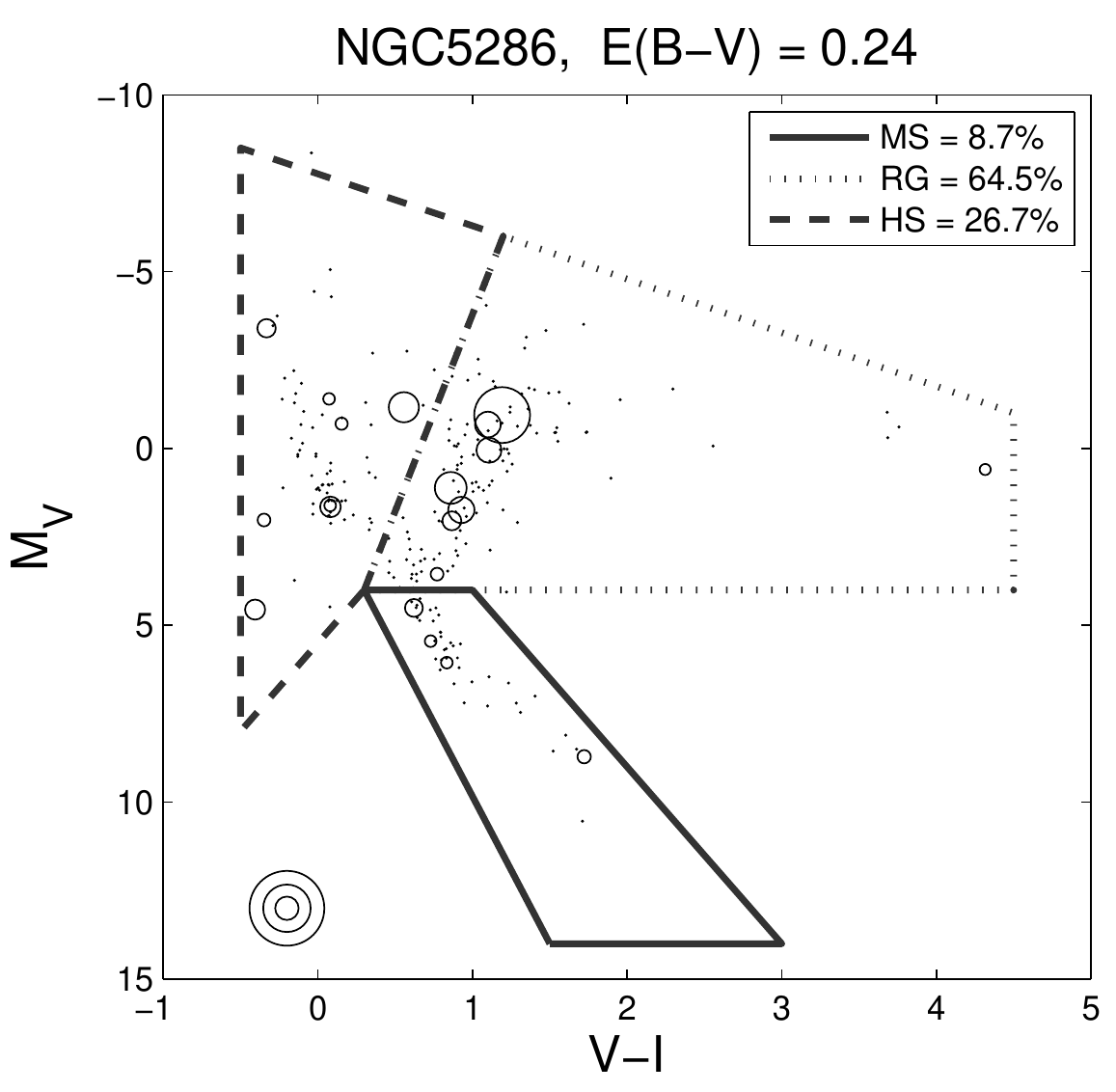}}
  \captcont{.../continued}
\end{figure}

\clearpage
\begin{figure}
  \centering
    \subfigure{
\includegraphics[width=0.75\textwidth]{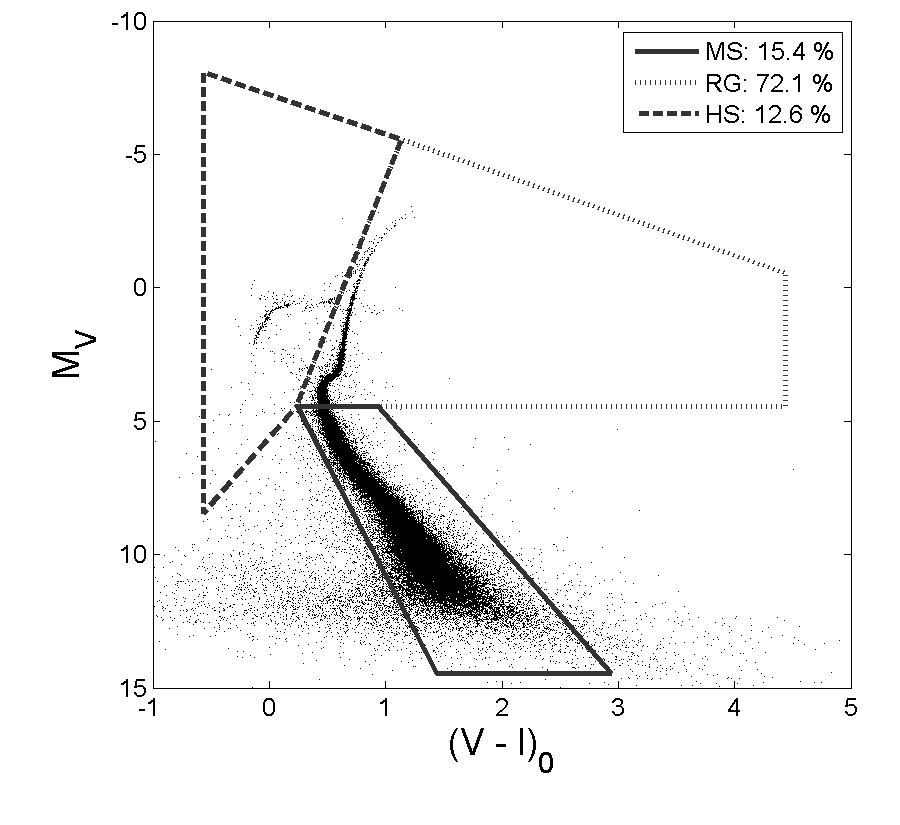}}
    \subfigure{
\includegraphics[width=0.7\textwidth]{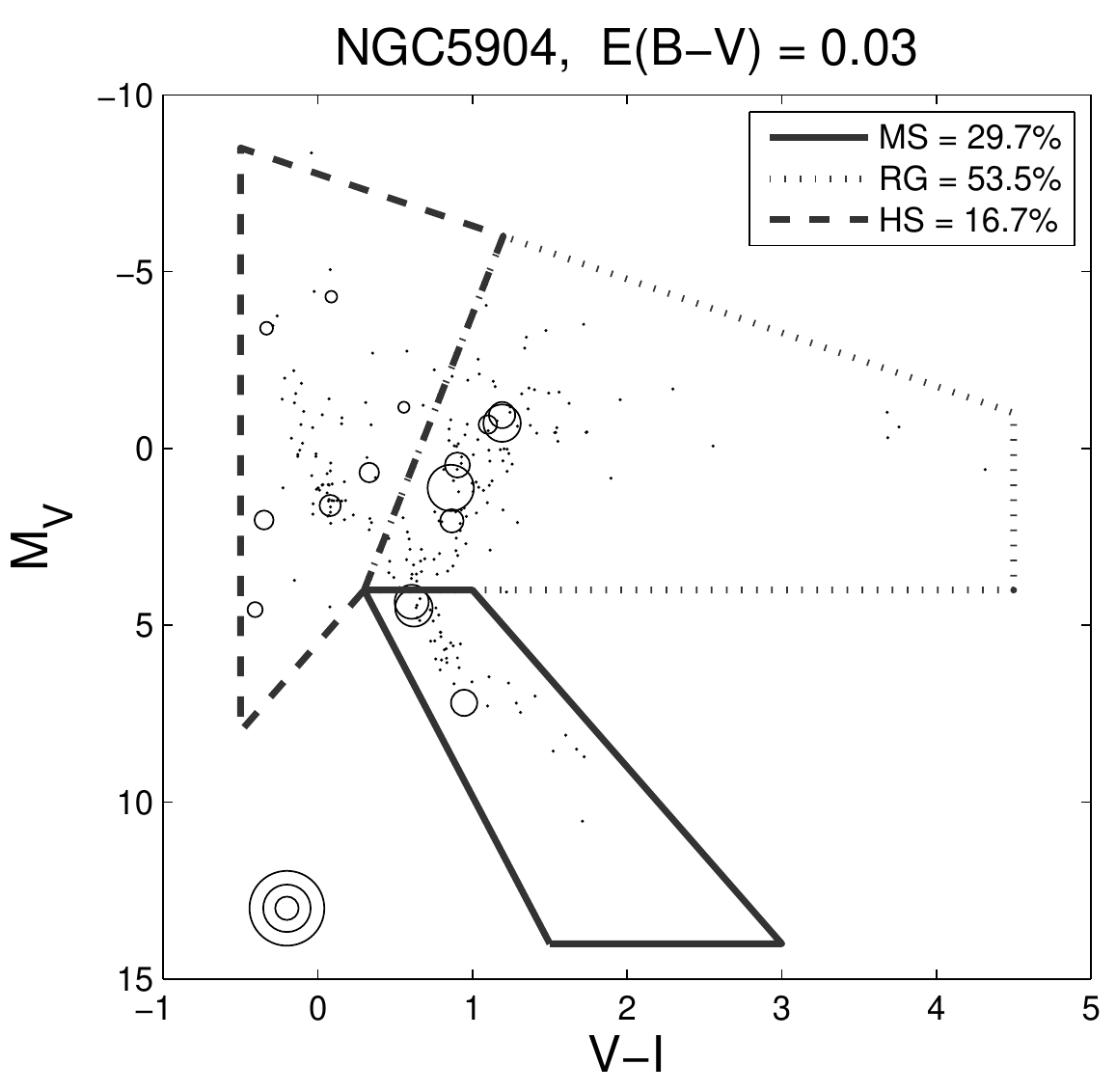}}
  \captcont{.../continued}
\end{figure}

\clearpage
\begin{figure}
  \centering
    \subfigure{
\includegraphics[width=0.75\textwidth]{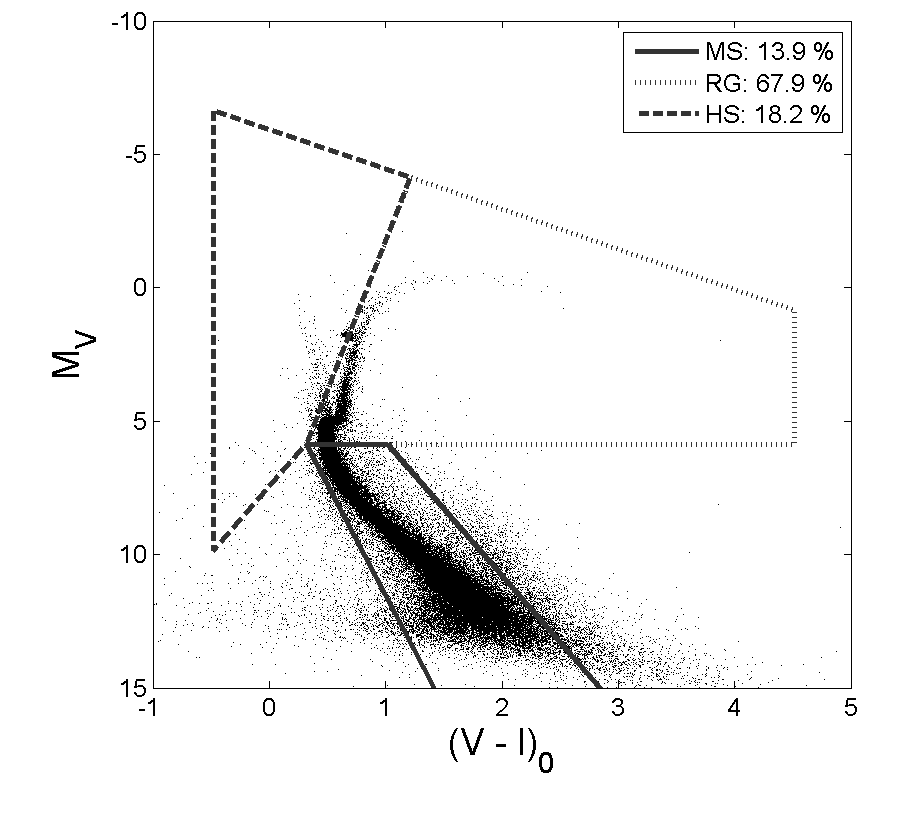}}
    \subfigure{
\includegraphics[width=0.7\textwidth]{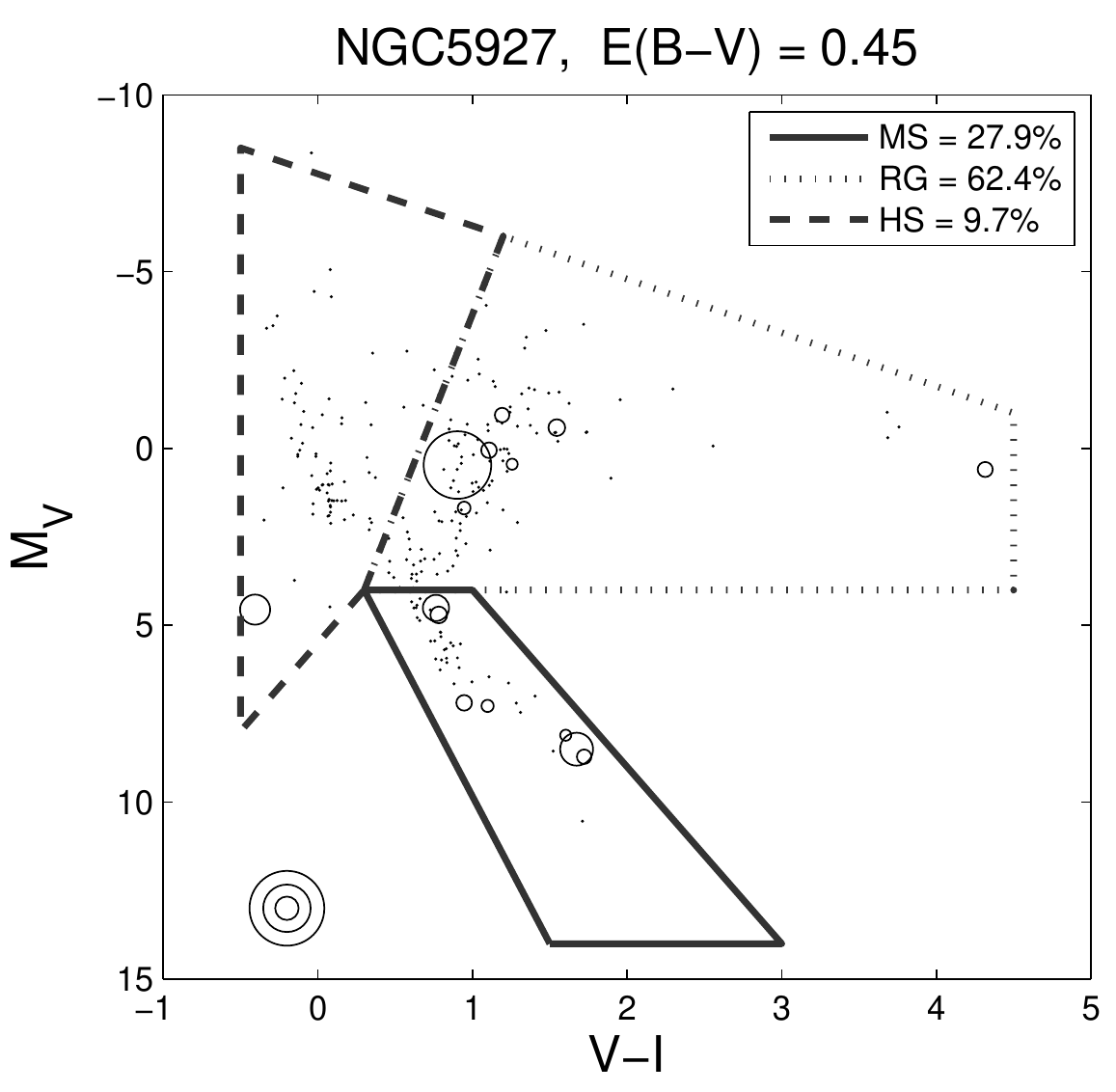}}
  \captcont{.../continued}
\end{figure}

\clearpage
\begin{figure}
  \centering
    \subfigure{
\includegraphics[width=0.75\textwidth]{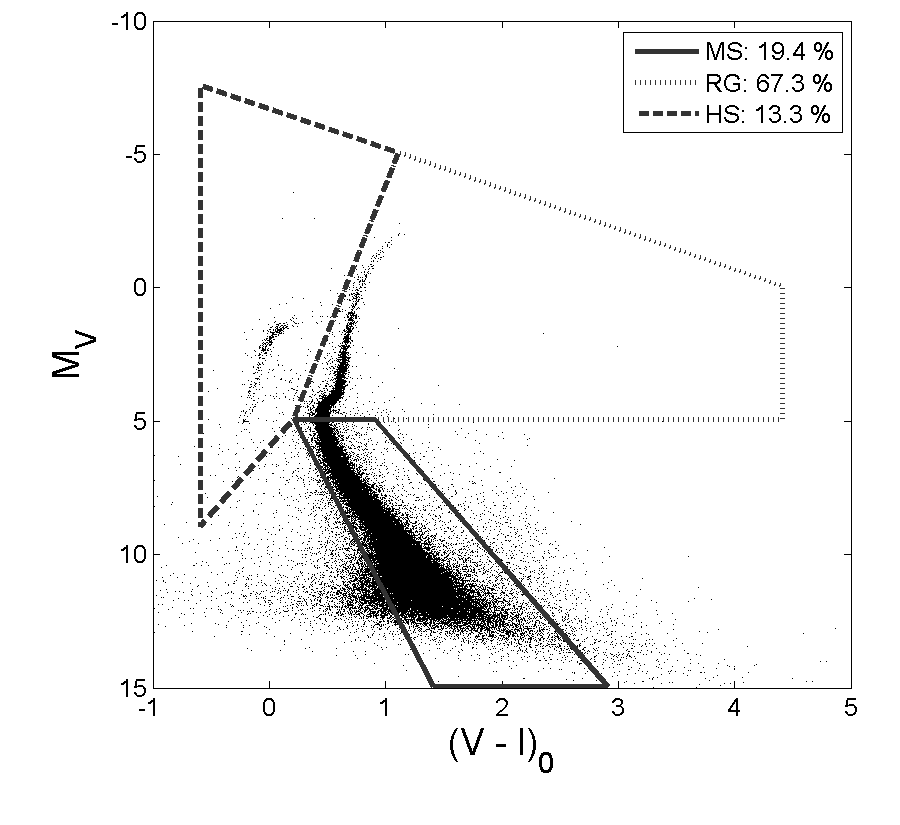}}
    \subfigure{
\includegraphics[width=0.7\textwidth]{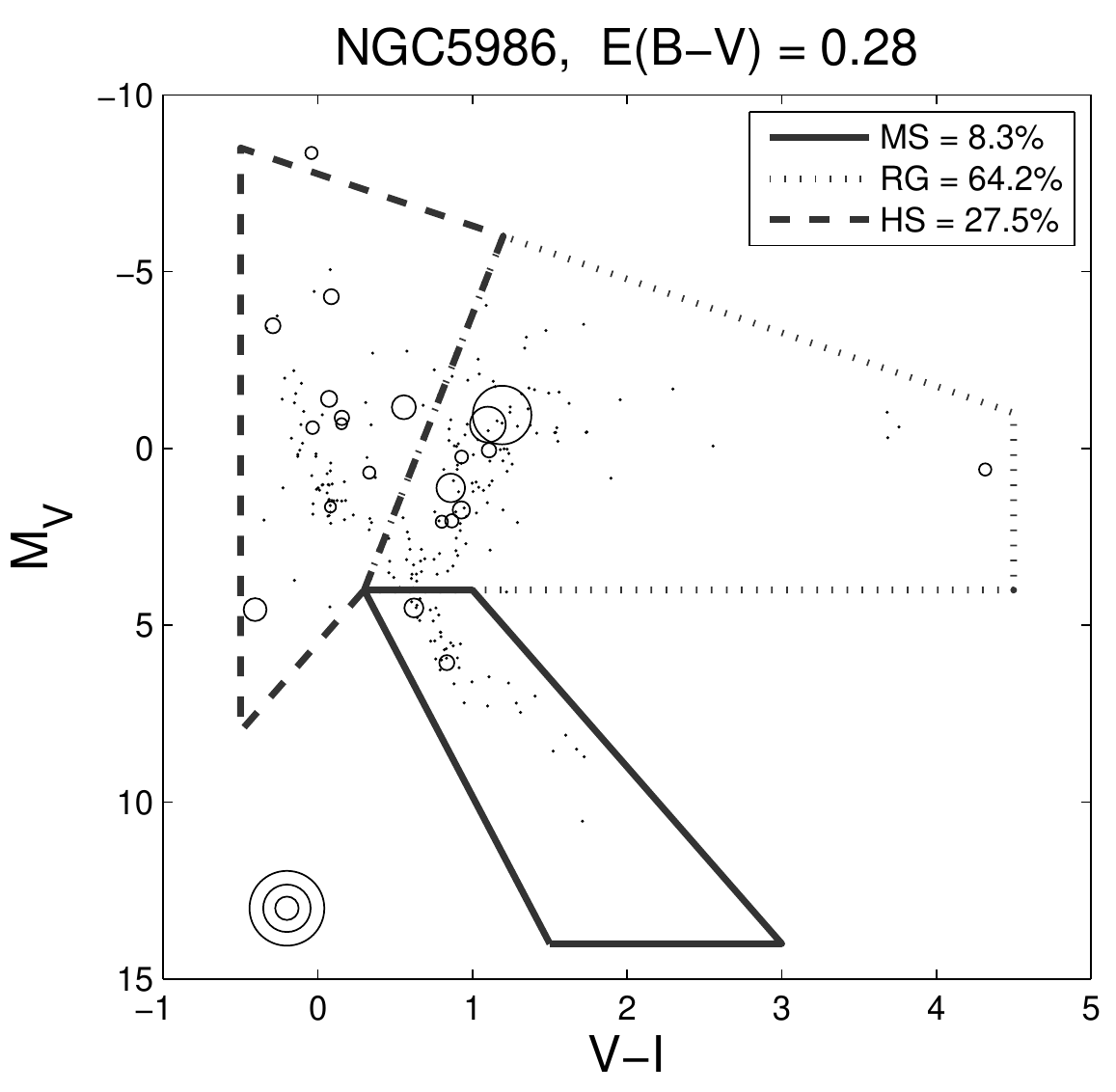}}
  \captcont{.../continued}
\end{figure}

\clearpage
\begin{figure}
  \centering
    \subfigure{
\includegraphics[width=0.75\textwidth]{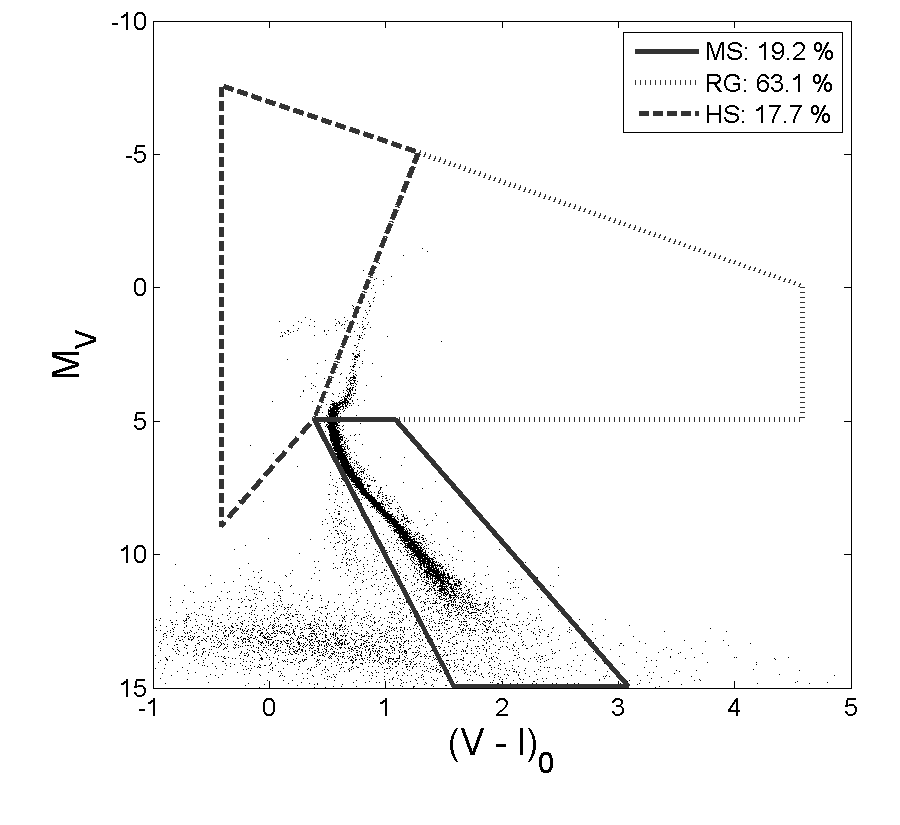}}
    \subfigure{
\includegraphics[width=0.7\textwidth]{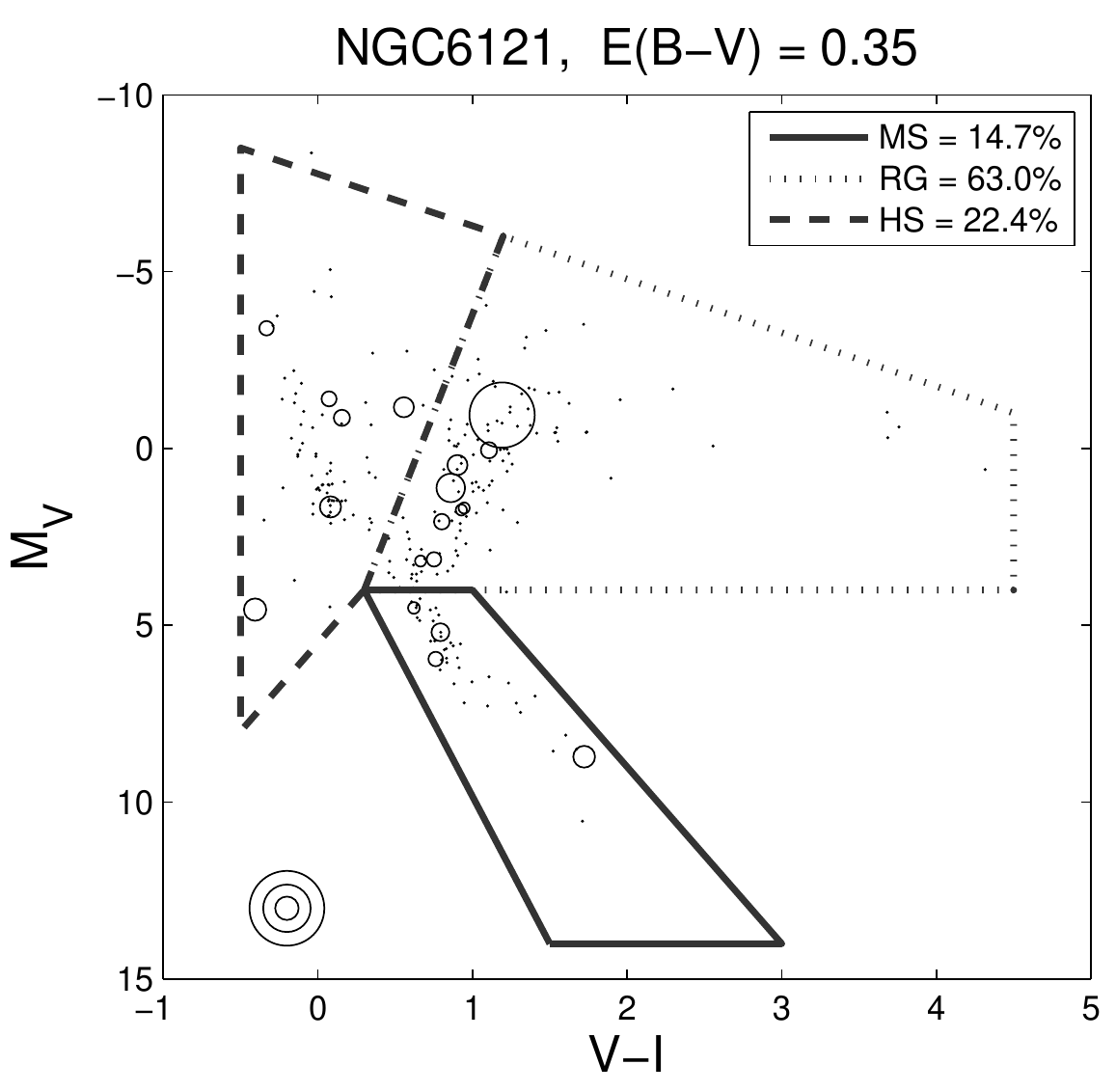}}
  \captcont{.../continued}
\end{figure}

\clearpage
\begin{figure}
  \centering
    \subfigure{
\includegraphics[width=0.75\textwidth]{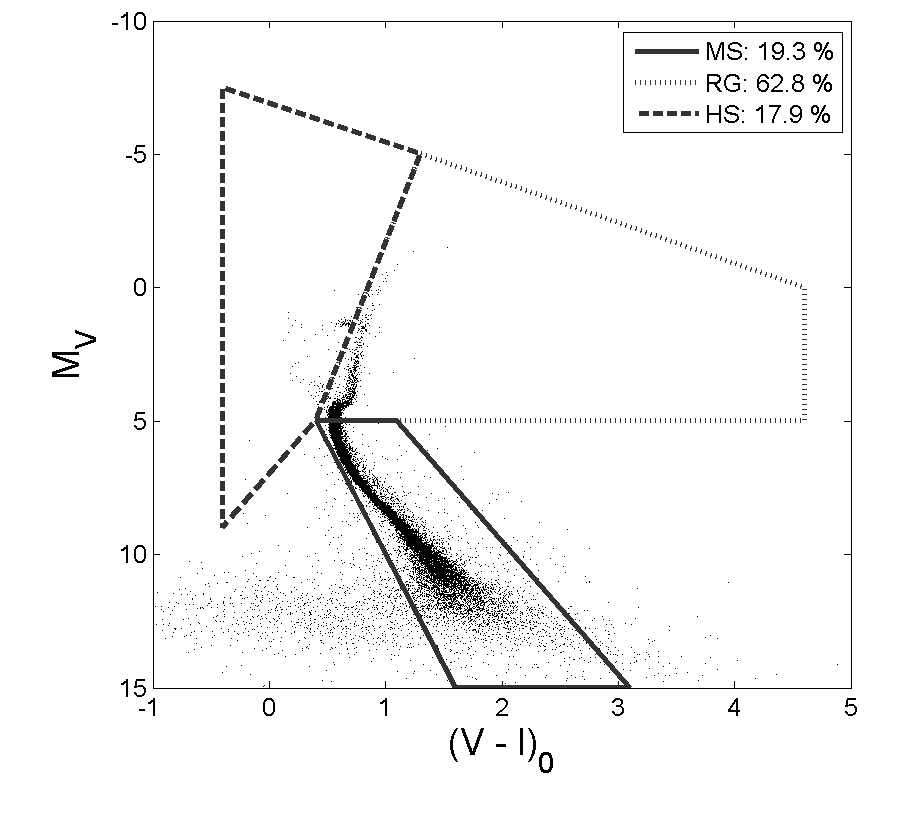}}
    \subfigure{
\includegraphics[width=0.7\textwidth]{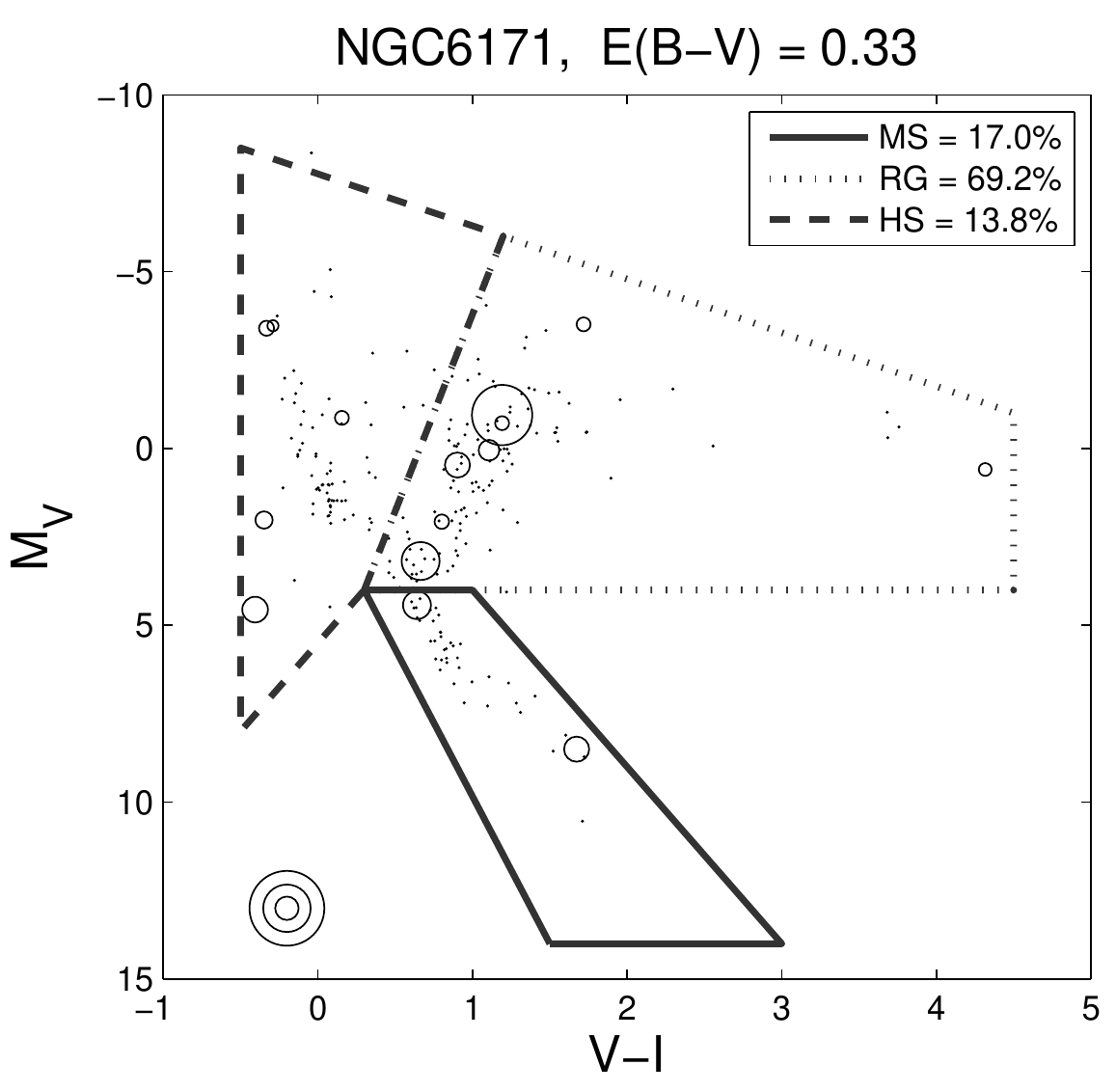}}
  \captcont{.../continued}
\end{figure}

\clearpage
\begin{figure}
  \centering
    \subfigure{
\includegraphics[width=0.75\textwidth]{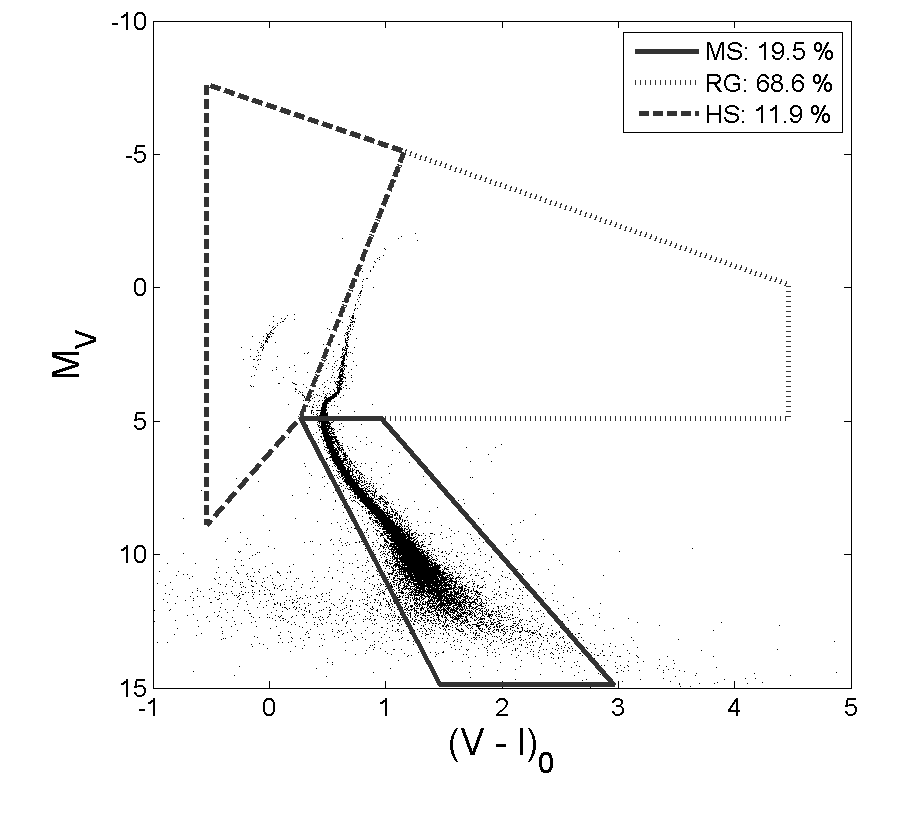}}
    \subfigure{
\includegraphics[width=0.7\textwidth]{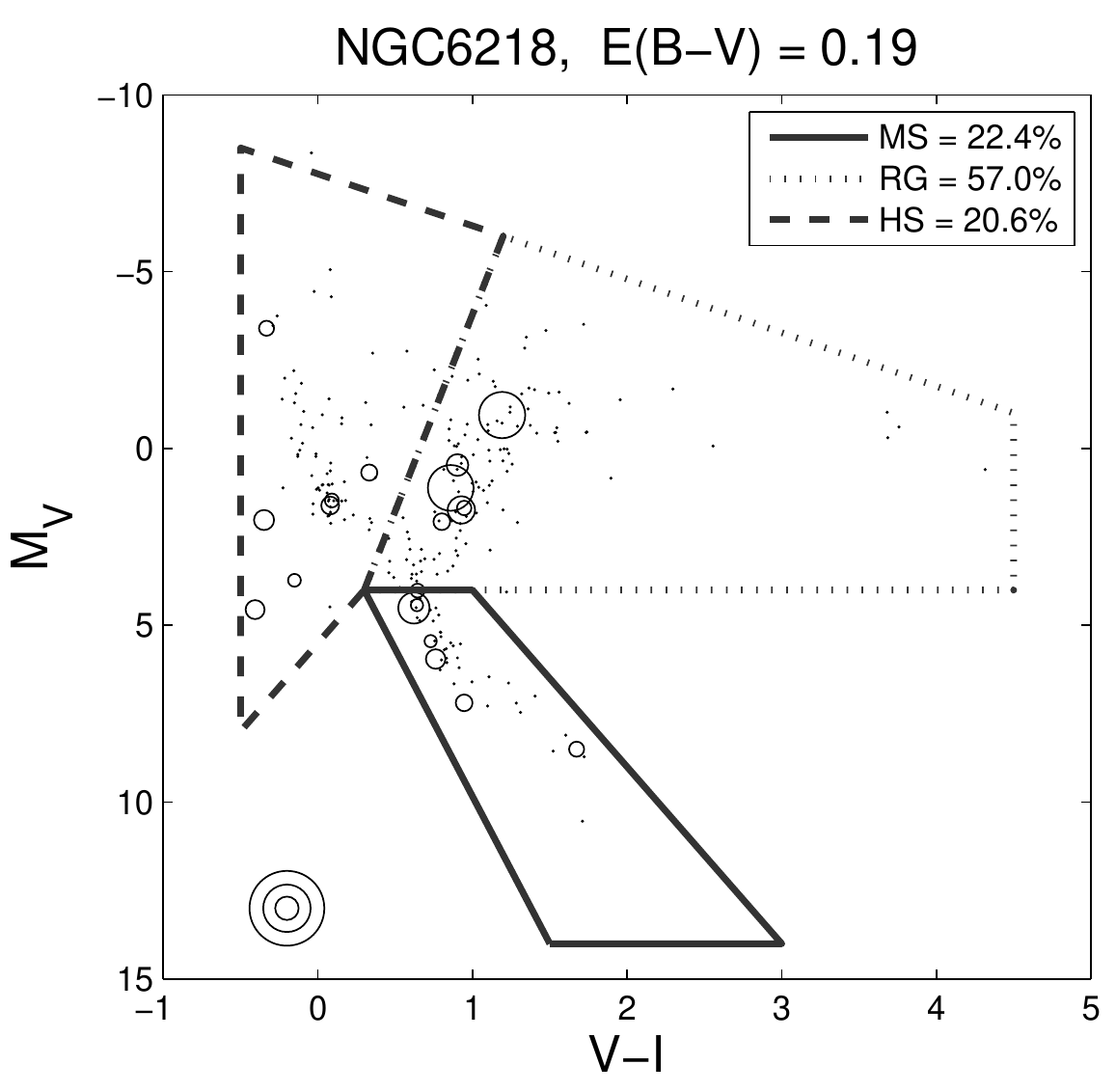}}
  \captcont{.../continued}
\end{figure}

\clearpage
\begin{figure}
  \centering
    \subfigure{
\includegraphics[width=0.75\textwidth]{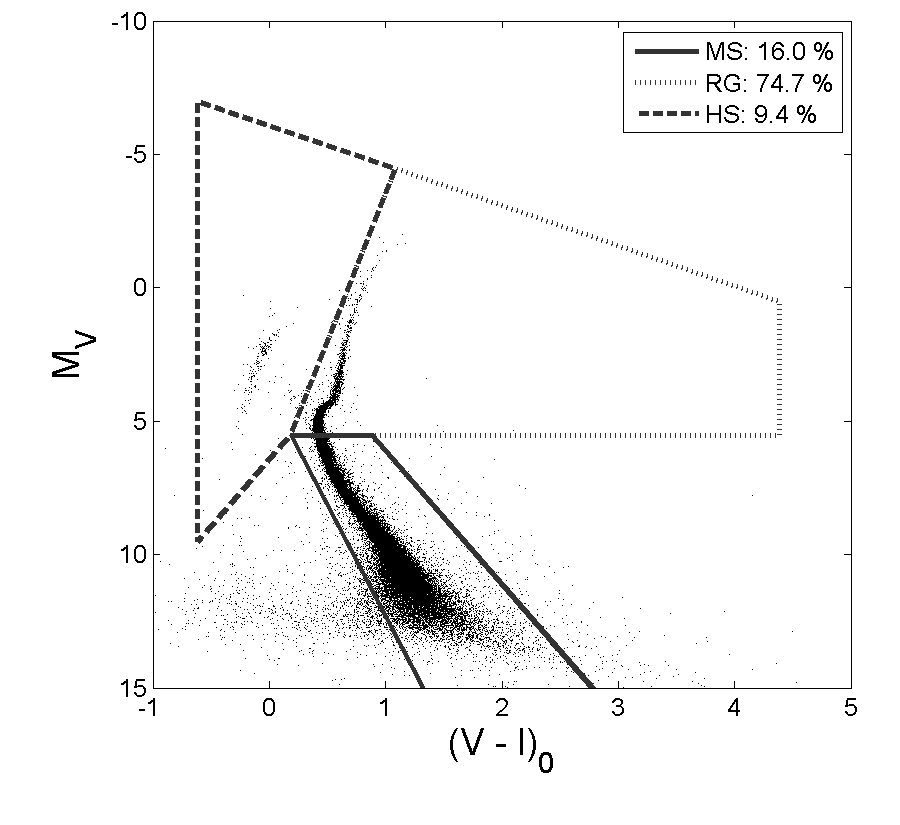}}
    \subfigure{
\includegraphics[width=0.7\textwidth]{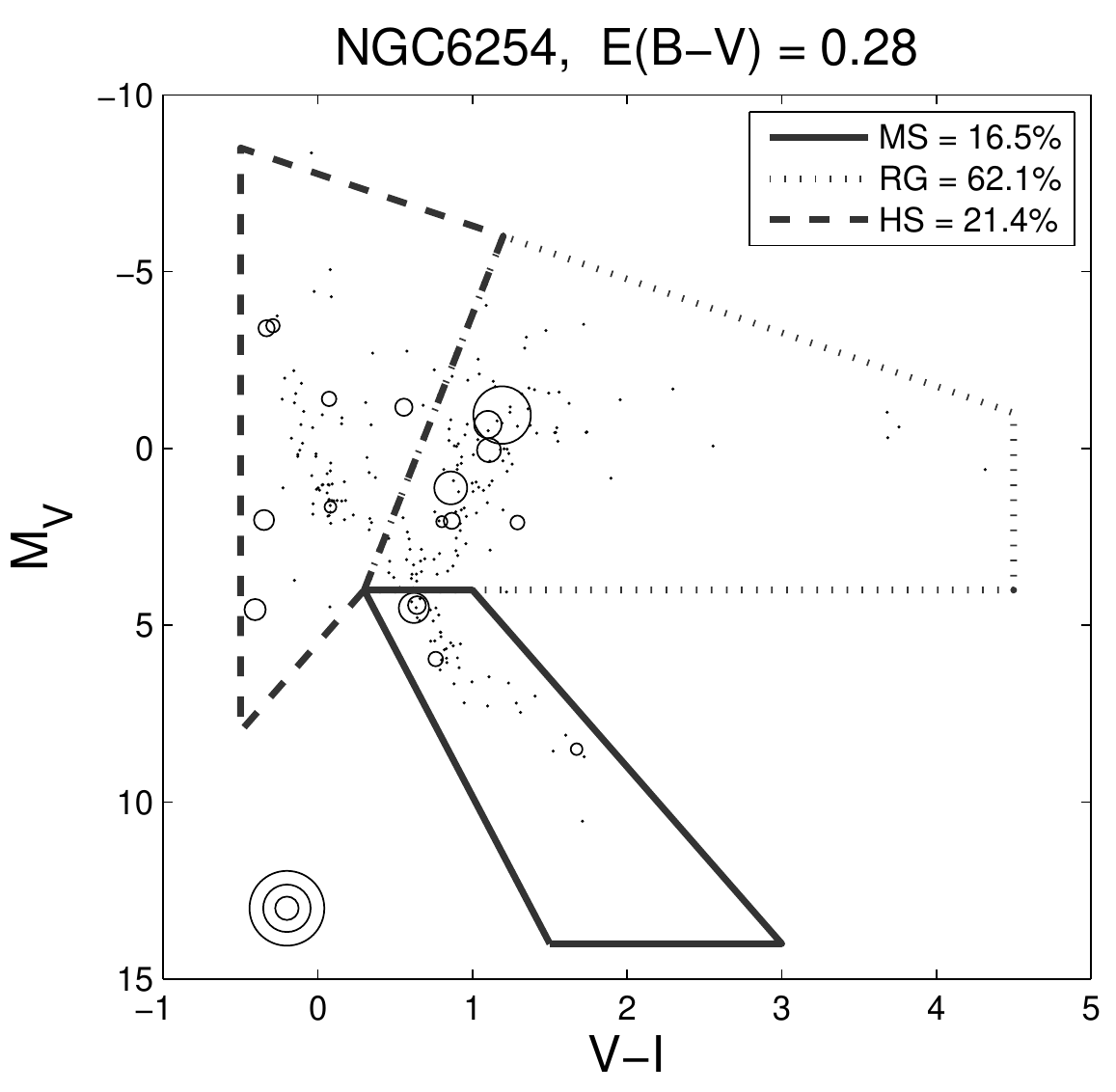}}
  \captcont{.../continued}
\end{figure}

\clearpage
\begin{figure}
  \centering
    \subfigure{
\includegraphics[width=0.75\textwidth]{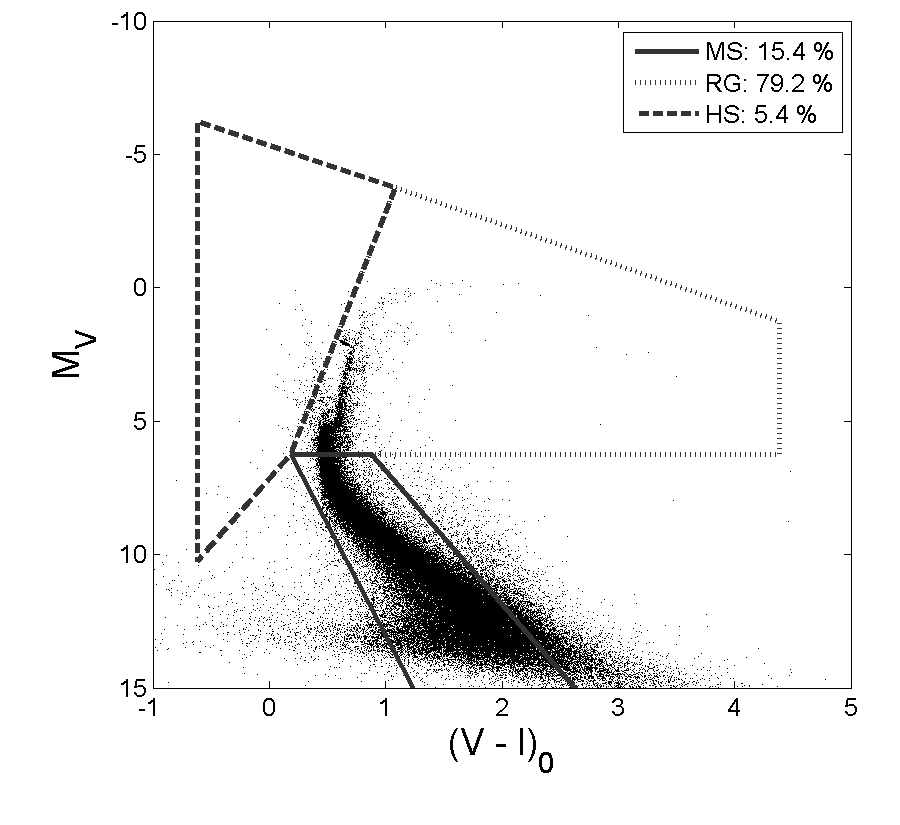}}
    \subfigure{
\includegraphics[width=0.7\textwidth]{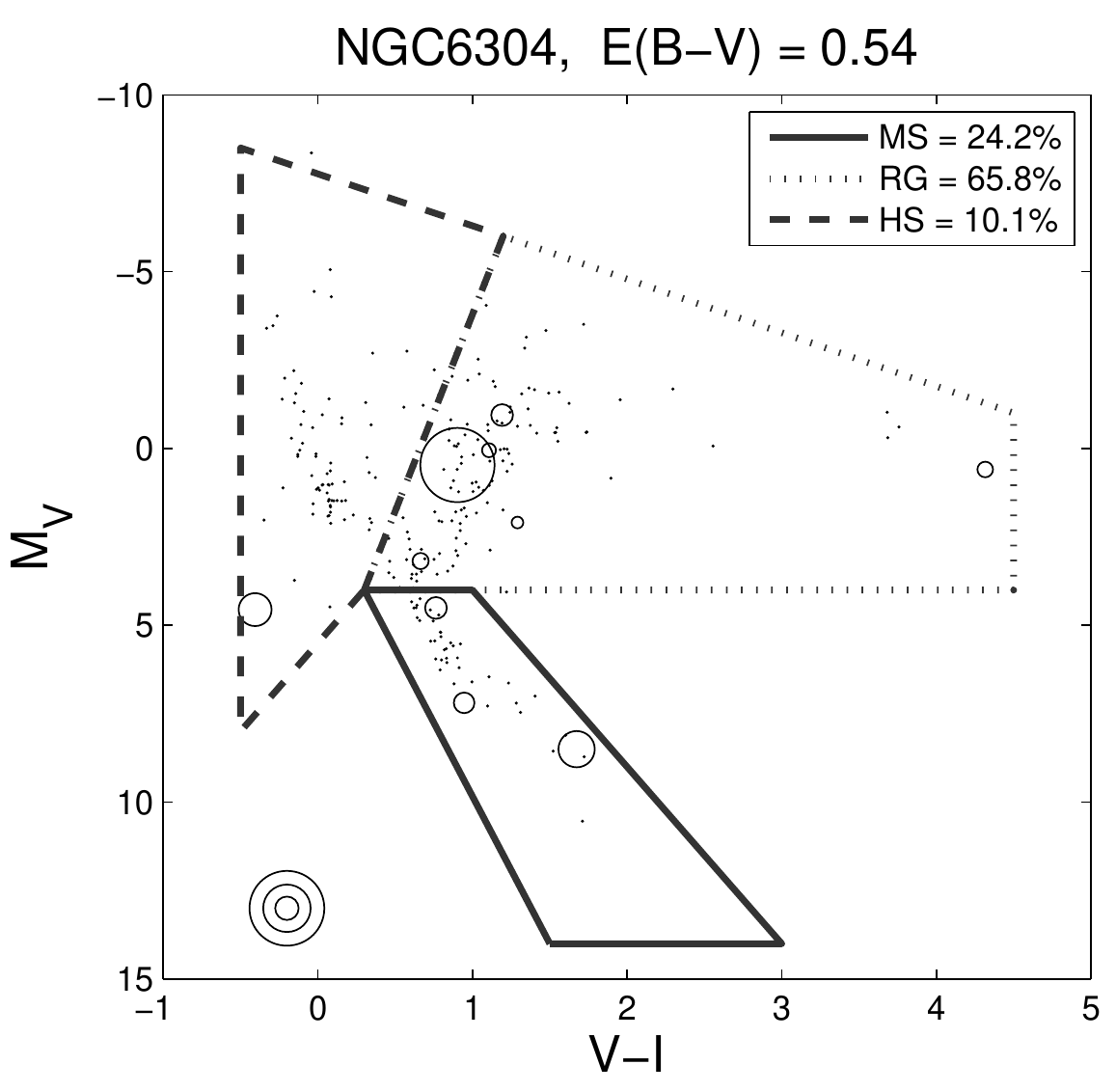}}
  \captcont{.../continued}
\end{figure}

\clearpage
\begin{figure}
  \centering
    \subfigure{
\includegraphics[width=0.75\textwidth]{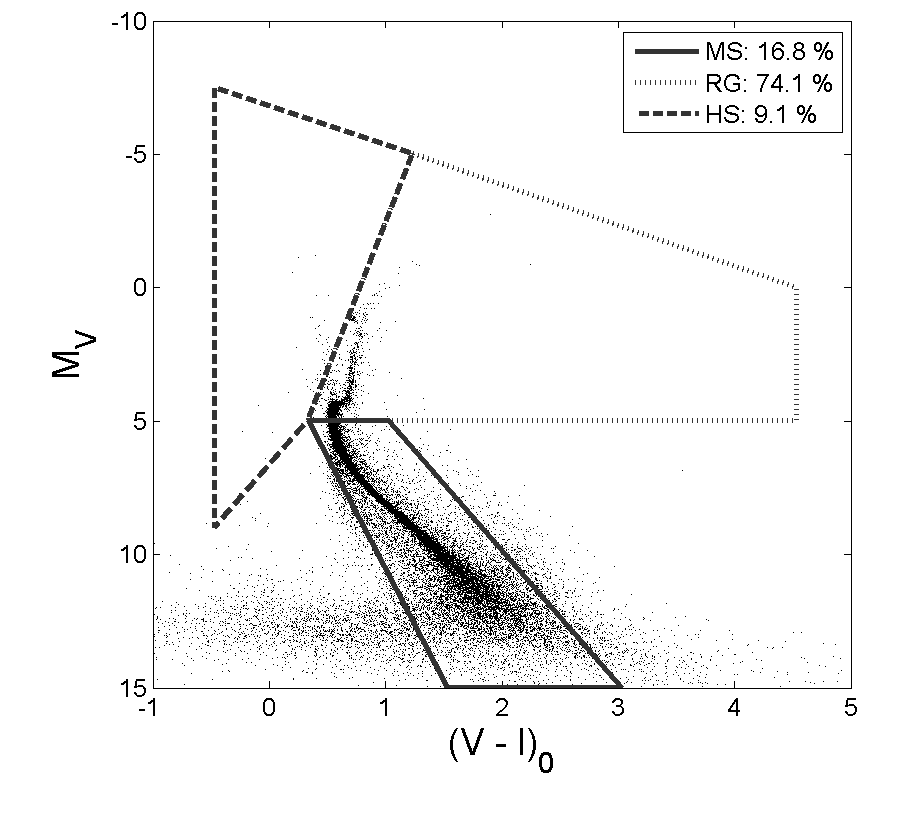}}
    \subfigure{
\includegraphics[width=0.7\textwidth]{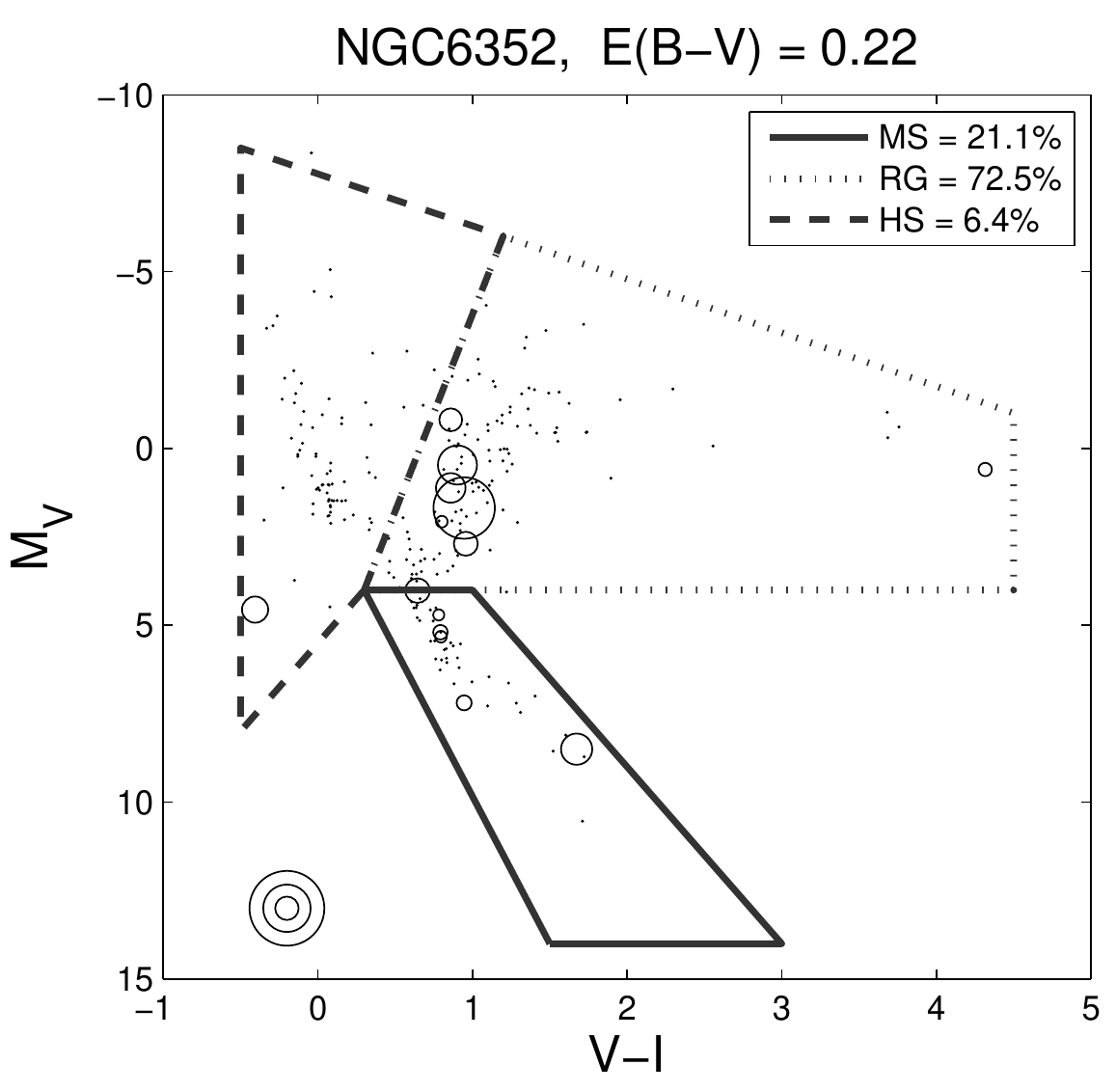}}
  \captcont{.../continued}
\end{figure}

\clearpage
\begin{figure}
  \centering
    \subfigure{
\includegraphics[width=0.75\textwidth]{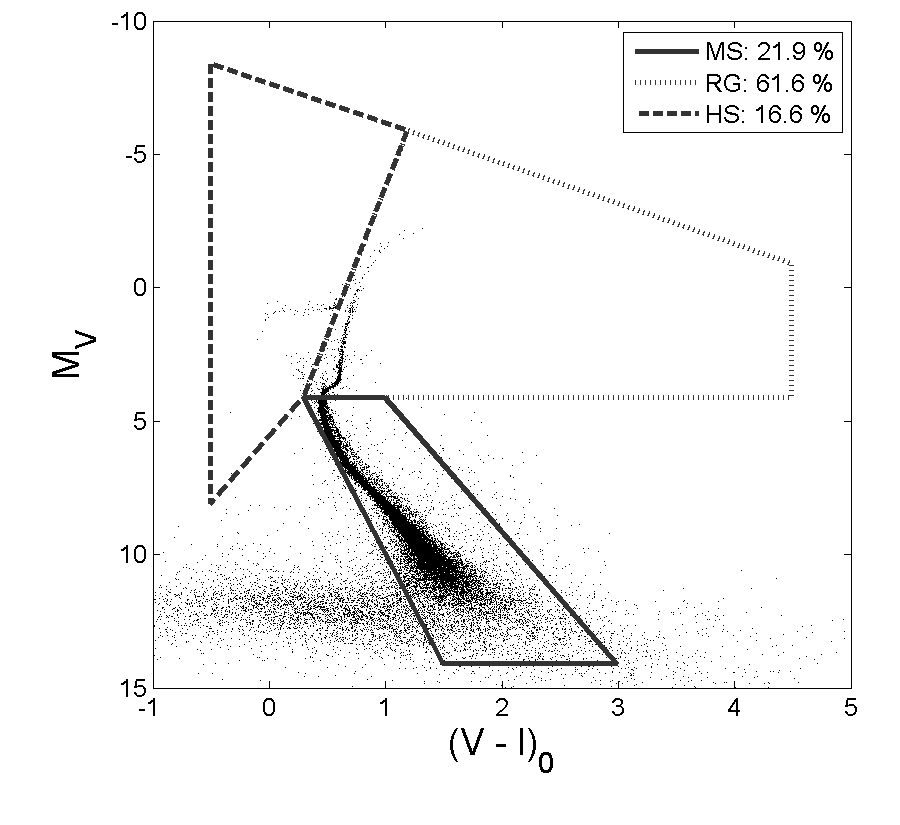}}
    \subfigure{
\includegraphics[width=0.7\textwidth]{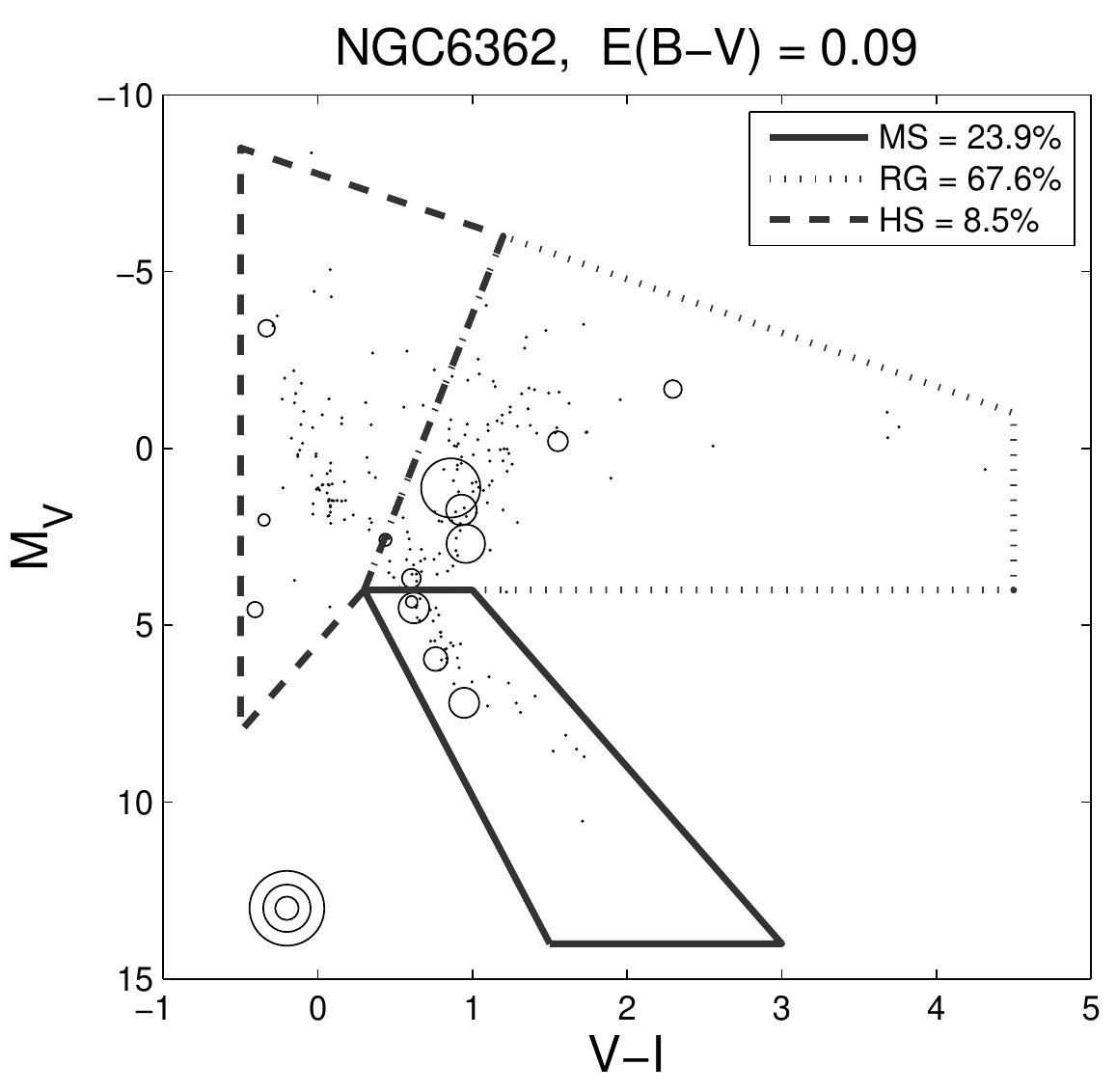}}
  \captcont{.../continued}
\end{figure}

\clearpage
\begin{figure}
  \centering
    \subfigure{
\includegraphics[width=0.75\textwidth]{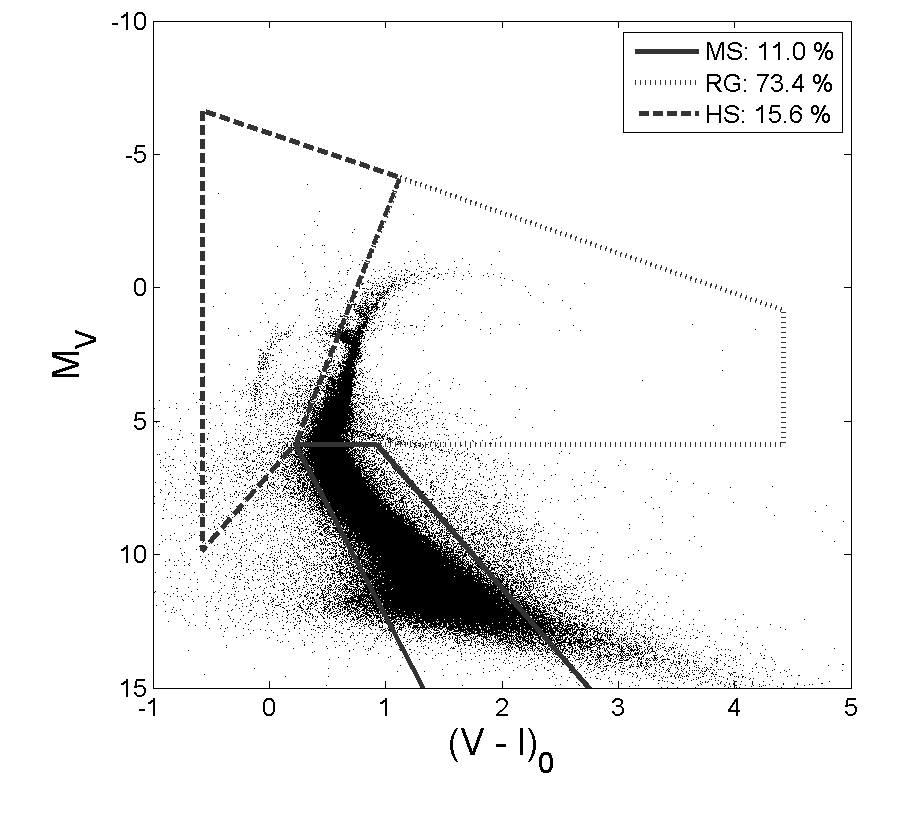}}
    \subfigure{
\includegraphics[width=0.7\textwidth]{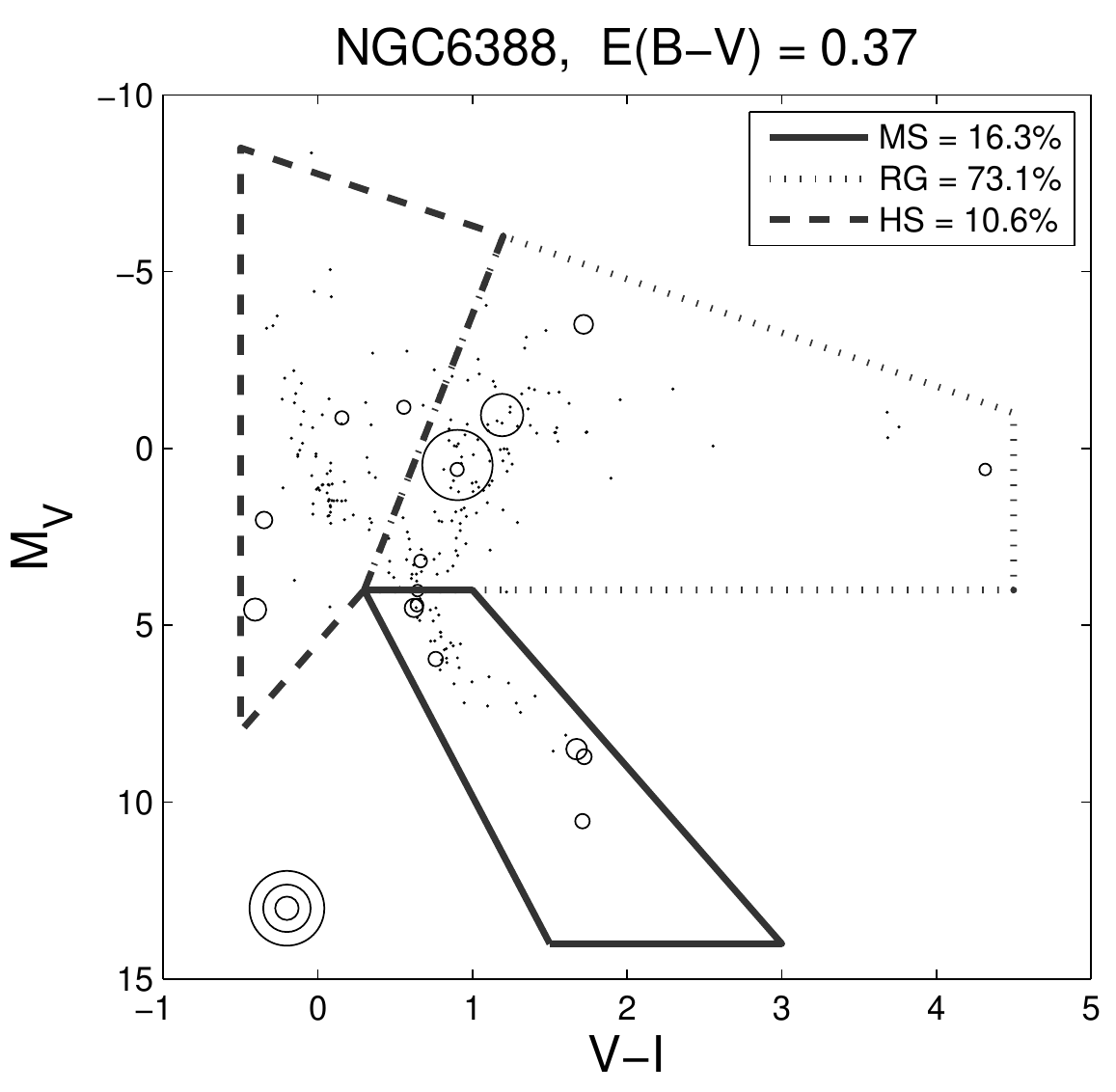}}
  \captcont{.../continued}
\end{figure}

\clearpage
\begin{figure}
  \centering
    \subfigure{
\includegraphics[width=0.75\textwidth]{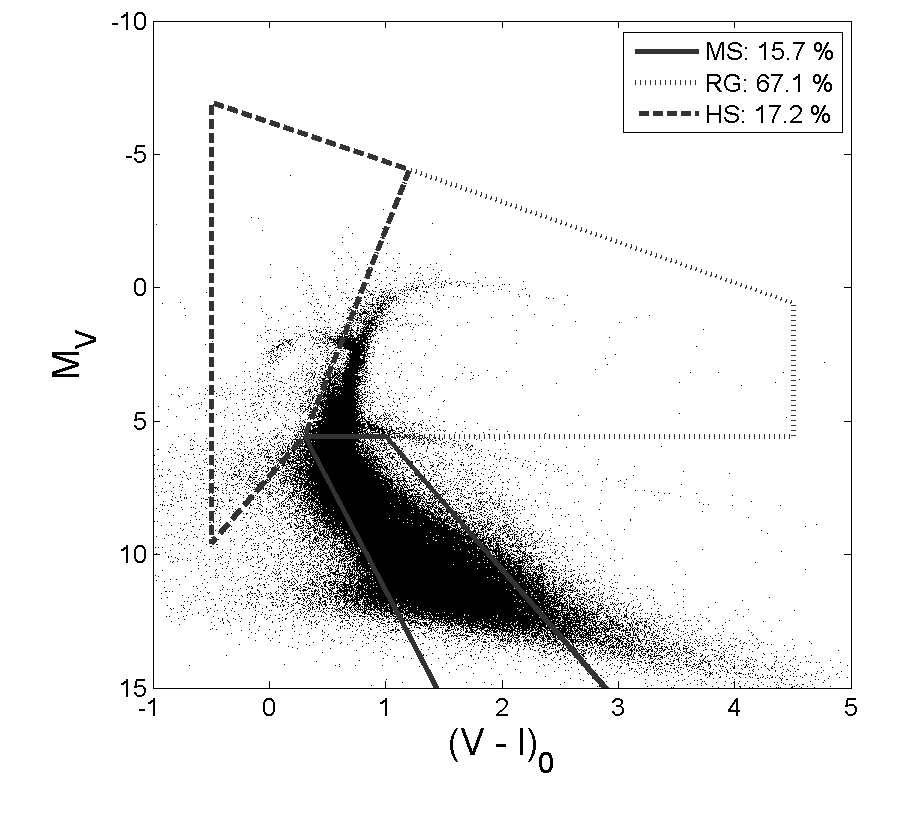}}
    \subfigure{
\includegraphics[width=0.7\textwidth]{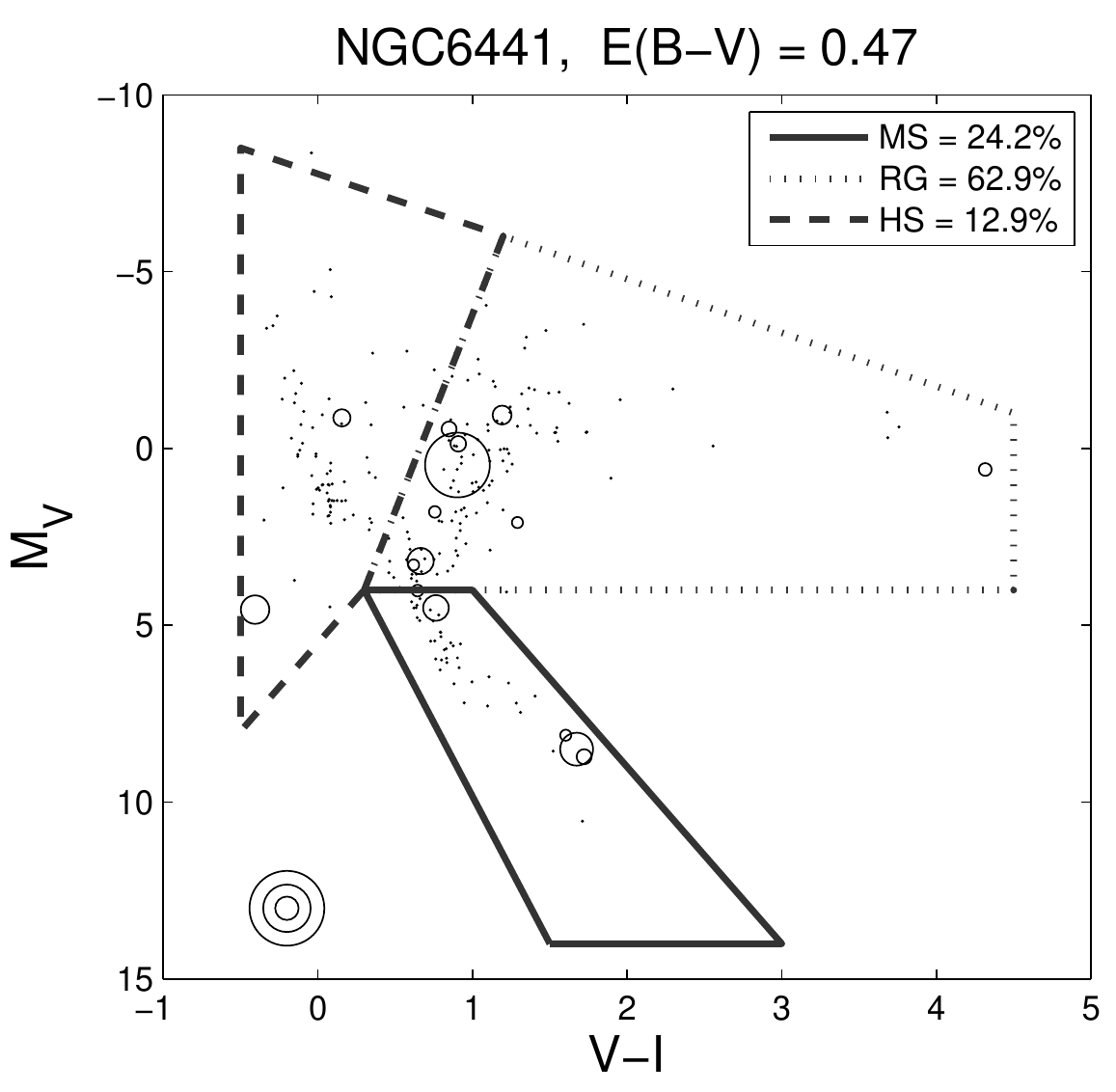}}
  \captcont{.../continued}
\end{figure}

\clearpage
\begin{figure}
  \centering
    \subfigure{
\includegraphics[width=0.75\textwidth]{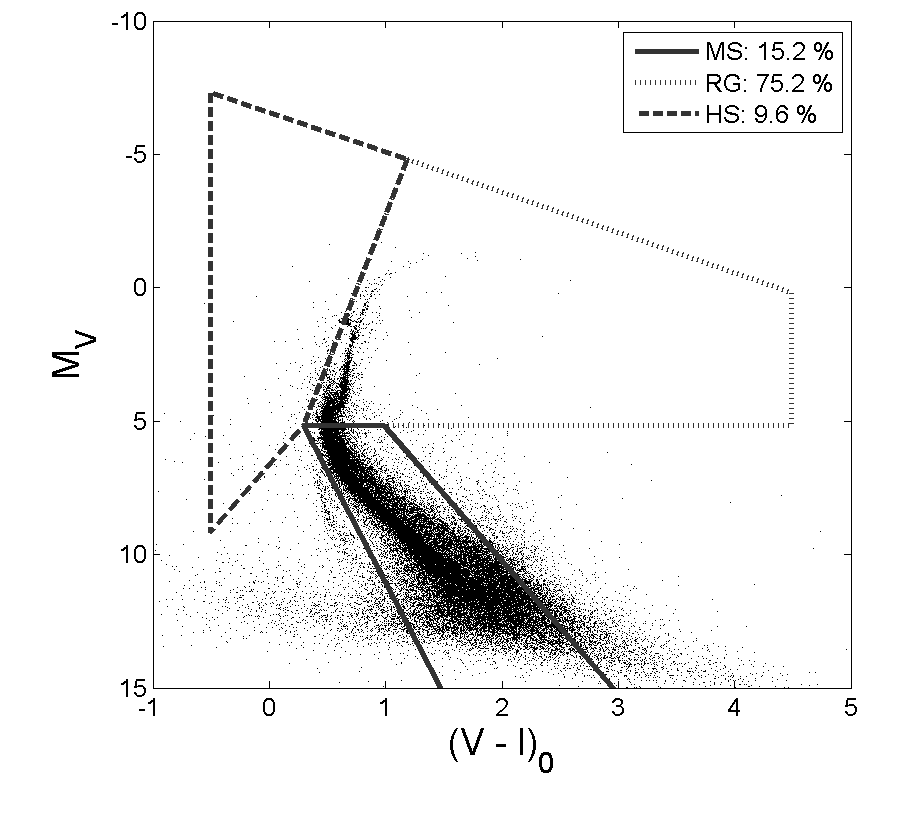}}
    \subfigure{
\includegraphics[width=0.7\textwidth]{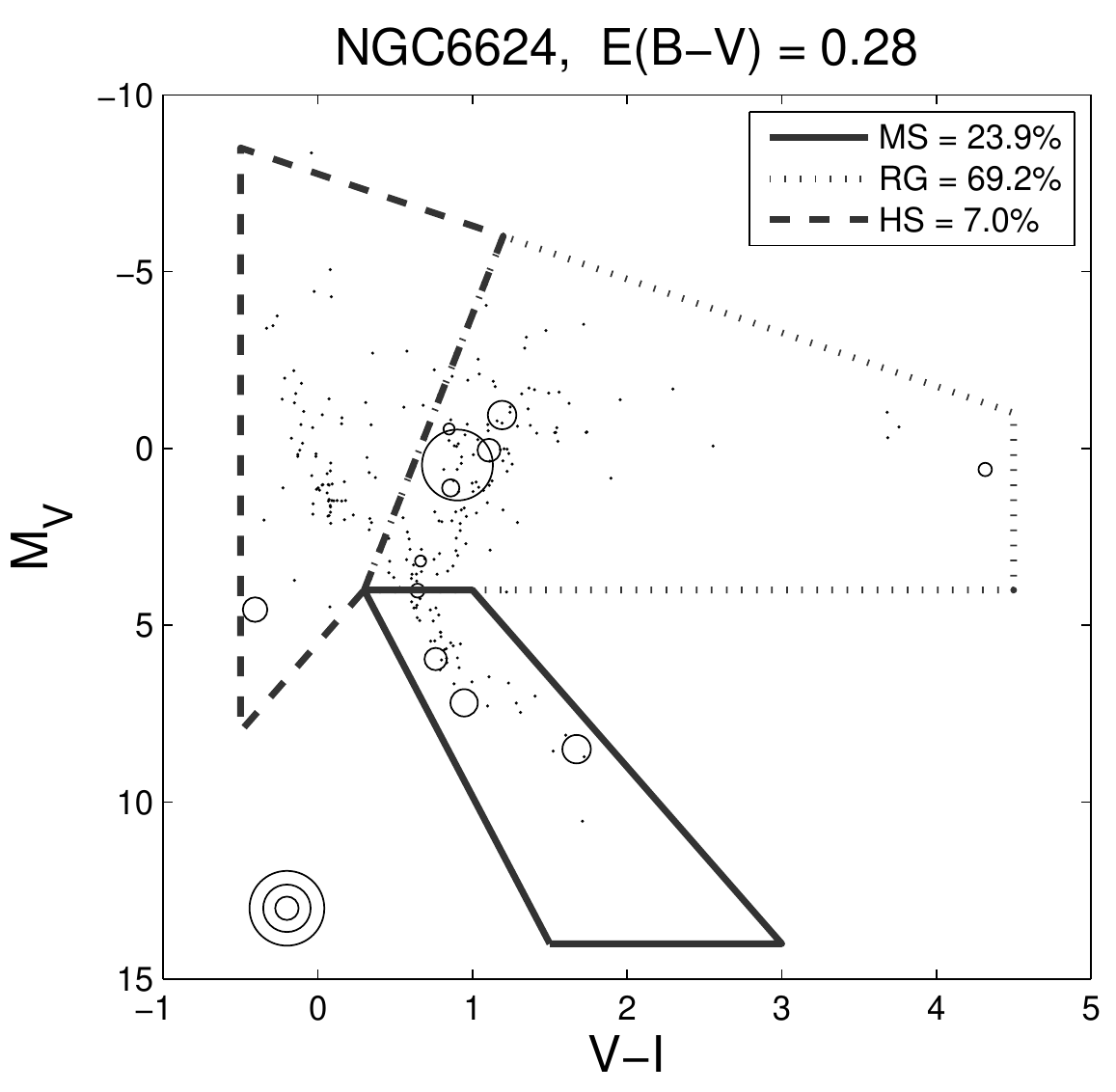}}
  \captcont{.../continued}
\end{figure}

\clearpage
\begin{figure}
  \centering
    \subfigure{
\includegraphics[width=0.75\textwidth]{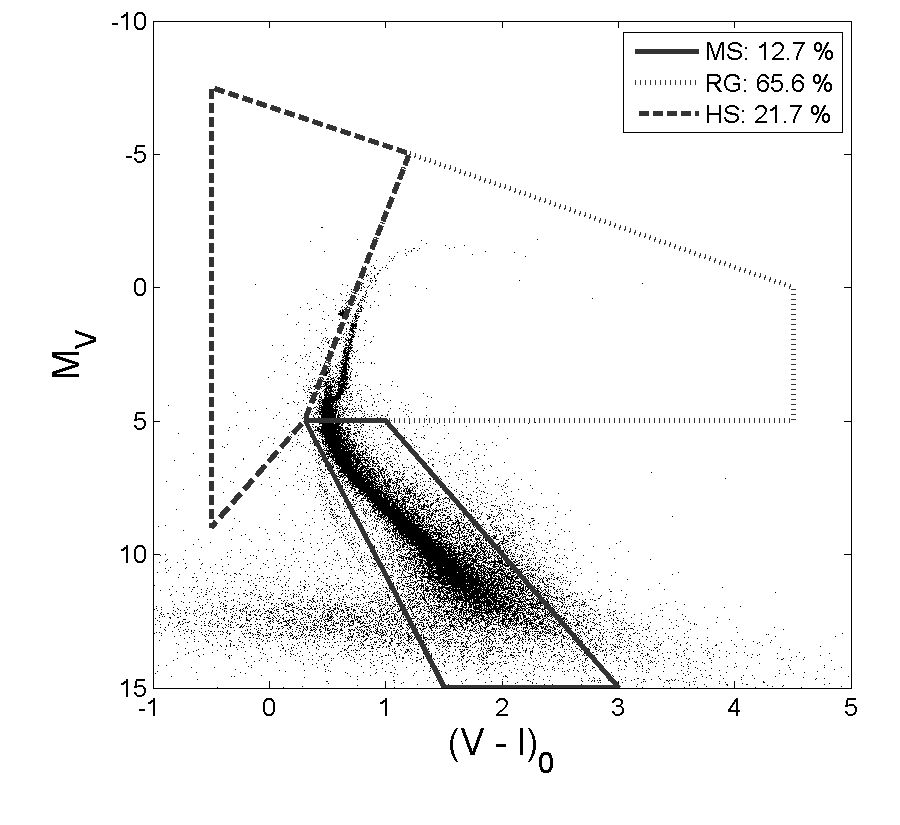}}
    \subfigure{
\includegraphics[width=0.7\textwidth]{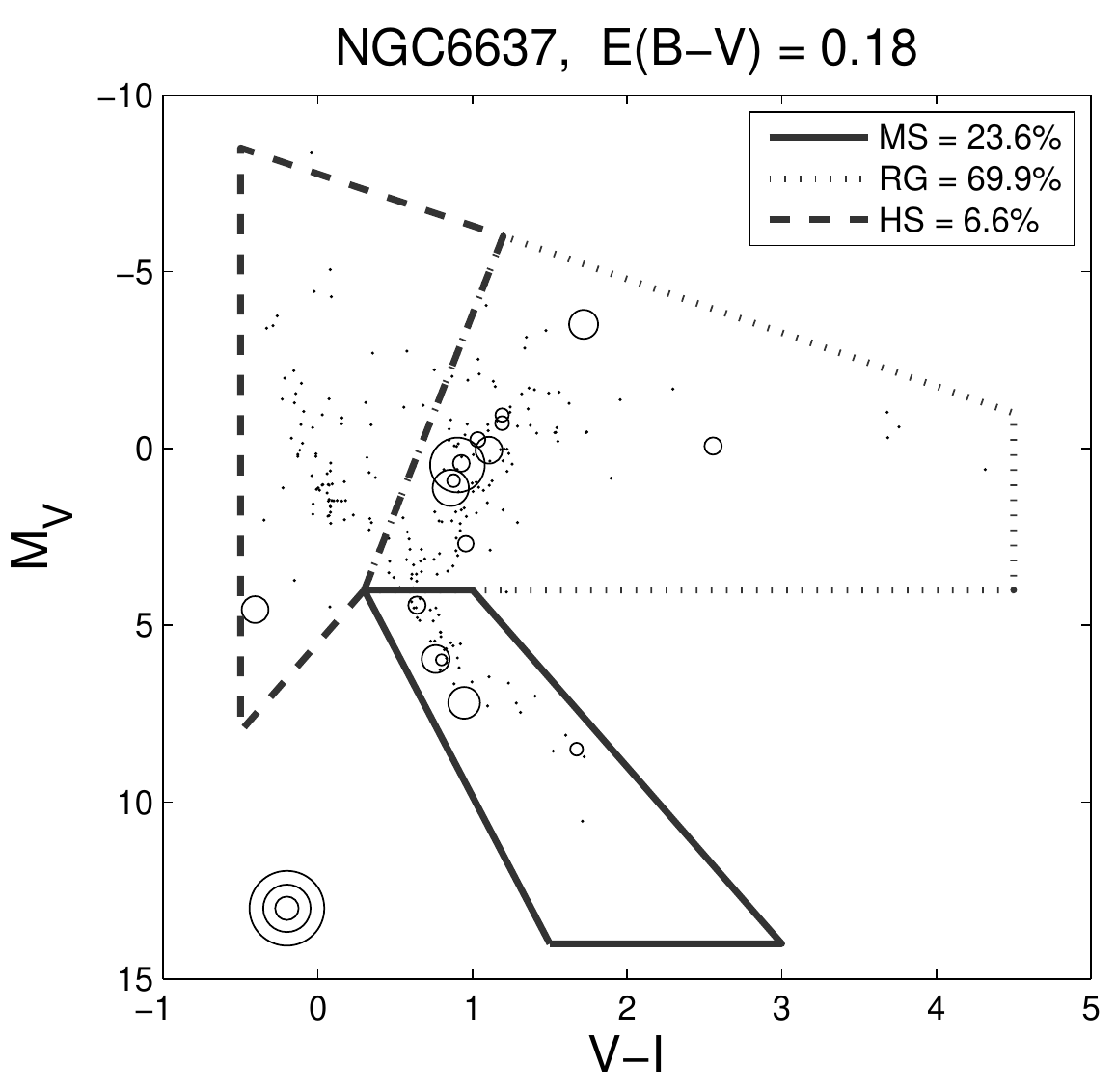}}
  \captcont{.../continued}
\end{figure}

\clearpage
\begin{figure}
  \centering
    \subfigure{
\includegraphics[width=0.75\textwidth]{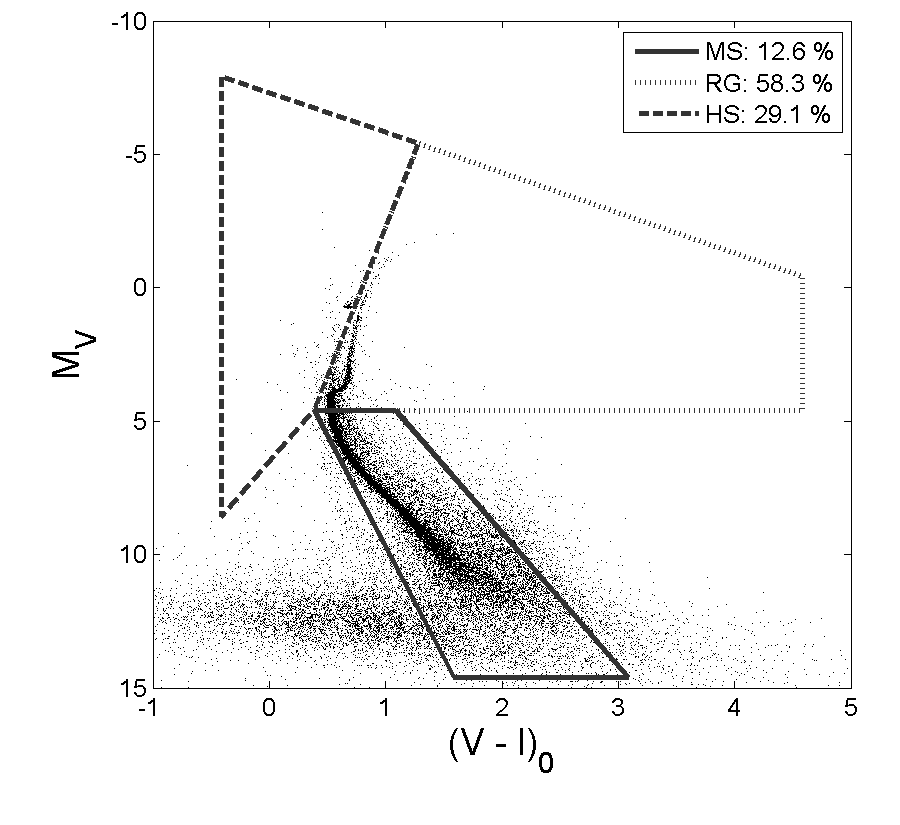}}
    \subfigure{
\includegraphics[width=0.7\textwidth]{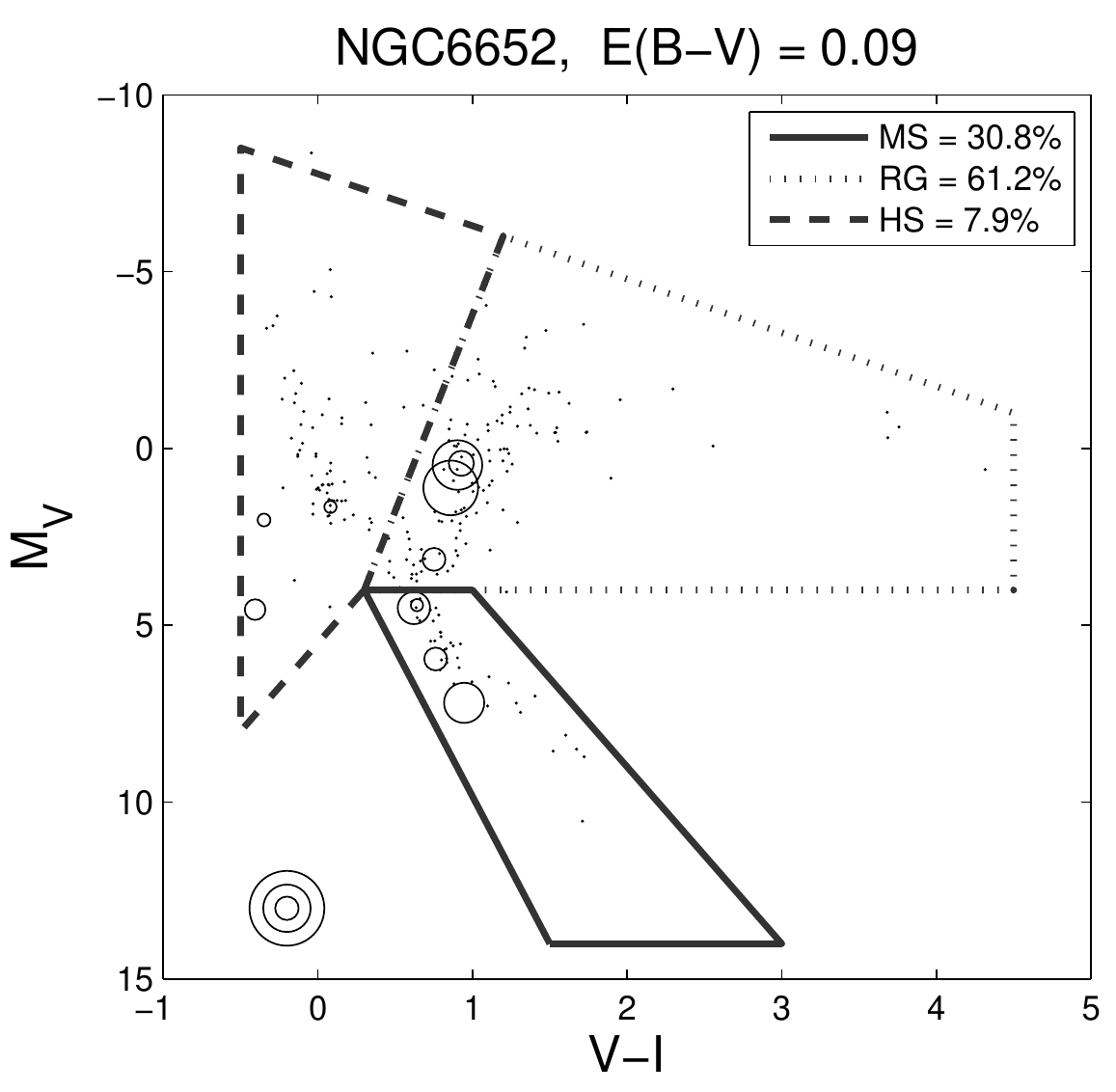}}
  \captcont{.../continued}
\end{figure}

\clearpage
\begin{figure}
  \centering
    \subfigure{
\includegraphics[width=0.75\textwidth]{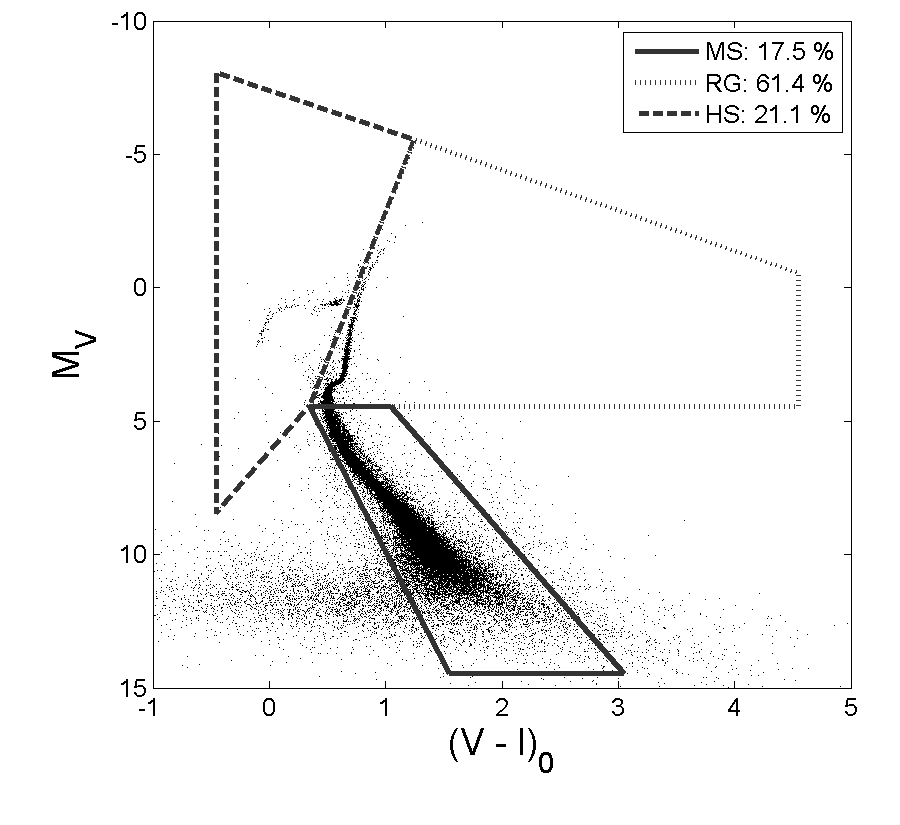}}
    \subfigure{
\includegraphics[width=0.7\textwidth]{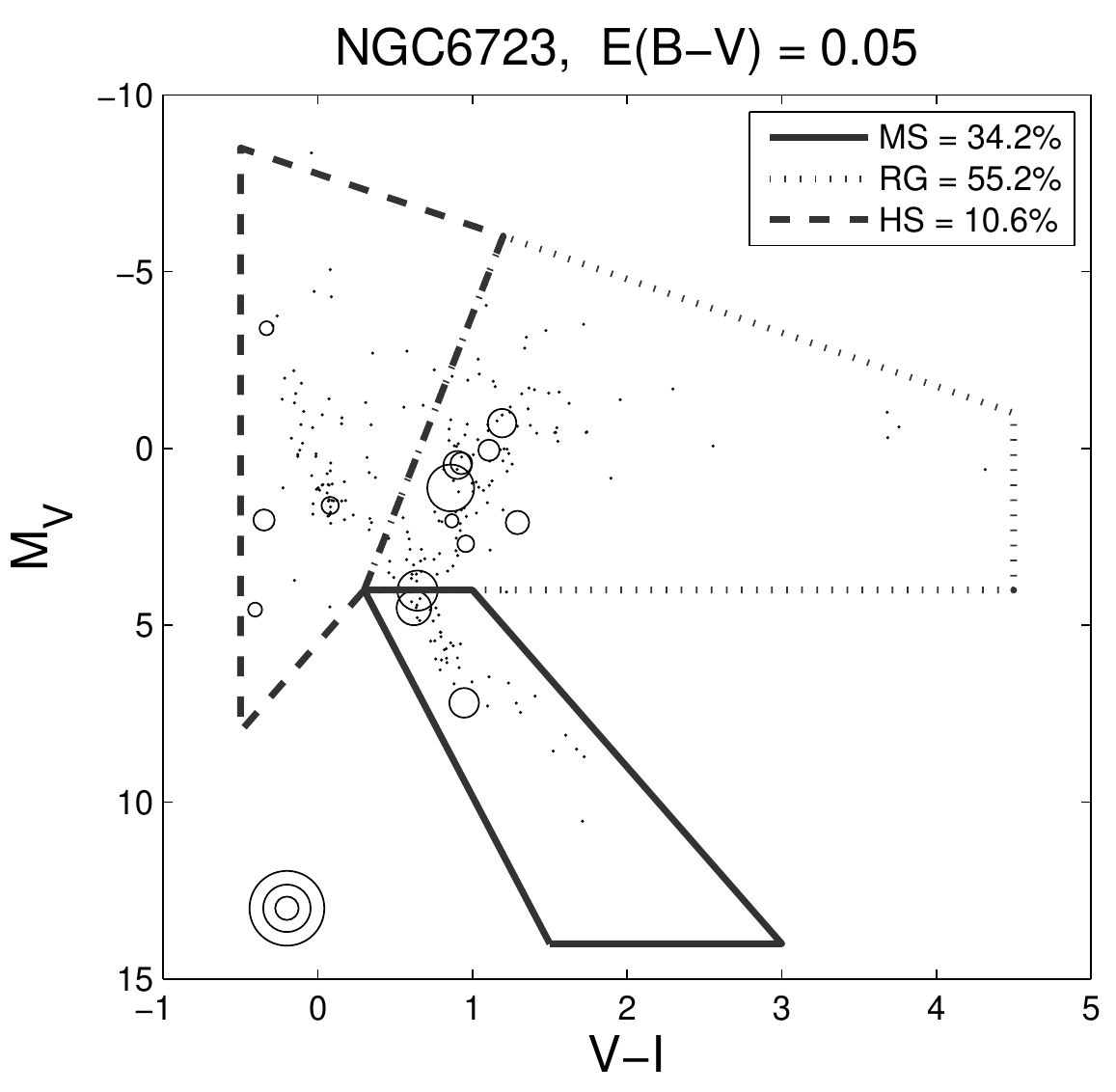}}
  \captcont{.../continued}
\end{figure}

\clearpage
\begin{figure}
  \centering
    \subfigure{
\includegraphics[width=0.75\textwidth]{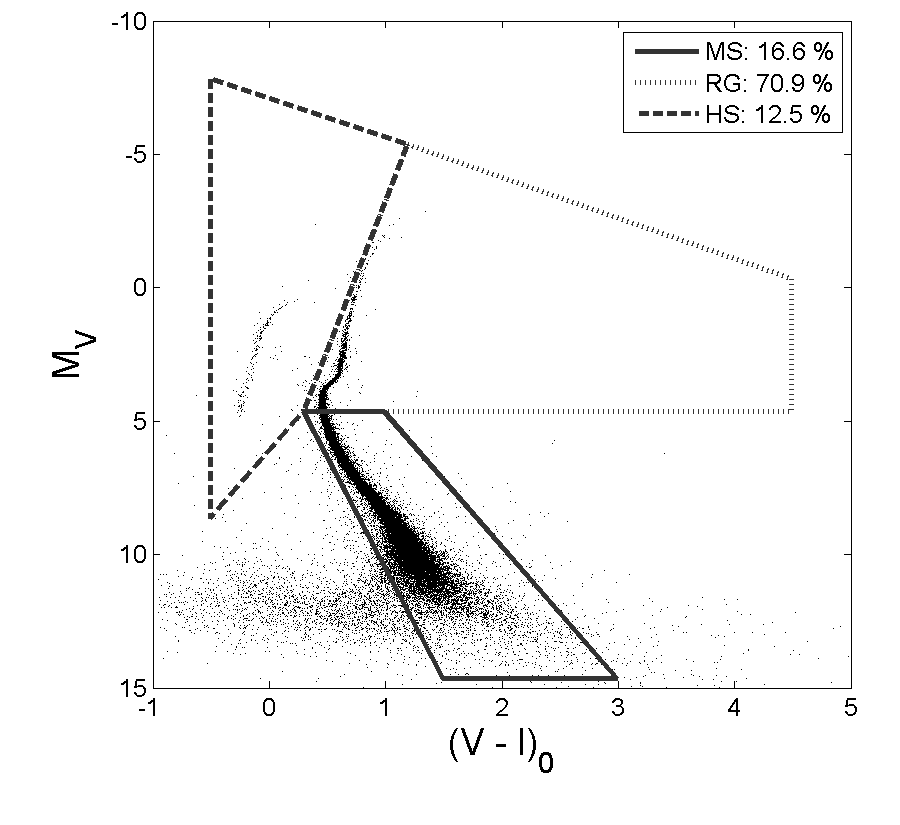}}
    \subfigure{
\includegraphics[width=0.7\textwidth]{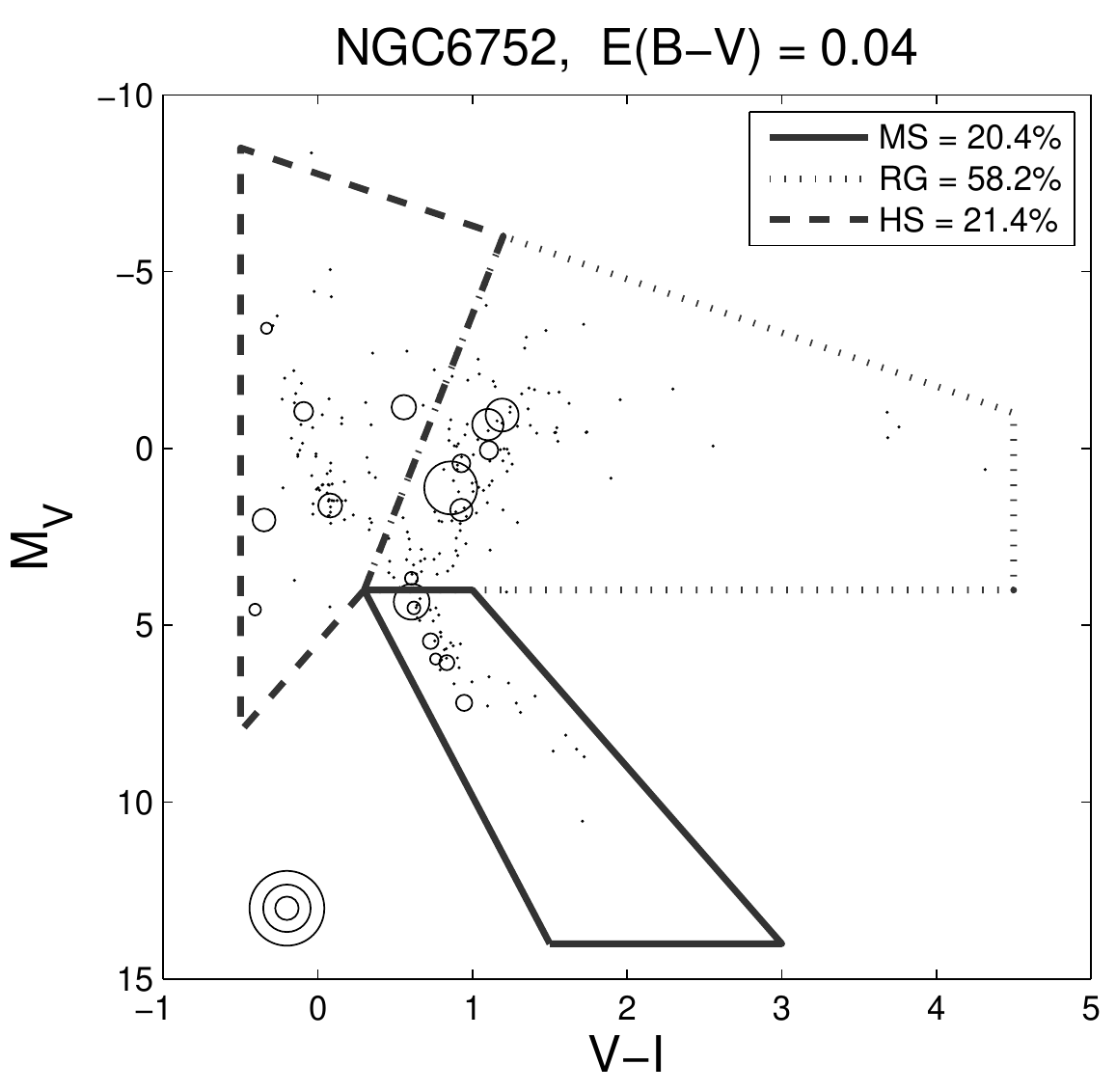}}
  \captcont{.../continued}
\end{figure}

\clearpage
\begin{figure}
  \centering
    \subfigure{
\includegraphics[width=0.75\textwidth]{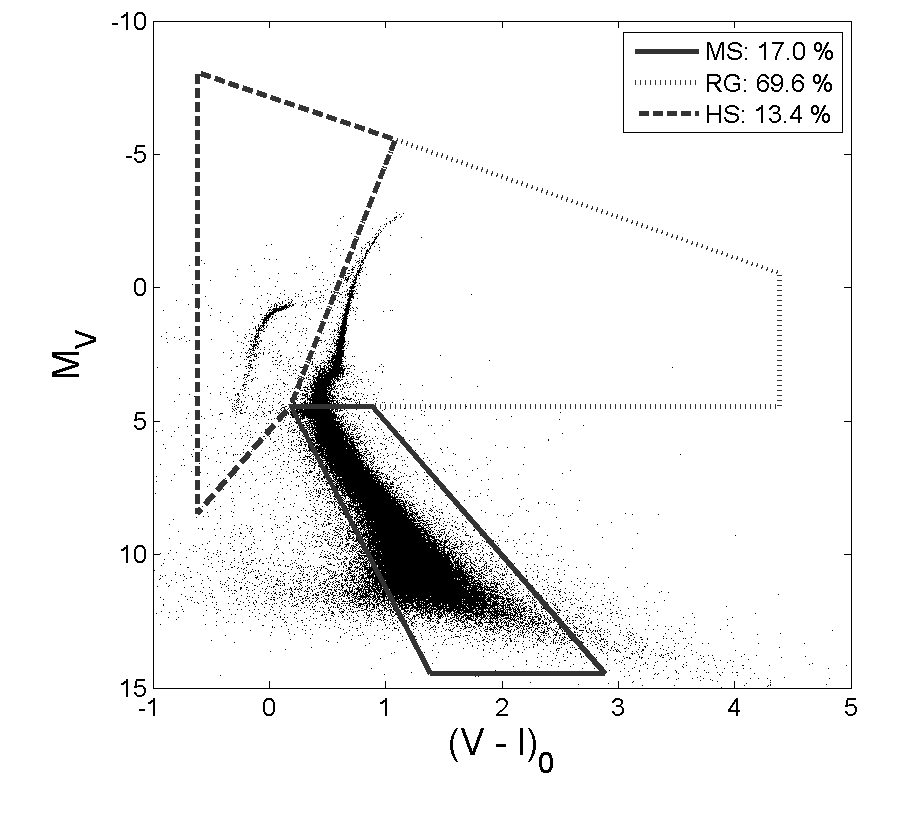}}
    \subfigure{
\includegraphics[width=0.7\textwidth]{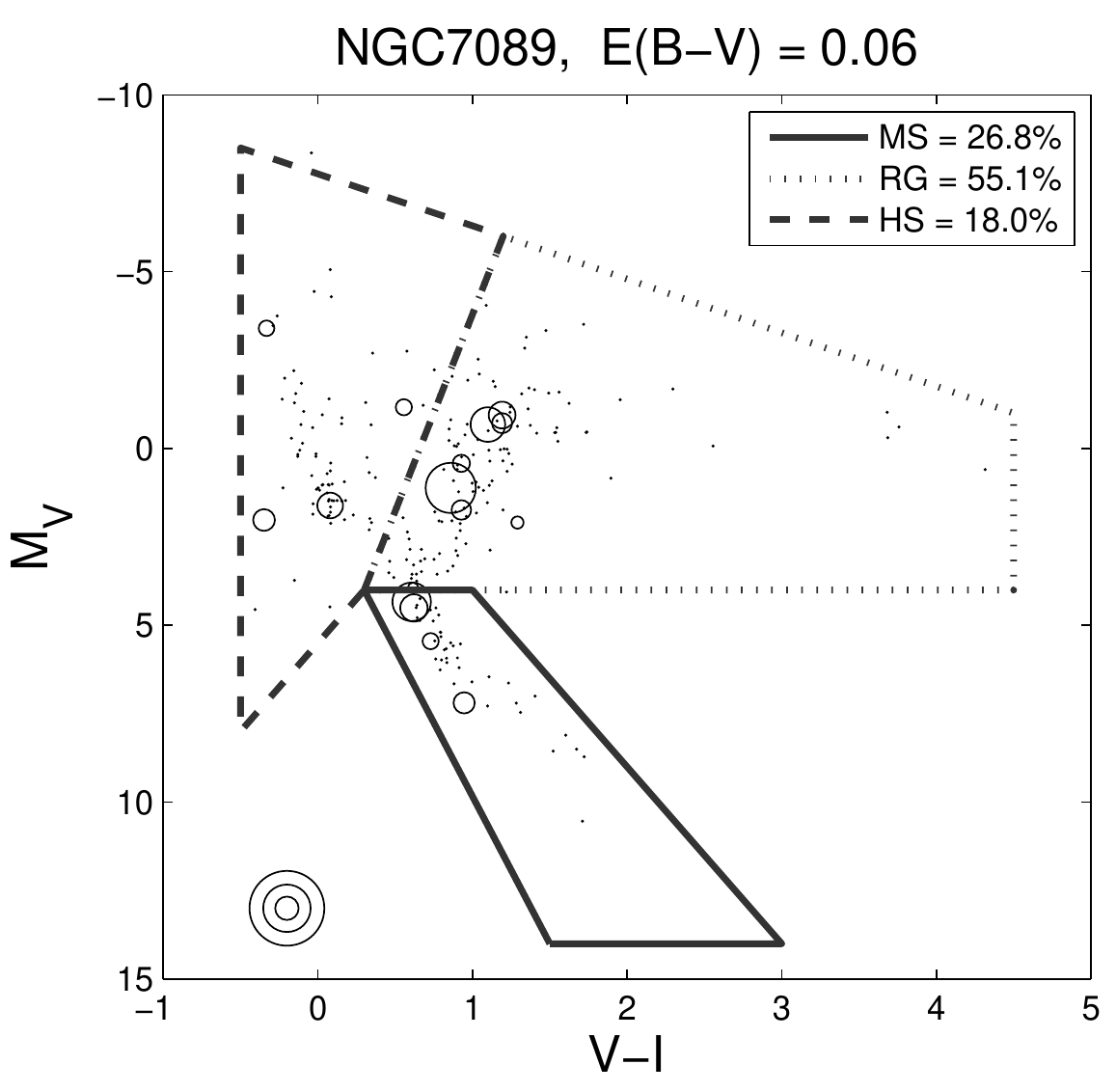}}
  \caption{.../continued}
\end{figure}

\clearpage

\section{APPENDIX D}\label{sec:AppendixD}

Here we present the ACS colour-magnitude diagrams for GGCs as
in \ap{AppendixC}, except that our evolutionary zones are now
more precisely defined.

\begin{figure}[H]
  \centering
\includegraphics[width=0.65\textwidth]{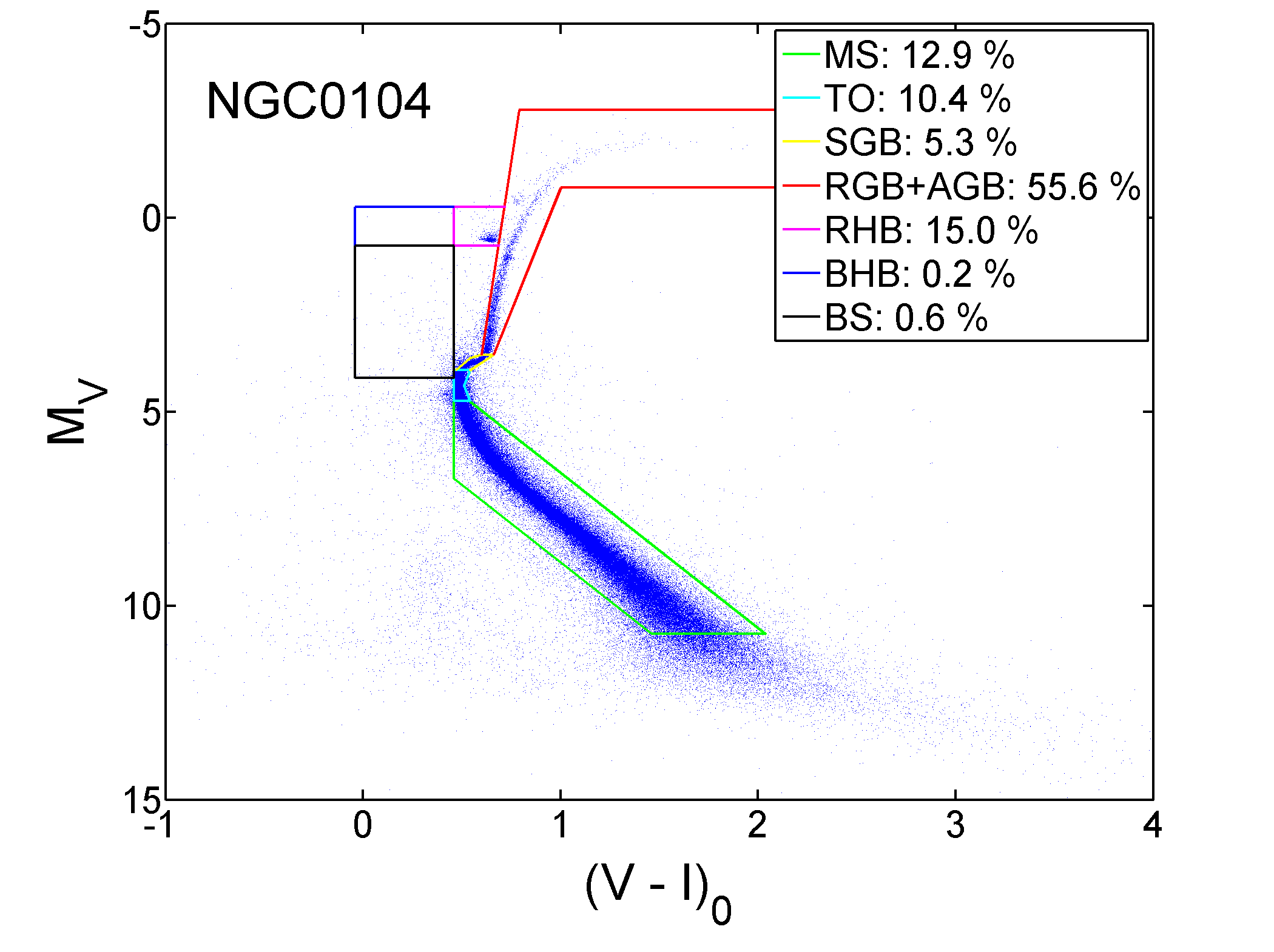}
  \captcont{Galactic globular cluster colour-magnitude diagrams.}
  \label{fig:CMDsReal}
\end{figure}

\begin{figure}[H]
  \centering
\includegraphics[width=0.65\textwidth]{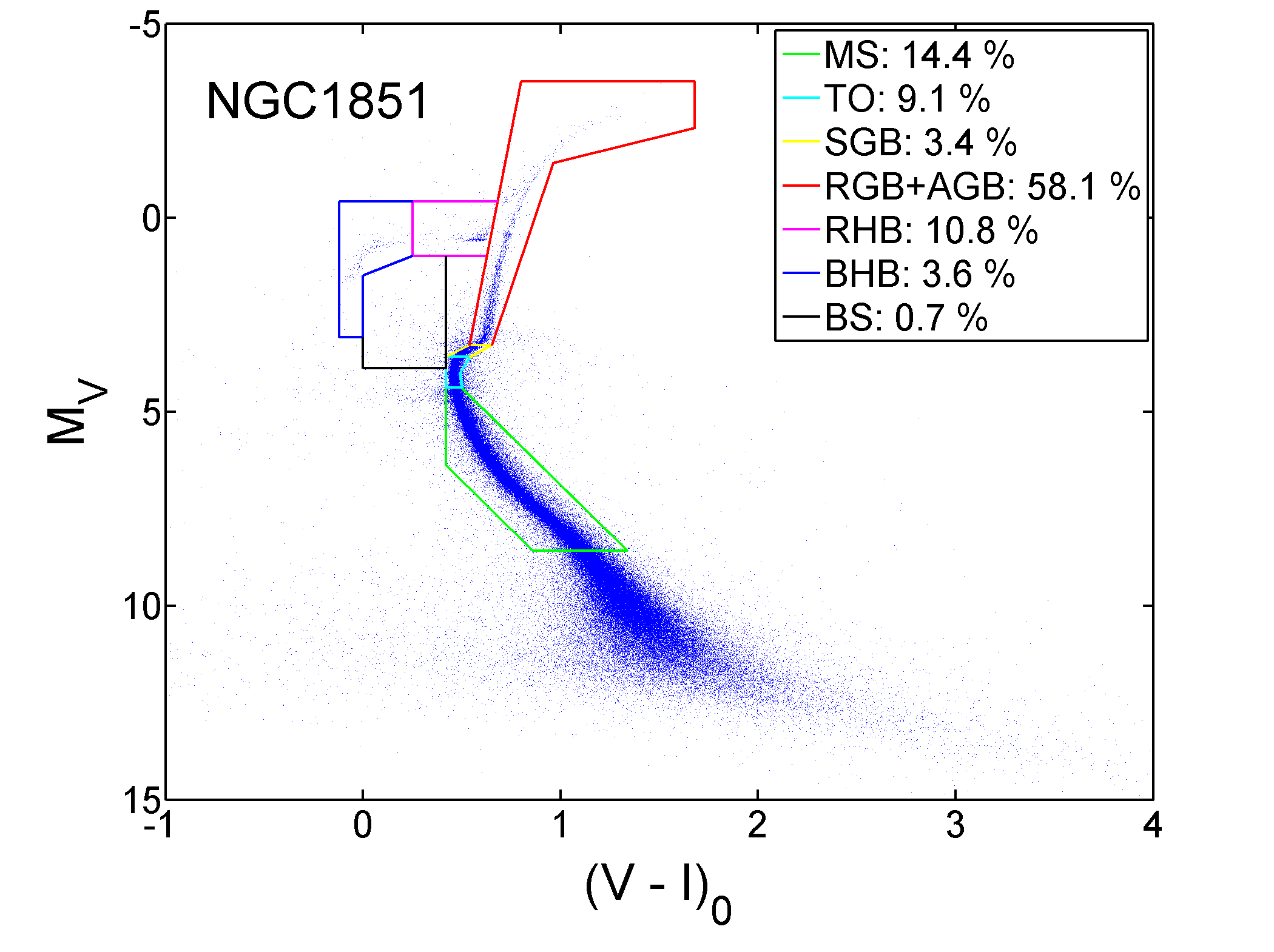}
  \captcont{.../continued}
\end{figure}

\clearpage
\begin{figure}
  \centering
\includegraphics[width=0.7\textwidth]{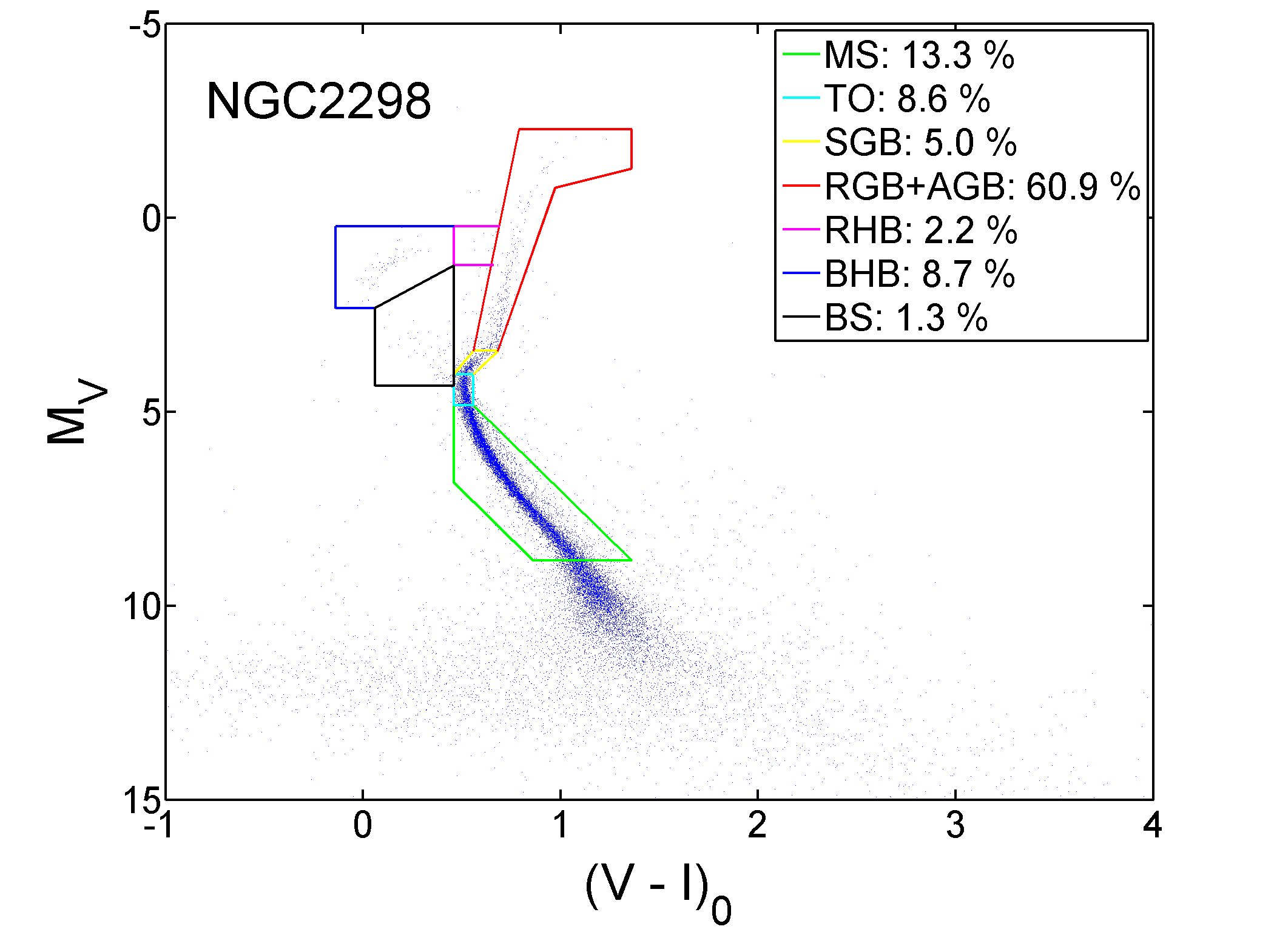}
  \captcont{.../continued}
\end{figure}

\begin{figure}
  \centering
\includegraphics[width=0.7\textwidth]{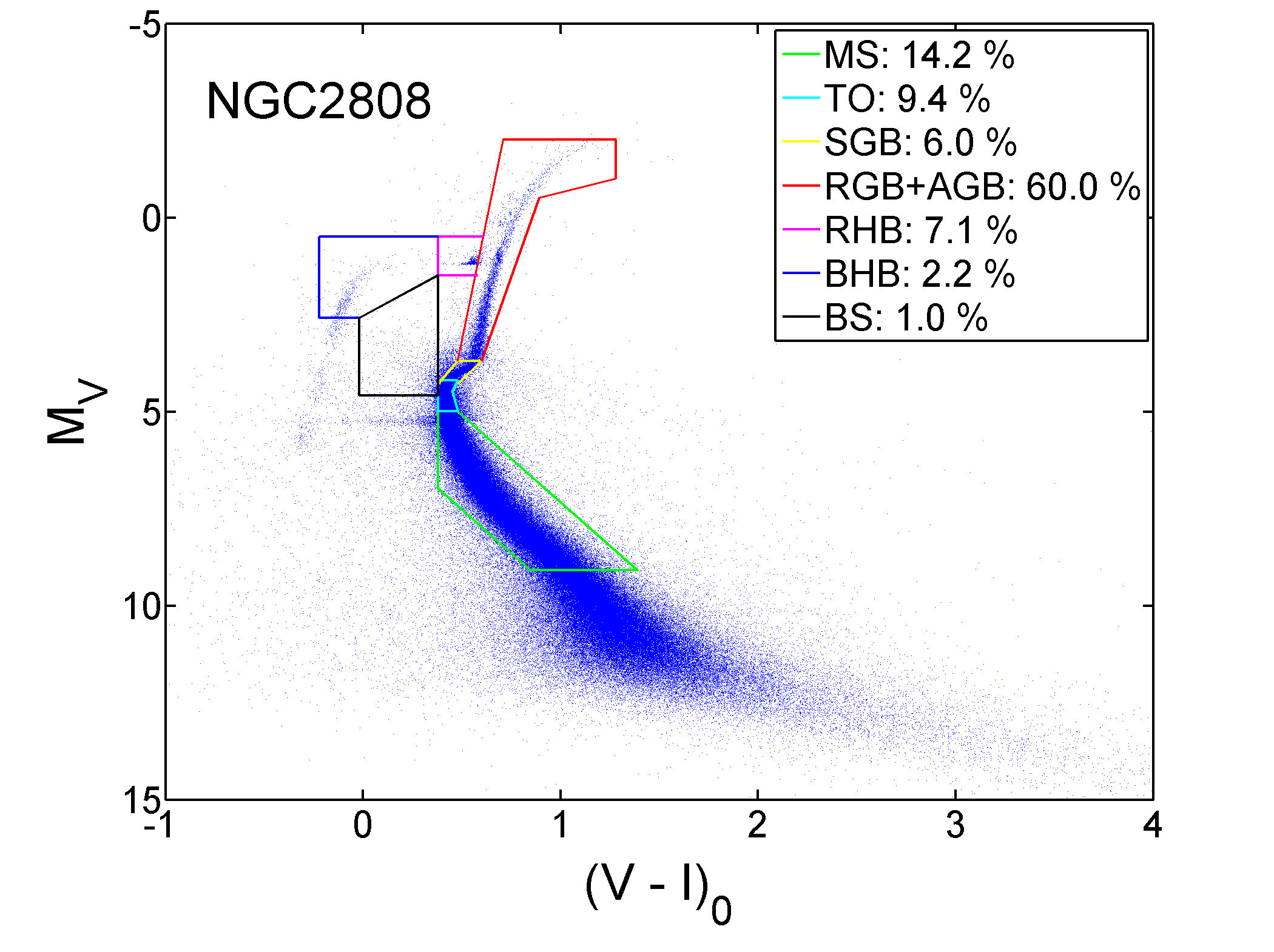}
  \captcont{.../continued}
\end{figure}

\clearpage
\begin{figure}
  \centering
\includegraphics[width=0.7\textwidth]{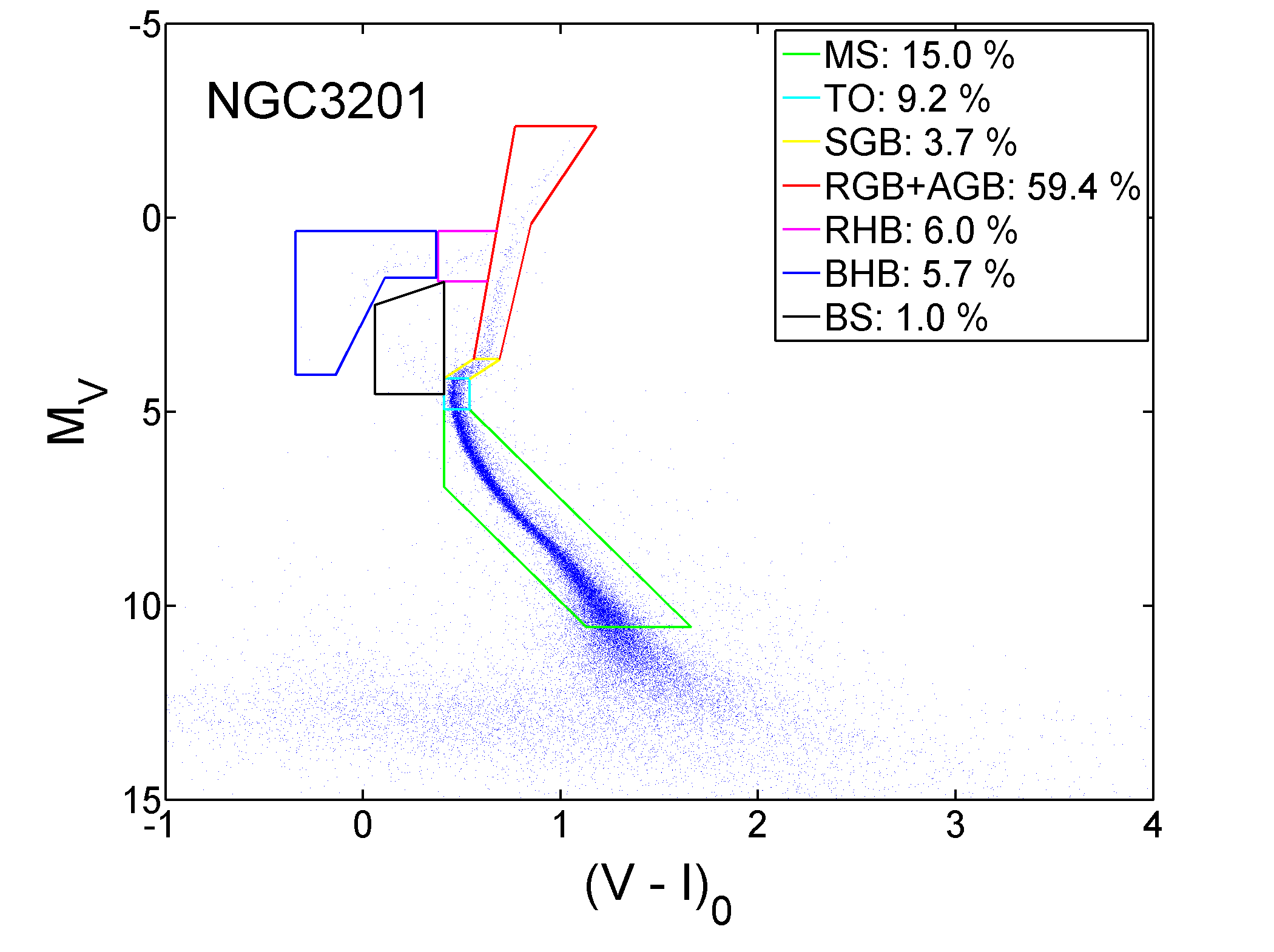}
  \captcont{.../continued}
\end{figure}

\begin{figure}
  \centering
\includegraphics[width=0.7\textwidth]{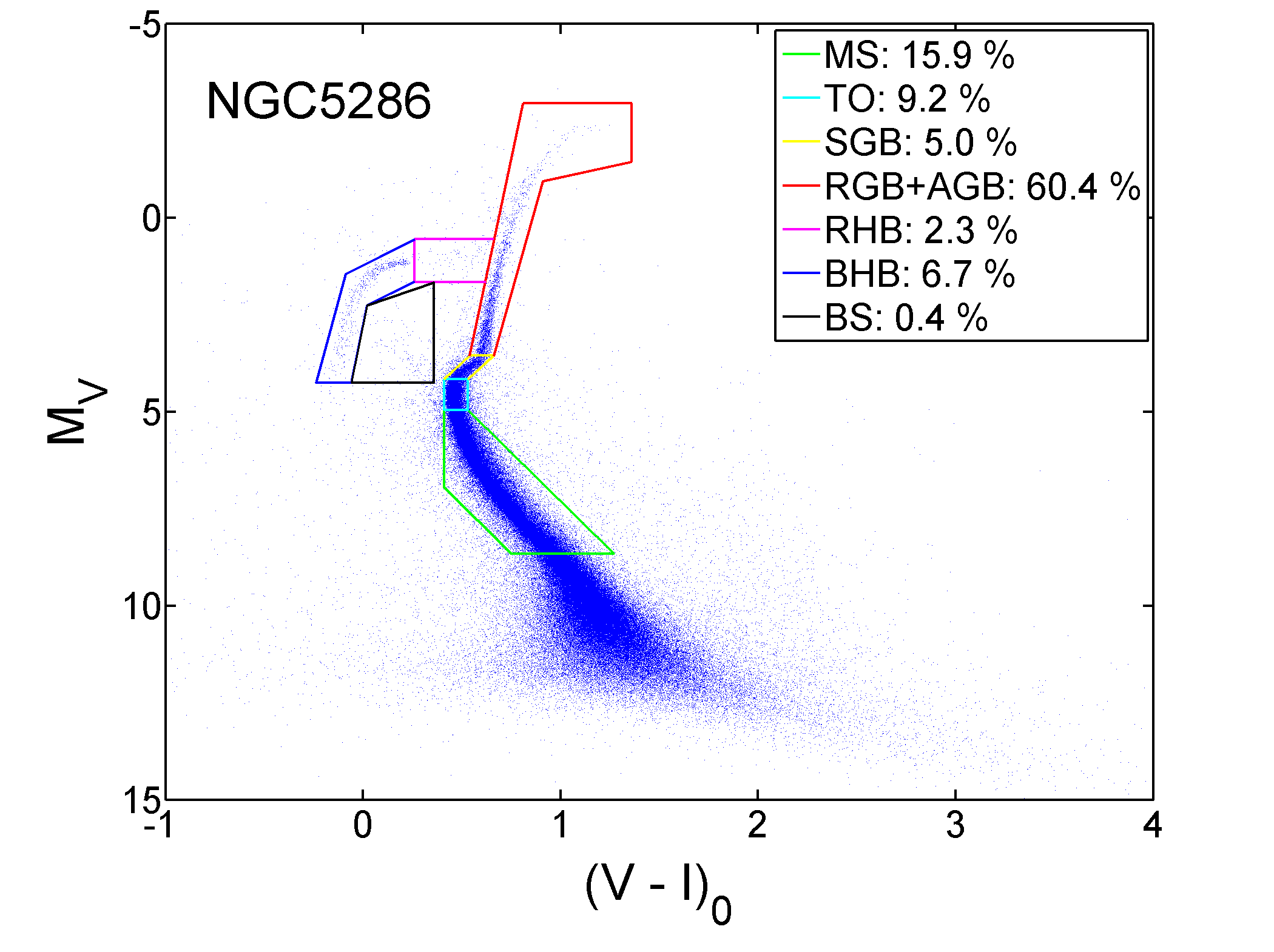}
  \captcont{.../continued}
\end{figure}

\clearpage
\begin{figure}
  \centering
\includegraphics[width=0.7\textwidth]{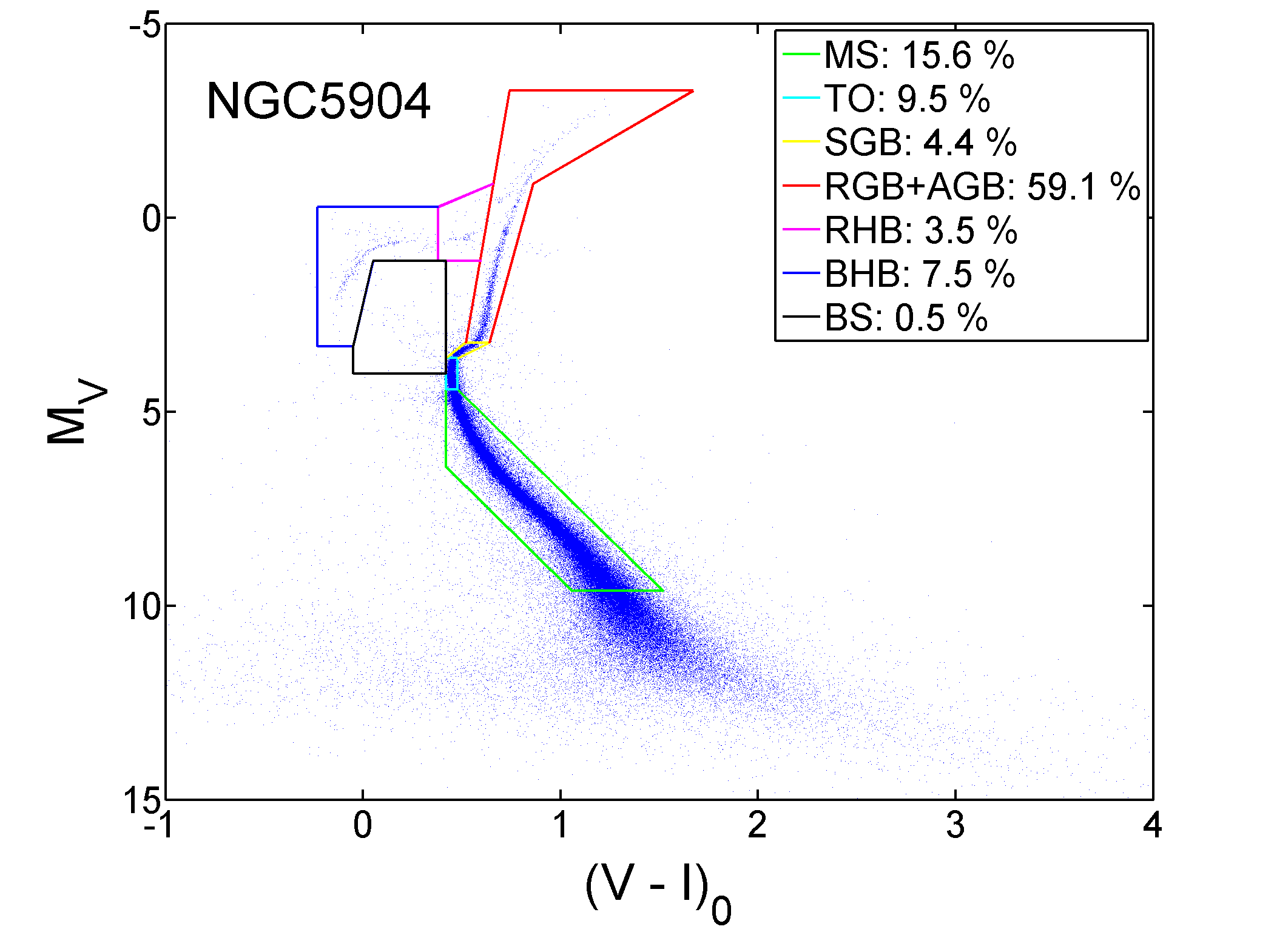}
  \captcont{.../continued}
\end{figure}

\begin{figure}
  \centering
\includegraphics[width=0.7\textwidth]{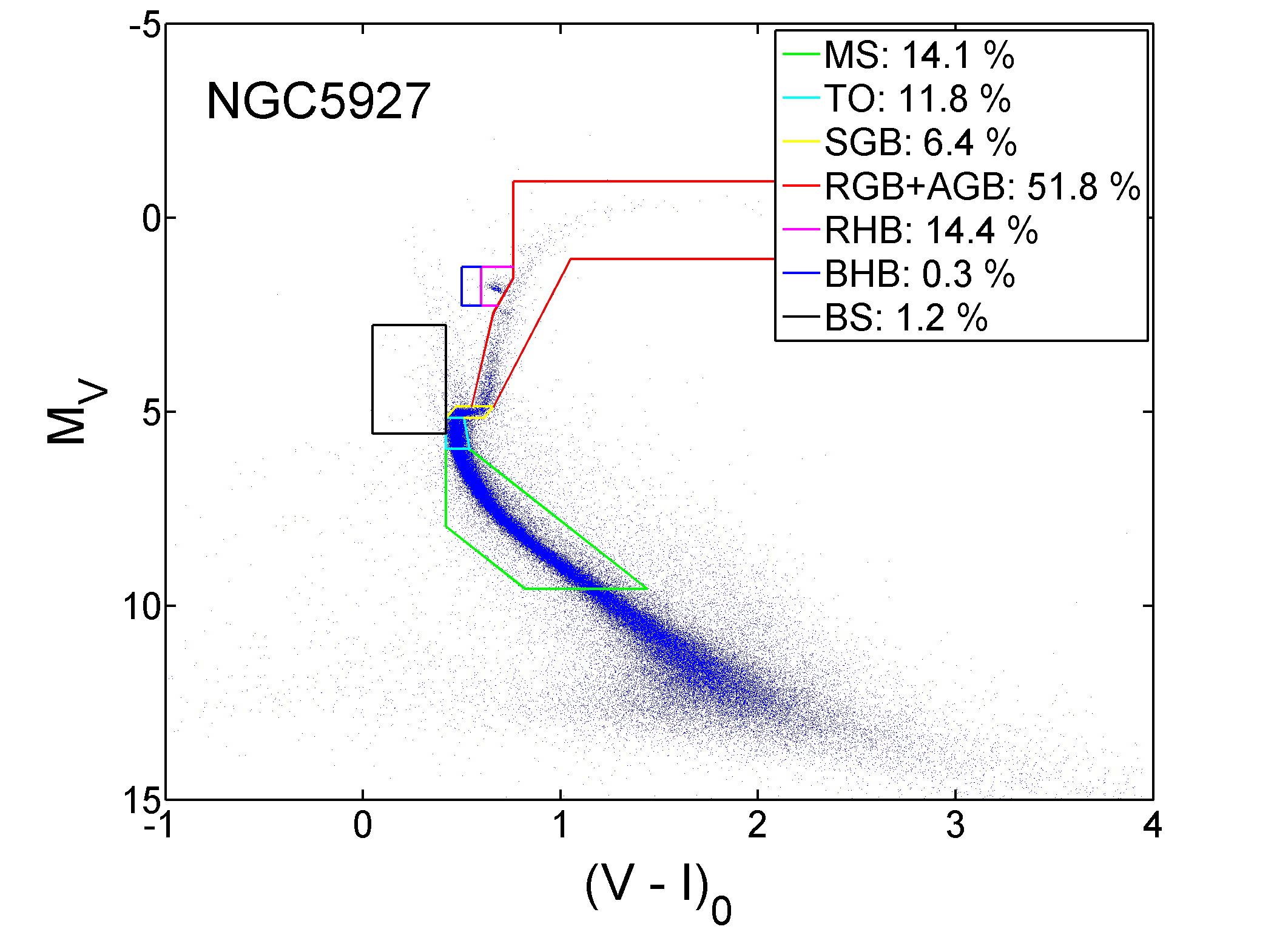}
  \captcont{.../continued}
\end{figure}

\clearpage
\begin{figure}
  \centering
\includegraphics[width=0.7\textwidth]{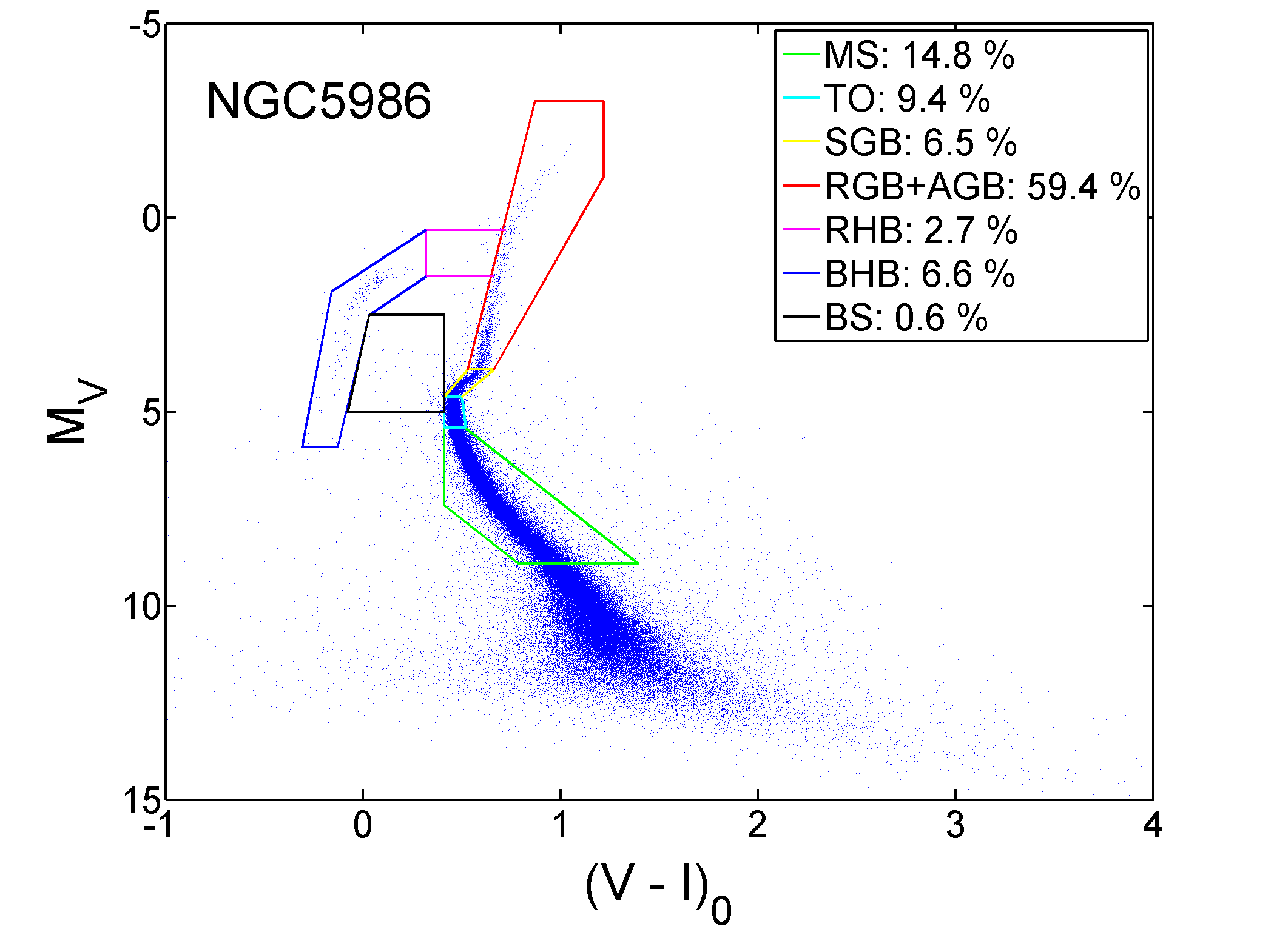}
  \captcont{.../continued}
\end{figure}

\begin{figure}
  \centering
\includegraphics[width=0.7\textwidth]{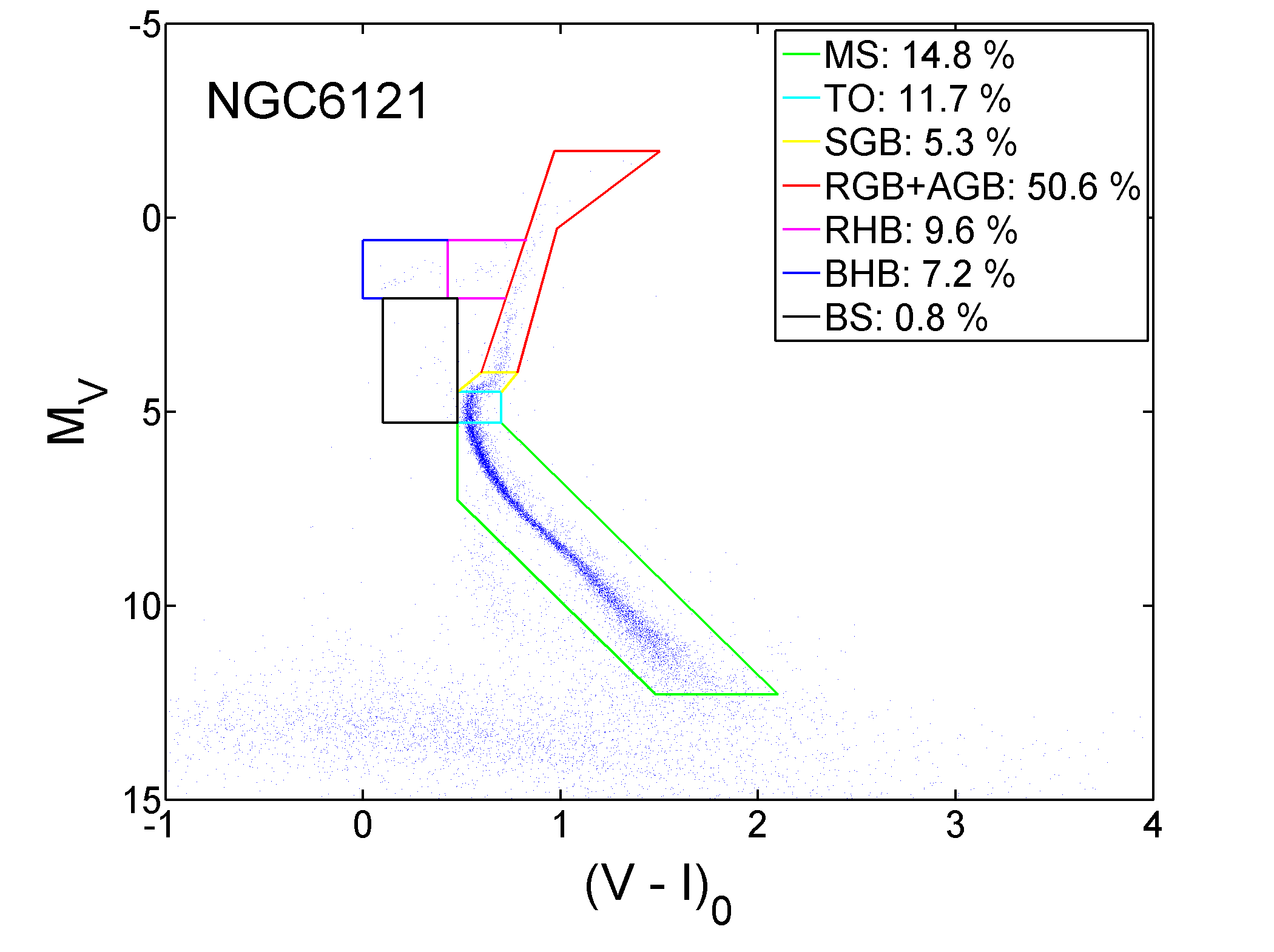}
  \captcont{.../continued}
\end{figure}

\clearpage
\begin{figure}
  \centering
\includegraphics[width=0.7\textwidth]{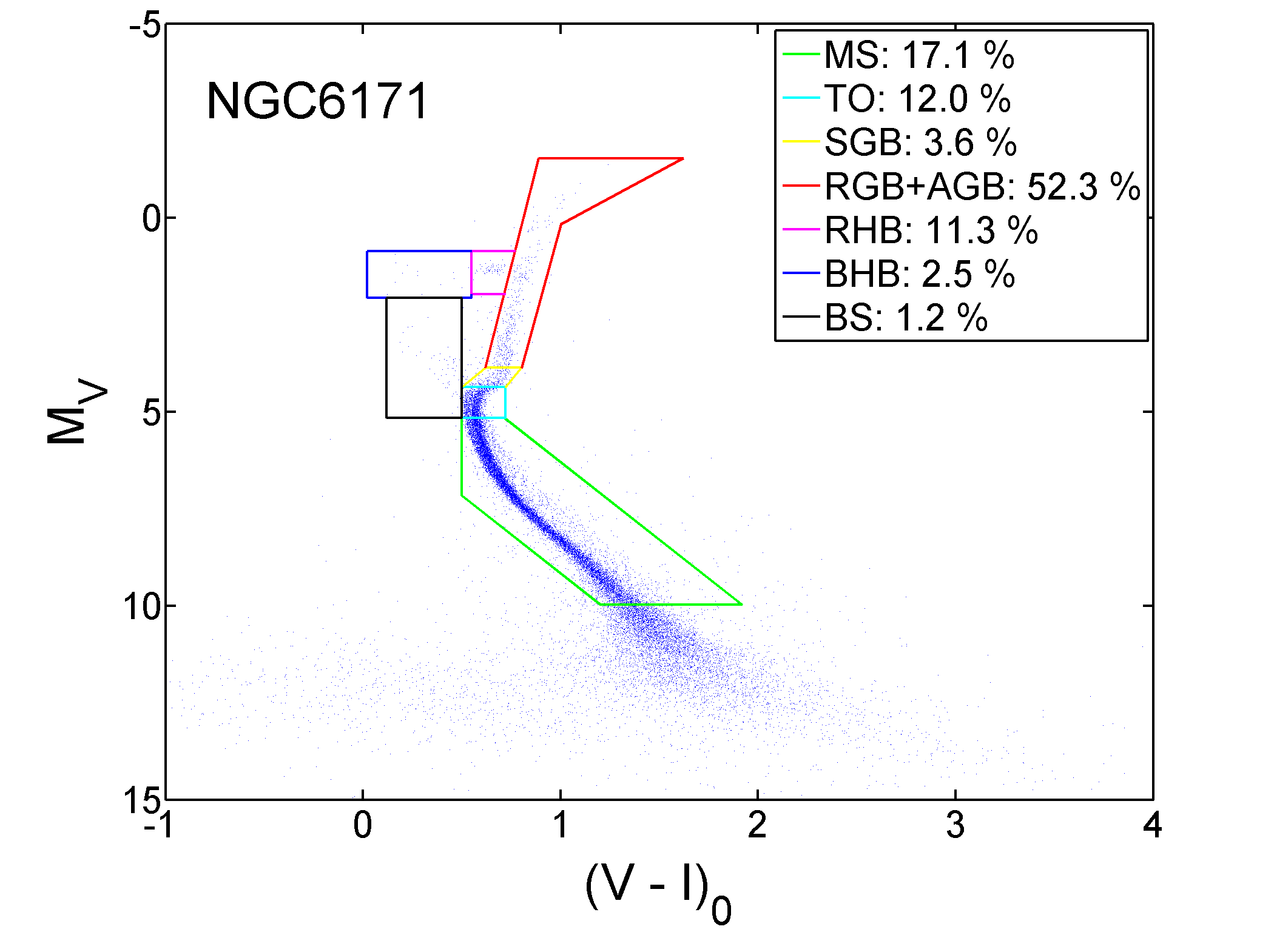}
  \captcont{.../continued}
\end{figure}

\begin{figure}
  \centering
\includegraphics[width=0.7\textwidth]{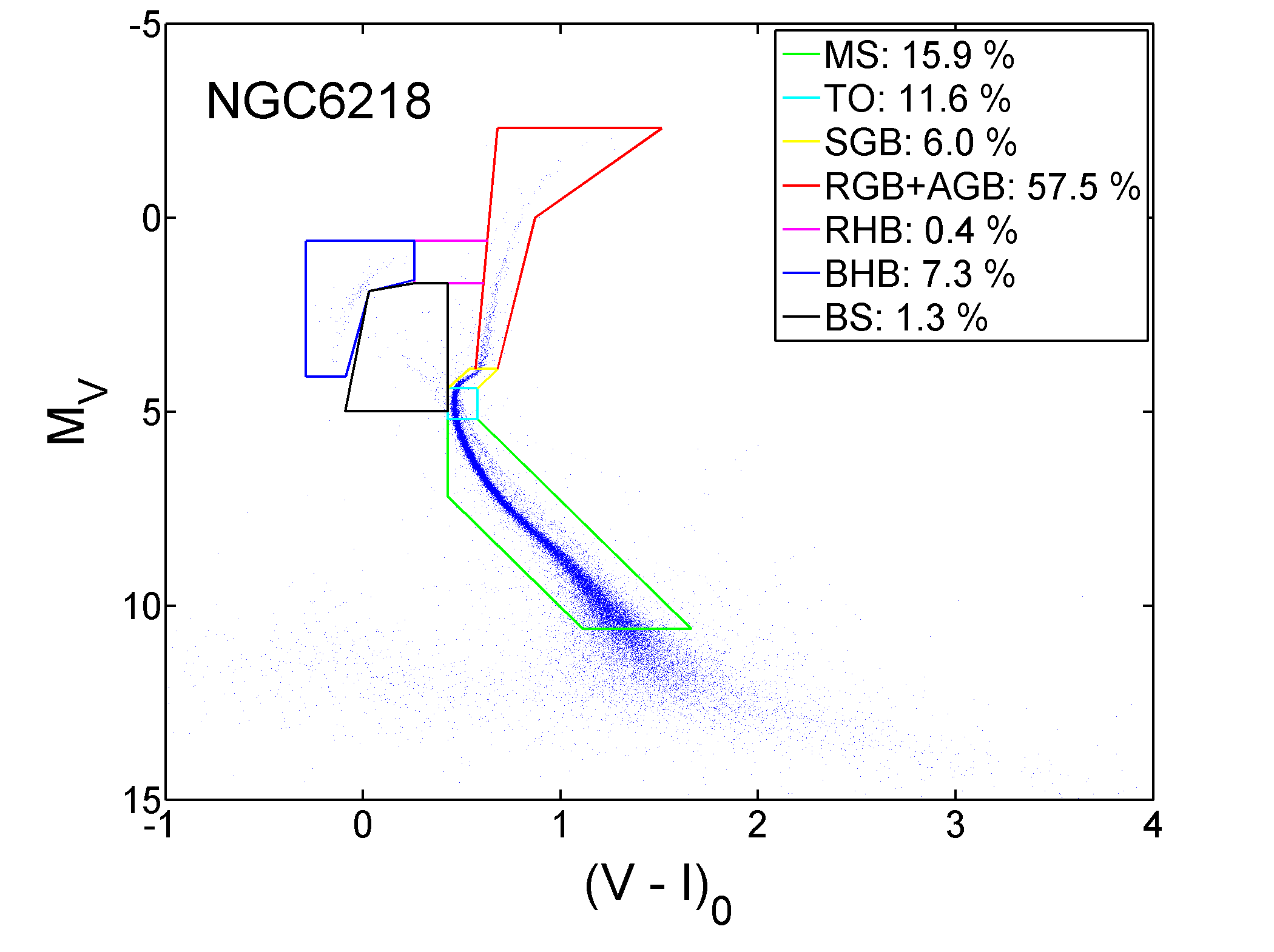}
  \captcont{.../continued}
\end{figure}

\clearpage
\begin{figure}
  \centering
\includegraphics[width=0.7\textwidth]{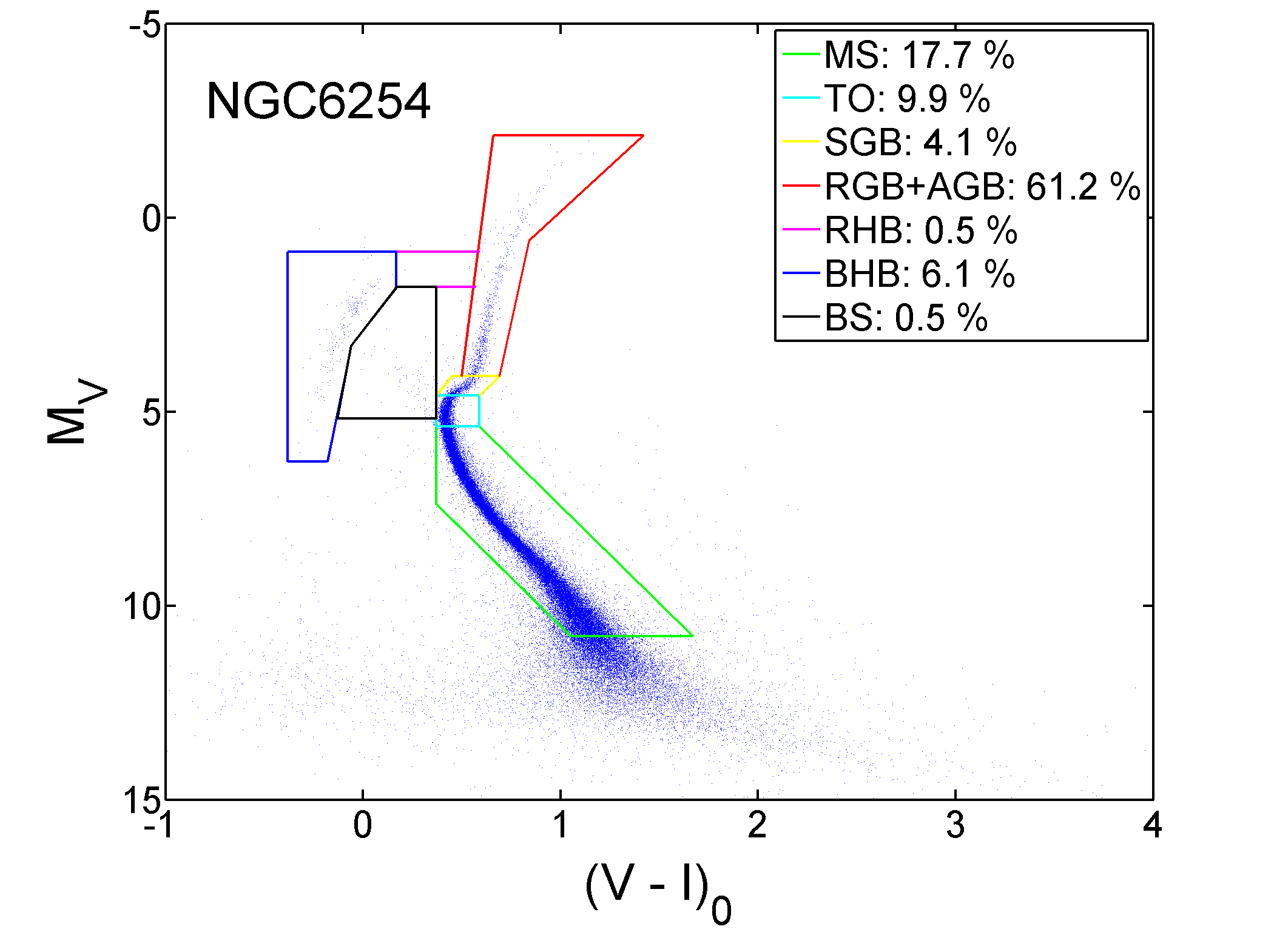}
  \captcont{.../continued}
\end{figure}

\begin{figure}
  \centering
\includegraphics[width=0.7\textwidth]{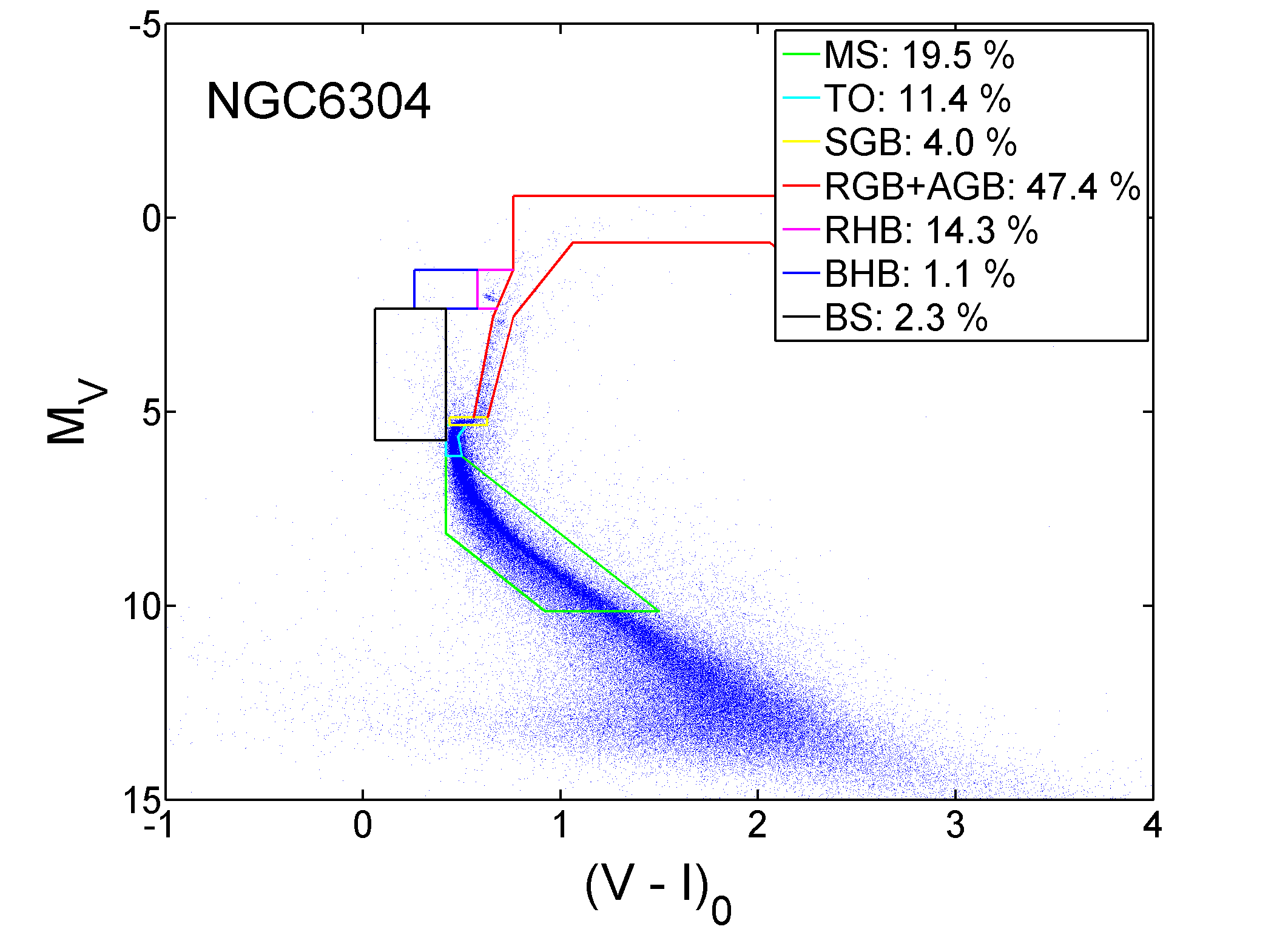}
  \captcont{.../continued}
\end{figure}

\clearpage
\begin{figure}
  \centering
\includegraphics[width=0.7\textwidth]{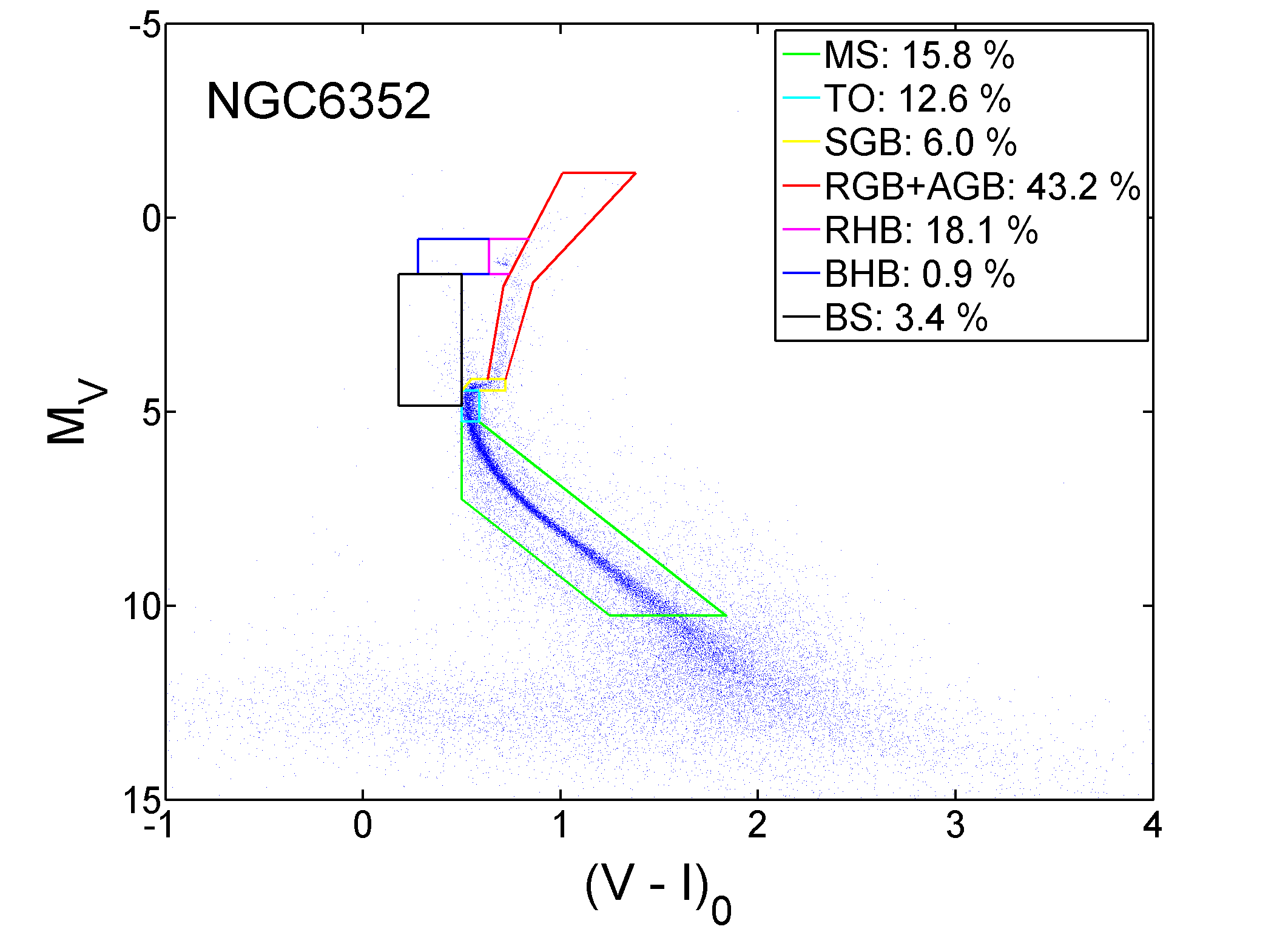}
  \captcont{.../continued}
\end{figure}

\begin{figure}
  \centering
\includegraphics[width=0.7\textwidth]{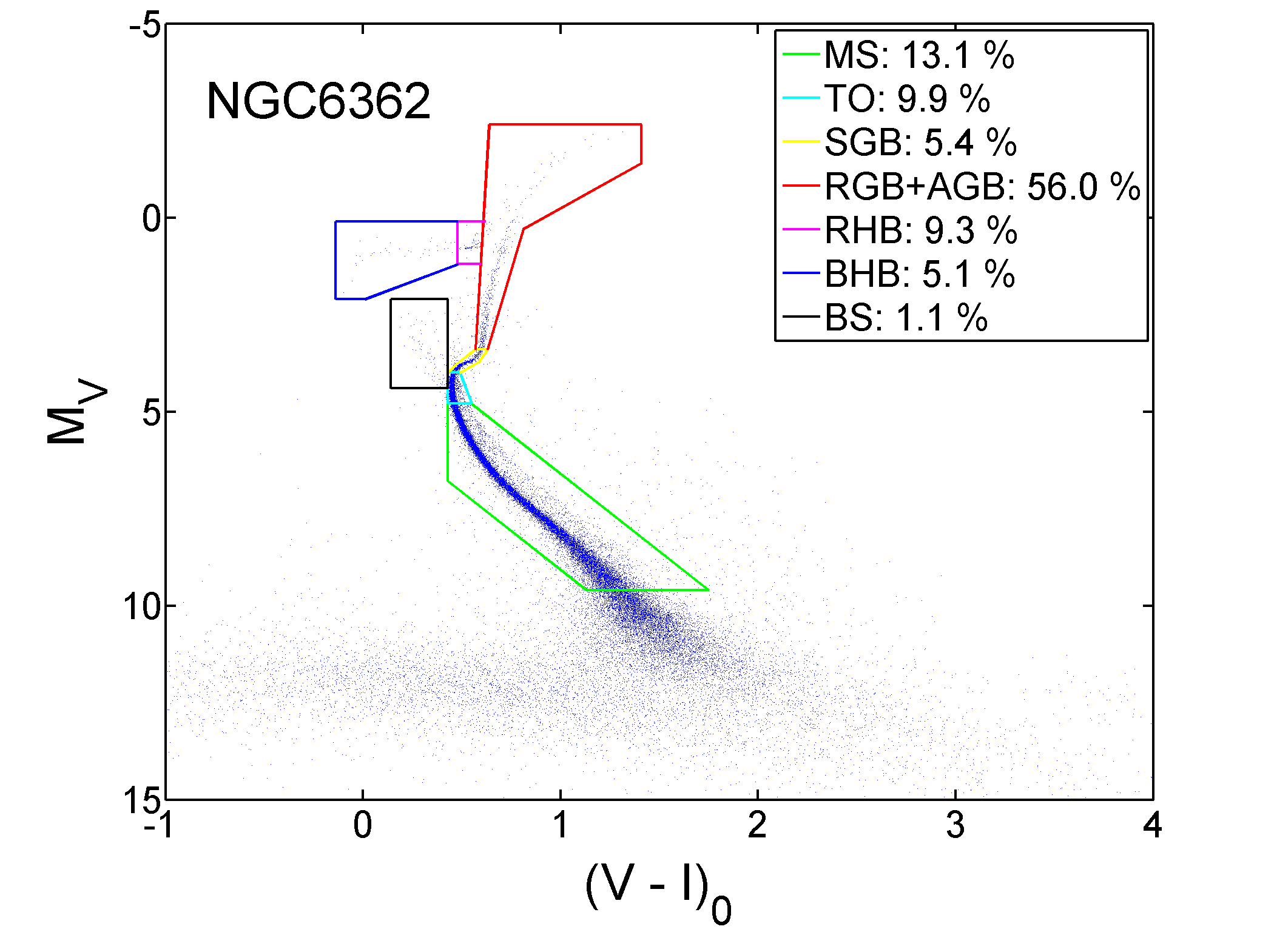}
  \captcont{.../continued}
\end{figure}

\clearpage
\begin{figure}
  \centering
\includegraphics[width=0.7\textwidth]{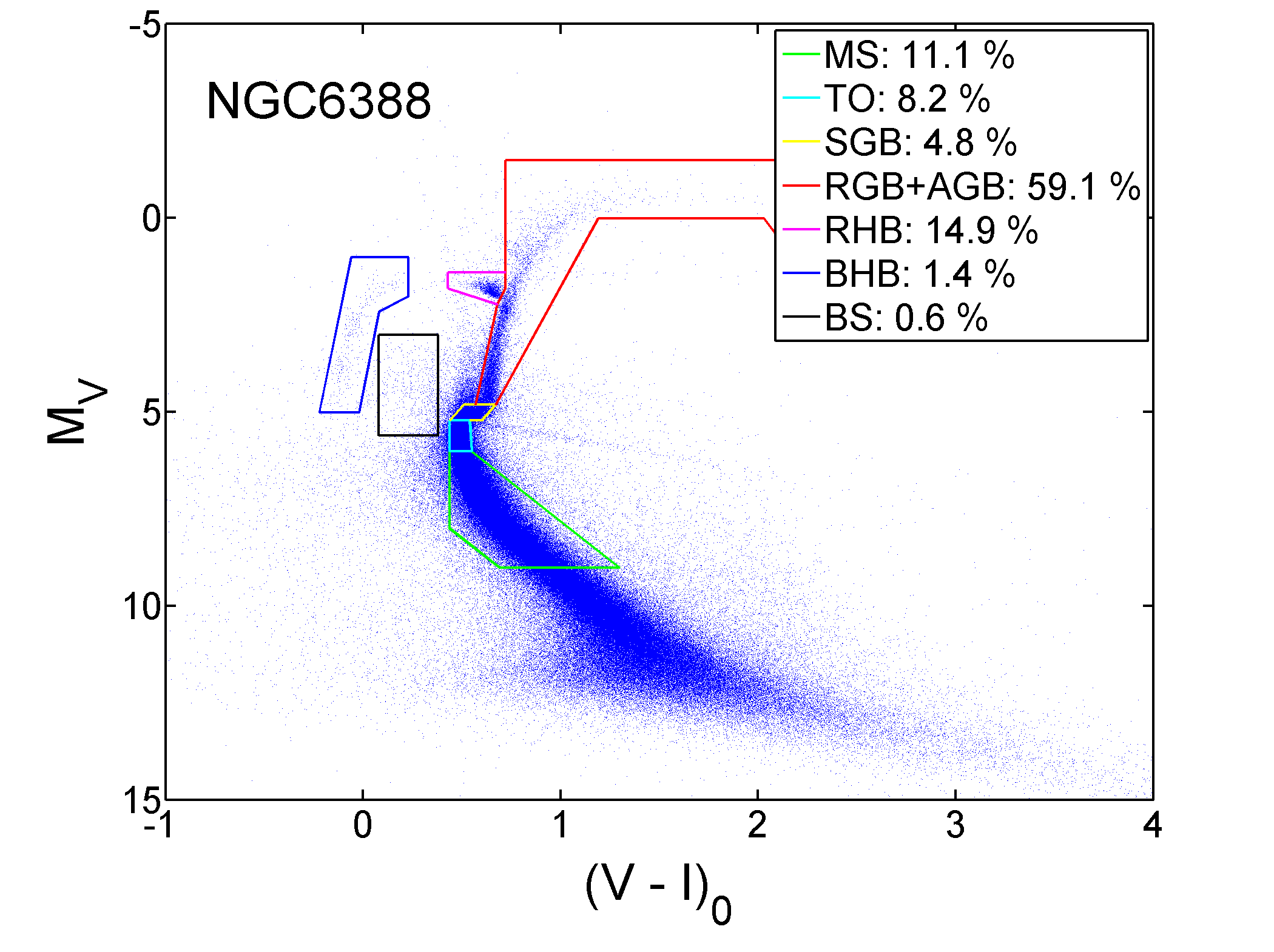}
  \captcont{.../continued}
\end{figure}

\begin{figure}
  \centering
\includegraphics[width=0.7\textwidth]{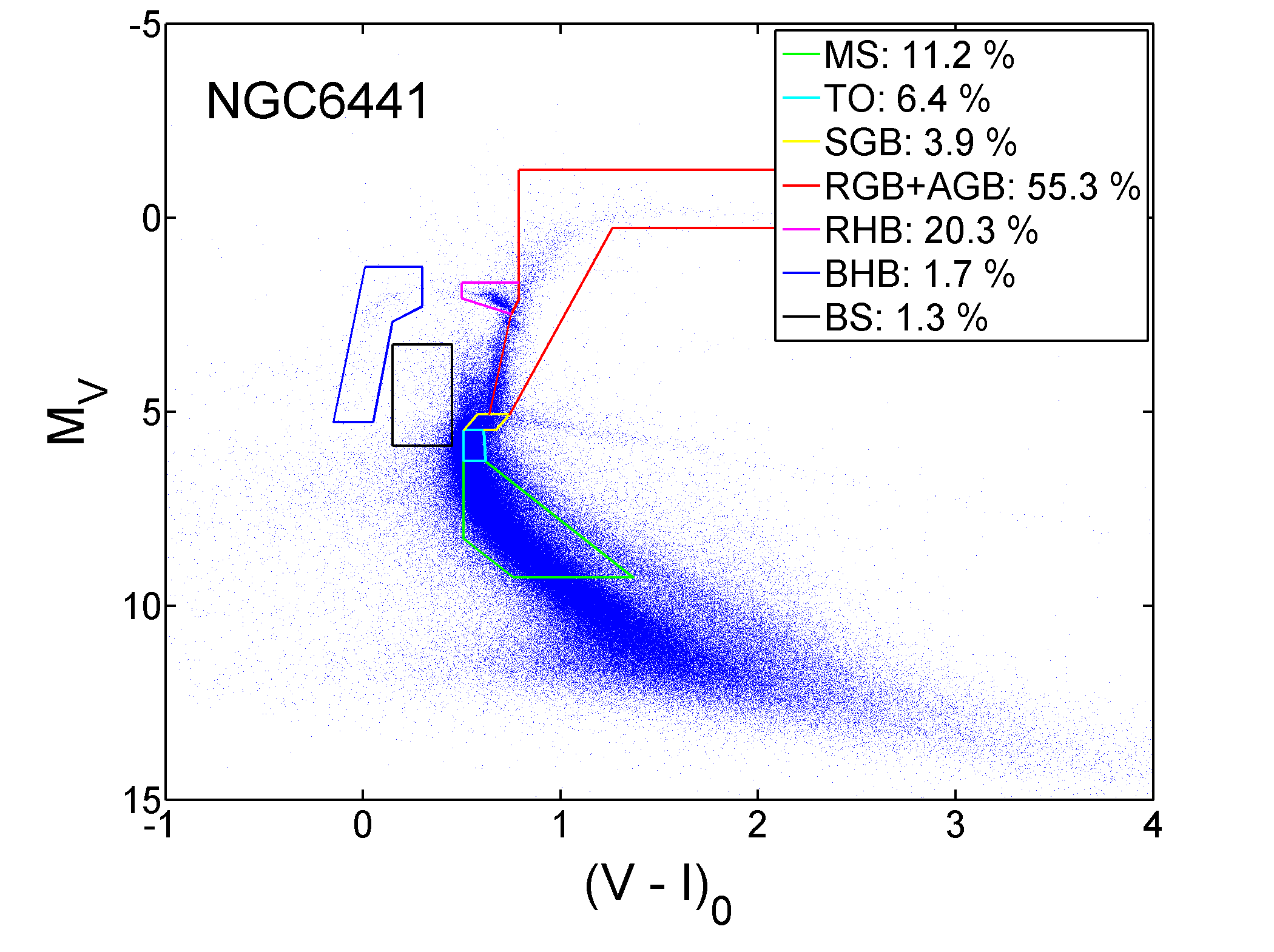}
  \captcont{.../continued}
\end{figure}

\clearpage
\begin{figure}
  \centering
\includegraphics[width=0.7\textwidth]{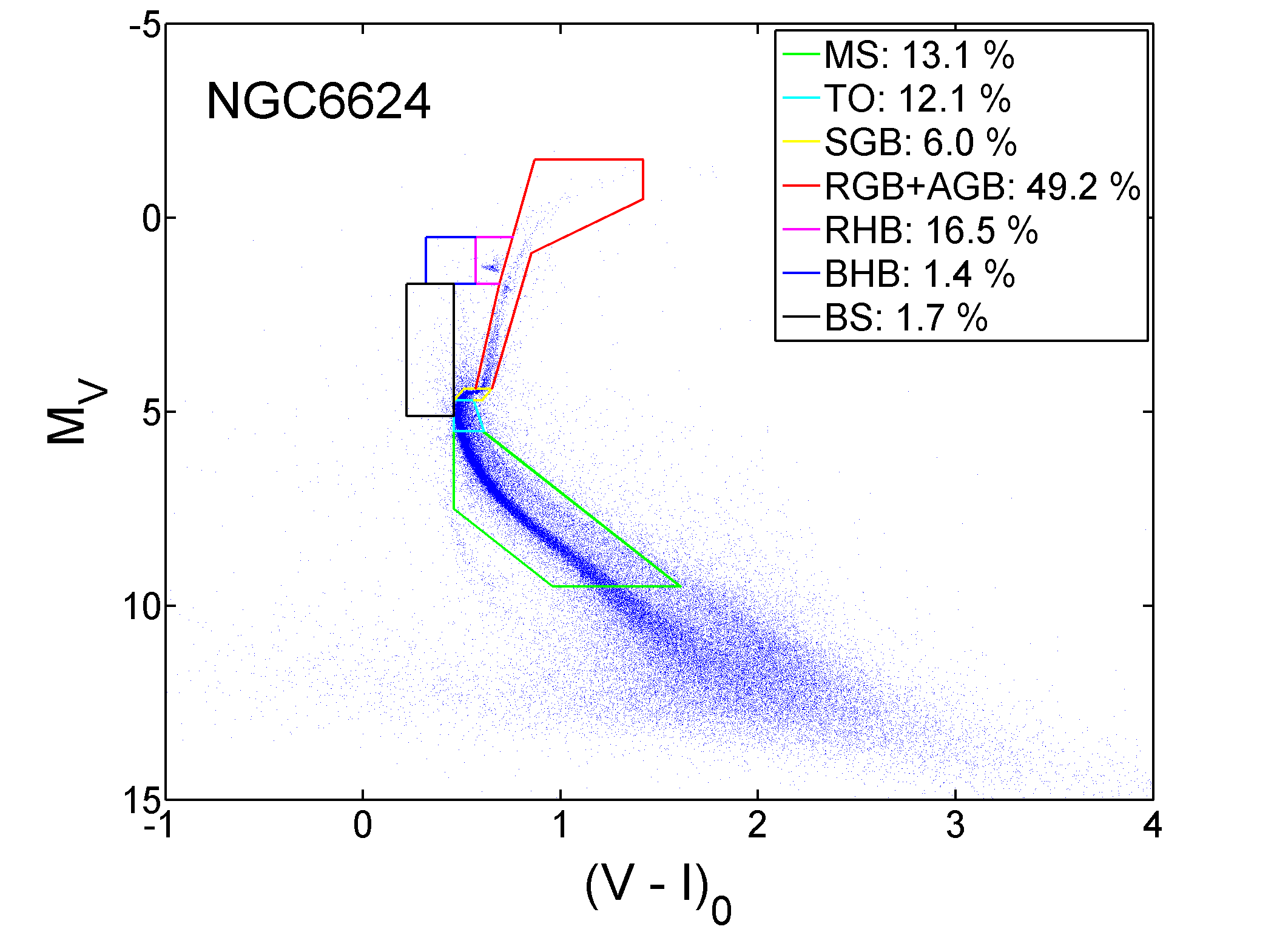}
  \captcont{.../continued}
\end{figure}

\begin{figure}
  \centering
\includegraphics[width=0.7\textwidth]{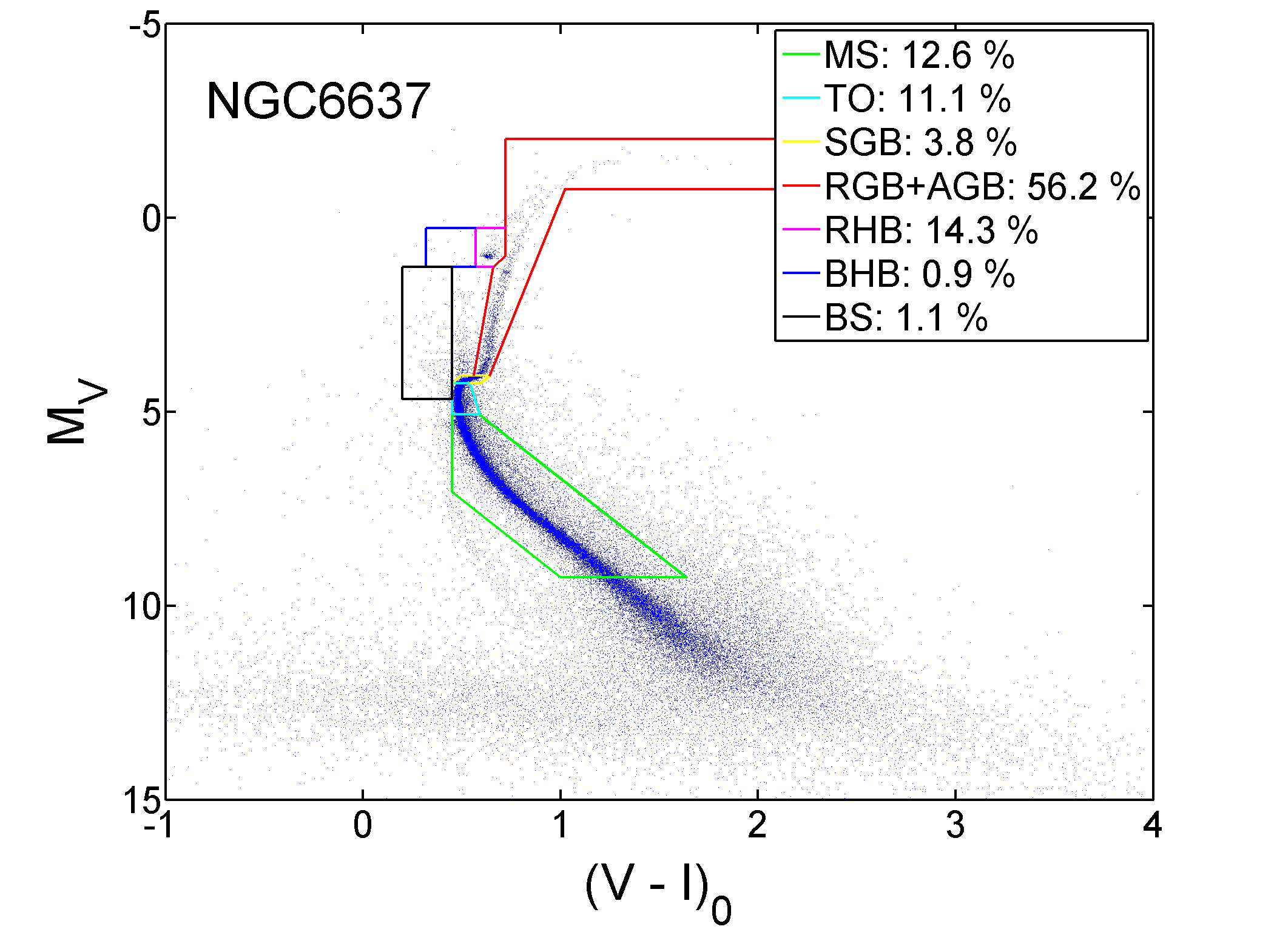}
  \captcont{.../continued}
\end{figure}

\clearpage
\begin{figure}
  \centering
\includegraphics[width=0.7\textwidth]{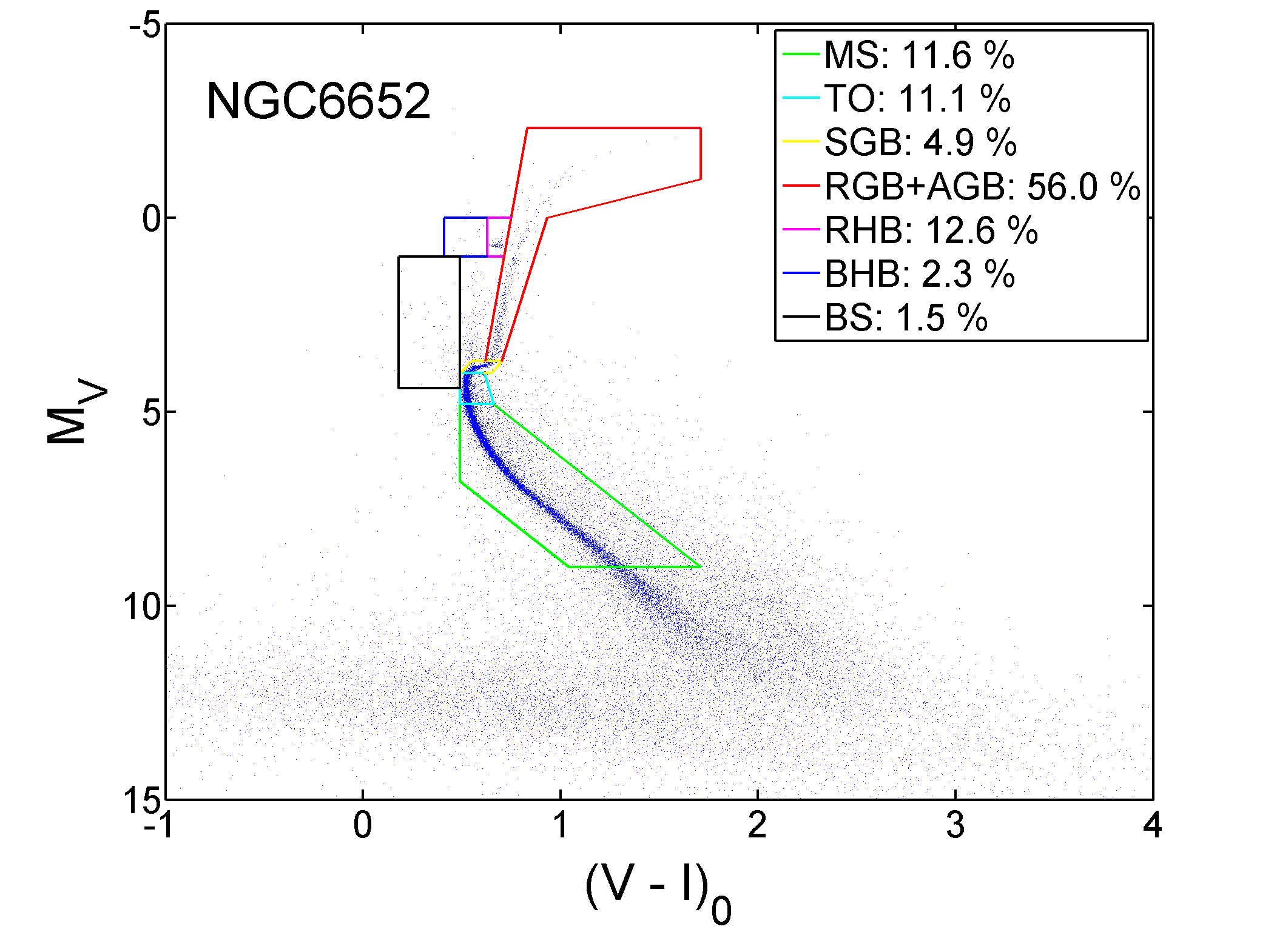}
  \captcont{.../continued}
\end{figure}

\begin{figure}
  \centering
\includegraphics[width=0.7\textwidth]{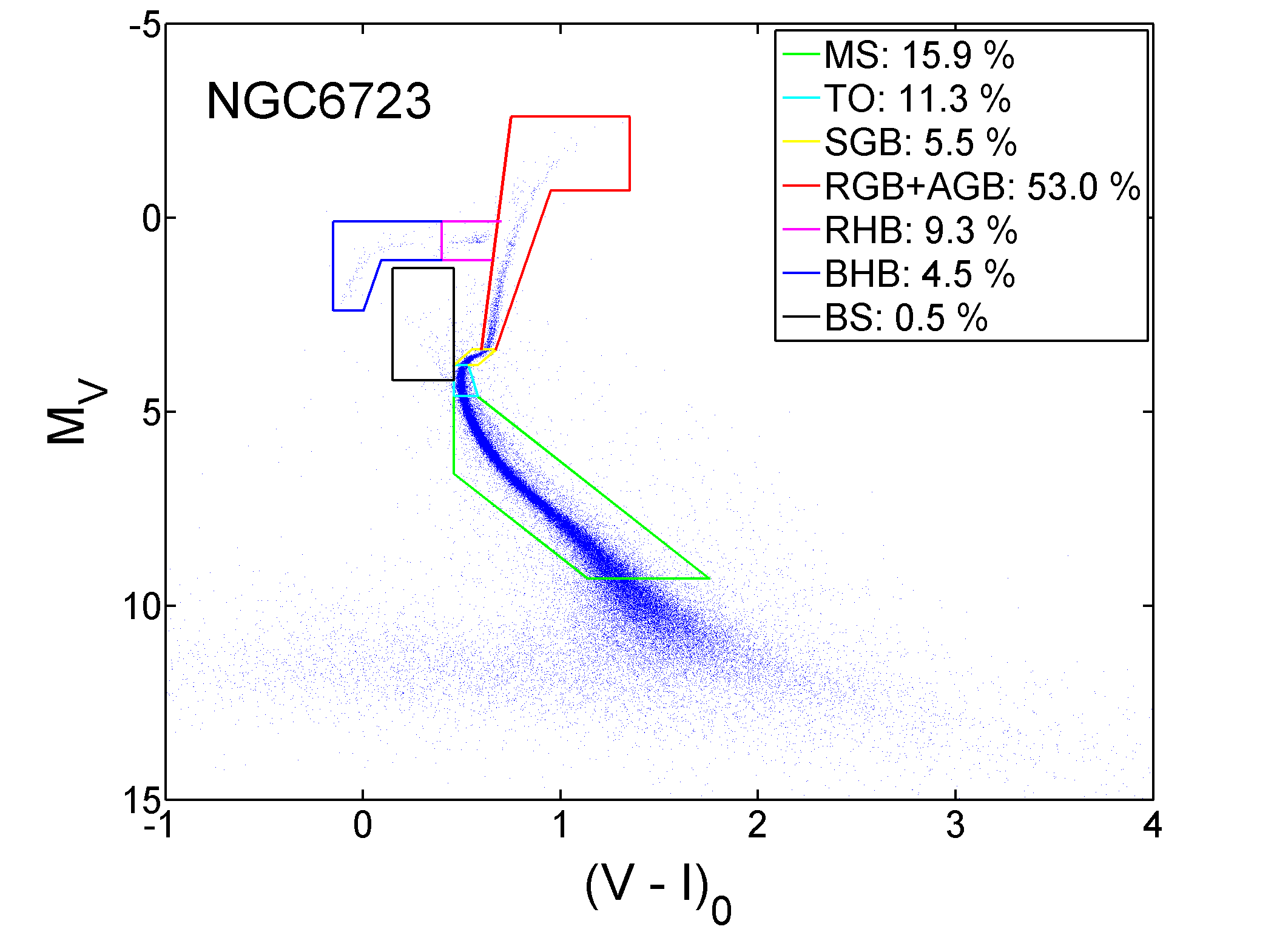}
  \captcont{.../continued}
\end{figure}

\clearpage
\begin{figure}
  \centering
\includegraphics[width=0.7\textwidth]{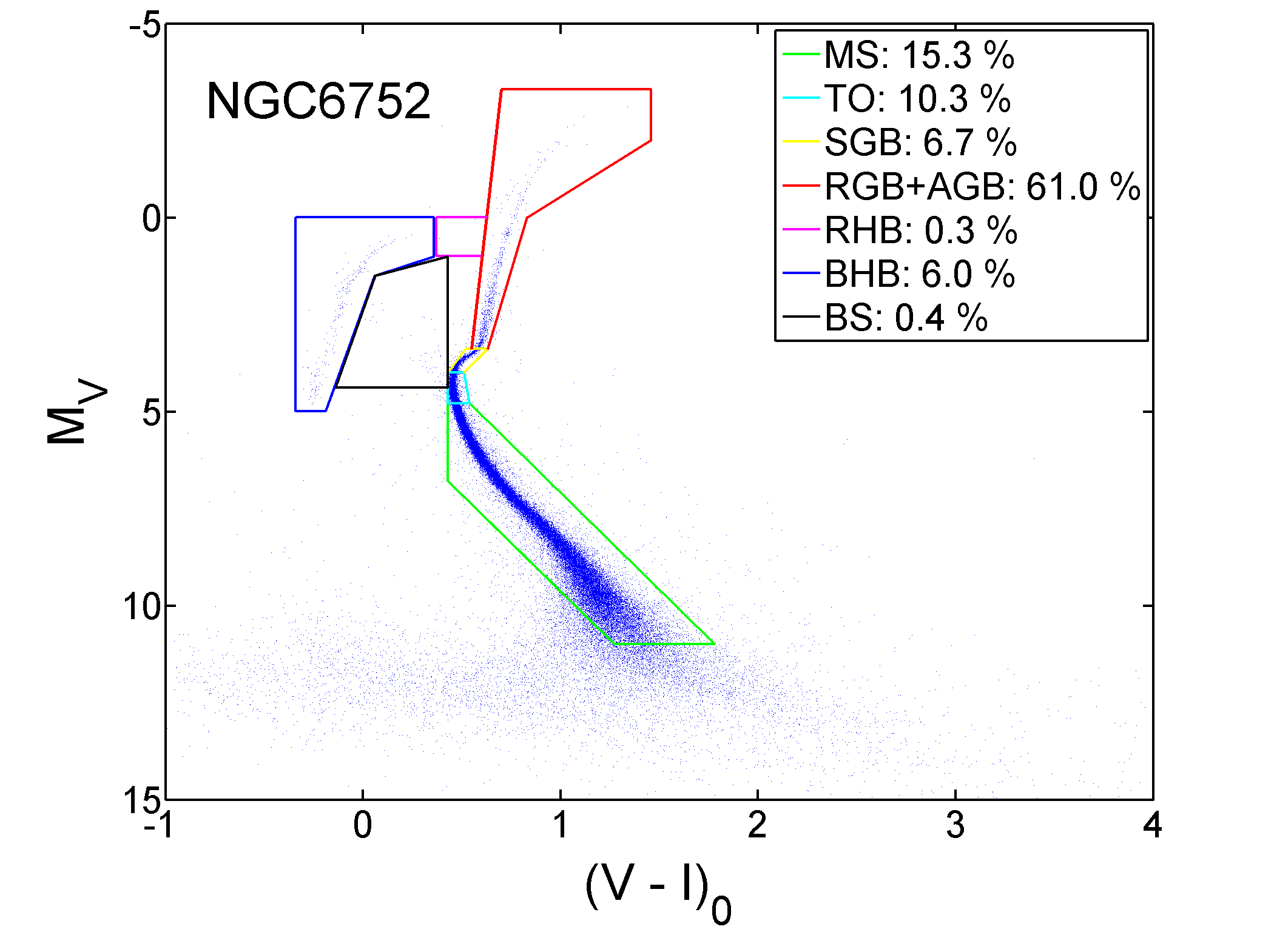}
  \captcont{.../continued}
\end{figure}

\begin{figure}
  \centering
\includegraphics[width=0.7\textwidth]{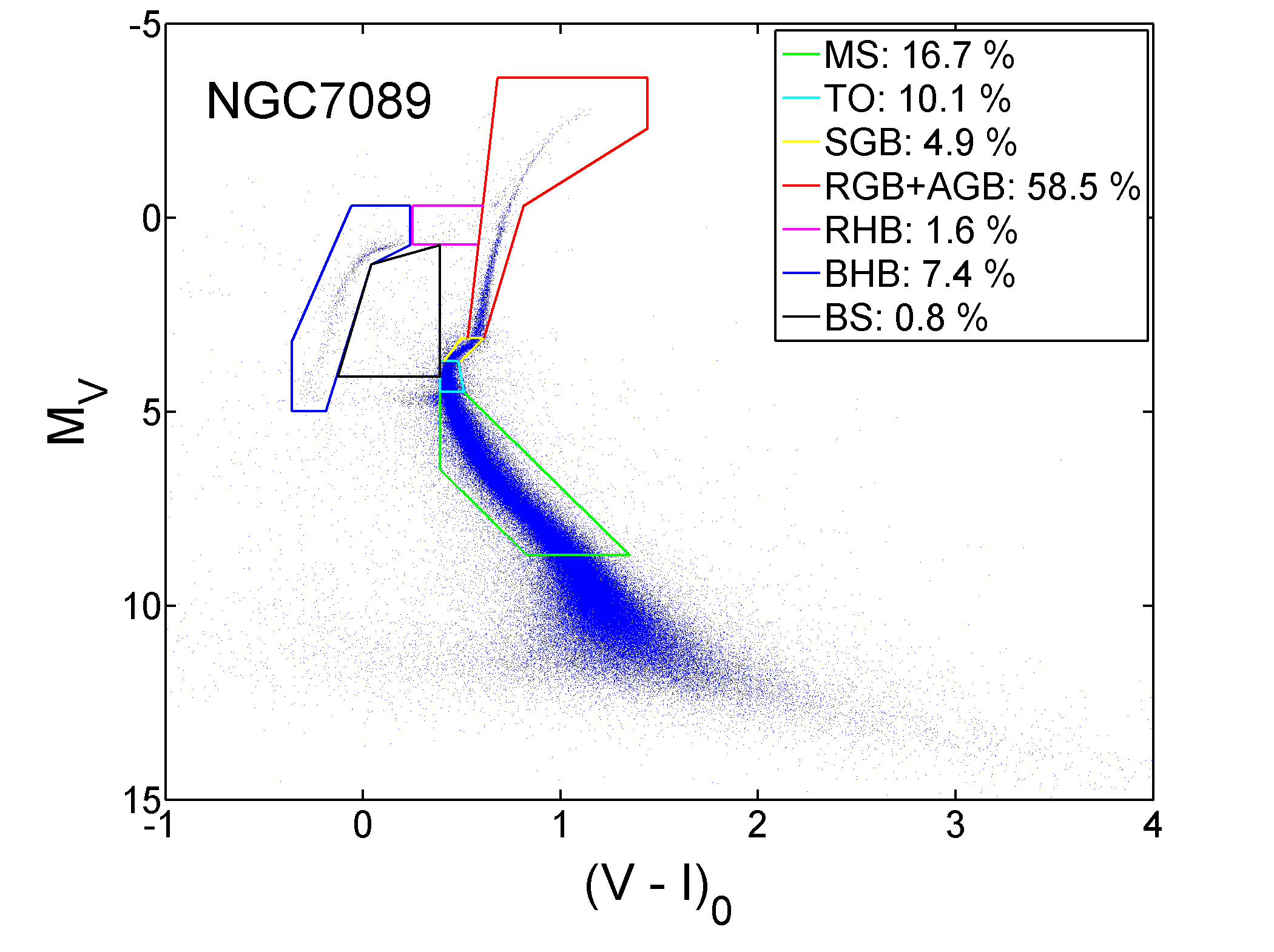}
  \captcont{.../continued}
\end{figure}

\clearpage

\end{document}